\documentclass[a4wide,twoside,titlepage]{scrartcl}

\usepackage[latin1]{inputenc}
\usepackage{eufrak}
\usepackage{a4wide}
\usepackage{amsmath}
\usepackage{graphicx}
\usepackage{epsfig}
\usepackage{color}

\usepackage{fancyhdr}
\font\bb=msbm10
\def\N{\mbox{\bb N}}
\def\Z{\mbox{\bb Z}}
\def\C{\mbox{\bb C}}

\begin{document}
\ifx\href\undefined\else\hypersetup{linktocpage=true}\fi

\newcommand{\bra}[1]{\langle #1|}
\newcommand{\ket}[1]{|#1 \rangle}
\newcommand{\braket}[2]{\langle #1|#2 \rangle}

%
%
%
\begin{titlepage}
\topmargin=55pt
\begin{center}

{\huge \bf Quasihole Tunneling}\\

\vspace{3mm}
{\huge \bf in the Fractional Quantum Hall Regime}\\ 

\vspace{6cm}
{\Large \bf Diplomarbeit} \\

\vspace{2cm}
{\Large vorgelegt von\\ \vspace{3mm} Moritz Helias}\\

\vspace{8cm}
{\large Quantentheorie der kondensierten Materie\\ Universität Hamburg \\ November 2003}
\end{center}
\end{titlepage}

\newpage
\thispagestyle{empty}
\phantom{Hallo}
\newpage

\setcounter{page}{1}

\pagenumbering{roman}
\tableofcontents
\newpage
\pagenumbering{arabic}


\section{Introduction}
\label{sec:intro}
Discovered in 1982 by Tsui, Stormer and Gossard \cite{TsuiStormerGossard}, the fractional Quantum Hall effect (FQHE) has opened a field of vivid activity until today. The reason for its continuing actuality can be attributed mainly to four fascinating features of this two-dimensional electron gas in a magnetic field.

As in the integer quantum Hall effect (IQHE) \cite{Laughlin1,Halperin} the measured Hall resistance in the FQHE is reproduced to very high accuracy due to a \emph{topological} origin of the \emph{quantization}. This was already pointed out by Laughlin \cite{Laughlin} short after the discovery of the effect. Quantization that can be traced back to topological features is not restricted to the IQHE and FQHE but was also found in superfluidity, superconductivity and in Josephson junctions. A collection on this subject by Thouless can be found in \cite{Thouless}.
Secondly, the FQHE is an effect caused by \emph{correlations}. It can be described by a Hamiltonian solely containing
two-particle interactions (see section \ref{sec:projection}). Thus, it belongs to the most strongly correlated systems studied so far.
Closely linked to the correlations is the appearance of fractionally charged \emph{quasiparticles} first introduced by Laughlin \cite{Laughlin}.
The prediction of these exotic particles inspired the efforts of experimentalists and theorists and the particles' existence became more and more manifest. Nevertheless investigating their properties -- such as in tunneling experiments -- is still a domain of recent experiments \cite{Heiblum1,Heiblum2} and accompanying theoretical work \cite{KaneFisher2}. These tunneling experiments raised the question what the dynamics of quasiparticles at a tunneling constriction is. It is also the motivation for this diploma thesis.
The last point is a more theoretical one: In some cases the edges of a sample in the FQHE -- like in the IQHE -- can be considered to be an embodiment of an one-dimensional interacting system for which analytically solvable models (\emph{Luttinger liquid}) are already known independently.

Motivated by experiments \cite{Saminadayer, Heiblum1, Heiblum2} and theory \cite{KaneFisher1,KaneFisher2} on single quasiparticle tunneling through a quantum point contact, a constricted fractional quantum Hall system will also be investigated in this work (section \ref{sec:InhomSystems}). To address the question of a single quasiparticle tunneling through this constriction, here we resort to quasiholes for which well approved trial wavefunctions are known.
Basing the work on the electronic many-particle Hamiltonian and these trial wavefunctions allows for a view on quasihole tunneling that is independent on the chiral Luttinger liquid model used so far for explaining the experiments.
Finite systems of few (4 - 6) electrons in the FQH regime will be investigated, which permits the electronic many-particle Hamiltionian to be dealt with by exact numerical diagonalization.

The outline of the main part of this work, which is divided into six sections, is as follows.
Section \ref{sec:overview} covers the major developments on the fractional quantum Hall regime concerning this work.
In doing so, the question of quasiparticle tunneling will be touched and embedded into past and ongoing research.
In section \ref{sec:one} the single particle basis for the chosen boundary conditions will be derived. The connection of the boundary conditions to an electric in-plane field will also be pointed out.
Section \ref{sec:two} covers the homogeneous (absent external potential) many-electron system.
A short-ranged interaction will be introduced and shown to be useful for the following work. 
The features of a system at a filling factor $\nu=\frac{1}{3}$ will be reviewed for Coulomb interaction and this short-range interaction.
Two different methods of inserting a quasihole into the system will be derived from Laughlin's trial wavefunctions \cite{Laughlin} 
and the stability of these excitations will be checked for both electron-electron interactions. Aside from that, a system providing bound quasihole states is found.
These bound states will be used in section \ref{sec:three} to create the most simple system in which tunneling of quasiholes can be observed.
In section \ref{sec:InhomSystems} an external potential is introduced to model a quantum point contact.
Corrections to the current operators arising from contributions of the next Landau level turn out to be crucial to obtain a consistent picture. The time evolution of an injected quasihole will be evaluated for both, weak and strong external potentials (compared to the excitation gap). Creating a tunneling constriction by a strong potential is counteracted by the incompressibility of the system. Ways to overcome this
competition are examined and lead to a realization of an effective tunneling barrier in which the time evolution of a quasihole can be studied.
The last section contains a summary of the main results and conclusions. Perspectives on possible further investigations and on questions
not exhausted in this work are given as well.

\newpage
\section{Quasiparticles in past and recent research: An overview on the FQH regime}
\label{sec:overview}
In the following, the major developments on the fractional quantum Hall regime from the perspective of this work will be mentioned.
This is not only intended to give a coarse overview but also to render more precise the question about quasiparticle tunneling and to relate
it to past and recent research.

As stated above, it was Laughlin \cite{Laughlin} who found the low energy excitations in a FQHE system to carry a fraction of an electron's charge. The finite creation energy of these quasiparticles explained the crucial physics like the incompressibility (section \ref{sec:gsmu}) which in turn helped explaining the vanishing longitudinal resistance $\rho_{x,x} = 0$.
His trial wavefunction also identified the origin of the gapped ground state (section \ref{sec:hardcore}) and revealed the uniqueness of the filling factors $\nu=\frac{1}{m}$, $m$ odd. Moreover, the one-to-one correspondence between flux quanta and zeros in the many-body wavefunctions (vortices) became apparent.
Numerical work by Yoshioka, Halperin and Lee on finite systems with rectangular geometry \cite{Yoshioka} was in agreement with Laughlin's proposal and also finite systems with spherical symmetry investigated by Haldane and Rezayi \cite{HaldaneRezayi} confirmed Laughlin's wavefunction for the ground state and the quasiparticle excitations.

The filling factors $\nu=\frac{p}{2 pm + 1} \quad p,m \in \N$, where the FQHE was observed as well, could not be explained by Laughlin's theory. Haldane explained it in a hierachichal scheme \cite{Haldane} as the FQHE of quasiparticles. Again the quasiparticles played an essential role. Another finding in this paper 
was the importance of the short-ranged part of the Coulomb interaction that was proven to be responsible for the appearance of the correlations in Laughlin's wavefunction. This was precised by Trugman and Kievelson \cite{Trugman} who even showed Laughlin's wavefunction to be an exact eigenstate for a special kind of short-ranged electron-electron interaction (section~\ref{sec:hardcore}).

These earlier developments which base directly on the many-particle Hamiltonian of the system (section~\ref{sec:ideas}) together with the trial wavefunctions of fractionally charged excitations (section~\ref{sec:QuasiHoles}) constitute the footing for the work at hand.
This is to point out the independence of our results on other theoretical work resting upon more elaborate theoretical models.
Although not used explicitly in this work, these developments shall be sketched here because they are mainly used in explaining the most recent experiments of quasiparticle tunneling.

Jain introduced a new quasiparticle -- a compound of an electron and an even number of vortices -- the composite fermion \cite{Jain}. This invention made a unification of the IQHE and FQHE possible: ``The FQHE is the IQHE of composite fermions''. This unification also includes an alternative to the hierachical picture in explaining the fractions $\nu=\frac{p}{2 pm +1}$.
According to Jain, the correlations are the key issue of this concept: The interaction between the electrons is incorporated into the system in the definition of the composite fermions. The system of strongly correlated electrons thus transforms into a system of weakly interacting composite fermions.
Their fermionic nature attended by weak interactions among them allow these compound particles to show the IQHE again.
Due to the weak interaction, this invention opened the field towards the well approved mean field descriptions and those that go beyond.
An overview on the wealth of developments basing on composite fermions can be found in \cite{Heinonen}.
The success of this theory proved the importance of two kinds of correlations for the FQHE: The Laughlin-like correlations that bind
vortices to electrons resident in the definition of the composite fermions and the fermionic Pauli correlations responsible for the rigidity of effect.

Recent experiments on quasiparticle tunneling \cite{Saminadayer,Heiblum1,Heiblum2} are closely related to a description of the fractional quantum Hall regime by edge states. This theory, used successfully in explaining the IQHE, was ported to the FQH regime. Its applicability in the latter case can intuitively be understood due to the unification of IQHE and FQHE by the analogy of electrons and composite fermions.

A more rigor treatment is based on the wavefunction picture. In an abruptly confined two-dimensional electron gas at integer filling the only excitations allowed by Pauli's principle appear at the edges of the sample where the Landau level crosses the Fermi energy. 
In the wavefunction picture for the IQHE these excitations can be shown to be bosonic. The same is true for the FQHE \cite{MacDonald}.
Creating an effective low energy theory of excitations near the two Fermi points of the one-dimensional edge makes the Luttinger liquid model
applicable in the case of the IQHE.
Similar low energy bosonic excitations in the FQHE are treated in the chiral Luttinger liquid model developed by Wen \cite{Wen}.
A subsequent work of Chamon and Wen \cite{ChamonWen} pointed out the importance of the confining potential to produce abrupt edges. ``Soft'' potentials were shown to produce more complicated edge structures.

Quasiparticle tunneling into the edges is considered to be a proper means of verifying the chiral Luttinger liquid nature and of investigating the structure of the edges.
Such tunneling experiments, first used for examining the quasiparticles themselves rather than the edges, 
were performed by Simmons \emph{et. al.} \cite{Simmons} already before Wen's theory. They measured fluctuations in the longitudinal conductivity $\rho_{xx}$ which were believed to origin from tunneling processes of quasiparticles between the current carrying edges as explained in a paper by Pokrovsky and Pryadko \cite{Pokrovsky}. However, the fractional filling of the system could not be ruled out completely as a source of these fluctuations. To exclude them, non-equilibrium processes had to be investigated.

\begin{center}
\begin{figure}
\input{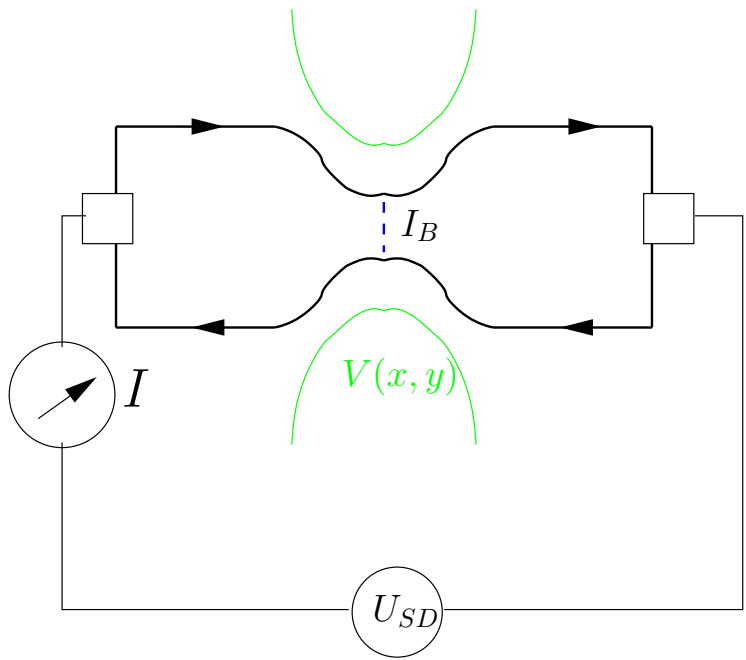}
\caption{Proposed geometry for quasiparticle tunneling between one-dimensional edges in the FQH regime. The source-drain voltage $U_{SD}$ causes a current $I$ through the sample. The edges are bent due to an external potential $V(x,y)$. Tunneling of quasiparticles is expected to occur where the edges are close to each other. The tunneling current $I_B$ causes fluctuations in the transmitted current $I$ which is measured.}
\label{fig:edgetunnel}
\end{figure}
\end{center}

Kane and Fisher \cite{KaneFisher1} developed a non-equilibrium theory that allowed the calculation of fluctuations in the current through a FQH system. It is based on the chiral Luttinger liquid model by Wen \cite{Wen} and describes the tunneling process as a point-like tunneling impurity between two edges of the sample. To sketch the idea, a schematics of the two edge channels that conduct the current through the sample is given in Fig. \ref{fig:edgetunnel}. These edge channels are bent due to an external potential $V(x,y)$ that creates a quantum point contact in the system. The quasiparticles flowing along the edges are believed to tunnel from one edge to the other and therefore cause fluctuations in time in the current $I$. These can be directly observed in the experiment.
An experiment of a quantum point contact in the FQH regime like in Fig. \ref{fig:edgetunnel} was realized by Saminadayer and Glattli \cite{Saminadayer}. Near the potential $V(x,y)$ of the constriction the filling factor was $\nu=\frac{1}{3}$. For small currents $I_B$ (weak backscattering limit) the measured shot-noise \cite{Shottky,KaneFisher1} in the current $I$ let them infer a fractional charge of $\frac{1}{3} e$ for the particles that are tunneling. 

The regime of strong backscattering $I_B \simeq I$ was investigated by Griffith \emph{et. al.} \cite{Heiblum1} and showed behavior as expected:
If the potential of the barrier is sufficiently high the system is effectively divided into two halves and only electrons can tunnel between them.

Just recently Chung \emph{et. al.} \cite{Heiblum2} made experiments at lower temperature on a system of two quantum point contacts.
The quasiparticles flowing along the edge were reduced in density by transmitting them through the first quantum point contact prior to hitting the second one. Thus being in a regime where quasiparticles arrived ``one by one'' at the constriction, single quasiparticle tunneling could be investigated in the absence of correlations between the particles.
The measurements showed that single-quasiparticle tunneling was only observed if the temperature was sufficiently high (73 mK) while at lower
temperature (23 mK) only electrons were found to tunnel even at quite transparent constrictions.
This very surprising result is supported by recent calculations by Kane and Fisher \cite{KaneFisher2} founded on the chiral Luttinger liquid theory.
Their calculations show ``strong evidence'' for only electrons being transmitted at $T=0$ and their proposed explanation is an Andreev reflexion.

\newpage
\section{Theoretical preparations}
\label{sec:one}
In this section the preparations for the numerical calculations on systems in the fractional quantum Hall regime are collected and derived. In section \ref{sec:ideas} the system to be treated will be introduced and the necessity of exact diagonalization will be pointed out.
The subsequent sections deduce the single particle basis used for the numerical calculations. In this context the periodic boundary conditions will be explained and a physical interpretation for the appearing phase factors will be given.
In section \ref{sec:efield} using these boundary conditions a possible realization of an in-plane electric field will be discussed.
\subsection{The model of the fractional quantum Hall system in this work}
\label{sec:ideas}
The model of the fractional quantum Hall systems treated in this work make the simplification
of electrons to be confined in a two dimensional plane.
Perpendicular to this x-y-plane there is a homogeneous magnetic field $\vec B = B \vec e_z$.
The electrons interact via a repulsive interaction potential $V_{int}$, for which we will use the Coulomb interaction as well as a short ranged
interaction and in some cases they are subjected to an external potential $V_{ext}$ which is used to model a constriction inside the system.
Additionally, there can be an in-plane electric field used as a driving force for a current.

Accordingly, the many particle Hamiltonian has the following form
\begin{eqnarray}
\label{eq:ManyPartHamiltonian}
H = \underbrace{\frac{1}{2 m} \sum_{i=1}^{N_e} \vec\Pi_i^2}_{H_{kin}} + \sum_{i=1}^{N_e} V_{ext}(\vec r_i) + \sum_{i < j} V_{Int}(\vec r_i - \vec r_j) .
\end{eqnarray}
Here, $\vec \Pi$ is the kinetic momentum that incorporates the vector potential caused by the magnetic field and the in-plane electric field.
We will calculate in the limit of high magnetic field, $B \rightarrow \infty$. 
So the spin of the electrons is assumed to be polarized and there is no spin degree of freedom in the Hamiltonian. 
The states of electrons in a magnetic field are known to
be quantized in macroscopically degenerate Landau levels $E_n$ of energy $E_n = \hbar \omega_c (\frac{1}{2} + n)$ with $\omega_c = \frac{e B}{m}$.

Here we are interested in the case of a partly filled first Landau level, mainly in the filling factor $\nu = \frac{N_e}{N_s} = \frac{1}{3}$, where $N_s$ denotes the number of states in the lowest Landau level.
In the limit of high magnetic fields the energy is dominated by the energy of the Landau level quantization. The degeneracy of the Landau levels forces us to use exact diagonalization to account for the interaction and for external potentials. Thus it is reasonable to determine the eigenbasis of the single particle kinetic Hamiltonian $H_{kin}$ and treat the interactions and external potentials by diagonalization in the degenerate space of the lowest Landau level.

This basis will be derived in the following sections.

%
\subsubsection{One electron in a magnetic field: Harmonic oscillator}
Consider electrons that are constrained to move in the $x$-$y$-plane and which are exposed to a homogeneous magnetic field $\vec B = B \vec e_z$ parallel to the $z$-axis.
The vector potential can be chosen in Landau gauge as
\begin{eqnarray}
  \label{eq:VekPot}
  \vec A(\vec r) &=& B \; x \; \vec{e_y},
\end{eqnarray}
where $\vec r = (x,y)$ is the coordinate in the plane.

The Hamiltonian describing one electron in this system is
\begin{eqnarray}
  \label{eq:Hamilton}
  H &=& \frac{1}{2m} \Big( \Pi_x^2 + \Pi_y^2 \Big) \\ \nonumber
\text{with} \quad
  \Pi_x &=& -i \hbar \frac{\partial}{\partial x} \\ \nonumber
  \Pi_y &=& -i \hbar \frac{\partial}{\partial y} + e B \; x \\ \nonumber
\end{eqnarray}
The operators $\Pi_x = P_x + e A_x$ and  $\Pi_y = P_y + e A_y$ are the kinetic momenta of the electron, $\vec P$ denotes the canonic momentum.
A natural energy scale of the system is given by the cyclotron frequency $\omega_c = \frac{e B}{m}$, the typical length unit is the classical cyclotron radius $l_0 = \sqrt{\frac{\hbar}{e B}}$.
By rescaling the kinetic momenta to these natural units two operators $Q$ and $S$ can be defined as
\begin{eqnarray}
  \label{eq:QSOperators}
  Q &=& \frac{1}{\sqrt{m \hbar \omega_c}} \Pi_y \\ \nonumber
  S &=& \frac{1}{\sqrt{m \hbar \omega_c}} \Pi_x. \\ \nonumber
\end{eqnarray}
Their commutator evaluates to
\begin{eqnarray}
  \label{eq:QSKommutator}
  [Q, S] &=& i,
\end{eqnarray}
which is the canonical commutation relation.

Following \cite{CT}, it is possible to express the Hamiltonian (\ref{eq:Hamilton}) using $Q$ and $S$. Knowing the commutator (\ref{eq:QSKommutator}) one recognizes this to be the problem of an one-dimensional harmonic oscillator, whose spectrum consists of equidistant discrete energy levels, commonly known as Landau-levels.
\begin{eqnarray}
  \label{eq:KanonHarmOsz}
  H &=& \hbar \omega_c ( Q^2 + S^2 ) \\ \nonumber
  E_n &=& \hbar \omega_c (\frac{1}{2} + n) ; \qquad n \in \N^0
\end{eqnarray}
In an analogous manner to the approach for solving the usual harmonic oscillator, raising and lowering operators $a^\dagger$ respectively $a$ can be defined as
\begin{eqnarray}
  \label{eq:aOps}
  a^\dagger &=& \frac{1}{\sqrt{2}} ( Q - iS ) \\ \nonumber
  a &=& \frac{1}{\sqrt{2}} ( Q + iS ). \\ \nonumber
\end{eqnarray}

With aid of the commutators $[Q, S^2] = 2 i S$ and $[S, Q^2] = -2 i Q$ following from (\ref{eq:QSKommutator}) the action of $a^\dagger$ or $a$ on an energy eigenstate of the system can be verified to raise and lower its energy by an energy quantum $\hbar \omega_c$.
%
%
\subsubsection{Magnetic translations}
\label{sec:magtrans}
Once chosen the Landau gauge, the Hamiltonian (\ref{eq:Hamilton}) given above commutes with the momentum operator $P_y$ in y-direction, since $\Pi_x$ and $\Pi_y$ do.
So it is possible to find a complete set of common eigenvectors $\ket{n, p_y}$ to both operators $H$ and $P_y$, where the label $n$ denotes the energy eigenvalue and
$p_y$ the eigenvalue of the momentum.
So we have $H \ket{n, p_y} = \hbar \omega_c (n+\frac{1}{2})\ket{E, p_y}$ and $P_y \ket{n, p_y} = p_y \ket{n, p_y}$.
Apart from the raising and lowering operators for the energy there must be a momentum-shift operator $B^\dagger(\delta p)$ affecting the eigenvalue of $P_y$ like
$p_y \rightarrow p_y+\delta p$. This implies the following commutator of $P_y$ and $B^\dagger$
\begin{eqnarray}
  \label{eq:PropBdagger}
  P_y \; B^\dagger(\delta p) \; \ket{E, p_y} &=& (p_y + \delta p) \; B^\dagger(\delta p) \; \ket{E, p_y} \\ \nonumber
  B^\dagger(\delta p) \; P_y \ket{E, p_y} &=& B^\dagger(\delta p) \; p_y \; \ket{E, p_y} \\ \nonumber
  \Rightarrow [P_y, B^\dagger(\delta p)] &=& \delta p \; B^\dagger(\delta p)
\end{eqnarray}
A simple momentum translation operator of the form $B^\dagger(\delta p) = \exp(\frac{i}{\hbar} \delta p \; y)$ does however not commute with the Hamiltonian,
 which is due to the non vanishing commutator between $y$ and $\Pi_x$.
But a generating operator $b^\dagger = y - i\frac{\hbar}{eB} \partial_x$ commutes with $\Pi_x$ as well as with $\Pi_y$ (compare Equ. (\ref{eq:Hamilton})) and is thus compatible to $H$.
A finite transformation constructed from this generator is a momentum shift by $\delta p$ accompanied by a coordinate shift in $x$-direction by $X_p = \frac{l_0^2}{\hbar} \delta p$.
\begin{eqnarray}
  \label{eq:MagTrans}
  b^\dagger &=& y - i l_0^2 \frac{\partial}{\partial x} \\ \nonumber
  B^\dagger(\delta p) &=& \exp( \frac{i}{\hbar} \delta p \; b^\dagger) \\ \nonumber
              &=& \exp \Big( \frac{i}{\hbar} \delta p \; y \Big) \exp \Big( \frac{\delta p l_0^2}{\hbar} \partial_x \Big) \\ \nonumber
  t_x(X_p) &:=& \exp \Big( \frac{i}{l_0^2} X_p \; y \Big) \exp \Big( X_p \partial_x \Big). \\ \nonumber
\end{eqnarray}
Due to the coupling of $x$ coordinate to the $y$-momentum, we define $t_x(\frac{\delta p l_0^2}{\hbar}) := B^\dagger(\delta p)$ as the ``magnetic translation operator in $x$-direction''. 
Since $t_x$ has the demanded properties of (\ref{eq:PropBdagger}) it is the (continuous)
 raising operator for the eigenvalue of $P_y$. This continuous symmetry of the system causes an infinite degeneracy of every Landau level.

Apart from the magnetic translation $t_x(X_p)$ in x-direction, the operator $P_y$ can be used as a generating operator of an ordinary translation in y-direction, that as well
commutes with $H$ 
\begin{eqnarray}
  \label{eq:MagTransY}
  t_y(Y) &=& \exp(Y \partial_y).
\end{eqnarray}
\subsubsection{Basis in direct space}
\label{sec:Basis}
By virtue of the operators ${a, a^\dagger, t(\delta p)}$ the eigenbasis can be determined in direct space.
Initially the ground state can be calculated by applying the lowering operator. We are looking for the simultaneous eigenvector of $H$ and $P_y$
for the eigenvalue 0 in both cases. This eigenvector has to satisfy 
\begin{eqnarray}
  \label{eq:GSOrt}
  a \ket{0,0} &=& 0 \\ \nonumber
  P_y \ket{0,0} &=& 0  \Rightarrow \phi_{0,0}(x,y) = \phi_{0,0}(x). \\ \nonumber
\end{eqnarray}
The second equation tells us that $\phi_{0,0}$ is only a function of $x$. Now using the direct space representation of the lowering operator, we gain a differential equation for $\phi_{0,0}(x)$
\begin{eqnarray}
  (\Pi_y - i \Pi_x) \ket{0,0} &=& 0 \\ \nonumber
  (-i\hbar \partial_y + eB x + \hbar \partial_x) \phi(x) &=& 0 \\ \nonumber
\Rightarrow \phi_{0,0}(x,y) &=& \frac{1}{\sqrt{\sqrt{\pi} l_0}} \exp(-\frac{x^2}{2 l_0^2}),
\end{eqnarray}
which, as a function of $x$, is normalized to $1$.
By applying $B^\dagger(p_y) = t_x(\frac{l_0^2}{\hbar}p_y)$ one obtains solutions with a different momentum $p_y$ in the same Landau-level. This leads to the explicit form of eigenfunctions in the lowest Landau-level
\begin{eqnarray}
  \label{eq:TransPy}
  \ket{0, p_y} &=& t_x(\frac{l_0^2}{\hbar}p_y) \ket{0,0} \\ \nonumber
  \phi_{0, p_y}(x,y) &=& \frac{1}{\sqrt{\sqrt{\pi} l_0}} \exp \Big( \frac{i}{\hbar} p_y \; y \Big) \exp \Big( \frac{-(x+X_p)^2}{2 l_0^2} \Big) \\ \nonumber
\mbox{with} \quad X_p &=& \frac{l_0^2}{\hbar} p_y
\end{eqnarray}
which are normalized to a delta distribution (like plane waves are).
The eigenfunctions of higher landau-levels can be obtained by the repeated application of the rising operator $a^\dagger$ which results in an additional factor
consisting of a Hermite polynomial $H_n(\frac{x+X_p}{l_0})$ of degree $n$ where n denotes the Landau-level. 
\subsubsection{Generalized periodic boundary conditions on the torus}
\label{sec:PRB}
The aim to describe an infinitely big system in a finite basis that can be handled for numerical diagonalization, can be achieved only if the system complies with certain symmetries, which reduce the degrees of freedom to a finite number and thus allows for a description in a finite basis.
Here, the system will be required to have a discrete translational symmetry where the periodicity is given by the size of a cell $(x,y) \in [0,L_1] \times [0, L_2]$. Thus all physical observables have to return to their original value when proceeding to another cell.

\begin{figure}
\includegraphics[scale=0.4]{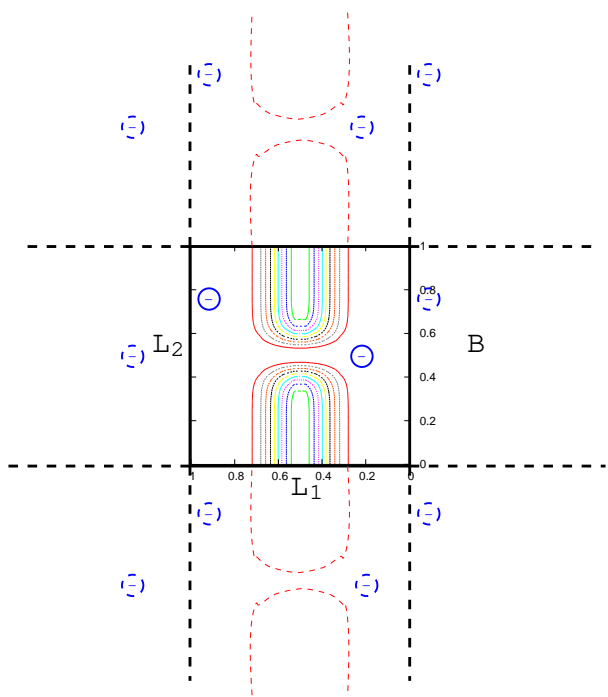}
\caption{Periodic tiling: The physical content of the unit cell  --i.e. external potentials (equipotential lines in the figure)-- is repeated over the entire system. A finite number of electrons (blue circles) in this cell is treated independently, the electrons in neighboring cells are mirrored. Although only depicted for the $y$-direction, the periodicity holds in $x$-direction as well.}
\label{fig:prbbarrel}
\end{figure}

To begin with, the number of electrons in the $x$-$y$-plane must be reduced to a finite number in order to have a finite number of degrees of freedom.
The idea is the following. Only a small number of electrons is put into one unit cell $[0,L_1] \times [0, L_2]$ and treated independently. All neighboring cells only contain images of these electrons at exactly the same position with respect to this cell's boundaries. The intention is to model a system that spreads out infinitely and doesn't need any confining potentials which would cause boundary effects. Since the interaction of electrons in distant cells is somewhat smaller than that between electrons in the same and directly neighboring cells, the mirroring of electrons is assumed not to destroy the desired effects. However, to gain results of a real infinite system with infinite degrees of freedom, a careful analysis for the size dependence of the observed effects would have to be made. A sketch of the repetition of the unit cell is found in figure \ref{fig:prbbarrel}. The equipotential lines are intended to depict a potential which is periodically repeated in every cell. 
Putting this more formally, we have a finite number of electrons in the system. Due to the cell wise periodicity of all potentials in the system,
we know by Bloch's theorem, that the eigenfunctions of the one particle Hamiltonian will be simultaneous eigenfunctions to the translations that
do a shift by one cell size. Since here we have a non-periodic vector potential in the kinetic part of the Hamiltonian, instead of ordinary 
translations we have to use the magnetic translations $t_x(L_1)$ and $t_y(L_2)$ from section \ref{sec:magtrans} in order to commute with $H_{kin}$.In complete analogy to Bloch's theorem, these translations are unitary operators which have eigenvalues of modulus 1, which is a phase factor the wavefunction picks up upon application of $t_x(L_1)$ or $t_y(L_2)$ respectively.
Thus, we arrive at generalized (because of \emph{magnetic} translations) periodic boundary conditions
\begin{eqnarray}
  \label{eq:PRB}
  t_x(L_1) \phi(x, y) = \exp(i \alpha) \phi(x,y)\\ \nonumber
  t_y(L_2) \phi(x, y) = \exp(i \beta) \phi(x,y),
\end{eqnarray}
where $\alpha$ and $\beta$ fix the eigenvalue for the respective translation.

The above procedure is of course only possible if additionally $t_x$ and $t_y$ commute. Generally this is not the case, since their commutator
evaluates to
\begin{eqnarray}
  \label{eq:KommTxTy}
  [\frac{i}{l_0^2}y + \partial_x, \partial_y] &=& - \frac{i}{l_0^2} \in \C \\ \nonumber
  t_y(L_2) t_x(L_1) &=& t_x(L_1) t_y(L_2) \exp \Big( \frac{i}{l_0^2} L_1 L_2 \Big).
\end{eqnarray}
The appearing phase factor $\exp \Big( \frac{i}{l_0^2} L_1 L_2 \Big)$ is the phase factor known from the Aharanov-Bohm effect \cite{AharanovBohm}.
The connection between these to phase factors can be understood in the following manner.
One has to keep in mind that the magnetic translations perform a coordinate shift with the additional property of transforming eigenstates of $H_{kin}$ to eigenstates with the same energy.
In the chosen Landau gauge, the vector potential $\vec A(x,y)$ is linear in $x$ and independent of $y$ and thus $\vec A(0,0) = 0$.
Assume we had an eigenvector $\Psi_0$ of $H_{kin}$ and want to construct another eigenvector which is shifted by $(a,b)$.  
In a first step we can just shift the reference frame by $(-a,-b)$ which changes $x\rightarrow x+a$, $y \rightarrow y+b$
\begin{eqnarray}
  \underbrace{\left. H_{kin} \right|_{x\rightarrow x+a, y\rightarrow y+b}}_{H_{kin}^\prime} \Psi_{0}(x+a,y+b) = E_0 \Psi_{0} (x+a,y+b) \nonumber
\end{eqnarray}
The coordinate shift in $H_{kin}$ only affects the vector potential which accordingly changes by $\vec A \rightarrow  \vec A^\prime = \vec A + eBa \vec e_x$.
The new Hamiltonian $H_{kin}^\prime$ can be written in a more complicated form
\begin{eqnarray}
  \exp (\frac{i e}{\hbar} \int_{c(a,b)} \vec A \; d\vec r) H_{kin}^\prime &=& H_{kin} \exp (\frac{i e}{\hbar} \int_{c(a,b)} \vec A \; d\vec r) \\ \nonumber
  H_{kin} \exp(\frac{i e}{\hbar} \int_{c(a,b)} \vec A \; d\vec r) \Psi_0(x+a,y+b) &=& E \underbrace{ \exp (\frac{i e}{\hbar} \int_{c(a,b)} \vec A \; d\vec r) \Psi_0(x+a,y+b)}_{\Psi_{a,b}(x,y)} \\ \nonumber
  \label{eq:AharBohmPhase}
\end{eqnarray}
The phase factor compensates for the offset $e B a$ in the vector potential and it is the same factor that appears in the
explanation for the Aharanov-Bohm effect.
Here $c(a,b)$ is a contour that connects the point $(0,0)$ with $(x+a,y+b)$. Since the vector potential is not
rotation free, $\Psi_{a,b}(x,y)$ is a functional of $c$ and depends on the chosen path to reach the endpoint.
Therefore, the exponential factor in \ref{eq:AharBohmPhase} also depends on this path.

This procedure can be regarded as an alternative derivation of the magnetic translations.
Comparing Equ. \ref{eq:AharBohmPhase} and Equ. (\ref{eq:MagTrans}), we obtain $t_x(L_1)$ from the first equation by choosing $c$ the path that is parallel to the x-axis from $(0,0)$ to $(x+L_1,y)$.

Analogously we can construct the translation $t_y$, where the phase factor is trivial.

\begin{center}
\begin{figure}
\includegraphics[scale=0.6]{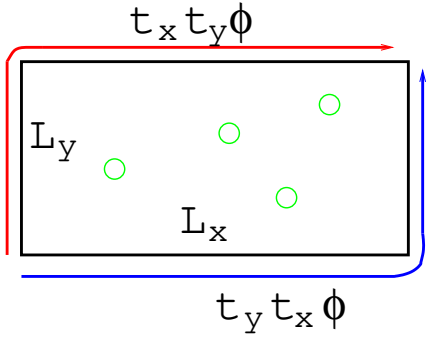}
\caption{Aharanov-Bohm effect. First applying $t_x(L_1)$ and then $t_y(L_2)$ shifts an eigenvector along the blue path, the reverse order moves it along red path. The solutions in the point $(L_1,L_2)$ differ by the Aharanov-Bohm phase related to the magnetic flux enclosed by the two paths. Commuting operators $t_x$ and $t_y$ imply an integer number of enclosed flux quanta (green circles).}
\label{fig:aharanovbohm}
\end{figure}
\end{center}

By successive application of $t_x(L_1)$ and then $t_y(L_2)$ or first $t_y(L_2)$ and then $t_x(L_1)$ to an eigenvector, move the solution around the unit cell from point $(0,0)$ to $(L_1,L_2)$ on two different traces. These two traces are sketched in figure \ref{fig:aharanovbohm}.
The Aharanov-Bohm phase by which the two solutions differ is just the phase factor in Equ. \ref{eq:KommTxTy}.
This phase factor can be written like in (\ref{eq:AharBohmPhase}) and --by  Stokes' theorem-- be related to the amount of magnetic flux enclosed by the contour.
To return to the commutation relation of $t_x(L_1)$ and $t_y(L_2)$, commuting operators demand a phase factor which is a multiple of $2 \pi$ in Equ. \ref{eq:KommTxTy}. Thus the wavefunction must pick up a phase factor $2 \pi N_s$ with $N_s$ integer. 
Writing the wavefunction as a function of the complex variable $z = x+iy$, this corresponds to a complex function with $N_s$ zeroes in the unit cell. This number $N_s$ on the other hand is the number of flux quanta $\Phi_0 = \frac{h}{e}$ piercing the unit cell, which can be seen from the expression of the total flux $\Phi$ through the area $L_1 L_2$ of the unit cell
\begin{eqnarray}
  \label{eq:FluxCondition}
  \Phi &=& L_1 L_2 B = N_s \frac{h}{e} = N_s \Phi_0 \\ \nonumber
  \Rightarrow L_1 L_2 &=& 2 \pi N_s l_0^2,
\end{eqnarray}
where the definition of the magnetic length $l_0 = \sqrt{\frac{\hbar}{e B}}$ was used.

From the second equation in (\ref{eq:PRB}), keeping in mind the form of the wavefunction in (\ref{eq:TransPy}), the quantization of the y-momentum follows
\begin{eqnarray}
  \label{eq:QuanPy}
  p_j = \hbar \frac{2 \pi}{L_2} j + \frac{\hbar}{L_2} \beta \qquad j \in \Z.
\end{eqnarray}
Keeping in mind that the magnetic translation $t_x(L_1) \stackrel{(\ref{eq:MagTrans})}{=} B^\dagger(\frac{\hbar}{l_0^2}L_1) \stackrel{(\ref{eq:KommTxTy})}{=} B^\dagger(N_s \hbar \frac{2 \pi}{L_2})$ has the meaning of a raising operator for the eigenvalue $p_y$ and provided the periodic boundary condition (\ref{eq:PRB}), this is an identification of the states $\ket{n, p_y}$ and $\ket{n, p_y+ k N_s \hbar \frac{2 \pi}{L_2}}$,
where $k$ is an integer.
Thus we are looking for a linear combination of those identified vectors satisfying the periodic boundary conditions.
These will be named $\ket{n, j}_{prb}$ and can be constructed from the non-periodic solutions of (\ref{eq:TransPy}) by summing up all identified wavefunctions
\begin{eqnarray}
  \label{eq:PerBasis}
  \ket{n, j}_{prb} &=& \sum_{k \in \Z} \big(t_x(L_1)\exp(-i \alpha) \big)^k \; B^\dagger \big(\frac{\hbar}{L_2} (2 \pi j + \beta) \big) \; \frac{1}{\sqrt{n!}}(a^\dagger)^n \; \ket{0,0} \\ \nonumber
  &=& \frac{1}{\sqrt{n}} \sum_{k \in \Z} \exp(-i k \alpha) \; t_x \big( (2 \pi j + \beta) \frac{l_0^2}{L_2} + k L_1 \big) \; (a^\dagger)^n \; \ket{0,0}. \\ \nonumber
\end{eqnarray}
To obtain the wavefunction in direct space, $a^\dagger$ and $t_x$ have to be expressed in this representation, which then results in
\begin{eqnarray}
  \label{eq:BasisOrt}
  \phi_{n, j}(x,y) &=& \frac{1}{\sqrt{\sqrt{\pi} n! 2^n L_2 l_0}} \sum_{k \in \Z} \exp \big( i \frac{X_j + k L_1}{l_0^2} y - i k \alpha \big) \exp \Big( -\frac{ \big( x + X_j + k L_1 \big)^2 }{2 l_0^2} \Big) \\ \nonumber
  &\times& H_n \Big( \frac{x + X_j + k L_1}{l_0} \Big) \\ \nonumber
  X_{j,\beta} &=& (j + \frac{\beta}{2 \pi}) \frac{L_1}{N_s}.
\end{eqnarray}
Here $H_n$ are the Hermite polynomials. This wavefunction is normalized to unity upon integration over the domain of the unit cell.
\subsubsection{Physical meaning of $\alpha$ and $\beta$}
\label{sec:AlphaBeta}
The two parameters $\alpha$ and $\beta$ were introduced in section \ref{sec:PRB} to fix the eigenvalues of the magnetic translation operators according to Equ. (\ref{eq:PRB}).
A physical meaning of magnetic fluxes that pierce the torus can be attributed to the parameters $\alpha$ and $\beta$ as shown by Tao and Haldane \cite{Tao}. The parameters introduced in their paper can be shown to be equivalent to ours by the following consideration.

\begin{figure}
\includegraphics[scale=0.8]{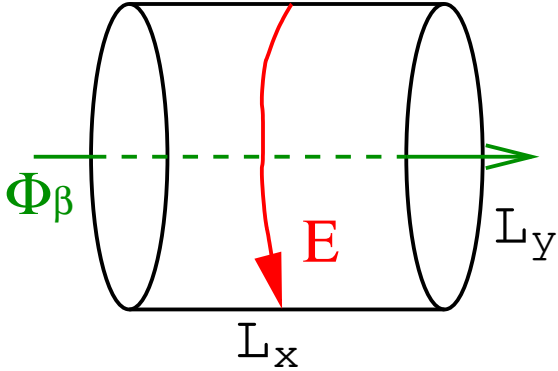}
\caption{The parameter $\beta$ can be interpreted as the magnetic flux $\Phi_\beta$ of a solenoid parallel to the y-axis,
if $x=0$ and $x=L_1$ is identified by the periodic boundary condition.}
\label{fig:BetaFlux}
\end{figure}

The periodicity in the tilings of one unit cell can alternatively be regarded as an identification of the left border of the unit cell with its right one and the upper with the lower one. This imposes a torus topology.
Because the $x$- and $y$- axis differ only by the selected gauge, it is sufficient to focus on one parameter, here $\beta$.
If we apply a unitary transformation
\begin{eqnarray}
  U(y) = \exp(-i \frac{\beta}{L_2}y)
\label{eq:UnitBeta}
\end{eqnarray}
on both, the Hamiltonian and the periodic eigenvectors, the new eigenvectors $\ket{\tilde k}$ satisfy ``simple'' boundary conditions with respect to $t_y(L_2)$, like
\begin{eqnarray}
  t_y(L_2) U(y) \ket{k} = \underbrace{U(y) \ket{k}}_{\tilde{\ket{k}}}.
\end{eqnarray}
Therefore, both borders can be identified by means of $t_y(L_2)$ like $(x,y) \stackrel{t_y(L_2)}{\simeq} (x,y+L_2)$.
The transformed Hamiltonian gains an additional term residing in the momentum operator $\Pi_y$, which transforms according to
\begin{eqnarray}
  \tilde{\Pi}_y = U(y) \Pi_y U^\dagger(y) = \Pi_y + \frac{\hbar}{L_2} \beta.
\label{eq:TrxPiy}
\end{eqnarray}
The additional term proportional to $\beta$ can be seen as belonging to a vector potential $\vec A_{\beta} = \frac{\hbar \beta}{e} \vec e_y$.
This again is related to a flux of $\Phi_{\beta} = \frac{\hbar}{e} \beta$ passing through the torus as sketched in figure \ref{fig:BetaFlux}.
If now $\beta$ is varied with time, an electric field $E_y = - \dot{\vec{A}}_{beta} = -\frac{\hbar \dot \beta}{e L_2} \vec e_y$ is induced in the plane along y-direction.

To stress, that this interpretation is reasonable, the influence of $\beta$ on the basis states from Equ. (\ref{eq:BasisOrt})
(which still have to be multiplied by $U(y)$) is reviewed.
An electron in the state $\ket{k}$ is localized around $-X_{k,\beta}$ in $x$-direction. Since $X_{k,\beta}$ depends on $\beta$, the electron moves
in $x$-direction is $\beta$ changes adiabatically. The velocity of this motion is
\begin{eqnarray}
v_x &=& -\dot X_{k,\beta} \\ \nonumber
    &=& -\frac{L_1}{2 \pi N_s} \dot \beta \\ \nonumber
    &=& \underbrace{\frac{L_2 L_1}{2 \pi N_s}}_{l_0^2} \frac{e}{\hbar} E_y = \frac{E_y}{B},
\end{eqnarray}
which coincides with the classical $\vec E \times \vec B$ drift.


\subsection{FQH system with an applied electric field}
\label{sec:efield}
The parameters $\alpha$ and $\beta$, formally introduced in the periodic boundary conditions in section \ref{sec:PRB}, turned out
to have a physical meaning of magnetic fluxes as shown in the previous section. It was shown that a time dependent change of $\beta$ can be used
 to generate an electric in-plane field in the system. The same is true for $\alpha$ since $x$- and $y$-direction only differ by the selected gauge.
In this section we will assume to describe a system in the basis of states in the lowest Landau level $(n=0)$  from equation (\ref{eq:PerBasis}) and use $\beta$ for creating an applied electric field in $y$-direction. In time dependent perturbation theory the time evolution of the system will be calculated.

A homogeneous magnetic field $\vec B = B \vec e_z$ and an electric field $\vec \epsilon = \epsilon \vec e_y$ in both
can be represented by a vector potential depending on space coordinates and time
\begin{eqnarray}
  \label{eq:Vekpot}
  \vec{\epsilon} &=& -\vec \nabla \phi - \frac{\partial \vec{A}}{ \partial t} \\ \nonumber
  \vec{B} &=& \vec \nabla \times \vec A \\ \nonumber
  \phi &=& 0 \\ \nonumber
  \vec A(\vec r, t) &=& (B x - \epsilon \; t) \; \vec e_y.\\ \nonumber
\end{eqnarray}
The part of $\vec A$ describing the magnetic field was chosen in Landau-gauge which makes  
it possible to use periodic boundary conditions (\ref{eq:PRB}) with fixed values of $\alpha$ and $\beta$, say 0. The electric field in y-direction can be achieved by a homogeneous term of $\vec A$ linear in time.
Therefore the many-particle Hamiltonian becomes time dependent and obtains the following form
\begin{eqnarray}
  \label{eq:Hamiltonian}
  H(t) &=& \sum_{j=1}^{N_e} \frac{1}{2 m} (\Pi_{j,x}^2 + \Pi_{j,y}(t)^2) + \sum_{j \neq k} V_{int}(|\vec r_j - \vec r_k|) + \sum_{j=1}^{N_e} V_{ext}(\vec r_j) \\ \nonumber
  \Pi_{j,x} &=& -i \hbar \frac{\partial}{\partial x_j} \\ \nonumber
  \Pi_{j,y}(t) &=& -i \hbar \frac{\partial}{\partial y_j} + e B \; x_j - e \epsilon \; t. \\ \nonumber
\end{eqnarray}
Here, $\Pi_x$ and $\Pi_y$ are the kinetic momentum operators, $V_{int}(|\vec r|)$ is the interaction potential and $V_{ext}(\vec r)$ is an external potential compatible with the periodic boundary conditions, needed to model a constriction.

It is possible to isolate the Hamiltonian's time dependence from (\ref{eq:Hamiltonian}) in a time dependent parameter $\beta(t)$.
The second parameter $\alpha$ influences the momentum in $x$-direction.
Both parameters --- $\alpha$ and $\beta$ --- can as well be cast into modified periodic boundary conditions by a gauge transformation $U_{\alpha,\beta}(x,y) = \exp(-\frac{i}{\hbar} ( x \alpha + y \beta))$. This will yield the generalized periodic boundary conditions from section (\ref{sec:PRB}).
Either the boundary conditions are kept fixed and the parameters are regarded as parts of the Hamiltonian or the Hamiltonian
will have $\alpha = \beta = 0$ and the periodic boundary conditions will catch up a phase factor of $\exp(i \alpha)$ upon a magnetic translation one unit cell to the right or a factor of $\exp(i \beta)$ when going one cell up, respectively.
However, now we will keep the boundary conditions fixed and concern the parameters as parts of the Hamiltonian
\begin{eqnarray}
  \label{eq:HamiltonAlpha}
  H_{\alpha, \beta} &=& \sum_{j=1}^{N_e} \frac{1}{2 m} (\Pi_{\alpha,j,x}^2 + \Pi_{\beta,j,y}^2) + \sum_{j \neq k} V_{coul}(\vec r_j - \vec r_k) + \sum_{j=1}^{N_e} V_{ext}(\vec r_j) \\ \nonumber
  \Pi_{\alpha,j,x} &=& -i \hbar \frac{\partial}{\partial x_j} + \frac{\hbar}{L_1} \alpha \\ \nonumber
  \Pi_{\beta,j,y} &=& -i \hbar \frac{\partial}{\partial y_j} + \frac{e}{c} B \; x_j + \frac{\hbar}{L_2} \beta. \\ \nonumber
\end{eqnarray}
Here, $L_1$ and $L_2$ again denote the extent of the unit-cell.
%
%
%
%
The time-dependent problem to be solved thus is the solution of the Schrödinger-equation of the time-dependent Hamiltonian
\begin{eqnarray}
  \label{eq:SGl1}
  H(t) = H_{\alpha=0, \, \beta=-\frac{L_2 \epsilon e}{\hbar}  t}.
\end{eqnarray}
\subsubsection{Time evolution in time dependent perturbation theory}
The time dependent perturbation theory allows to calculate the state's time evolution in the case where the time dependent terms in the Hamiltonian can be considered to be small with respect to the unperturbed Hamiltonian $H_0$, where 'small' compares the typical energy of the perturbation to the spacing of the eigenenergies of $H_0$.
In the adiabatic limit the perturbing term is switched on smoothly in an infinitely long period of time while the total change of $H(t)$ is yet finite.
Kato has shown \cite{Kato} that even in the case of crossing levels the time evolution will follow the stationary eigenstates of $H(t)$, for fixed $t$,
 except for a phase factor (see next section).
Applied to our system, when starting in an eigenstate of $H(0) = H_{\alpha=0, \; \beta=0}$ in $t=0$ the system will follow this state on changing
$\beta$ adiabatically thus staying in the corresponding eigenstate of the operator $H(t) = H_{\alpha=0, \; \beta(t)}$ for all $t$.
Now we want to consider in addition the phase factor which is lost when using the adiabatic theorem. This phase factor is crucial for the task
of constructing the time evolution operator of the system from its eigenstates.

For every time $t$ (and consequently every value of $\alpha$ and $\beta$) we will therefore consider a spectral decomposition of $H$ according to
\begin{eqnarray}
  \label{eq:SpektralZerl}
  H_{\alpha, \beta} = \sum_k \ket{k_{\alpha, \beta}} \bra{k_{\alpha, \beta}} E_{k, \alpha, \beta}.
\end{eqnarray}
Starting in an eigenstate $\ket{k_{0,0}}$ for $t=0$, the adiabatic evolution would correspond to following the state $\ket{k_{0,\beta(t)}}$ for all $t$.
If however the electric field is not switched on adiabatically, we can still use this time dependence as a starting point to govern the true time evolution by means of
perturbation theory using the adiabatic states as a basis.
Since the basis is complete (in the subspace of the lowest Landau level), an arbitrary state in the lowest Landau level can be written as
\begin{eqnarray}
  \label{eq:Psit}
  \Psi(t) &=& \sum_k a_k(t) \exp(-\frac{i}{\hbar} \int_0^t E_{k, \alpha, \beta(t')} \; dt') \; \ket{k_{\alpha, \beta(t)}} \\ \nonumber
          &=& \sum_k a_k(t) \exp(\frac{i}{L_2 \epsilon e} \int_0^{\beta=-\frac{L_2 \epsilon e}{\hbar}t} E_{k, \alpha, \beta^\prime} \; d\beta^\prime) \; \ket{k_{\alpha, \beta}}.
\end{eqnarray}

Using this ansatz in the time dependent Schrödinger-equation we obtain the following result (we omit here the dependence on $\alpha$ for the sake of simplicity)
\begin{eqnarray}  
  i \hbar \frac{\partial \Psi}{\partial t} &=& H(t) \Psi(t) \\ \nonumber
  i \hbar \frac{\partial \Psi}{\partial t}(t) &=& 
  i \hbar \sum_k \Bigg( \Big( \dot a_k(t) + \frac{1}{i \hbar} a_k(t) E_{k,\beta(t)} \Big) \ket{k_{ \beta(t)}} + a_k(t) \frac{\partial}{\partial t} \ket{k_{ \beta(t)}} \Bigg) \times \\ \nonumber
    && \times \exp ( -\frac{i}{\hbar} \int_0^t E_{k,\beta(t')} \; dt') \\ \nonumber
  H(t) \Psi(t) &=& \sum_k a_k(t) E_{k,\beta(t)} \exp ( -\frac{i}{\hbar} \int_0^t E_{k,\beta(t')} \; dt' ) \ket{k_{ \beta(t)}} .
\end{eqnarray}
By projecting on the state $\bra{k_{ \beta(t)}}$ and using the basis' orthonormality we arrive at a differential equation for the coefficients 
$a_k(t)$
\begin{eqnarray}
  \label{eq:DglAk}
  i \hbar \dot a_k(t) = i \hbar \sum_l a_l(t) \bra{k_{\beta(t)}} \frac{\partial \ket{l_{ \beta(t)}}}{\partial t} \exp( -\frac{i}{\hbar} \int_0^t (E_{l,\beta(t')} - E_{k,  \beta(t')}) \; dt' ).
\end{eqnarray}
In the first place, the sum over $l$ includes off-diagonal elements as well as diagonal terms. But it turns out however that only the diagonal terms contribute, which
can be shown with help of Hellmann-Feynman's theorem \cite{Lange}, by calculating
$\bra{k_\beta} \frac{\partial \ket{l_\beta(t)}}{\partial t} = - \frac{L_2 e \epsilon}{\hbar} \bra{k}\frac{\partial \ket{l_\beta}}{\partial \beta}$.

Hellmann-Feynman's theorem is applicable on parameter dependent eigenvectors of a Hermitian operator dependent on the same parameter. Normalized eigenvectors assumed,
we have
$\frac{\partial}{\partial \beta} \braket{k_\beta}{l_\beta} = 0$, from which follows $\frac{\partial \bra{k_\beta}}{\partial \beta} \ket{l_\beta} = - \bra{k_\beta} \frac{\partial \ket{l_\beta}}{\partial \beta}$. In our case we yield for $H_\beta$
\begin{eqnarray}
  \label{eq:FeynmanHellman1}
  \bra{k_\beta}H_\beta\ket{l_\beta} &=& E_{k,\beta} \delta_{k, l} \\ \nonumber
  \frac{\partial E_{k,\beta}}{\partial \beta} \delta_{k, l} &=& \bra{k_\beta}\frac{\partial \ket{l_\beta}}{\partial \beta} (E_{k, \beta} - E_{l, \beta}) + \bra{k_\beta} \frac{\partial H_\beta}{\partial \beta} \ket{l_\beta}
\end{eqnarray}
Here $\delta_{k,l}$ denotes the Kronecker symbol. With (\ref{eq:HamiltonAlpha}) the derivative of $H_\beta$ can be expressed as $\frac{\partial H_\beta}{\partial \beta} = \frac{\hbar}{m L_2} \sum_j \Pi_{\beta,j,y}$. 
As we restrict the state to lie in the sub-space of the lowest Landau-level, the matrix elements of the kinetic momentum operators $\Pi$ are zero. This is because $\Pi$ can be expressed as a linear combination of $a^\dagger$ and $a$. It will be shown explicitly in section \ref{sec:PiZero}.

From equation (\ref{eq:FeynmanHellman1}) for $k=l$ we would conclude that the eigenenergies should be constant with $\beta$ and thus with time.
This however is only true for homogeneous systems. The reason for that is the projection to the lowest Landau-level.
In the language of perturbation theory, the eigenstates are given to lowest order, while the eigenenergies are calculated to one order above.
A more precise treatment will be given in section \ref{sec:corrections}. For the moment it is enough to assume $E_{k,\beta}$ to depend on $\beta$.

So we have as a final form the differential equation for the coefficients $a_k(t)$
\begin{eqnarray}
  \label{eq:AkDglfinal}
  i \hbar \dot a_k(t) &=& a_k(t) i \hbar \bra{k_{\beta(t)}}\frac{\partial \ket{k_{\beta(t)}}}{\partial t} \\ \nonumber
  a_k(t) &=& a_k(0) \exp \Big( \int_0^t \bra{k_{\beta(t')}} \frac{\partial \ket{k_{\beta(t')}}}{\partial t} dt' \Big) \\ \nonumber
         &=& a_k(0) \exp \Big( \int_0^{\beta=-\frac{L_2 \epsilon e}{\hbar}t} \bra{k_{\beta'}} \frac{\partial \ket{k_{\beta'}}}{\partial \beta} \; d\beta' \Big).
\end{eqnarray}
This is a phase factor as the argument in the exponential function in (\ref{eq:AkDglfinal}) is an imaginary number, since 
$\bra{k}\frac{\partial \ket{k}}{\partial \beta} = -\frac{\partial \bra{k}}{\partial \beta}\ket{k} = -\bra{k}\frac{\partial \ket{k}}{\partial \beta}^* $.

Given an arbitrary initial state of the system, we can calculate the time evolution of the system by projecting the state onto the eigenstates for $t=0$ which yields
the coefficients $a_k(0)$. These then evolve according to (\ref{eq:AkDglfinal}). The state of the system for the time $t$ is the superposition of these contributions given by equation (\ref{eq:Psit}).
Comparing this result to the adiabatic limit in the next section, here we have an additional phase factor of equation (\ref{eq:AkDglfinal}).
\subsubsection{Adiabatic limit of time evolution}
To appreciate the results from the previous section, they will be compared to the adiabatic limit of the time evolution.
In the limit of an infinitesimally weak electric field the time evolution of a state can be calculated by means of the adiabatic theorem, first used by Fock and Born, later generalized by Kato \cite{Kato}. Here we use it in a similar way as in \cite{Avron}. This theorem facilitates the calculation of the time evolution of a system described by a parametric dependent Hamiltonian, where the parameter is varied infinitely slowly by a finite amount or ---put differently--- the total change of the Hamiltonian from $t=-\infty$ to $t=0$ is finite.
Following the nomenclature in \cite{Kato} the time dependent Schrödinger equation is rewritten by rescaling the time variable by $t = \tau s$.
This makes it possible to separate the parameter $s$ running through a finite range from $\tau$ which goes to infinity
\begin{eqnarray}
  \label{eq:rescaledSchroedinger}
  i \hbar \partial_t \Psi(t) &=& H(t) \Psi(t) \\ \nonumber
  \rightarrow i \hbar \partial_s \Psi_\tau(s) &=& \tau H(s) \Psi_\tau(s).
\end{eqnarray}
The aim is to arrive at an expression for the time evolution of the system in the limit of $\tau\rightarrow \infty$, namely $\Psi(s)$.
The adiabatic theorem as proven in the paper of Kato states that for an eigenfunction $\phi(0)$ of $H(0)$ there exists the following approximate expression for
the real time evolution formally described by the unitary operator $V_\tau(s)$
\begin{eqnarray}
  \label{eq:AdiabatTheorem}
  V_\tau(s) \phi(0) &=& \exp \Big(\frac{-i \tau}{\hbar} \int_0^s E(s) \; ds \Big) \; \phi(s) + O(\tau^{-1}) \\ \nonumber
  \mbox{where} \quad H(s) \phi(s) &=& E(s) \phi(s)
\end{eqnarray}
In the limit of $\tau \rightarrow \infty$ this expression becomes exact, indicated by the vanishing deviation-term.

Applied to the problem of an additional electric field, we can determine the time evolution for infinitely weak electric fields.
According to equation (\ref{eq:SGl1}) $\tau$ can be identified with the parameters introduced earlier as
\begin{eqnarray}
  \label{eq:ident}
  \beta = \underbrace{-\frac{L_2 \epsilon e}{\hbar}}_{-\tau^{-1}} t.
\end{eqnarray}
Using (\ref{eq:ident}) in equation (\ref{eq:AdiabatTheorem}) we arrive at
\begin{eqnarray}
  \label{eq:AdiabatFinal}
  \Psi(\beta) &=& \lim_{\epsilon \rightarrow 0} \exp \Big(\frac{i}{L_2 \epsilon e} \int_0^\beta E(\beta) \; d\beta \Big) \; \phi(\beta).
\end{eqnarray}
So the phase factor in equation (\ref{eq:AkDglfinal}) would be lost by this adiabatic treatment.
\newpage
\section{The homogeneous system}
\label{sec:two}
\subsection{Laughlin's wavefunction, interaction and correlations}
\label{sec:hardcore}
In this section some well known features of the homogeneous system are collected and partly adapted to our geometry, since they are crucial
for understanding and motivating the quasihole excitations.

For the circular gauge of the vector potential in the absence of any background potential, Laughlin has proposed in \cite{Laughlin} a
Jastrow-type variational wavefunction for the case of filling factors $\nu = \frac{1}{m}$ where $m$ is an odd integer,
\begin{eqnarray}
  \label{eq:Laughlin}
  \Psi_m(z_1,...,z_{N_e}) = \prod_{i<j} (z_i-z_j)^m \; \exp \left(-\frac{\sum_{i=1}^{N_e} |z_i|^2}{4 l_0^2} \right).
\end{eqnarray}
Picking one electron's coordinate, say $z_1$, while keeping the others fixed, the wavefunction exhibits $(N_e-1)m$ zeros, $m$ of which
appear whenever $z_1$ approaches another electron's position. This $m$-fold zero produces strong correlation holes around each electron which in turn reduces the Coulomb energy. Since the number of zeros of the wavefunction is fixed by the number of flux quanta through the system,
the only freedom is where to put these zeros. In equation (\ref{eq:Laughlin}) the maximum number $(N_e-1)m$ of zeros is used to create the most effective correlation holes. These are the reason for this ground state to be separated by a gap from the system's excited states.
Numerical comparisons showed for different repulsive interaction potentials that this trial wavefunction has a big overlap with wavefunctions found by exact diagonalization \cite{Laughlin,HaldaneRezayi}. Independently Yoshioka, Halperin and Lee \cite{Yoshioka} found these correlations in their numerical work.

Along with the ground state Laughlin also found the elementary excitations of the system to be quasielectrons and quasiholes.
Laughlin mapped this system by an analogy to an one component plasma and thus identified these quasiparticles to have fractional charge $\pm \frac{1}{m}e$.
These excitations are created whenever the filling factor $\nu$ deviates from its value $\frac{1}{m}$: If it increases, quasielectrons are created; on decrease there are additional quasiholes.
The finite amount of energy $\epsilon_{\pm}$ needed to create one of these particles makes the system incompressible, because an infinitesimal change in the system's area (which of course changes $\nu$) causes quasiparticles to be created each of which rising the energy by $\epsilon_{\pm}$. Thus the compressibility is infinite. This incompressibility will be verified in section \ref{sec:gsmu}.

\subsubsection{Short-range interaction}
\label{sec:ShortRange}
Although the overlap of Laughlin's wavefunction with the exact one is high for Coulomb interaction it is not an eigenstate of the system.
 Haldane and Rezayi \cite{HaldaneRezPseudo} treated the interaction by means of introducing pseudopotentials with different ranges. They
showed that the ground state mainly depends on the pseudopotential parameter of shortest range and that it is quite robust against changes
of the other parameters. They found out that the Laughlin wavefunction is an eigenfunction in the case of a short range interaction which only
has the one nonzero pseudopotential parameter for the shortest distance. This treatment was specifically used on the spherical geometry and relied on conservation of angular momentum, which is not true for our system.

This short range interaction was generalized by Trugman and Kievelson \cite{Trugman} who wrote down an analytical form for it without resorting to angular momentum conservation. Although they used the open plane geometry we can follow their idea here and show that it is also applicable to rectangular geometry. In the paper cited above, an expansion of the interaction potential is considered as
$V_{int}$ in terms of its range $b$ as
\begin{eqnarray}
  \label{eq:RangeExpansion}
  V_{int} &=& \sum_{j=0}c_j b^{2 j} \nabla^{2 j} \delta(\vec r).
\end{eqnarray}
Going to the Fourier-space, this expansion is seen to be a Taylor series of a symmetric function in $|q|$ (only even powers of $|q|$ appear),
which is true for every real valued isotropic potential which has an analytic Fourier transform in $q=0$.
The requirement for the potential to have no singularity in $q=0$ is fulfilled if the potential has a finite mean value, since
the Fourier component $\tilde V(0) = \int_0^{\infty} dr \; r V(r)$ is the mean value of the potential.
%
%
%
%
Thus, if $V(q)$ is analytic in $q=0$, it is expandable into a power series in $|q|^2$ and we can determine the coefficients $C_j$ of Equ.
(\ref{eq:RangeExpansion}). This for example is true for a Yukawa-potential. The Coulomb potential however does not have this property.

The first term $c_0 \delta(\vec r)$ in (\ref{eq:RangeExpansion}) vanishes for spin-polarized antisymmetric wavefunctions.
The leading term in the limit of small ranges $b$ thus is
\begin{equation}
  V_{short}(\vec r) = b^2 \nabla^2 \delta(\vec r).
  \label{eq:ShortRange}
\end{equation}
This interaction was shown \cite{Trugman} to reveal Laughlin's wavefunction as the unique ground state for $\nu = \frac{1}{3}$.
Therefore, using this interaction instead of the Coulomb potential, we can assume the Laughlin-like correlations in the wavefunction, which are manifest in the relative zeros of two electrons' coordinates, to be more pronounced than for Coulomb interaction.

The Fourier transform of this short-range interaction is
\begin{eqnarray}
  \label{eq:FourierShort}
  \tilde V_{short}(\vec q) &\propto& -|\vec q|^2.
\end{eqnarray}
In the case of inhomogeneous systems (section \ref{sec:InhomSystems}) the work of Krause-Kyora \cite{Krause-Kyora} showed
that using this short-ranged interaction is advantageous due to the absence of the long-range part of the Coulomb potential which
causes oscillations in the density profile even far off the actual constriction potential.
Therefore, we will make use of this interaction in parts of the following calculations.
The term ``short-range interaction'' and ``hard-core interaction'' will be used equivalently in what follows.

Apart from introducing Laughlin's wavefunction in the next section, the formal reason for this interaction to make Laughlin's trial wavefunction an eigenstate will also be investigated.
\subsubsection{Laughlin's wavefunction in rectangular geometry}
\label{sec:Laughlin}
Haldane and Rezayi \cite{HR} ported Laughlin's wavefunction the rectangular geometry with periodic boundary conditions and they
investigated impurity effects on this ground state. From this paper we will cite the many particle wavefunction here since it will be used 
in section \ref{sec:HardcoreHR} to show that the expectation value of the short-range
interaction vanishes for this state and also to construct the quasihole creation operator in section \ref{sec:QHOpIdea}.

The starting point is a Jastrow-ansatz which implies a many-particle wavefunction that separates into a product of a relative- and center-of-mass-function, where the relative part factorizes to a product of functions $f(z_i-z_j)$ each depending only on the difference of two particles' coordinates $z_i,z_j$ with $z_i=x_i+iy_i$.
The center of mass coordinate is given by $Z=\sum_i z_i$.
Thus we can write
\begin{eqnarray}
  \label{eq:Jastrow}
  \Psi(z_1, \ldots, z_{Ne}) &=& \prod_{j < k} f(z_j - z_k) F_{N_s}(Z) \exp \left(- \pi \frac{N_s}{L_1 L_2} \sum_i x_i^2 \right).
\end{eqnarray}
The wavefunction $F_{N_s}(Z) \exp(- \pi \frac{N_s}{L_1 L_2} \sum_i x_i^2)$ has the same form as a single-particle wavefunction of an electron in a magnetic field containing $m = \frac{N_s}{N_e}$ flux quanta.
Its explicit form \cite{HR} is
\begin{eqnarray}
  \label{eq:Fcm}
  F_{N_s}(Z) &=& \exp(K Z^*) \prod_{\nu=1}^m \vartheta_1 \left( \pi \frac{Z^* - Z_\nu^*}{L_2}| i\frac{L_1}{L_2} \right) \\ \nonumber
  \mbox{where} \; \exp(-i K L_2) &=& (-1)^{N_s} \exp(i \beta) \\ \nonumber
  \mbox{and} \; \exp \left( \frac{2 \pi}{L_2} \sum_{\nu} Z_\nu^* \right) &=& (-1)^{N_s} \exp(i \alpha + K L_1),
\end{eqnarray}
where $\vartheta_1$ is the odd elliptic theta function of first kind \cite{Gradstein}. It is important to note that $\vartheta_1(z) \propto z$ for small $|z|$ such that $F_{N_s}$ has $m$ zeros.
The solution of (\ref{eq:Fcm}) is uniquely defined by giving the real wavevector $K$ and these $m$ zeros $Z_\nu$. The number of linear independent solutions of (\ref{eq:Fcm}) were shown \cite{HR} to be equal to the number of zeros $Z_\nu$, thus there are $m$. They cause an $m-$fold degeneracy of the ground state of a homogeneous system. Alternatively, according to Tao and Haldane \cite{Tao}, this degeneracy can be regarded as originating from the translational invariance of the center of mass wavefunction.

The function $f(z)$ in \ref{eq:Jastrow} has to be chosen such that the periodic boundary conditions are fulfilled and they have to be
odd functions in order to obey the Pauli principle. Like in the open plane, demanding an $m$-fold zero whenever two electrons approach each other leads to the single solution
\begin{eqnarray}
  \label{eq:GroundstateHR}
  \Psi_{GS}(z_1,...,z_{Ne}) &=& \prod_{j < k} \vartheta_1 \Big( \pi \frac{-i (z_j^* - z_k^*)}{L_2} | i\frac{L_1}{L_2} \Big)^m F_{N_s}(Z) \exp(-\frac{\pi N_s \sum_{j=1}^{Ne} x_j^2}{L_1 L_2}).
\end{eqnarray}
Fixing all but one electron's coordinate and counting the zeros of the wavefunction as a function of the last free coordinate, we clearly should find $N_m$ zeros within the unit cell in accordance with the Aharanov-Bohm effect, as stated in section \ref{sec:PRB}. It is important to note, that we will find an m-fold zero whenever approaching one of the fixed electrons' coordinate, giving $(N_e-1)m$ zeros in total. This is seen from (\ref{eq:GroundstateHR}). The remaining $m$ zeros can be found in the center of mass wavefunction (see Equ. (\ref{eq:Fcm})). Their positions obviously depend on the positions of the other electrons. This is an important point to make, because it shows that we cannot fix these zeros with respect to one of the electrons at a certain point. 
Instead, if we want to fix a zero of the wavefunction at some given point with respect to one electron's coordinate independently of the other electrons' positions, we have to insert an additional flux quantum to gain the freedom of localizing the respective zero arbitrarily. In section \ref{sec:QuasiHolesDelta} this will be used to create a quasihole.
\subsubsection{Vanishing short-range interaction for Laughlin's wavefunction in rectangular geometry}
\label{sec:HardcoreHR}
In the previous section the equivalent to Laughlin's wavefunction in a system with periodic boundary conditions was cited.
What Trugman and Kievelson \cite{Trugman} did for the open plane geometry is possible to show for periodic boundary conditions as well: This wavefunction yields a vanishing expectation value $\langle V_{short} \rangle$ for the short-range interaction, $V_{short}$ given by Equ. (\ref{eq:ShortRange}).
Furthermore, also the trial wavefunction for a quasihole which will be given in Equ. (\ref{eq:HoleEx}) minimizes this interaction to zero.

To proof these two statements, we can consider a trial-wavefunction of the following form
\begin{eqnarray}
  \Psi(z_1,\ldots,z_{N_e}) &=& \prod_{i<j} \tilde \vartheta(z_i - z_j)^m f(z_1,\ldots,z_{N_e})
\end{eqnarray}
Here $\tilde \vartheta(z) = \vartheta_1(-i \pi \frac{z^*}{L_2} | i\frac{L_1}{L_2})$ is used as a short hand and $f(z_1,\ldots,z_{N_e})$ is a symmetric function. $\vartheta_1$ is the odd elliptic theta function of first kind. 
In the case of Laughlin's ground state wavefunction (\ref{eq:Jastrow}) the function $f$ is given by $f = F_{N_s}(Z) \exp(- \pi \frac{N_s}{L_1 L_2} \sum_i x_i^2)$, with $F_{N_s}(Z)$ being the center of mass function given in (\ref{eq:Fcm}).
If we take the trial wavefunction of a quasihole excited state, $f = \Pi_{k=1}^{N_e} \tilde \vartheta(z_k-z_0) F_{N_s+1}(Z+\frac{z_0}{m}) \exp(- \pi \frac{N_s+1}{L_1 L_2} \sum_i x_i^2)$ (as can be seen from Equ. (\ref{eq:HoleEx})).
The exact form of $f$ however does not matter here; only the fact that it is 2 times differentiable and that it doesn't have any singularities will be used in the following.

Writing the coordinates $z = x + i y$ as $\vec r = x \vec e_x + y \vec e_y$,
the expectation value of the two body interaction can be calculated as
\begin{eqnarray}
	\langle V_{short} \rangle &=& \int d^2 r_1 \ldots d^2 r_{N_e} \; |\Psi(z_1,\ldots,z_{Ne})|^2 \; \sum_{i<j} V(r_i - r_j) \\ \nonumber
	&=& \frac{N_e(N_e-1)}{2} \int d^2 r_1 \ldots d^2 r_{N_e} \; |\Psi(z_1,\ldots,z_{Ne})|^2 \; V_{short}(r_1 - r_2).
\end{eqnarray}
Introducing new coordinates $r_{-} = r_1 - r_2$ and $r_{+} = r_1 + r_2$, this integral can be recast into
\begin{eqnarray}
	\langle V_{short} \rangle &=& \frac{N_e(N_e-1) b^2}{4} \int d^2 r_3 \ldots d^2 r_{N_e} \int d^2 r_{+} \times \\ \nonumber
	&& \times \underbrace{ \int d^2 r_{-} \; \left| f(z_1,\ldots,z_{Ne}) \prod_{i<j} \tilde \vartheta(z_i - z_j)^m \right|^2 \; \nabla ^2 \delta(\vec r_{-}) }_{=: I}, \nonumber
\end{eqnarray}
where the short-range interaction potential was substituted. The last integral $I$ can be integrated by parts twice, resulting in
\begin{eqnarray}
	\label{eq:evalI}
	I &=& \int d^2 r_{-} \; \nabla ^2 \left( \left|f(z_1,\ldots,z_{Ne}) \prod_{i<j} \tilde \vartheta(z_i - z_j)^m \right|^2 \right) \; \delta(\vec r_{-})\\ \nonumber
	&=& \left.\nabla^2 \left( \left|f(z_1,\ldots,z_{Ne}) \prod_{i<j} \tilde \vartheta(z_i - z_j)^m \right|^2 \right) \right|_{\vec r_{-} = 0}.
\end{eqnarray}
The boundary terms vanish for both steps of partial integration. The boundary term after the first integration contains a factor 
$\nabla \delta(\vec r_{-})$ with $r_{-,x} = \pm L_1$ or $r_{-,y} = \pm L_2$ as limits. After the second integration, the factor $\delta(\vec r_{-})$ in the same limits makes the boundary term vanish.
Also the last expression (\ref{eq:evalI}) vanishes, because after differentiating we can factor out $|\tilde \vartheta(z_1 - z_2)|^{2m-2}$ that behaves like $|z_1-z_2|^{2m-2}$ if $\vec r_{-} \rightarrow 0$. Thus, the positive semi-definite interaction energy is zero for both ---the ground state and the quasihole state---  and, since this is the only contributing operator in the Hamiltonian (apart from the frozen kinetic energy), those states must be ground states of the homogeneous system.

Going to higher powers in $j$ in Equ. (\ref{eq:RangeExpansion}) and applying the same argument as before, the expectation value vanishes as long as $j < m$. However, since we are interested in a filling factor $\nu = \frac{1}{3} = \frac{1}{m}$, the order $j=1$ in the short-range interaction is already sufficient to force a threefold zero to reside on every pair of relative coordinates in the ground state. The reason for that is the Pauli principle which demands an odd number of zeros on every relative coordinate.
So, the interaction potential given in (\ref{eq:ShortRange}) has the property of producing Laughlin's wavefunction (\ref{eq:GroundstateHR}) in the case of a filling factor $\nu=\frac{1}{3}$.


\subsubsection{Projection to the lowest Landau level}
\label{sec:projection}
\label{sec:Projector}
So far we have seen trial wavefunctions for the homogeneous system. Now the many particle Hamiltonian given in Equ. (\ref{eq:ManyPartHamiltonian}) of section \ref{sec:ideas} shall be treated by numerical diagonalization first in absence of external potentials $V_{ext}$.
Treating a quantum mechanical system in a finite basis needs as a prerequisite a reasonable way to reduce the infinite dimensional Hilbert space to some subspace of finite dimension while keeping enough degrees of freedom to describe the essential physics. As shown above, an electron subjected to a magnetic field exhibits an equidistant energy spectrum of Landau-levels. The properties of a fractional quantum Hall ground state were shown to be reproducible when restricting the state to live in the lowest Landau level, neglecting admixture of higher Landau levels. Analytically this was done by Laughlin in \cite{Laughlin}. From perturbation theory this limit corresponds to the case where the Landau level spacing $\hbar \omega_c$ is much bigger than the perturbation of the system, namely the Coulomb interaction and the impurity potential.

This restriction to the lowest Landau level will also be applied here. But although it facilitates numerical calculations, some other problems are encountered when projecting to the lowest Landau level as described below in section \ref{sec:corrections}.

The projection operator $P_{LLL}$ can be constructed easily once the single-particle basis states $\ket{0,k} = a^\dagger_k \ket{0}$ of the lowest Landau level from section \ref{sec:Basis} are known. Here $a_k^\dagger$ denotes the creation operator of the single particle state (\ref{eq:PerBasis}) with the momentum quantum number $k \in \{0,\ldots, N_s-1 \}$.
We can build many-particle states out of them as slater determinants. Alternatively we can write the state in second quantization
\begin{equation}
\label{eq:ManyPartBasis}
\ket{j_1,\ldots,j_{N_e}} = a^\dagger_{j_1} \ldots a^\dagger_{j_{N_e}} \ket{0}.
\end{equation}
The dimension of the basis is ${N_s \choose N_e}$ and we can number the states by ascending multi-indices $j_1 < \ldots < j_{N_e}$, where $N_e$ is the number of electrons, $N_s$ the number of single particle states in the lowest Landau level.
The projection operator can be constructed simply as
\begin{eqnarray}
  \label{eq:PLLL}
  P_{LLL} &=& \sum_{j_1<\ldots<j_{N_e}} \ket{j_1,\ldots,j_{N_e}}\bra{j_1,\ldots,j_{N_e}}.
\end{eqnarray}
The Hamiltonian from Equ. (\ref{eq:ManyPartHamiltonian}) can now be projected by this operator. This yields
\begin{eqnarray}
  \label{eq:HamiltonProj}
  P_{LLL} H P_{LLL} &=& \frac{\hbar \omega_c}{2}N_e P_{LLL} + P_{LLL} \left( \sum_{i=1}^{N_e} V_{ext}(\vec r_i) + \sum_{i < j} V_{Int}(\vec r_i - \vec r_j) \right) P_{LLL}.
\end{eqnarray}
Due to the projectors left an right to the operators, this Hamiltonian only causes mixing in the subspace of the lowest Landau level.
Therefore, this is equivalent to diagonalizing the unprojected Hamiltonian in the many particle basis (\ref{eq:ManyPartBasis}). Stated differently, we just have to write the operators $V_{ext}$ and $V_{int}$ in second quantization where the single- and two-particle matrix elements are calculated in the basis $\ket{0,j}$. 
The kinetic energy of $\frac{\hbar \omega_c N_e}{2}$ is just a constant and does not affect the eigenstates at all. Omitting it is equivalent
to changing the origin of the energy scale.
This also makes clear why we have to diagonalize at all, since all many-particle states are degenerate with respect to the kinetic part of the
Hamiltonian and the interaction cannot be treated as a small perturbation.
To arrive at the eigenstates of this projected Hamiltonian we therefore have to diagonalize
\begin{eqnarray}
  \label{eq:ProjH}
  H_{LLL} &=& \sum_{j_1,j_2=1}^{N_s} V_{ext j_1,j_2} a^\dagger_{i} a_j + \sum_{j_1,j_2,j_3,j_4} \frac{1}{2} V_{int j_1,j_2,j_3,j_4} a^\dagger_{j_1} a^\dagger_{j_2} a_{j_4} a_{j_3}.\\ \nonumber
\end{eqnarray}
Actually, there is an additional diagonal single particle operator in the case of Coulomb interaction. It is the constant interaction energy of one electron with its images in neighboring cells \cite{Yoshioka}. The matrix elements for the interaction and the various external
potentials used in the inhomogeneous system are found in the appendix \ref{app:MatEls}.

However, in this section we will first work with the homogeneous system, thus $V_{ext} \equiv 0$.
\subsubsection{Groundstate energies and chemical potential}
\label{sec:gsmu}
%
%
The purpose of these calculations is to model a constriction in a fractional quantum Hall system and investigate how quasiholes
near this barrier behave. More concrete, the question whether single quasiholes can pass through the barrier should be answered.
The dependence on the barrier's parameters will be surveyed. To get a measure of the potentials needed to create a tunneling barrier (for electrons), the chemical potential of the electrons in the system must be calculated. If the height of the constriction is higher than the chemical potential, the electrons can only pass it by tunneling.
The chemical potential is calculated for different system sizes from 4 to 6 electrons at a filling factor $\frac{1}{3}$.
Another purpose of this calculation is to confirm the system's incompressibility, which is recognizable as a jump in the chemical potential
when going from slightly lower to slightly higher filling factors than $\frac{1}{3}$.
\begin{table}[htbp]
\begin{tabular}{|l|l|l|l|l|}
\hline
System &  GS energy $(\mbox{enu})$ & energy per electron $(\mbox{enu})$ & degeneracy & $\Delta_{gap} (\mbox{enu})$\\
\hline
4/12 C & -1.660765 & -0.415191 & 3 & 0.047174\\
4/11 C & -1.687588 & -0.421897 & 11 & 0.063354\\
4/13 C & -1.589986 & -0.397497 & 13 & 0.064071\\
5/12 C & -2.194106 & -0.438821 & 12 & 0.052964\\
3/12 C & -1.072924 & -0.357641 & 12 & 0.011904\\
\hline	
5/15 C & -2.06322 & -0.41264 & 3 & 0.06313\\
5/14 C & -2.081965 & -0.41639 & 14 & 0.05010\\
5/16 C & -1.98999 & -0.39800 & 16 & 0.05698\\
6/15 C & -2.61189 & -0.43532 & 5 & 0.04643\\
4/15 C & -1.49802 & -0.37451 & 15 & 0.03072\\
\hline
6/18 C & -2.471389 & -0.41190 & 3 & 0.06301\\
6/17 C & -2.488045 & -0.41467 & 17 & 0.055052\\
6/19 C & -2.396557 & -0.39943 & 19 & 0.058339\\
7/18 C & -2.994021 & -0.42772 & 18 & 0.032446\\
5/18 C & -1.890334 & -0.37807 & 18 & 0.016655\\
\hline
4/12 H & 0.0000 & 0.0 & 3 & 0.18049\\
4/11 H & 0.200229 & 0.050057 & 11 & 0.205743\\
4/13 H & 0.0000 & 0.0 & 13 & 0.192552\\
5/12 H & 0.609253 & 0.121851 & 12 & 0.176295\\
3/12 H & 0.0000 & 0.0 & $>$ 30 & n.c.\\
\hline	
5/15 H & 0.0000 & 0.0 & 3 & 0.20662\\
5/14 H & 0.21559 & 0.043118 & 14 & 0.198772 \\
5/16 H & 0.0000 & 0.0 & 16 & 0.199536\\
6/15 H & 0.449654 & 0.074942 & 5 & 0.235736\\
4/15 H & 0.0000 & 0.0 & $>$ 30 & n.c.\\
\hline
6/18 H & 0.000 & 0.0 & 3 & 0.195\\
6/17 H & 0.215296 & 0.035883 & 17 & 0.133884\\
6/19 H & 0.000 & 0.0 & 19 & 0.197\\
7/18 H & 0.5257 & 0.0751 & 18 & 0.1304\\
5/18 H & 0.0000 & 0.0 & $>$ 30 & n.c.\\
\hline
\end{tabular}
\caption{Ground state energies, degeneracies and gap of a system of 4 $\ldots$ 6 electrons for a filling factor near $\nu = \frac{1}{3}$. The first column specifies the number of electrons, the number of flux quanta and the kind of interaction potential (H for hard-core, C for Coulomb).}
\label{tab:GShomogeneous}
\end{table} 
%
%
%
Table \ref{tab:GShomogeneous} summarizes the ground state energies, their degeneracies and the gap of the system for both, Coulomb and hard-core interaction for different filling factors around $\nu = \frac{1}{3}$. Note that all energies are given in units of 
\begin{equation}
\mbox{enu} := \frac{e^2}{\epsilon l_0(N_s)}
\end{equation}
where $N_s$ is the respective number of flux quanta of the system. This energy unit will be used throughout the document. 
The energy per electron of the systems with $N_e/N_s=$4/12, 5/15 and 6/18 with Coulomb interaction are in agreement with the values calculated by Yoshioka in \cite{Yoshioka}.
%
%
%
%
%

From the values in table \ref{tab:GShomogeneous} the chemical potential of electrons could be calculated just by taking the difference between ground state energies of a system with $N_e$ and $N_e \pm 1$ electron. However, this approach has some drawbacks which are partly due to the small size of the system, partly because of the other quantities that are affected by changing $N_e$.
If we take a system with only four or five electrons and change the number of electrons by $\pm 1$ this change can hardly be treated as infinitesimal with respect to the total number of electrons, as demanded by a reasonable definition of $\mu$. In fact, in the case of $\frac{5}{15} \rightarrow \frac{6}{15}$ changing the number of electrons by $+1$ results in a filling of $\frac{2}{5}$, which should itself show a fractional quantum Hall effect.  
The second reason not to calculate the chemical potential this way is of more practical nature. When calculating the Coulomb interaction between electrons, to conserve the
overall charge neutrality, the interaction with a uniform positive background of density $\sigma_{+} = \frac{N_e e}{L_1 L_2}$ is taken into account. Thus, changing $N_e$ also alters this background charge and affects the eigenenergy.

To overcome the problems mentioned above, Yoshioka uses in \cite{Yoshioka84} a different approach to calculate the chemical potential for electrons in his numerical finite size studies by tracing it back to the dependence of the ground state energy on the number of flux quanta.
To show the step in the chemical potential, around the FQH-effect fillings, the chemical potential is calculated for slightly lower ($-$) and slightly higher ($+$) fillings of the system. This $\mu^\pm$ is calculated in the following way. Let $E_0(\nu)$ be the energy per electron at the filling factor $\nu$,
\begin{eqnarray}
  \mu^+_{electron} &=& E_{GS}^{N_e+1} - E_{GS}^{N_e} \\ \nonumber
		 &=& (N_e+1) E_0(\nu(Ne+1)) - N_e E_0(\nu)\\ \nonumber
  		 &=& (N_e+1) E_0(\nu) + N_e \frac{\partial E_0}{\partial \nu} \frac{1}{N_s} - N_e E_0(\nu) \\ \nonumber
		 &=& E_0(\nu) + \frac{N_e}{N_s} \frac{\partial E_0}{\partial \nu} \\ \nonumber
		 &\simeq& E_0(\nu) + \nu \; \frac{E_0(\nu_+) - E_0}{\nu_+ - \nu} \\
\mbox{analogously} \\ \nonumber
  \mu^-_{electron} &\simeq& E_0(\nu) + \nu \; \frac{E_0(\nu_-) - E_0}{\nu_- - \nu} \\
\mbox{where} \quad \nu &=& \frac{N_e}{N_s} \quad \nu_+ = \frac{N_e}{N_s-1} \quad \nu_- = \frac{N_e}{N_s+1}. \nonumber
\label{eq:muChak}
\end{eqnarray}
Since here only the number of flux quanta is varied, the problem of changing the neutralizing background is not present.
This calculation yields the  values from table \ref{tab:MuPMHomogeneous} for the chemical potential for adding an electron at slightly higher and slightly lower fillings than $\frac{1}{3}$.
\begin{table}[htbp]
\begin{tabular}{|l|l|l|l|}
\hline
System & $\mu^+_{electron} \; (\frac{e^2}{\epsilon l_0(N_s=1)})$ & $\mu^-_{electron}  \; (\frac{e^2}{\epsilon l_0(N_s=1)})$ \\
\hline
4/12 C & 0.27732 & -1.1745891\\
5/15 C & 1.2132899 & -2.091855\\
6/18 C & 2.1070906 & -2.4859929\\
\hline
4/12 H & 7.304929 & 0.0\\
5/15 H & 11.277894 & 0.0\\
6/18 H & 15.528201 & 0.0\\
\hline
\end{tabular}
\caption{Chemical potential for electrons and quasiholes for Coulomb (C) and hard core (H) interaction}
\label{tab:MuPMHomogeneous}
\end{table}
To address the calculation for hard core interaction, first note that according to section \ref{sec:HardcoreHR},
the ground state energy in the system with hard-core interaction vanishes whenever the wavefunction has a threefold zero if two electrons come close to each other. Therefore, at least $m N_e$ zeros or flux quanta are needed. If there are less, we arrive at a finite ground state energy.
On the other hand, if there is one flux quantum more than needed to establish these correlations, the hard-core interaction vanishes, too. So, adding a flux quantum is a gapless excitation of the system with hard-core interaction, while removing one needs a finite creation energy, since the correlations have to be changed.  

In the case of Coulomb interaction, there is a finite but yet small amount of energy to pay in either case.
The negative chemical potential $\mu_-$ means that the system wants to absorb electrons until it has reached a filling of $\frac{1}{3}$,
when the chemical potential jumps to a positive value and stops this process.

Since the elementary excitations of the system are quasiparticles according to Laughlin \cite{Laughlin},
we can relate this jump in the chemical potential to the creation energy of quasiparticles. Increasing the density amounts to adding quasielectrons, decreasing it can be understood as adding quasiholes. Since the quasiparticles only have $\frac{1}{m}$ of an electron's charge,
the creation of a quasihole needs $\epsilon_{h} = -\frac{1}{m} \mu_-$, for a quasielectron we need $\epsilon_{e} = \frac{1}{m} \mu_+$.
Therefore, for both interactions we can conclude that the system is incompressible, because enlarging the area is equivalent to inserting quasiholes, while compressing it amounts to inserting quasielectrons, which in both cases costs a finite amount of energy for infinitesimal changes of the density.

\subsubsection{Correlations in homogeneous systems}
\label{sec:correlations}
For electrons interacting with each other by the hard-core pseudopotential, the ground state should exhibit Laughlin-like correlations.
This can not only be confirmed by vanishing hard-core interaction but also by looking at the two particle correlation function
\begin{eqnarray}
\label{eq:CorrAvgd}
 g(z) &=& \frac{L_1 L_2}{N_e(N_e-1)} \bra{\Psi}\sum_{i \neq j} \delta(\vec r + r_i - r_j) \ket{\Psi}.
\end{eqnarray}
%
This correlation or pair distribution function was already used by Yoshioka in \cite{YoshiokaCorr}. Since we are calculating in a
basis with quasi-periodic boundary conditions (see section \ref{sec:PRB}) this operator's matrix elements are a bit different from those used by
Yoshioka. They are given in the appendix \ref{app:MatEls}.

\begin{figure}
\vspace{-1cm}
\includegraphics[scale=0.4]{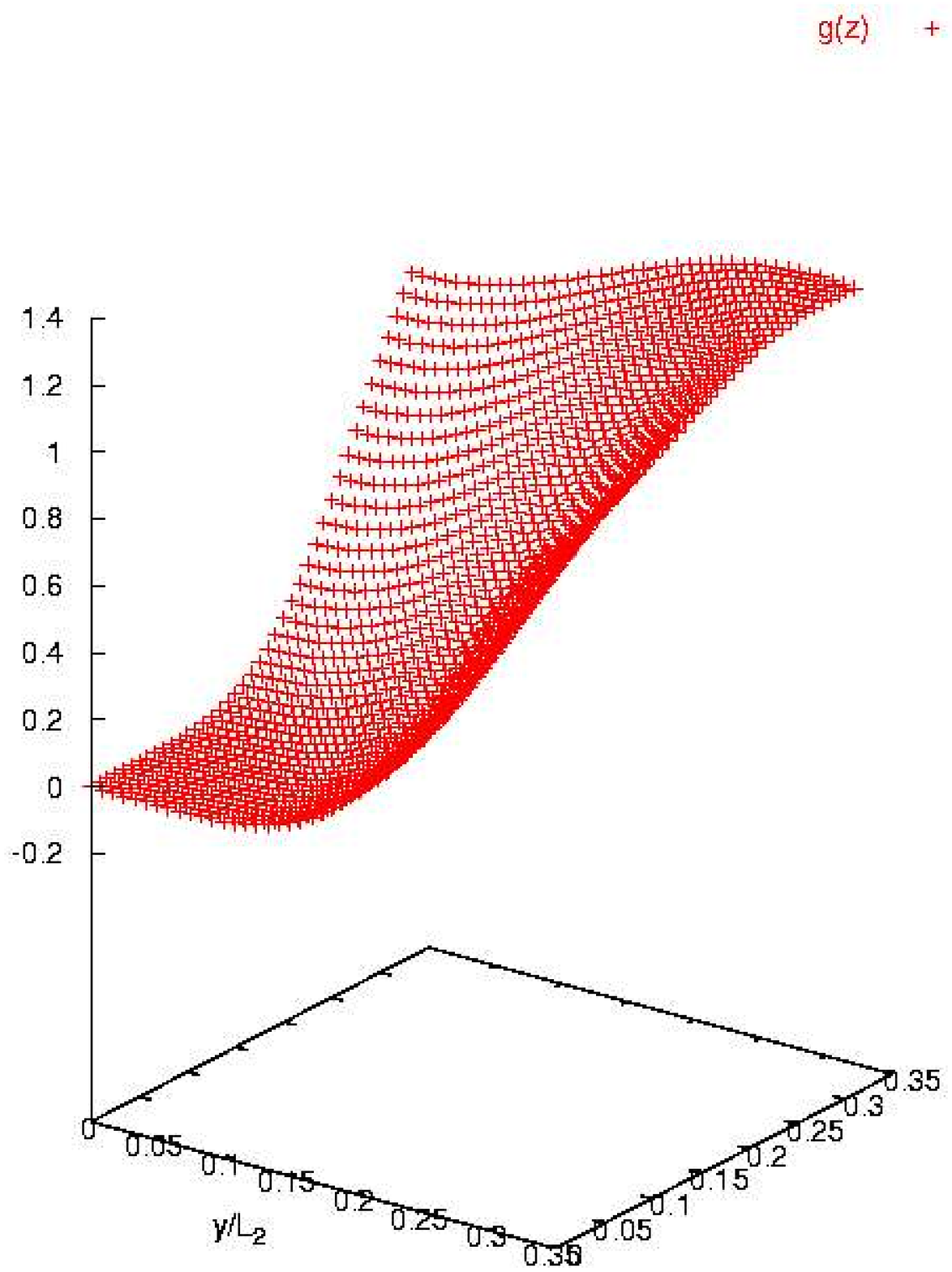} \includegraphics[scale=0.4]{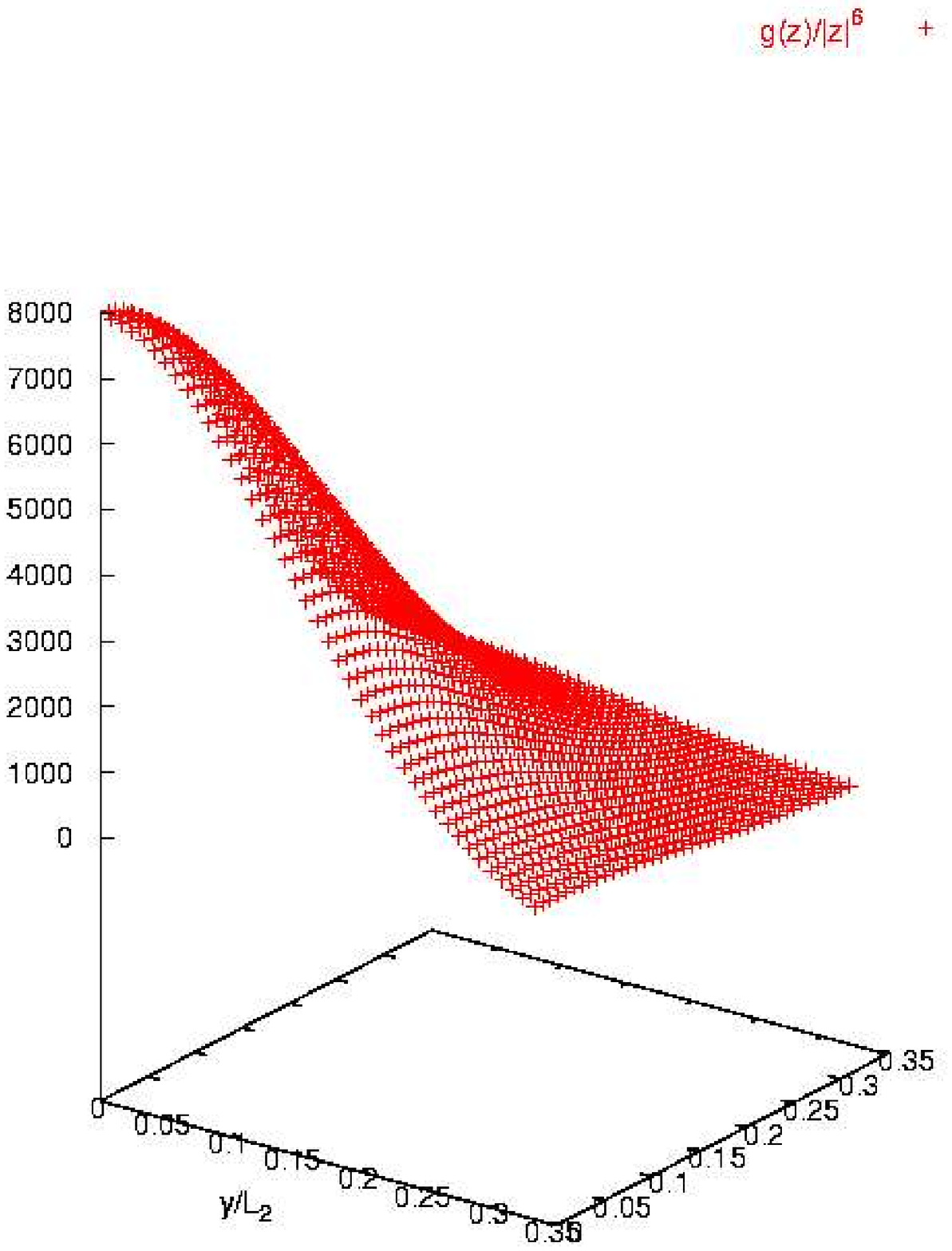}
\caption{Two particle correlation function for a system with $N_e/N_s = 5/15$ and hard-core interaction. The left plot shows the correlation function itself, the right plot is the quotient $g(z)/|z|^6$. The finite value for $|z| \rightarrow 0$ confirms that $g(z) \propto |z|^6$ for small $|z|$ as expected for Laughlin-like correlations.}
\label{fig:corr5-15hard}
\end{figure}

The Laughlin wavefunction has a relative factor of $(z_i-z_j)^m$ for every pair of electrons. In the case of $\nu=\frac{1}{3}=\frac{1}{m}$
it should lead to a correlation function $g(z)$ that behaves like $|z|^{2 m}$.
Evaluating $g(z)$ for each of the 3 degenerate ground states in a system with 5 electrons and 15 flux quanta yields figure \ref{fig:corr5-15hard}. The right plot shows the quotient $\frac{g(z)}{|z|^6}$. The finite value it approaches for small $z$ directly confirms the Laughlin-like correlations.
\subsection{Quasihole excitations}
\label{sec:QuasiHoles}
A quasihole excitation is an excitation of the fractional quantum Hall system to which we can attribute particle like properties such as charge and position. The charge of this object was shown to be a fraction of the charge of an electron by Laughlin in \cite{Laughlin}. Another argument than the plasma analogy used by Laughlin will be invoked here to make this fractional charge more plausible and to
get a coarse idea and intuitive understanding of the connection of flux quanta and quasiholes. It is the following picture adapted from Shankar \cite{Shankar}.

\begin{center}
\begin{figure}
  \input{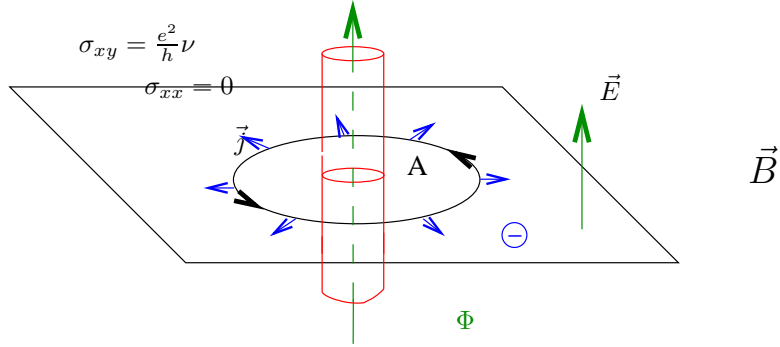}
  \caption{Gedankenexperiment to create a quasihole: A solenoid is pierced through the plane of the 2DEG in the gapped groundstate at filling $\nu=\frac{1}{m}$ and its magnetic flux is increased adiabatically by one flux quantum $\Phi_0$. The induced electric field $\vec E$ doesn't cause transitions but the charge of $-e \nu$ is repulsed out of the area $A$.}
  \label{fig:FracChargeRep}
\end{figure}
\end{center}

Imagine electrons confined in an infinite plane with a perpendicular homogeneous magnetic field $\vec B$. Let the filling factor be the fraction
$\nu = \frac{1}{m}$, $m$ being an odd integer. The system is then in a fractional quantum Hall state and shows a Hall conductivity of
$\sigma_{xy} = \nu \frac{e^2}{h}$ while the longitudinal conductivity is zero, $\sigma_{xx} = 0$. The ground state of the system is separated by a gap from the excited states. This gap prevents the system from undergoing transitions in the presence of an adiabatic perturbation.
A thin solenoid is pierced through the surface as depicted in Fig. \ref{fig:FracChargeRep}.
The flux $\Phi$ through this solenoid --- initially being 0 --- will be increased adiabatically to one flux quantum $\Phi_0=\frac{h}{e}$. Since this is an adiabatic change, the system's fractional state will not be destroyed because it is gapped and the conductivity remains as stated above. 
The time-dependent change of $\Phi$ produces an induced field $\vec E$ by Maxwell's equation $\nabla \times \vec E = - \partial_t \vec B$.
This field is directed tangential to a circular boundary $\partial A$ enclosing the solenoid (see fig. \ref{fig:FracChargeRep}). Due to the form of the conductivity tensor in the fractional quantum Hall state, this electric field produces a current flowing radially away from the solenoid through the boundary $\partial A$. This current can be integrated over time to yield the amount of charge $Q_{rep}$ that is repulsed by the insertion of the flux quantum. This calculation yields
\begin{eqnarray}
	Q_{rep} &=& \int_{-\infty}^0 dt \; \int_{\partial A} d\vec s \underbrace{\vec j}_{\sigma_{xy} E \vec{e_r}} \\ \nonumber
		&\stackrel{d\vec s \parallel \vec{e_r}}{=}& \sigma_{xy} \int_{-\infty}^0 dt \; \int_{A} \; d\vec S \underbrace{\nabla \times \vec E}_{-\partial_t \vec B} \\ \nonumber
		&=& -\sigma_{xy} \int_{-\infty}^0 dt \; \frac{\partial \Phi}{\partial t} \\ \nonumber
		&=& -\sigma_{xy} \Phi_0 \\ \nonumber
		&=& -e \nu.
\end{eqnarray}
The total charge that flows through the boundary while increasing the flux in the solenoid in this process is $Q_{rep} = \nu e$, so there remains a positive charge deficit of $Q_{qh} = \nu e$ inside the boundary. This is the charge that is attributed to the quasihole.
As Shankar points out \cite{Shankar}, thinking in this picture, the fractional charge of a quasihole is a consequence of the quantized conductivity of the system and not vice versa.
\subsubsection{Variational wavefunction for a quasihole excitation}
\label{sec:QHOpIdea}
Following the paper of Haldane and Rezayi $\cite{HR}$ we can construct a wavefunction that exhibits a quasihole excitation. Starting from a fractional quantum Hall system's ground state, a quasihole can be regarded as a zero in the wavefunction relative to all electrons' coordinates. As already stated in section \ref{sec:Laughlin}, to fix a zero of the trial wavefunction at a specified position with respect to every electron, an additional flux quantum must be introduced into the system to gain the desired freedom and in accordance with the Aharanov-Bohm effect.

Such a wavefunction of a state with a localized quasihole at $z_0 = x_0 + i y_0$ was given along with the system's ground state in this paper \cite{HR}. It can be understood as a modification of the ground state wavefunction (\ref{eq:GroundstateHR}).
\begin{eqnarray}
  \label{eq:HoleEx}
  \Psi_{hole}(z_1,...,z_{Ne}; z_0) &=& \prod_{j=1}^{Ne} \vartheta_1 \Big( \pi \frac{-i (z_j^*-z_0^*)}{L_2} | i\frac{L_1}{L_2} \Big) \times\\ \nonumber
  &\times& \prod_{j < k} \vartheta_1 \Big( \pi \frac{-i (z_j^* - z_k^*)}{L_2} | i\frac{L_1}{L_2} \Big)^m \times \\ \nonumber
  &\times& F_{N_s+1}(Z+\frac{z_0}{m}) \exp \left(-\frac{\pi (N_s+1) \sum_{j=1}^{Ne} x_j^2}{L_1 L_2} \right)
\end{eqnarray}
If we compare the ground state wavefunction (\ref{eq:GroundstateHR}) with (\ref{eq:HoleEx}), the procedure to create the quasihole
excitation can formally be seen as a multiplication with a relative zero
(since $\vartheta(z) \simeq z$ for small $|z|$) for every electrons' coordinate $z_i$ with respect to the location of the hole $z_0$. 
Additionally, the center of mass function $F_{N_s+1}(Z)$ satisfies the boundary conditions for $N_s+1$ flux quanta and is translated by $\frac{z_0}{m}$. In the Gaussian the number of flux quanta $N_s$ is replaced by $N_s+1$.
Physically this can be interpreted as inserting an additional flux quantum at $z_0$, manifest in an Aharanov-Bohm phase of $2 \pi$ when encircling $z_0$ with one of the electrons.
The increase of the number of flux quanta is also reflected in the change of the number of flux quanta in the Gaussian.
Finally the center of mass coordinate has to be translated which will appear to be crucial for conserving the boundary conditions.

These operations will be incorporated into a quasihole creation operator $O_{Hole}(z_0)$. It will be derived such that
acting on the ground state function (\ref{eq:GroundstateHR}) results in the quasihole wavefunction (\ref{eq:HoleEx}).
\begin{equation}
  \label{eq:OperDef}
  \Psi_{hole}(z_1,\ldots,z_{N_e}; z_0) = O_{hole}(z_0) \Psi(z_1,\ldots,z_{N_e})
\end{equation}
After deriving this operator we can let it act on a many particle basis state of a system with $N_s$ flux quanta.
These many particle basis states are linear combinations of slater determinants of the single particle basis states from Equ. (\ref{eq:BasisOrt}) in section \ref{sec:PRB}. Since we are adding a flux quantum by this process, the resulting vector can be expressed in the basis of a system with $N_s+1$ flux quanta. If we once know how these basis states of the $N_s$ system translate to those of the $N_s+1$ system with an inserted quasihole, the operator can as well be applied to arbitrary states expressed in the ``$N_s$''-basis. From the operator's definition it is clear that it always produces
only trial wavefunctions for quasihole states which may be good or not. For homogeneous systems with short-range interaction or with Coulomb interaction this approach will be shown to produce excitations that are identical or close to the true quasihole excitation, respectively.

The rather lengthy calculation of the operator's derivation can be found in Appendix \ref{app:DerivQHOp}.
\subsubsection{Creating a quasihole by pinning of a vortex}
\label{sec:QuasiHolesDelta}
Regarding the trial wavefunction for a quasihole excitation in Equ. (\ref{eq:HoleEx}) from the previous section, it is possible to make two points.
First, due to the $m-$fold zero in the relative coordinates this wavefunction must have a vanishing expectation value of the short-range interaction energy as was shown in section \ref{sec:HardcoreHR}. Second, because of this vanishing ground state energy, all states
only differing in the quasihole's position $z_0$ will be degenerate. Since hard-core interaction is a positive definite pseudo-potential (in spin polarized systems), the ground state of a system with $N_s = m N_e + 1$ flux quanta cannot have lower energy than zero. Thus, Equ. (\ref{eq:HoleEx}) must be a vector that lies in the space of the degenerate ground state of a homogeneous system.
So, it must be possible to find a ground state with vanishing energy, even if the position of one vortex is fixed as a quasihole at $z_0$.

This constraint can be imposed to the system by adding a delta potential $\sum_{i=1}^{N_e}\delta(x_i-x_0)\delta(y_i-y_0)$ at $z_0 = x_0 + i y_0$
to the Hamiltonian. 
In the case of a filling factor of $\nu=\frac{1}{3}$, diagonalizing the Hamiltonian with short-range interaction (\ref{eq:ShortRange})
must then result in wavefunction (\ref{eq:HoleEx}), since this is the only wavefunction that
at the same time suffices all constraints: The periodic boundary conditions, a zero at $z_0$ and 3 relative zeros on every electron.
The part $F_{N_s+1}$ of the wavefunction \ref{eq:HoleEx} still is defined by a wavevector $K$ and $3$ zeros $Z_\nu$ (see eq. (\ref{eq:Fcm})) and hence produces a threefold degeneracy as in the homogeneous case.


\subsubsection{Quasiholes in homogeneous systems with hard-core interaction}
\label{sec:QuasiHolesHard}
In this section quasihole excitations will be investigated for a homogeneous system of electrons interacting via hard-core interaction.
The filling factor is set to $\nu = \frac{1}{3}$, realized by $N_e=5$ electrons and $N_s=15$ flux quanta.

\begin{center}
\begin{figure}[htbp]
  \includegraphics[scale=0.4]{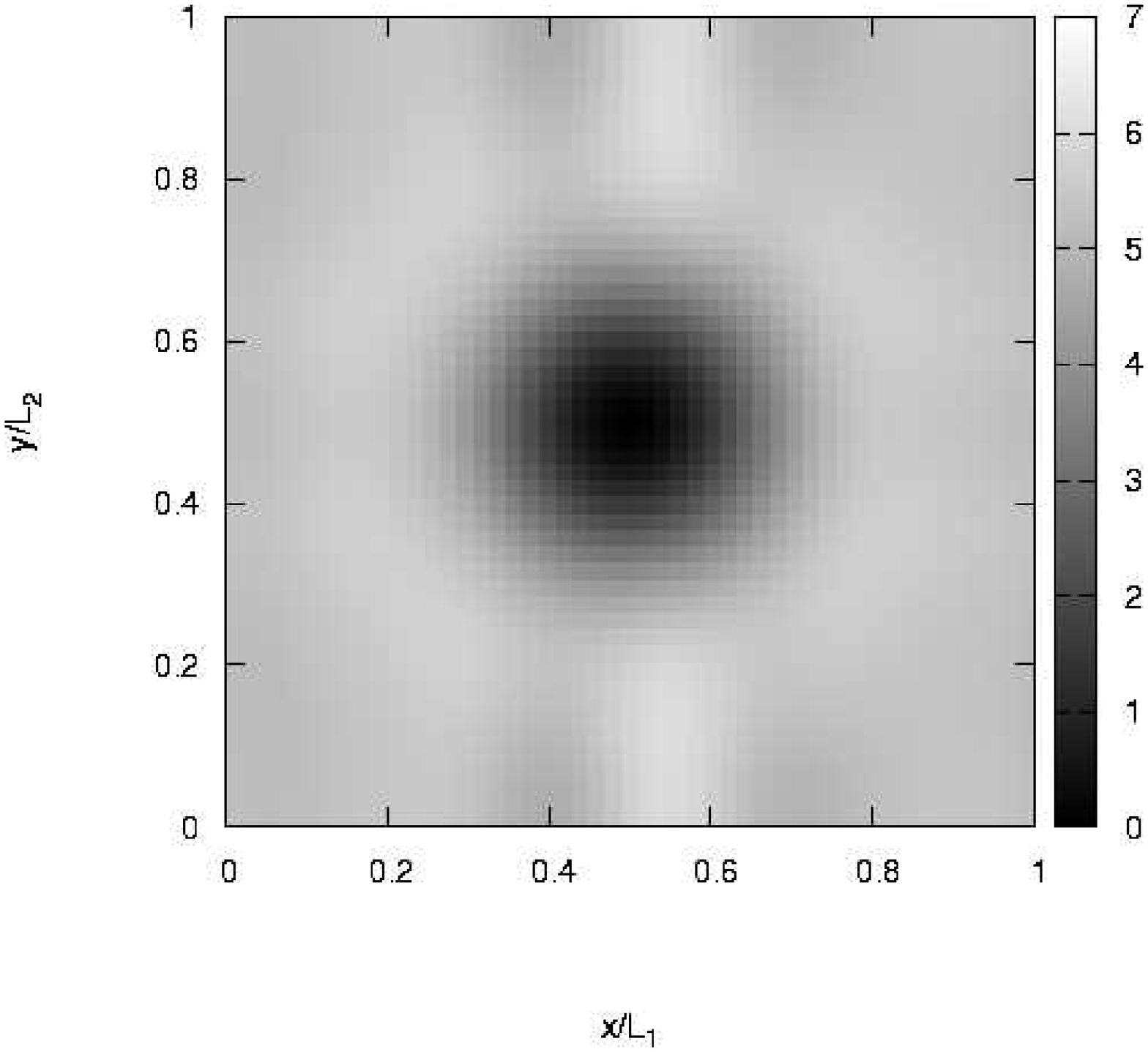}\includegraphics[scale=0.4]{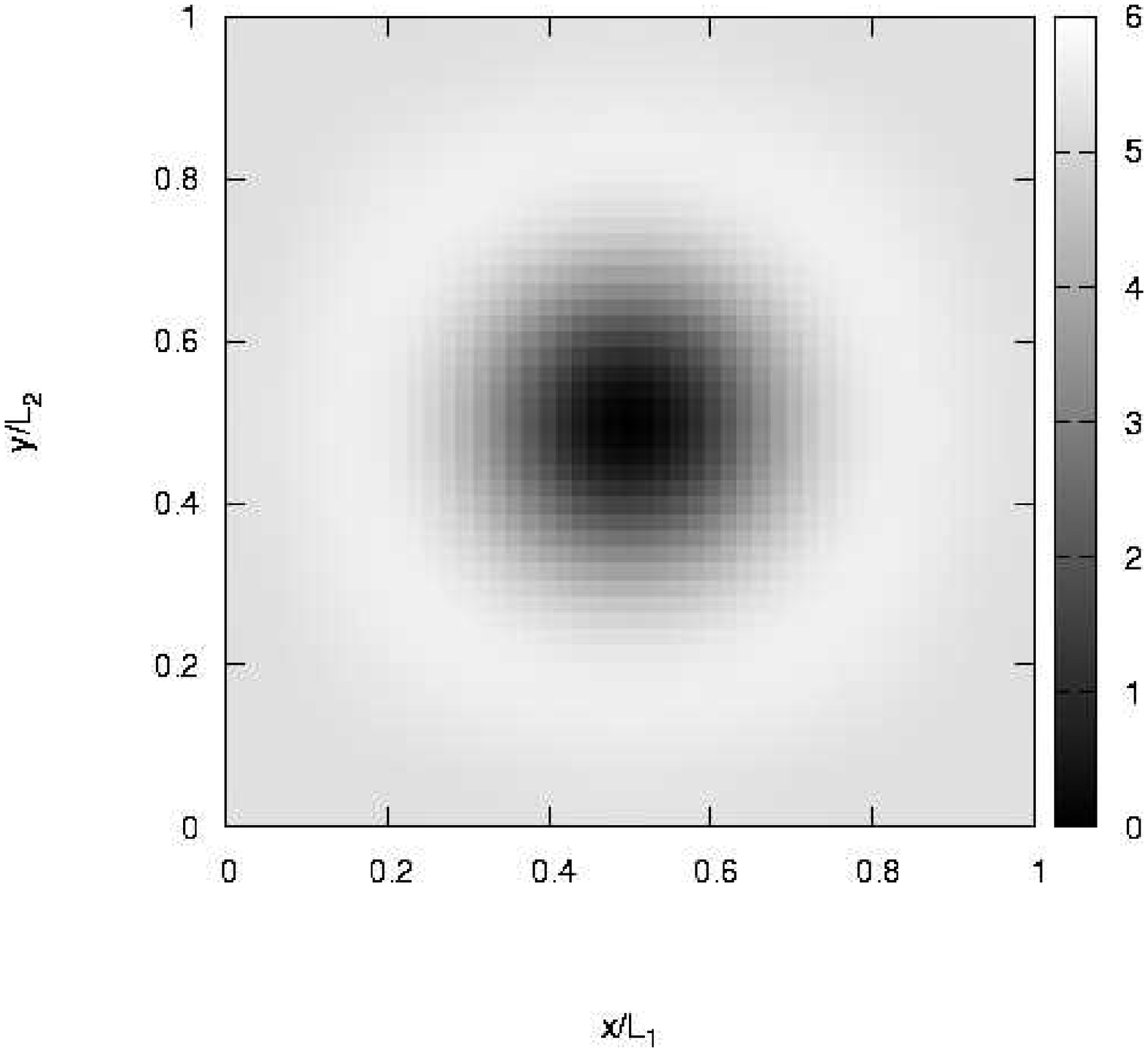} 
  \caption{Electronic density of a quasihole-state created by application of the quasihole creation operator on the ground state of a homogeneous system with $N_e/N_s=5/15$ and hard-core interaction. Left plot: Density after application on one of the three deg. ground states; Right plot: Densitiy obtained by superposition of the three degenerate states.}
  \label{fig:QuasiHoleHom}
\end{figure}
\end{center}

\begin{figure}[htbp]
  \includegraphics[scale=0.6]{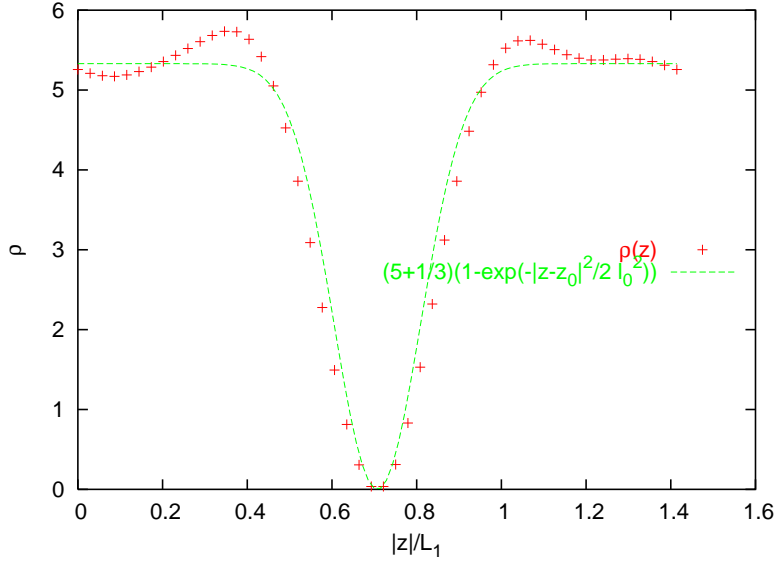}
  \caption{Cut along $x=y$ through the electronic density of the $N_e/N_s=5/16$ system with inserted quasihole. A Gaussian of width $l_0$ covers the area of one flux quantum and fits quite well the profile. For asymmetry see text.}
  \label{fig:FracCharge}
\end{figure}

\begin{figure}[htbp]
  \includegraphics[scale=0.45]{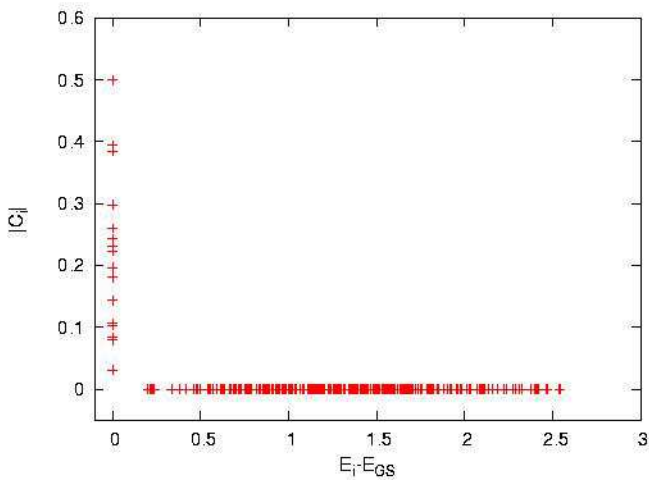}
  \caption{Absolute values of projection coefficients of a quasihole state created by application of the quasihole creation operator on one of the ground states of the $N_e/N_s=5/15$-system by projecting on the eigenstates of a homogeneous $N_e/N_s=5/16$-system.}
  \label{fig:ProjHoleHom}
\end{figure}

Applying the creation operator for a quasihole, deduced in section \ref{sec:QuasiHoles}, to each of the three degenerate ground states of the $N_e/N_s=5/15$ system results in three quasihole states having the hole at the same position. 
The electronic density of one of these states is shown in Fig. \ref{fig:QuasiHoleHom}, left plot. The states are not identical and only after superposition of the three densities a circular symmetric result is achieved (apart from finite size effects, see below), as shown in Fig. \ref{fig:QuasiHoleHom}, right plot.
To get a measure of the charge $q_{qh}$ repulsed by inserting the quasihole, a Gaussian of width $l_0$ is fitted to the density. The area below this Gaussian is $2 \pi l_0^2$ which takes exactly $\frac{1}{N_s}$ of the system's area. In a homogeneous system, this Gaussian would therefore envelop $\frac{1}{3}$ of an electron's charge. Fig. \ref{fig:FracCharge} shows a quite good agreement of the the Gaussian $( 5 + \frac{1}{3} ) (1-\exp(-\frac{|z|^2}{2 l_0^2}))$ with the density profile which is an indicator of a fractional amount of charge $q_{qh} = \frac{1}{3}e$ that was repulsed by insertion of the vortex (compare section \ref{sec:QuasiHoles}). The slight asymmetry encountered in the density profile is probably a finite size effect. In an infinite system the cut must be symmetric.
Still unclear is the stability of this excitation. The projection of the quasihole-state to the eigenstates of the homogeneous $N_e/N_s=5/16$-system gives the answer: The excitation lies completely inside the space of the 16-fold degenerate ground state (see figure \ref{fig:ProjHoleHom}), hence it is an eigenstate itself and a gapless excitation as in the case of the addition of a flux quantum in section \ref{sec:gsmu}.
By application of the quasihole creation operator on each of the three degenerate ground states of the homogeneous $N_e/N_s=5/15$-system, three linearly independent degenerate states with eigenenergy zero and a quasihole at the desired position could be created. Stated differently, the space of quasihole excitations is a three-dimensional subspace of the ground state sector of the $N_e/N_s=5/16$-system. This is true for every point in the unit cell the quasihole is created. 
The vanishing eigenenergies in systems with hard-core interaction are an indicator for Laughlin-like correlations between the electrons. Bearing the construction of the quasihole operator in mind, it is clear that these correlations ---established in the ground state of the homogeneous $\nu=\frac{1}{3}$ system--- are conveyed by the operator to the state containing a quasihole.

\begin{center}
\begin{figure}[htbp]          
  \includegraphics[scale=0.45]{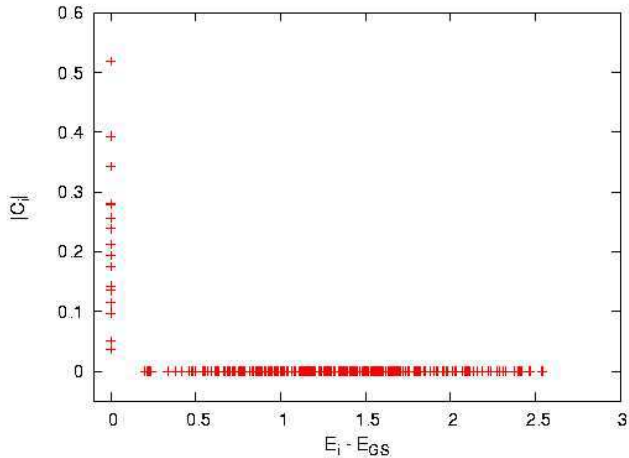}
  \caption{Absolute values of projection coefficients of the quasihole-state created by diagonalizing a $N_e/N_s=5/16$-system with hard-core interaction and a delta potential by projecting on the eigenstates of a homogeneous $N_e/N_s=5/16$-system.}
 \label{fig:ProjHardDeltaHole}
\end{figure}
\end{center}

As described in section \ref{sec:QuasiHolesDelta}, there is an alternative way to insert a quasiholes by pinning a flux quantum through a delta-shaped potential at a fixed position. In the absence of other external potentials and using hard-core interaction this results in a threefold degenerate  ground state with vanishing energy, as expected.
The density of one of these states looks very similar to the result gained earlier by means of the operator (see left plot in Fig. \ref{fig:QuasiHoleHom}).
Projecting this quasihole state to the eigenstates of the homogeneous $N_e/N_s=5/16$-system shows that this excitation lies completely inside the space spanned by the 16-fold degenerate ground state vectors (figure \ref{fig:ProjHardDeltaHole}). Therefore the stability of this excitation is obvious since it is an eigenstate again. This is of course independent of the position where the hole was created.
Another interesting feature of the method using a delta potential to create quasihole excitations are the excited states of such a system.
As already known from Tab. \ref{tab:GShomogeneous}, in a 5/16-system there is a 16-fold degenerate ground state. Placing a delta
potential into such a system, which is chosen to have a much smaller integral energy than the gap of $0.200 \mbox{enu}$,
 should cause a mixing mainly within the ground state sector. Compared to the gap of a homogeneous 5/15-system, the gap of this system is almost 
unaltered, which confirms that there is negligible mixing between states below and above the gap.
The diagonalization with a delta-potential $V_{\mbox{delta}} = 0.01 \mbox{enu} L_1 L_2 \sum_{i=1}^{N_e} \delta(\vec r_i - \vec r_0)$
at the origin ($\vec r_0=0$) yields the spectrum in Fig. \ref{fig:spec5-16delta}. The threefold degenerate ground state with zero energy is confirmed. The 16-fold ground state splits into three different energy bands, two of which are made up out of three, one out of ten almost degenerate states. Evaluating the densities for these states reveals their nature: The 10-fold quasi-degenerate subspace consists partly of states looking like superpositions of quasiholes located at positions that are different from $r_0$. The best example is state 11, whose density is given in Fig. \ref{fig:dnExcitedHole} (right plot). The quasihole seems to be localized mainly at the origin. The slight dip in the density at $r_0$ is a common feature of these 10 states, but most of the other 9 eigenstates have a more complicated density profile. Nevertheless, another sign that supports the interpretation of these states to be quasihole ground states at different positions will be given in section \ref{sec:twodeltas}.
  
A qualitatively different picture is obtained for the three higher states 13,14,15.
The superposition of their densities shows a ring like structure around the delta's position, as found in Fig. \ref{fig:dnExcitedHole} (left).
This looks very much like an excited state of the quasihole. The inner radius of this ring is about $2 l_0$, the outer one about $3 l_0$.
\begin{center}
\begin{figure}[htbp]          
  \includegraphics[scale=0.45]{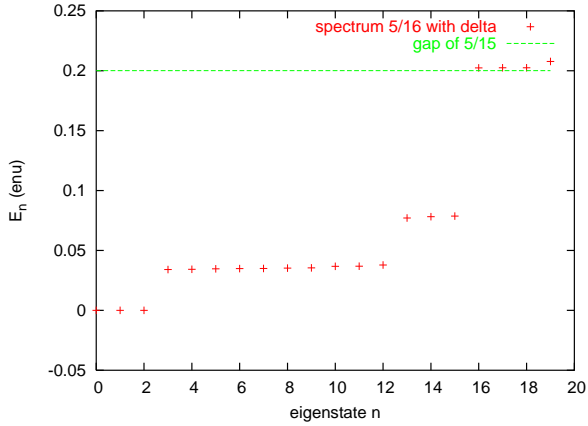}
  \caption{Lowest 20 eigenenergies of the 5/16-system with hardcore interaction and a delta potential at $\vec r_0 = 0$. The gap of a homogeneous 5/15-system is given for comparison.}
  \label{fig:spec5-16delta}
\end{figure}
\begin{figure}[htbp]                 
  \includegraphics[scale=0.4]{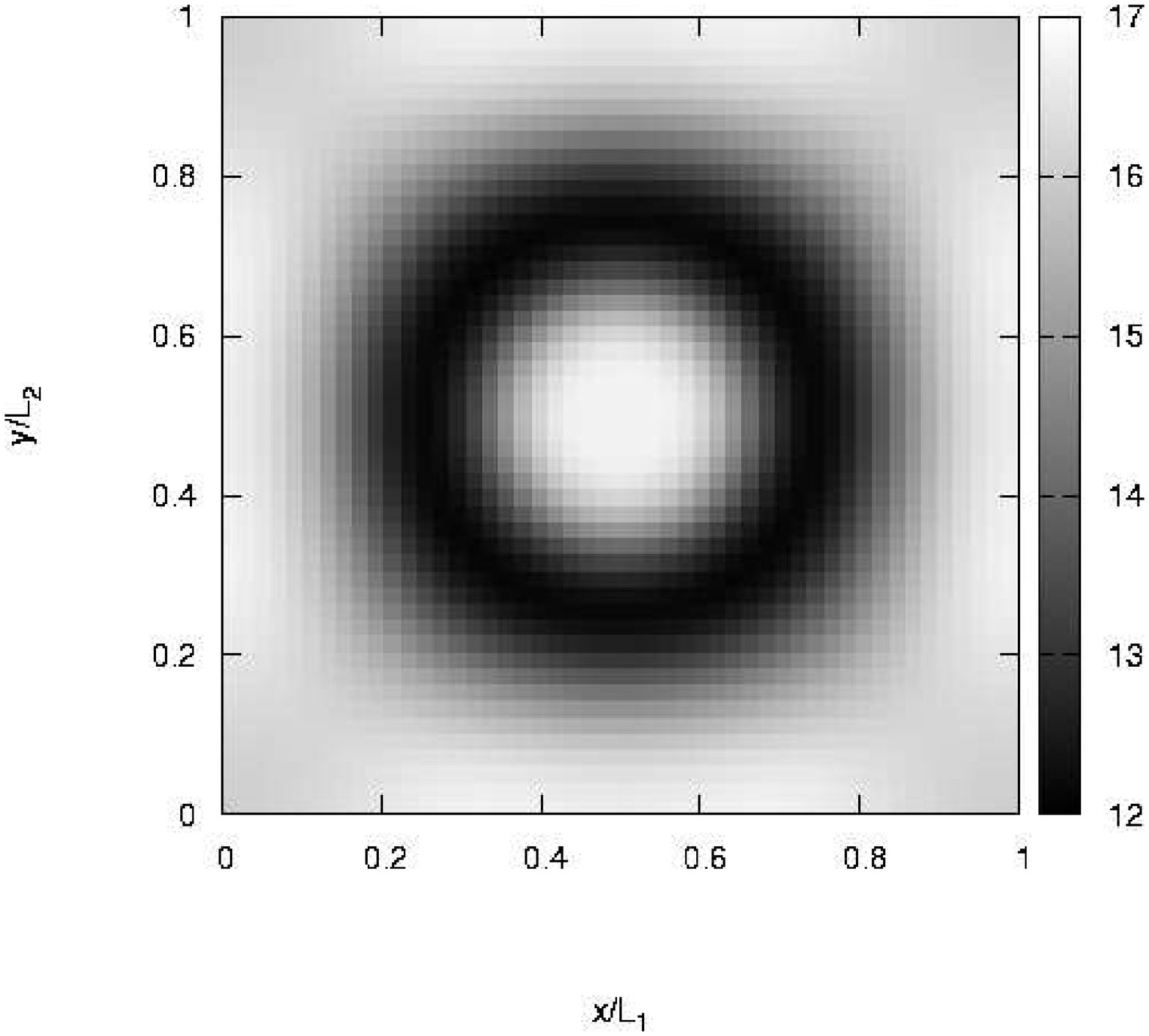}\includegraphics[scale=0.4]{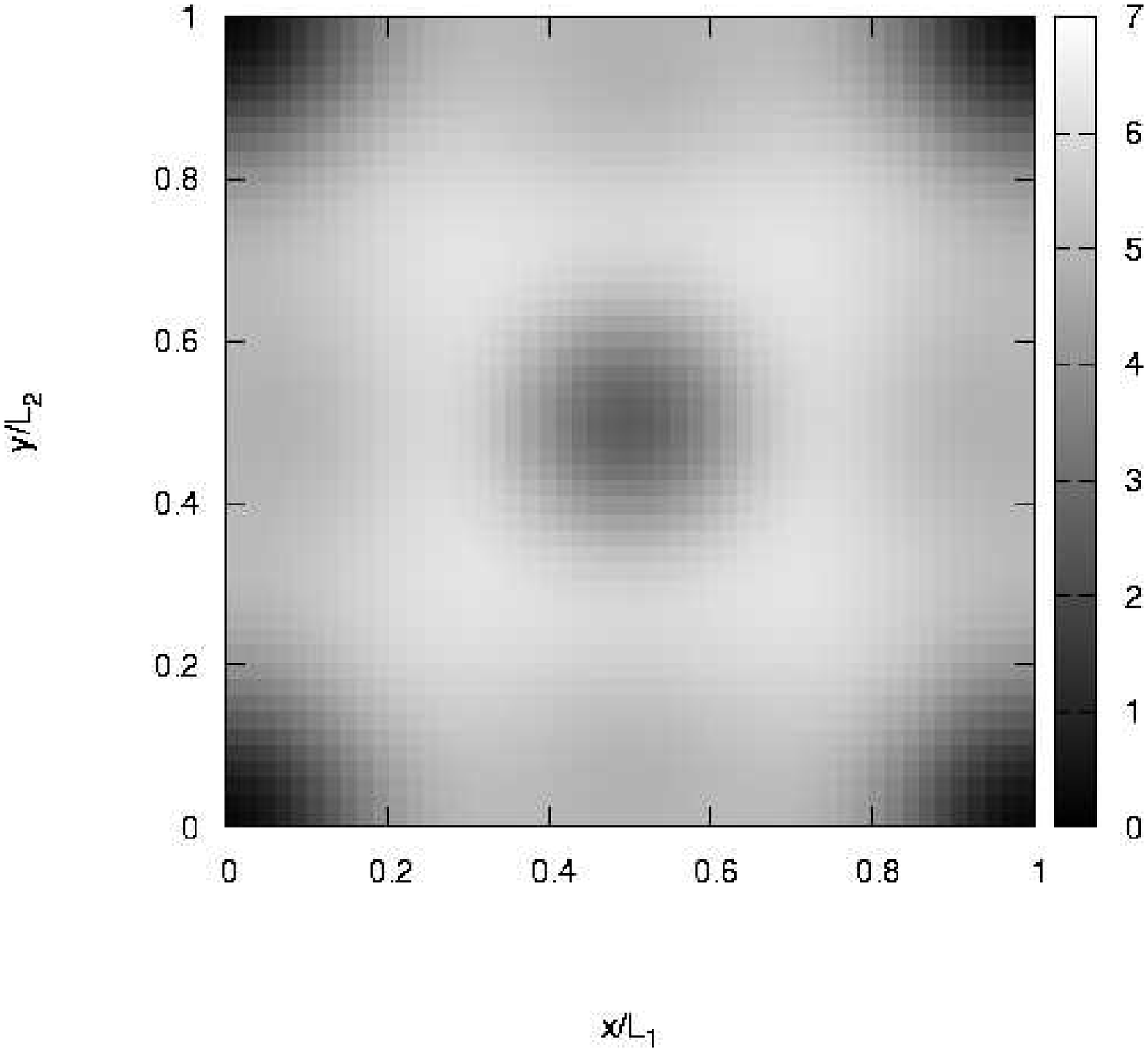}
  \caption{Left: Superposition of the densities of the threefold quasi-degenerate states 13,14,15 in a 5/16-system with a delta potential at $\vec r_0=(0.5,0.5)$. Right: Density of eigenstate 11 of the same system. The magnetic length is about $l_0 \simeq \frac{L_1}{10}$.}
  \label{fig:dnExcitedHole}
\end{figure}
\end{center}
\subsubsection{Quasiholes in homogeneous systems with Coulomb interaction}
\label{sec:QHcoulomb}
\begin{figure}[htbp]
\vspace{-1cm}
  \includegraphics[scale=0.4]{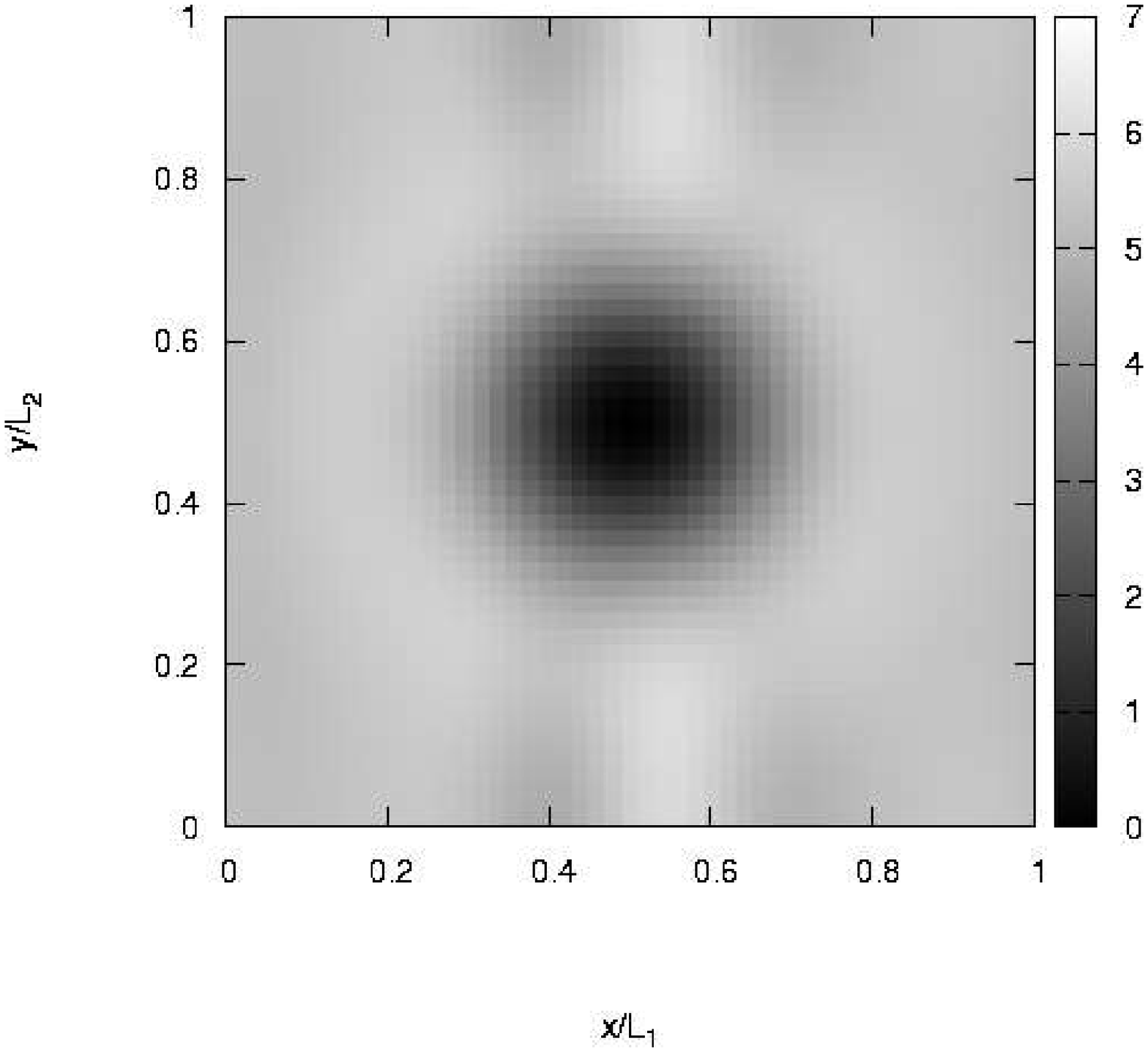} \includegraphics[scale=0.4]{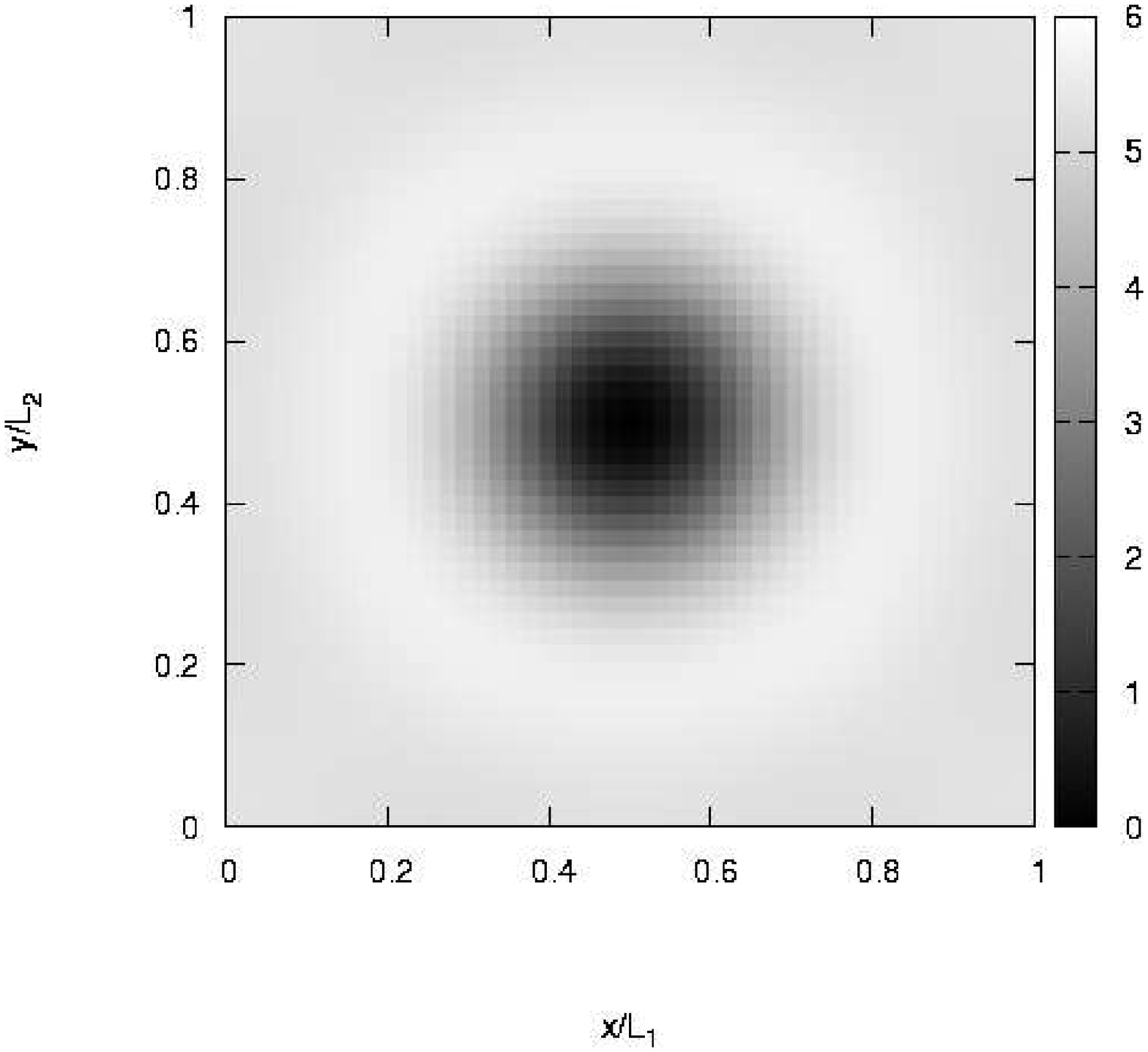}
\vspace{-1cm}
  \caption{Density of a quasihole in a 5/16-system with Coulomb interaction created by application of the creation operator to a ground state of the homogeneous 5/15-system. Left plot: One single state; Right: Superposition of the densities of three states obtained from the tree deg. ground states.}
  \label{fig:DensityHoleHomCoulombOperator}
\end{figure}

\begin{figure}[htbp]
  \includegraphics[scale=0.4]{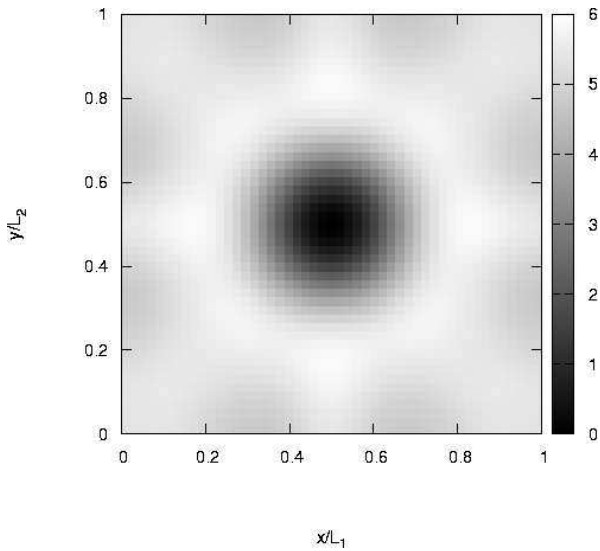}
  \caption{Density of a quasihole in a 5/16-system with Coulomb interaction and a delta potential obtained as the ground state by diagonalization.}
  \label{fig:DensityHoleHomCoulombDelta}
\end{figure}

Again a system of 5 electrons is chosen, the magnetic field causes a flux of 15 flux quanta through the unit cell, but the electron interact via the Coulomb potential.
The same quantities as in the case of hard-core interaction are investigated. Beginning with the density, an obvious difference between 
the quasihole created by delta potentials in contrast to those obtained by application of the operator (compare Fig. \ref{fig:DensityHoleHomCoulombOperator} (left) and Fig. \ref{fig:DensityHoleHomCoulombDelta}), is, that the former ones show a four fold symmetry axis.
\begin{figure}[htbp]
  \includegraphics[scale=0.45]{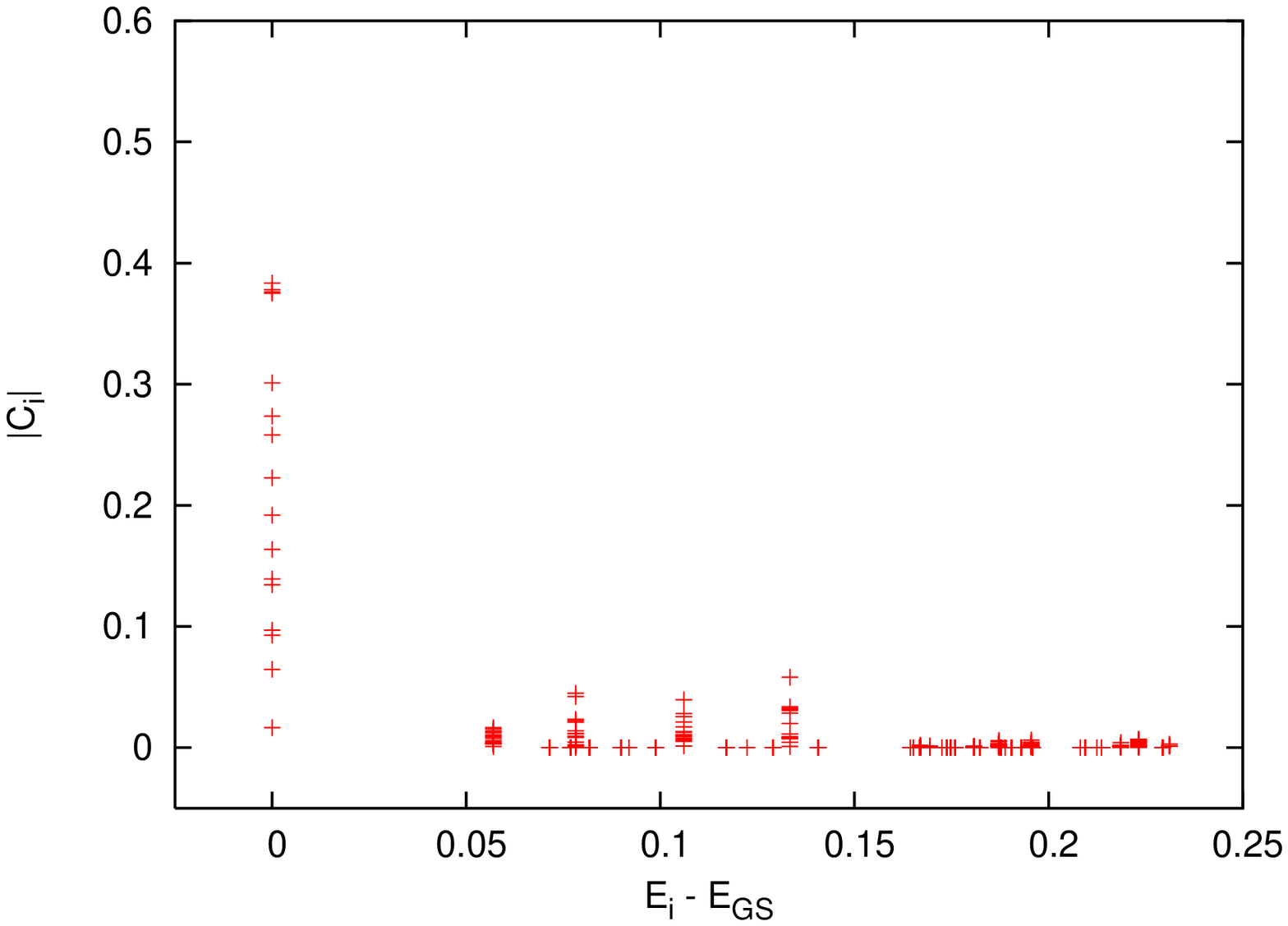}\includegraphics[scale=0.45]{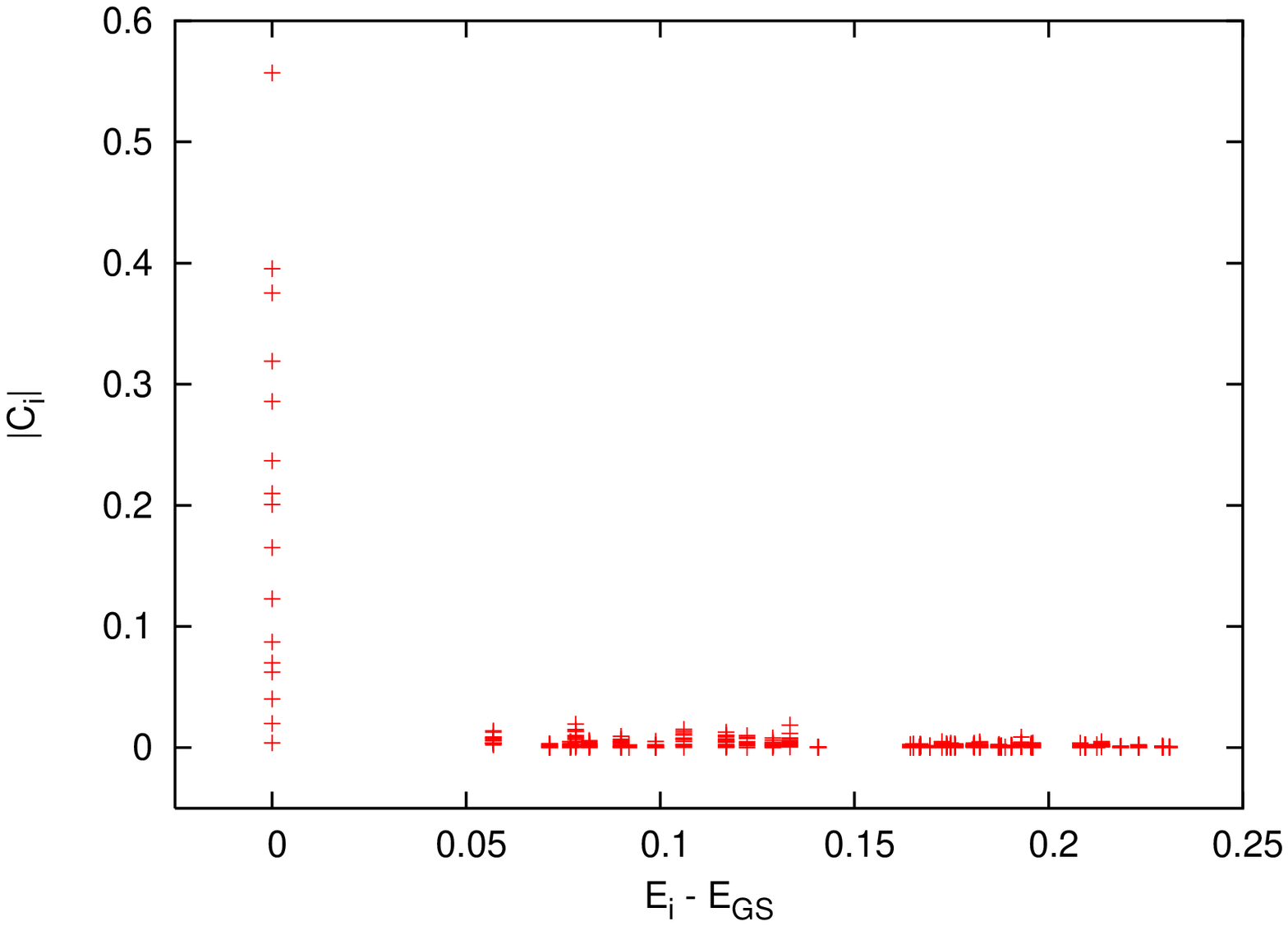}
  \caption{Absolute values of projection coefficients for a quasihole-state with Coulomb interaction obtained by projecting on the eigenstates of a homogeneous 5/16-system. Left plot: Quasihole is generated by the creation operator. The excitation energy is $0.0023975 \mbox{enu}$ above the GS.
Right plot: Quasihole obtained as ground state from diagonalization with a delta potential. Its excitation energy is $0.00079831 \mbox{enu}$ above the GS.}
  \label{fig:ProjHoleHomCoulomb}
\end{figure}
Another way to compare the both methods of creating a quasihole is by looking at their spectral decomposition. The states are projected onto the eigenstates of a homogeneous 5/16-system.
This reveals another difference: In the case of Coulomb interaction the quasihole generated by a delta potential has smaller contributions of the excited states than the one generated by the operator. This can be attributed to the fact that the operator was derived from Laughlin's trial wavefunction, which is exact for hard-core interaction, but which in turn is only a good approximation in the case of Coulomb interaction.
Aside from the projection coefficients $|C_i|$ themselves, the energy of the quasihole states can be computed by using Equ. (\ref{eq:ExcEn}) and the known eigenenergies $E_i$ of the homogeneous system
\begin{equation}
E = \sum_i |C_i|^2 E_i .
\label{eq:ExcEn}
\end{equation}
Tab. \ref{tab:EnergiesCoulombHomHole} shows the results for both --- Coulomb and hard-core interaction.
Obviously, quasihole states are no longer eigenstates of the system when Coulomb interaction is in use. Furthermore,
the delta potential created quasiholes have excitation energies which are lower by approximately a factor of three compared to those obtained by means of the operator.
\begin{table}[htbp]
\begin{tabular}{|l|l|l|}
\hline
System & Method & $\Delta$ $(\frac{e^2}{\epsilon l_0})$\\
\hline
5/15 C & operator & 0.0023975\\
5/15 C & delta potential & 0.0007983\\
5/15 H & operator & 0.0\\
5/15 H & delta potential & 0.0\\
\hline
\end{tabular}
\caption{Comparison of the excitation energy (above GS) of a quasihole created by application of the operator to a homogeneous 5/15-systems's ground state and by diagonalizing a 5/16-system with a delta potential for hard-core (H) and for Coulomb (C) interaction.}
\label{tab:EnergiesCoulombHomHole}
\end{table}
The stability of a hole excitation in the system with Coulomb interaction is not obvious in advance.
To clarify this, the time evolution of the occupation numbers $n_j(t)$ is surveyed. This quantity can be calculated according to Equ. (\ref{eq:OccNumTDev}), once the spectral decomposition of the quasihole-state into eigenstates of the homogeneous 5/16-system and thus its time evolution $\Psi(t)$ is known,
\begin{equation}
n_j(t) = \langle \Psi(t)|a^+_j a_j|\Psi(t) \rangle.
\label{eq:OccNumTDev}
\end{equation}
\begin{figure}[htbp]
  \includegraphics[scale=0.4]{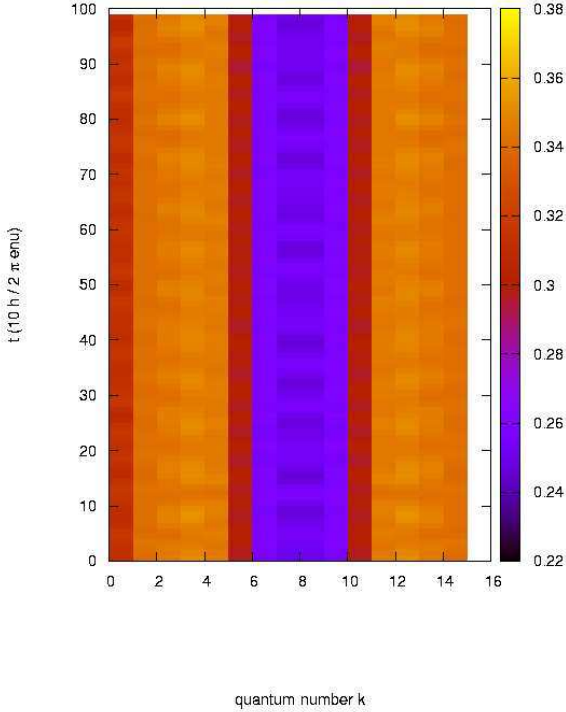}
  \caption{Time evolution of the occupation numbers $n_j(t)$ for the quasihole obtained by diagonalization with a delta potential at $\vec r_0 = (0.5,0.5)$ and Coulomb interaction.}
  \label{fig:LandauMatrixHomCoul}
\end{figure}
Fig. \ref{fig:LandauMatrixHomCoul} shows the time dependence of $n_j(t)$. The color code indicates the occupation probability of the single particle state $j$ ($x$-axis) at time $t$ ($y$-axis).
In a homogeneous 5/15-system the occupation of a single particle state is equal $\frac{1}{3}$.
In contrast in Fig. \ref{fig:LandauMatrixHomCoul} the states around $j = 6 \ldots 9$ have a lower occupation probability due to the inserted quasihole. This dip in the occupation is nearly constant in time and only shows some minor oscillations. 
Since the quantum number $j$ is coupled to the point $X_j = \frac{j}{L_1 N_s}$ around which the single electron state is localized, we can conclude from the constance of $n_j(t)$ that the hole does not move in $x$-direction. As both axes - $x$ and $y$ - of the unit cell in this particular system differ only by the selected gauge, we can infer that there can neither be a motion of the quasihole in $y$-direction. Instead it rests in the middle of the cell where it was created and shows some breathing mode like oscillations. These can be observed by calculating the density as a function of time.
\subsection{Conclusions on homogeneous systems}
Laughlin's trial wavefunctions for the ground state and quasihole-excited state in a homogeneous system at $\nu=\frac{1}{3}$ were introduced
and shown to be eigenfunctions for a special short-range interaction.

In a homogeneous system known results were reproduced thereby verifying our calculations. This yields eigenenergies from which the chemical potential $\mu$ for electrons is calculated and the incompressibility of the system is confirmed for both --- Coulomb and short-ranged --- interaction. The correlations were found to be identical to those of Laughlin's wavefunction and the short-range interaction energy $\langle V_{\mbox{short-range}} \rangle_{GS}$ is understood to be a suitable indicator for this.
Based on the trial wavefunction of a quasihole excitation adapted to our geometry, a quasihole creation operator could be derived.
Its validity was verified by looking at the properties of the excitation it produces upon application to the ground state of a homogeneous $\nu=\frac{1}{3}$ system: The excitation carries a fractional charge of $\frac{1}{3}e$ and lies in the ground state sector (as expected) for short-ranged interaction, thus having excitation energy $0$. For Coulomb interaction a finite excitation energy is found.

An alternative method of creating a quasihole is by fixing one vortex of the many-particle wavefunction in a system with one flux quantum in excess ($N_s = 3 N_e + 1$); this was possible by diagonalizing this system with a delta potential to pin one zero of the wavefunction.
Comparing the ground state obtained this way to the excitation created by the quasihole-operator reveals both approaches to be equivalent for short-ranged interaction.
In the case of Coulomb interaction an energetically lower (factor of $4$) quasihole excitation is obtained by the second procedure.
Small contributions above the ground state of the homogeneous system to this state cause a breathing mode oscillation, but the quasihole is still stable.

Another interesting feature of a delta potential in a system with $N_s = 3 N_e + 1$ flux quanta are localized quasihole states at the potential's position. These will be of use in the following section to create a setup where tunneling can be observed.
Furthermore, excited states of the quasihole at the delta potential with a ring-like structure were found.
To some extent, results from section \ref{sec:tunnel} indicate these states to be qualitatively different from the lower lying ones.
Analyzing them more thoroughly could be the matter of further work.

\newpage
\section{Tunneling of quasiholes between delta potentials}
\label{sec:three}
\label{sec:tunnel}
One aim of this work is to investigate the ability of quasiholes to tunnel through a constriction in the system.
Before turning to this more complex problem, a much simpler case will be investigated here. The question is whether there is evidence
for quasihole tunneling in our system at all. A setup that is helpful to shed light on this was already found in
section \ref{sec:QuasiHolesHard}: A system of $N_e$ electrons interacting via hard-core interaction, $N_s = 3 N_e + 1$ flux quanta and a delta potential. The ``additional'' flux quantum was found to have a threefold degenerate localized state at the delta potential's position.
Introducing into this system a second delta potential at a different position should result in tunneling of the quasihole between the localized states at the one delta and the other. 
Analogous to the picture of a single particle tunneling through a potential barrier, the role of the barrier is played here by the space between the delta potentials where the quasihole has to pay much energy to live. 
If tunneling occurs, symmetric or antisymmetric linear combinations made up out of the localized states should be observed to form the ground state and lowest excited states, respectively.

The idea behind this construction is to assume the system of $N_e$ electrons and $N_s = 3 N_e + 1$ flux quanta to be effectively treatable as a single-quasihole system. One has however to remember that this single quasihole lives on the background of the homogeneous system's fractional quantum Hall state. This state can be thought of as the vacuum with respect to quasiholes.
Returning to the single quasiparticle picture, let us denote the $l$-th localized state of the quasihole at the delta potential $i$ as $\ket{l}_i$, where $i=1,2$ and $l=1,2,3$. Of course, in a system with two delta potentials these states are no longer eigenstates. Instead we use them
as a basis and try to obtain a reasonable estimate for the ground state and lowest excited states within this subspace.
The restriction to this space can be justified by perturbation theory, since due to the symmetry of the problem with respect to exchanging the delta potentials, the states $\ket{l}_i$ must be degenerate. They have some energy $\epsilon = \bra{l}_i H \ket{l}_i$ for $i=1,2$ and $l=1,2,3$, $H$ being the Hamiltonian of the system with both delta peaks. The mixing due to tunneling must therefore be most dominant for these states and we will
take it into account by a Bardeen-type tunneling Hamiltonian $H_t = \sum_{l, l^\prime} t_{l,l^\prime} \ket{l}_1 {}_2\bra{l^\prime} + t^*_{l,l^\prime} \ket{l^\prime}_2 {}_1\bra{l}$.

Now, there needs to be an assumption to set up the tunneling coefficients $t_{l,l'}$.
To render a connection between the quantum number $l$ and some physical observable, one has to keep in mind that the threefold degeneracy of the quasihole state emerged from the (threefold) degeneracy of the ground state of the homogeneous system at $\nu = \frac{1}{3}$. The latter degeneracy originated from conservation of the center-of-mass momentum $P_y$.
Returning to the system with a quasihole and one delta potential, there are three orthogonal quasihole states at every point $z_0$ in the unit cell.
So there must be another operator --- call it $G(z_0)$ (like the center of mass $P_y$ in the homogeneous case) --- that commutes with $H$ and whose eigenstates discriminate three orthogonal eigenvectors within this subspace.
If $G$ is independent of the point $z_0$, we will be able to separate the quasihole state sector into three eigenspaces of $G$ and label them by $l=1,2,3$. Of course, it is possible that $G(z_0)$ depends on the position $z_0$ of the quasihole.
But however it is tempting to assume that it does not.
Then it would be reasonable to suppose that the tunneling coefficients $t_{l,l'}$ only depend on whether the two coupled states $\ket{l}_1$ and $\ket{l^\prime}_2$ lie in the same eigenspace of $G$, $l=l'$, or if they lie in different ones $l \neq l'$.
This assumption yields tunneling coefficients $t_{l,l'} = \delta_{l,l'}t + (1-\delta_{l,l'}) t_1 \quad t,t_1 \in \C$

The Hamiltonian $H = \epsilon + H_t$ can be written in the basis $b$ of the direct sum of the ground state spaces of a quasihole at the one delta or the other:
$b = \{ \ket{1}_1,  \ket{2}_1,  \ket{3}_1,  \ket{1}_2,  \ket{2}_2,  \ket{3}_2 \}$, where the states are labeled by the
eigenstates of the suppositional operator G.
Hence the Hamiltonian has the following matrix form
\begin{equation}
H = \left(
\begin{matrix}
 \epsilon & 0 & 0 & t & t_1 & t_1\\
  0 & \epsilon & 0 & t_1 & t & t_1\\
  0 & 0 & \epsilon & t_1 & t_1 & t\\
  t^* & t_1^* & t_1^* & \epsilon & 0 & 0\\
  t_1^* & t^* & t_1^* & 0 & \epsilon & 0\\
  t_1^* & t_1^* & t^* & 0 & 0 & \epsilon\\      
\end{matrix}\right).
\label{eq:HamTunnel}
\end{equation}
Its eigenspectrum is
\begin{eqnarray}
eig (H) &=& \{ \epsilon - \lambda, \epsilon - \kappa, \epsilon - \kappa, \epsilon + \kappa, \epsilon + \kappa, \epsilon + \lambda \}\\ \nonumber
\lambda &=& \sqrt{2 |t_1|^2 - |t|^2 + 2|t+t_1|^2 } \\ \nonumber
\kappa &=& |t-t_1|\\ \nonumber
\label{eq:HamTunnelEig}
\end{eqnarray}
If $|t+t_1| > |t|$, then $\lambda > \kappa$ and the spectrum has the structure
consisting of one lowest eigenstate $\epsilon-\lambda$ followed by a two-fold eigenstate $\epsilon-\kappa$, another two-fold $\epsilon+\kappa$ and
a highest one with $\epsilon+\lambda$. This structure will be recognized in the numerically obtained spectrum in the next section.
\subsection{System with two delta potentials and one flux quantum in excess}
\label{sec:twodeltas}
\begin{figure}
\includegraphics[scale=0.5]{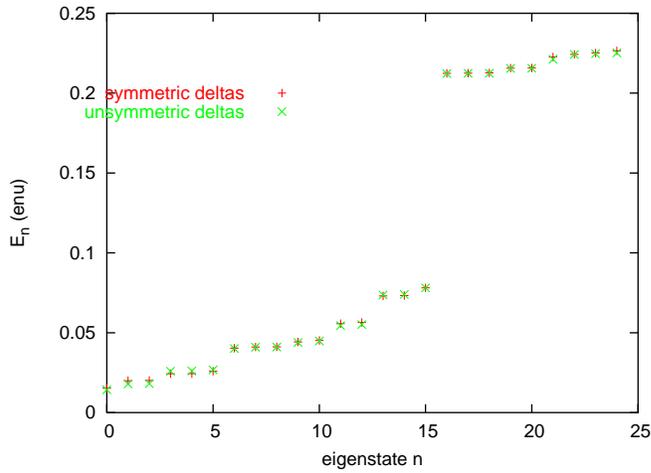}
\caption{Spectrum of a 5/16-system with two delta potentials $0.5 L_1$ apart;
red plot: Both delta potentials have a prefactor of $0.005 L_1 L_2 enu$; green plot: The prefactor is $0.004$ and $0.006 L_1 L_2 enu$ respectively.}
\label{fig:Spec2Delta}
\end{figure}

\begin{figure}
\includegraphics[scale=0.4]{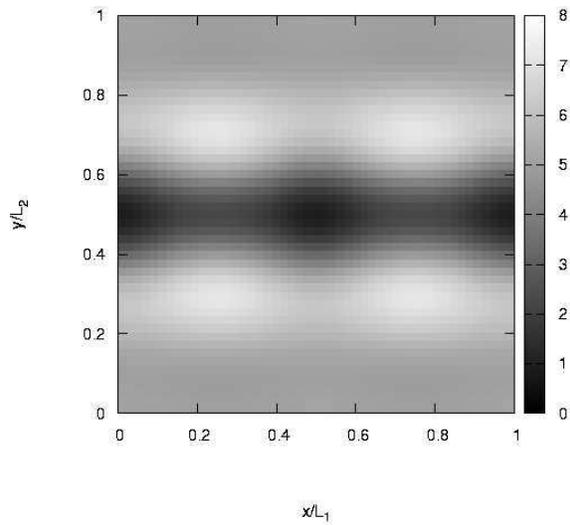}
\caption{Density of a 5/16-system with two delta potentials (strength $0.005 L_1 L_2 \mbox{enu}$) being $0.5 L_1$ apart. The two dips in the density (location of the potentials) are connected by a ``trench'' of low density.}
\label{fig:dn2delta0.5}
\end{figure}

\begin{figure}
\includegraphics[scale=0.5]{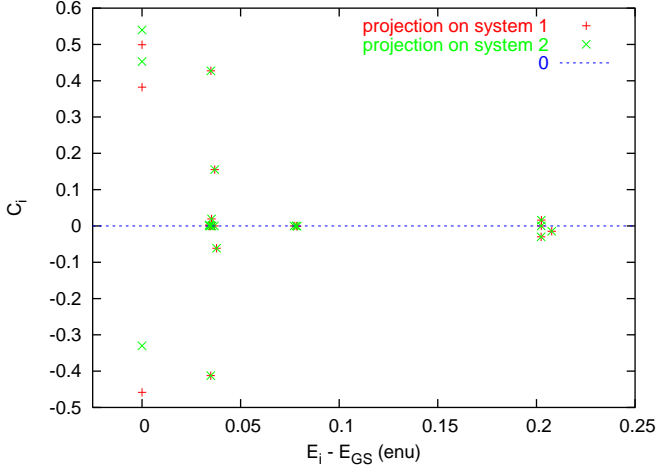}
\caption{Projection coefficients for the state with two delta potentials onto the eigenstates of the system with only one delta potential at the left or the right position. The distance between the potentials is $0.5 L_1$. The gap of the homogeneous system is $0.2 \mbox{enu}$.}
\label{fig:ProjBoth}
\end{figure}

\begin{figure}
\includegraphics[scale=0.5]{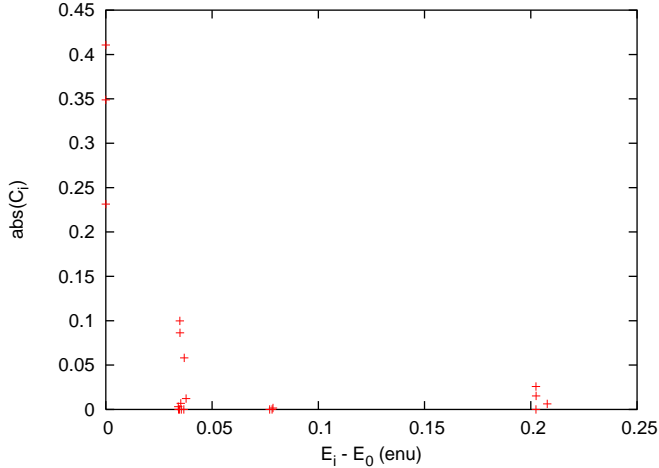}
\caption{The contributions of excited states of system 1 to the ground state projected onto the eigenstates of system 2. A high overlap with the ground state is found. The states 14,15,16 around $E_i-E_{GS} \simeq 0.76$ do not hardly contribute.}
\label{fig:ProjExc1on2}
\end{figure}

In a system of 5 electrons and 16 flux quanta there is a 16-fold degenerate ground state in the homogeneous system, as seen from Tab. \ref{tab:GShomogeneous}. Two delta potentials shall be introduced into this system. Their integral potential energy is chosen very small, such that they will not cause strong mixing between states below and above the gap. Introducing only one delta of height $0.01 L_1 L_2 enu$ produces the spectrum shown in Fig. \ref{fig:spec5-16delta} of section \ref{sec:QuasiHolesHard}. The splitting of the 16-fold degenerate ground state is below half of the gap's size. If we are to introduce the second delta and want the
mixing above the gap to be negligible, the amplitude of both deltas has to be yet a bit smaller to be on the save side. Here we chose a height of $0.005 enu L_1 L_2$ for
both deltas at the positions $(0.0,0.5)$ and $(0.5,0.5)$ in the unit cell.
Diagonalization yields the spectrum seen in Fig. \ref{fig:Spec2Delta}, red plot. The ground state sector exhibits a splitting, where the lowest six energies form a band that is supposed to originate from the coupling of the two times three degenerate ground states for the quasiholes localized at the two deltas, as described before. These states should therefore be symmetric or antisymmetric superpositions of eigenstates
for the quasihole at the left or the right delta. The structure of this spectrum is the same as expected from the simple model (Equ. \ref{eq:HamTunnelEig}).
A single ground state, two pairs of two fold degenerate states and a single highest state.
The density of the ground state is given in Fig. \ref{fig:dn2delta0.5}. It shows two dips at the positions of the delta potentials which in turn are located in a ``trench'' of low density connecting the two potential peaks. This picture suggests, that at this distance ($\simeq 5 l_0$) between the delta potentials, the delta peaks still can not be treated as decoupled. Therefore a splitting of the degenerate states is already visible at this distance, as seen from the spectrum in Fig. \ref{fig:Spec2Delta}, red plot.
To determine if this state $\ket{\Psi}$ is a superposition of a quasihole ground state located left and one located right, the projection coefficients
of $\ket{\Psi}$ onto the eigenstates of a system with one delta at $(0.0,0.5)$ (called system 1 in the following) and onto those of a system with a delta potential at $(0.5,0.5)$ (called system 2 in the following) are calculated.
Due to real vectors (for this special choice of positions), the coefficients in Fig. \ref{fig:ProjBoth} are real, too.
A first point to make is that contributions of states above the gap ($0.2 \mbox{enu}$) are small.
If the ground state $\ket{\Psi}$ of the system with both deltas would be a symmetric or antisymmetric linear combination of the ground states of the systems with a single delta and if the eigenspaces of localized quasihole states at different positions were orthogonal, one would expect half of the state to lie in the ground state sector of system 1, the other half to lie in the ground state space of system 2.
Unfortunately these spaces are not orthogonal. This non-orthogonality results from the fact that a quasihole excitation
created at an arbitrary positions in the unit cell always lies in the 16-fold degenerate ground state space of the homogeneous system.
Due to the symmetry upon exchanging the two delta potentials, the projection coefficients however must have the same magnitude for system 1 and system 2 (if the corresponding state on which is projected is non-degenerate).
To take into account the overlap between the threefold degenerate ground state spaces of system 1 and system 2, the
amount of the wavefunction that resides in the product space can be calculated as follows.
First the projection of $\ket{\Psi}$ onto the eigenspace of system 1 is performed. The result is given in Fig. \ref{fig:ProjBoth}. The part of the wavefunction that is projected by $P_{GS1}$ into the ground state sector will be called $\ket{\Psi_{GS1}} = P_{GS1} \ket{\Psi}$. It makes up $60.5$ per cent of the state. The ``rest'' of the state not lying within this space, $(1-P_{GS1}) \Psi$, is of course orthogonal to the ground state space 1. It will be projected in a second step onto the eigenspace of system 2. The resulting absolute values are given in Fig. \ref{fig:ProjExc1on2}. A big part $\ket{\Psi_{GS2 \perp GS1}} = P_{GS2}(1-P_{GS1}) \ket{\Psi}$ lies in the ground state space 2. It amounts to $34.5$ per cent of the whole state. The total amount of $\ket{\Psi}$ lying in the sector spanned by both ground state eigenspaces of system 1 and system 2, respectively, is just given by the squared norm  $N^2 = \braket{\Psi_{GS1}}{\Psi_{GS1}} + \braket{\Psi_{GS2 \perp GS1}}{\Psi_{GS2 \perp GS1}}$ and results in $94.4$ per cent. Thus the assumption for the state to be mainly an equally weighted linear combination of the ground states of the quasihole localized at the left or at the right delta potential is confirmed.

Another interesting finding is that there are nearly no contributions from the states $14,15,16$ around $E_i-E_{GS} = 0.76 \mbox{enu}$ in Fig. \ref{fig:ProjBoth}. This is another indication that these three states are qualitatively different from the ten states around $E_i-E_{GS} \simeq 0.035 \mbox{enu}$, as already proposed in section \ref{sec:QuasiHolesHard}.   
%
%
%
%

To determine whether the ground state is a symmetric or antisymmetric linear combination of the ground states of system 1 and system 2,
looking at the phases of the projection coefficients cannot give an answer to this question. The reason is simply that the phase of an eigenvector $\ket{l}_1$ in system 1 is arbitrary. The phase --- or the sign in our case of real vectors --- of the coefficients in Fig. \ref{fig:ProjBoth} thus depends on the arbitrary choice of the respective eigenvector's phase. The only meaningful definition of a symmetric or antisymmetric wavefunction must take into account the form of the contributing eigenstates of system 1 and system 2 in x-space. It has to be checked if the superposition is an odd or an even function with respect to interchanging the delta potentials.
For this purpose Fig. \ref{fig:dn2delta0.5}, which is the electronic density can be regarded as the density of the quasihole wavefunction. To get a quasihole-density from this plot, one has to subtract from the mean electronic density (5 in the units used here) this profile: The minima in Fig. \ref{fig:dn2delta0.5} become maxima and vice versa (inversion of the gray scale). What was called ``trench'' earlier now becomes a ``ridge''. This shape of the wavefunction without a node between its two maxima one would call a symmetric linear combination or a binding orbital in atomic physics. This qualitative shape of the quasihole wavefunction is more easily recognizable in Fig. \ref{fig:dncut}, where a cut through the density along the $x$-axis is shown.

\subsection{Distance dependence of the tunneling}
\label{sec:DistanceDepTunnel}
\begin{figure}
\includegraphics[scale=0.5]{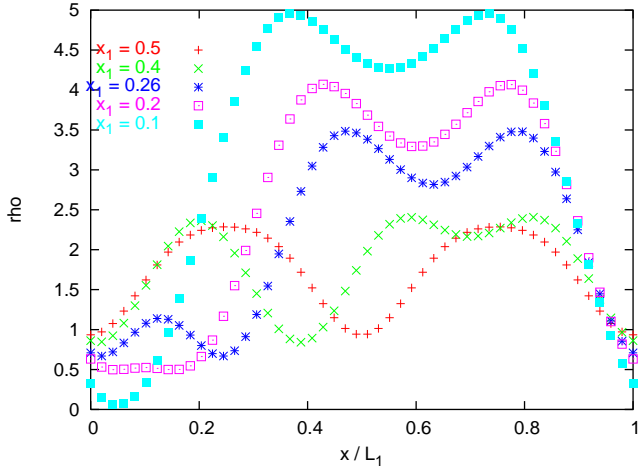}
\caption{Cut through the density along the line between two delta potentials in a 5/16-system for various positions $x_1$ of the second delta. The first delta is fixed at $x_0 = 0$. A qualitatively change occurs below $x_1 \simeq 0.2 L_1 \simeq 2 l_0$ where the two dips merge into one.}
\label{fig:dncut}
\end{figure}

\begin{figure}
\includegraphics[scale=0.45]{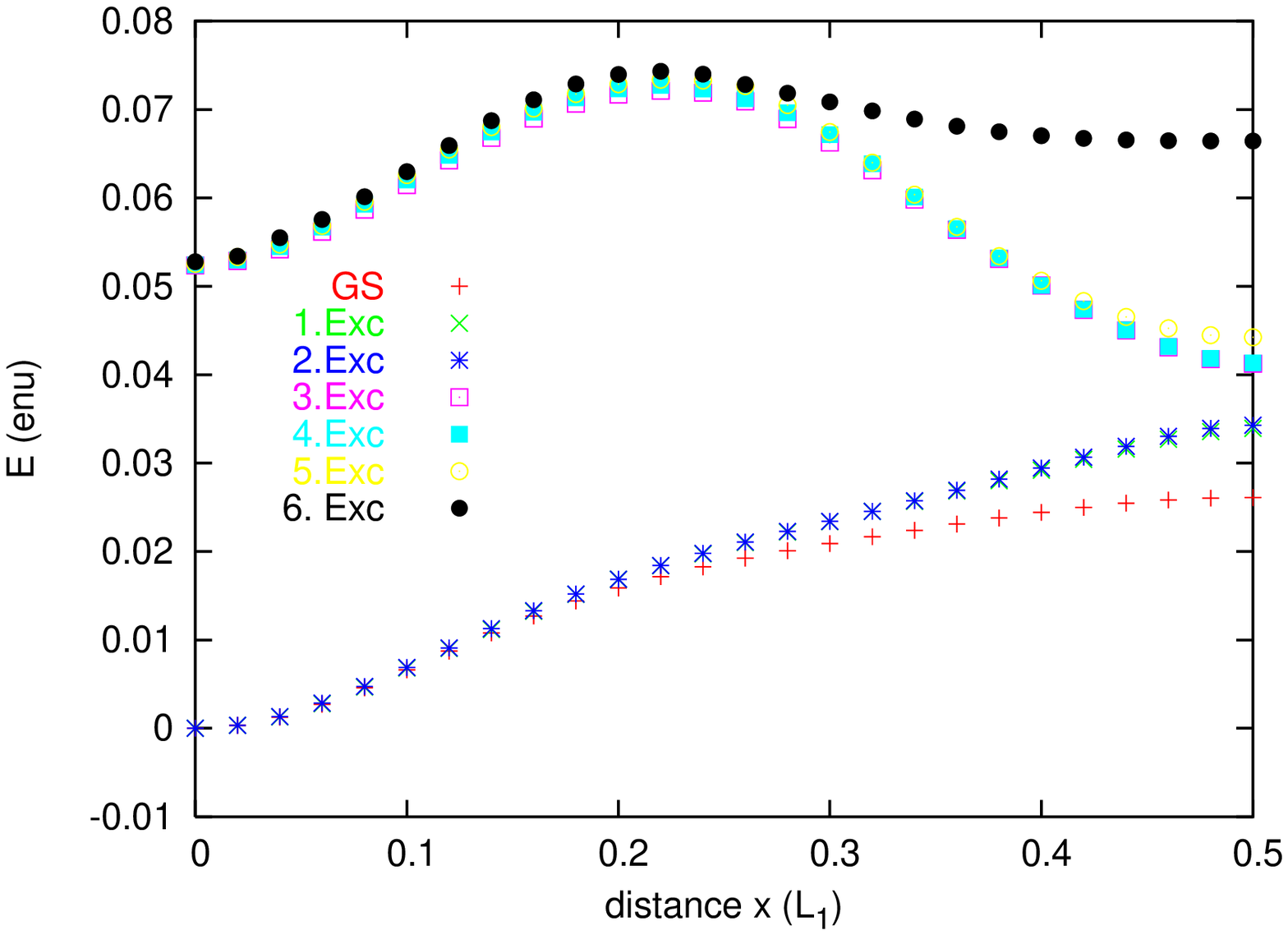}\includegraphics[scale=0.45]{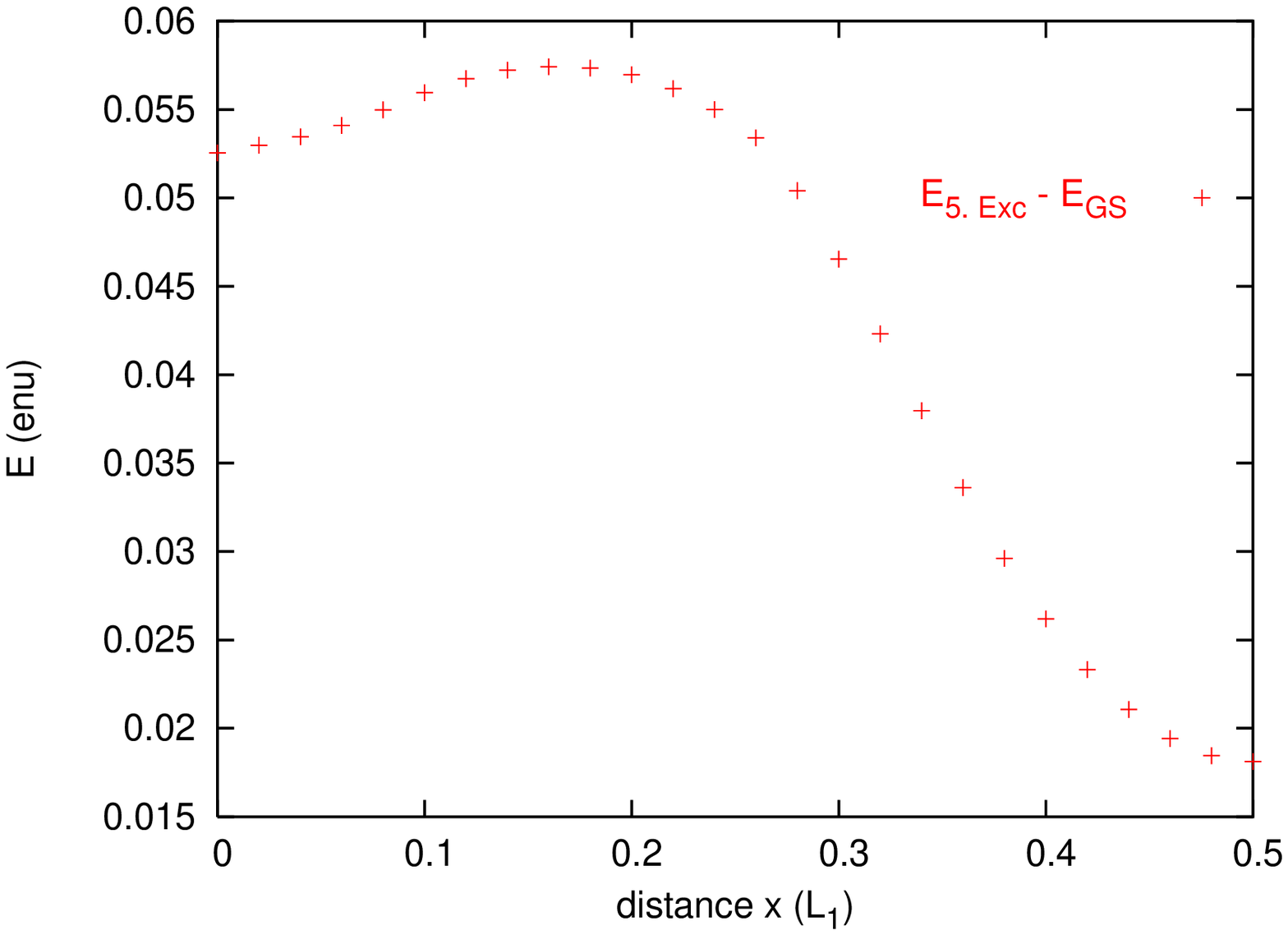}
\caption{Spectrum of a 5/16-system depending on the spacing between two delta potentials (strength = $0.005 L_1 L_2 enu$ each). At $x=0$ the two delta potentials effectively act as a single one. Left: lowest 7 eigenstates. Right: Splitting between the 5. exc. and the ground state.
A qualitative change for $x < 0.2 l_0$ can be observed.}
\label{fig:TunnelSplit}
\label{fig:sp2deltas}
\end{figure}

For smaller distances between the delta peaks the coupling due to tunneling is expected to increase because the width of the ``barrier'' decreases.
This dependence will be investigated here. In Fig. \ref{fig:dncut} a cut through the density 
parallel to the x-axis at $y = 0.5 L_2$ is plotted for different positions $x_1$ of the second delta peak. The first one is kept fixed at $(0.0,0.5)$. For distances $\ge 0.26$ the plots reproduce the two dips at the respective positions of the deltas. For smaller separations, there is a qualitative change and the curve for $x_1 = 0.1 L_1$ only shows one dip located in the middle of the two potentials.
This qualitative change occurs at a separation of $2 l_0 \simeq 0.2 L_1$. It is understandable,
since the spacial extent of a quasihole is of the order $l_0$ (width of the Gaussian fitted to the density, see Fig. \ref{fig:FracCharge}).
If the delta potentials come closer than twice this extention the states' overlap of localized quasiholes left and right becomes very large. The state that is preferred then is a quasihole situated in between both delta potentials. This can be infered from the density Fig. \ref{fig:dncut} as well as from the spectrum depending on the potentials' separation Fig. \ref{fig:sp2deltas}. For $x = 0$ the spectrum evolves into the state of a localized quasihole at the origin. For small separations between the potentials (i.e. $\le 2 l_0$) the energy of all states rises, but the structure of the spectrum remains the same. The rise in energy can simply be attributed to the potential energy caused by the delta potentials sitting on the slope of the quasihole's state where the wavefunction has small modulus but is not zero. At a separation around $ \ge 2 l_0$ the symmetric or antisymmetric state of the tunneling begins to be favored.

The dependence of the eigenspectrum on the distance between the two delta potentials, Fig. \ref{fig:sp2deltas}, depicts at the point $x = 0.5 L_1$ the situation from Fig. \ref{fig:Spec2Delta}, red plot. The six lowest eigenenergies are very close to each other. Decreasing the distance
increases the splitting between them. The comparison to the model at the beginning of this section, which lead to Eq. \ref{eq:HamTunnelEig}, shows that the spectrum evolves a bit more complex than expected. The highest state does not split off as much with respect to
the higher pair of degenerate states as the lowest does to the lower pair (as it should be according to the model). Therefore, the model made above is not applicable. This might be caused by mixing with higher states but the lowest six, which was neglected before. These higher states correspond -- as stated above -- either to quasiholes being localized at different positions than those of the two deltas under consideration or to excited states of the quasihole at the deltas' positions. Another reason might be the influence of the delta potentials on the ``vacuum'' state, i.e. the fractional quantum Hall state without any quasihole, which is the background on which the single quasiparticle lives.
It could cause deviations via influencing the tunneling matrix elements. If there are more than the proposed two different coefficients $t$ and $t_1$, the calculated spectrum might be reproduced.
Although the simple model in the beginning of this section does not apply unmodified, it is evident that the splitting of the six lowest states
is due to the tunnel-coupling. The dependence of the energy difference between the highest of these states and the ground state energy produces Fig. \ref{fig:TunnelSplit}. The splitting increases with reduced distance between the two deltas until a distance of $0.2 L_1 \simeq 2 l_0$ at which, as said before, the quasihole begins sitting in between the two delta potentials. This distance sets the lower limit for the tunneling regime on the spacing between the potentials.
%
%
%
%
\subsection{Asymmetric delta potentials}
\label{sec:AsymmDelta}
\begin{figure}
\includegraphics[scale=0.45]{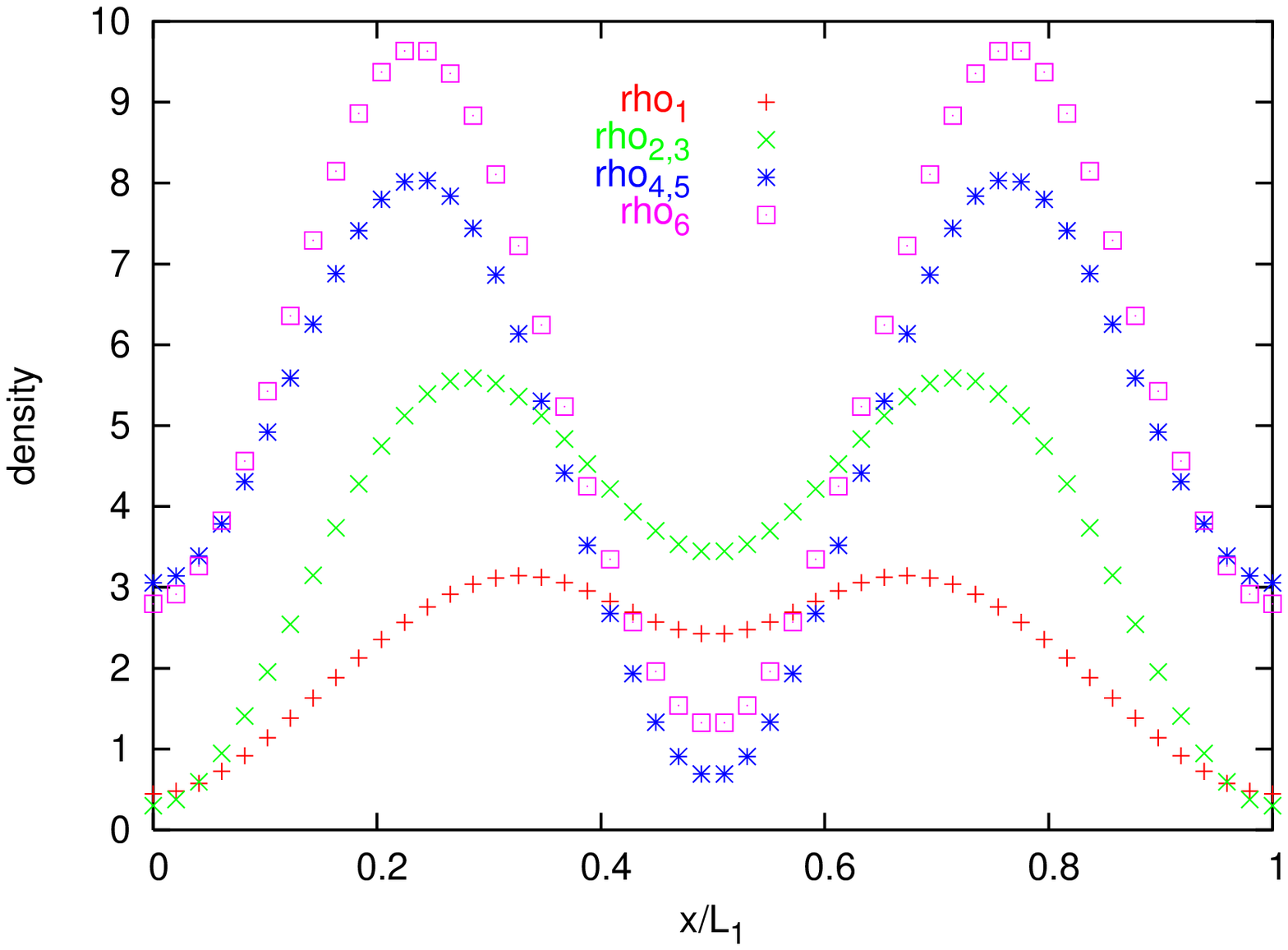}\includegraphics[scale=0.45]{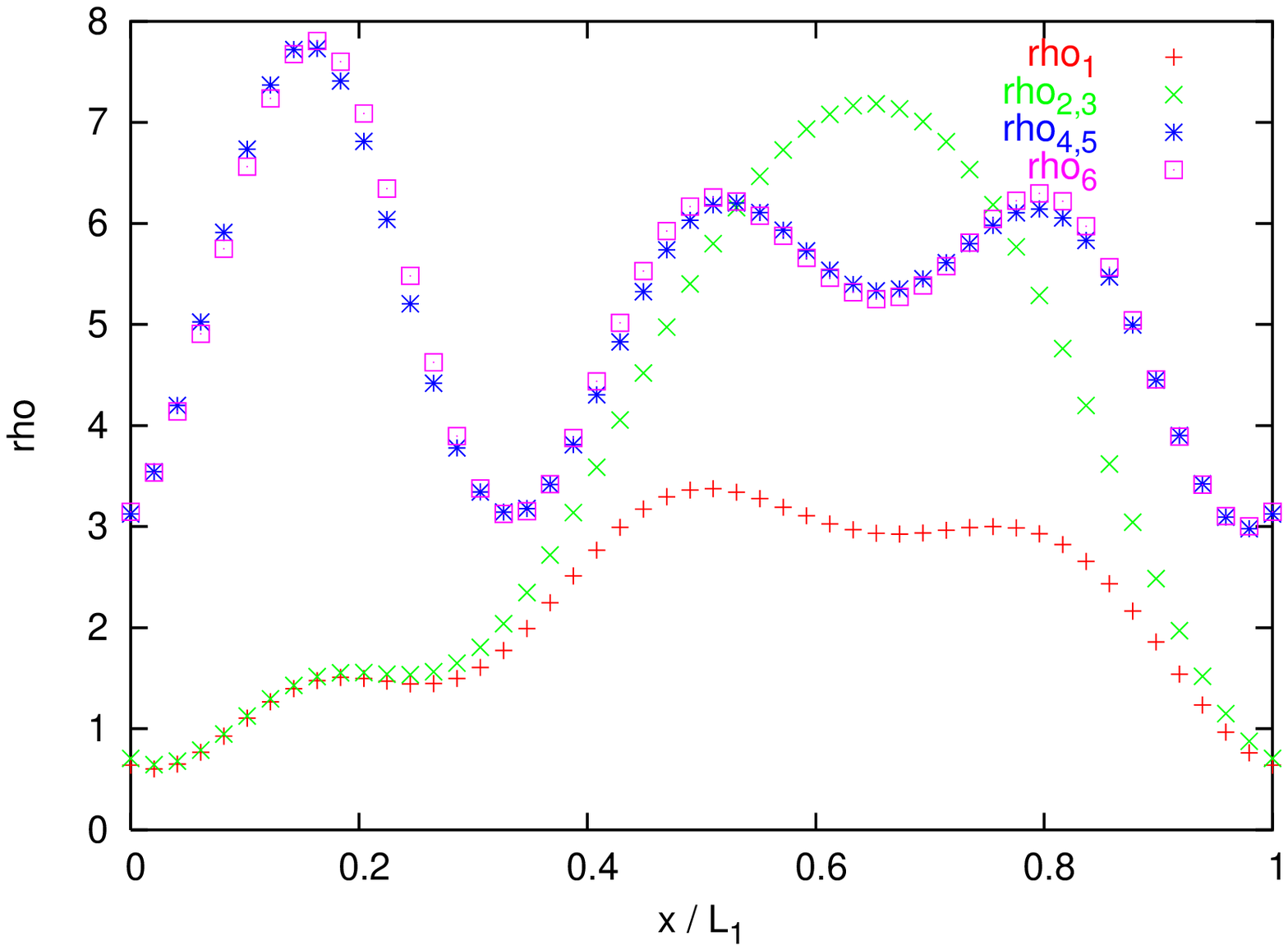}
\caption{Cut through the density along the $x$-axis for a 5/16-system with two delta potentials, one is fixed at $(0.0,0.0)$. Its strength is $0.006 L_1 L_2 \mbox{enu}$. The second delta has a strength of $0.004 L_1 L_2 enu$. Left plot: The second delta is situated at $(0.5,0.0)$. Right plot: The second delta sits at $(0.3,0.0)$.}
\label{fig:dnCut0.5}
\label{fig:dnCut0.3}
\end{figure}

%
%
%
%
Even at a distance of $5 l_0$ between the delta potentials there is still a coupling of the states with a localized quasihole
on the left delta and those where the quasihole is pinned by the right one. To confirm, that this tunnel coupling is distance dependent without
resorting to the spectrum, the following modification is made. The symmetry of the system is broken by increasing the height of one delta, while decreasing the other. The higher one, located at $(0.0,0.0)$ will have an amplitude of $0.006 L_1 L_2 enu$, the lower one at $(0.5,0.0)$ has a height of $0.004 L_1 L_2 enu$. These values were chosen such that the difference in the peak height is a bit larger than the splitting of the lowest six energies observed in Fig. \ref{fig:Spec2Delta}, red plot.
The asymmetry in the ``binding energy'' of $0.002 enu$ between a quasihole localized at the left and one at the right delta potential was chosen such that at a spacing of $x=0.3$, the tunnel-splitting of about $0.045 \mbox{enu}$ (Fig. \ref{fig:TunnelSplit}) should dominate the process.
The spectrum for the asymmetric case is given by the green points in Fig. \ref{fig:Spec2Delta} and is very similar to the symmetric case, but the 
splitting between the six lowest states should now correspond to states where the quasihole is either localized at the left, stronger delta, giving lower energy or at the right one, resulting in a higher eigenenergy.
A cut through the density (averaged over degenerate states 2,3 and 4,5) can be found in Fig. \ref{fig:dnCut0.5}. The asymmetry originating from the different peak heights is quite pronounced. The three lower lying states 1,2,3 have the quasihole stronger localized at the larger delta potential at $x = 0$, while the three higher ones -- 4,5,6 -- show a stronger dip in the density at the delta at $x = 0.5 L_1$.
Decreasing the distance between the deltas must increase the tunneling and therefore favor the symmetric or antisymmetric state if the tunnel-coupling exceeds the asymmetry in the deltas' peak heights. For a distance of $x=0.3 L_1$ this results in density profiles like in Fig. \ref{fig:dnCut0.3}. The asymmetry between the deltas is still strongly visible in the lower three states, while the upper three states are already very symmetric. A qualitative difference between the lower and the higher three states now can be attributed less to the place where the quasihole is localized, but to the symmetry of the states: The lower three ones favor a low density between the deltas, while the upper three ones have a peak in the density in between.
This situation resembles the case of a coupled two level system, where we have symmetric and antisymmetric linear combinations of the single systems' eigenfunctions. But still the asymmetry in the density between both delta potentials is visible. In the limit of much stronger tunneling, the state must become symmetric.
In the ideal case, from the calculation with asymmetric delta potentials, it is possible to extract information about the distance dependence of the tunneling matrix element $t$. Starting at large separations, tunneling is negligible and the quasihole state localized at the higher delta is favorable. Decreasing the distance until the tunneling becomes dominant over the asymmetry we should --- at sufficient strong tunneling --- end up in a symmetric density profile (as seen in the symmetric case).
The reason why this cannot be seen clearly in our system is due to its finite size: Even at a separation of half a unit cell, the tunneling is still quite strong. This was already anticipated from the density profile in Fig. \ref{fig:dn2delta0.5} (``trench'' in the density between the potentials), from the finite splitting of the ground state sector in Fig. \ref{fig:sp2deltas} at $x=0.5$ and finally also confirmed by the density profile in Fig. \ref{fig:dnCut0.5}. If the tunneling was negligible, the latter density plot would have to show the quasihole localized at the higher delta for the ground state, while it should reside at the lower delta in the excited state. Although this tendency is visible, it is already superposed by the effect of tunneling. The energy splitting between these states must be comparable to the difference in the peak height, which is however true (compare Fig. \ref{fig:Spec2Delta}, green plot). This lower limit on the tunnel coupling given by the finite size of the system demands a stronger asymmetry in the peak height. On the other hand, going to small distances between the potentials, the tunneling regime was seen to be limited at a separation of $\simeq 2 l_0$ (saturation in Fig. \ref{fig:TunnelSplit}, localized quasihole between the deltas in Fig. \ref{fig:dncut}).

\subsection{Conclusions on tunneling of a quasihole between delta potentials}
Two delta potentials in a system with an additional flux quantum are found to be a good model system to get a view on tunneling of quasiholes.
The ground state of this system could be shown to be to a high degree ($94.4$ per cent) a linear combination with equal coefficients of the quasihole ground states localized at the potentials, thus giving an evidence that the description in terms of a single-quasiparticle picture is valid.
Although the spectrum becomes complicated for strong tunneling, for weak tunneling its lowest energies resemble the structure obtained by the simple model just considering tunneling between the lowest localized quasihole states.

As a side effect, the classification of the excited states below the gap (see section \ref{sec:QuasiHolesHard}) could be substantiated: States with a ring-like structure were found to be absent in the spectral decomposition of the tunnel setup's groundstate, whereas all states with lower energy contributed.

The expectation was confirmed that tunneling should increase with decreasing distance between the potentials. It would be interesting to compare this distance dependence with the overlap of the quasihole wavefunctions. The saturation of the tunnel splitting for distances smaller than $2 l_0$ however tells us, that the dependence of the tunneling on the overlap can only hold for sufficient large separations.

The finite size of the system was found to give a lower limit on the strength of the tunneling, whereas the finite spacial extent of the quasihole wavefunction on the other side dictates an upper bound for it.
These limitations prevented the calculations with asymmetric potentials to yield quantitatively usable results.
However, these limitations were not exhausted yet. As seen from Fig. \ref{fig:dnExcitedHole} in section \ref{sec:QuasiHolesHard} there is an excited eigenstate of a quasihole localized mainly at the position diametrically opposed to the delta potential. Owing to its orthogonality to the quasihole ground state, choosing two diametrically opposed points within the unit cell to place the delta potentials will probably result in weakest possible tunneling (as also intuitively expected). Therefore one could start with a smaller asymmetry in the peak height and possibly extract the distance dependence of $t$ from the point of crossover between the asymmetry-dominated and the tunneling regime.

Possible further work could also include a comparison of the results obtained here to those for electrons (or holes). The same calculations at a filling factor $N_e/N_s$, where $N_s = N_e + 1$ would describe single-hole tunneling in an almost filled Landau level. The program \cite{VybornyMuellerHelias} used for the calculations performed here is also capable of calculations near the filling factor $\nu=1$.
Another interesting proof for the tunneling would be to prepare the system in a state where the quasihole is localized at one delta potential and to evaluate its time evolution. An oscillation of the quasihole between the localized states should be observable. 
\newpage
\section{Inhomogeneous systems: Constrictions in the FQH regime}
\label{sec:InhomSystems}
\subsection{Modeling the constriction}
\label{sec:gaussian}
After having seen in section \ref{sec:tunnel} that the quasiholes indeed show tunneling between two delta potentials, the 
aim is now to investigate their behavior if they are not bound to point-like potentials but are allowed to move in a system with a constriction.
We want to infer their transport properties in a fractional quantum Hall system with a constriction directly from their motion in the time domain.

The system to be considered therefore not only subjects the electrons to a homogeneous magnetic field,
 but also to some external potential $V(x,y)$. This potential serves to model a constriction for the electrons.
Its shape is that of a wall that crosses the system parallel to the y-axis with a notch of controllable width and 
depth inside this wall which forms a sort of a passage through it.

Two different kinds of potentials are used here. The first one is a Gaussian shaped barrier, the second type is a ``delta'' barrier in k-space.
The first type is just a potential in x-space. Since the calculations are to be performed in the
periodic basis derived in section \ref{sec:Basis}, the potential as well has to maintain the discrete
translational symmetry in order to be treatable in this basis. Put differently, $V(x,y)$ has to commute with the magnetic translations found in equations (\ref{eq:MagTrans}) and (\ref{eq:MagTransY}). Since $V(x,y)$ only contains the operators $x$ and $y$, every periodic potential with a period of one unit cell (in x- and in y-direction) commutes with these translations.
The part of the potential creating the ``wall'' parallel to the y-axis therefore has to be made periodic in the x-direction. This is done by a
superposition of Gaussian profiles, each of which is located in one unit cell,
\begin{eqnarray}
  \label{eq:VwallGauss}
  V_{Wall}(x) &=& s \frac{e^2}{\epsilon l_0^2} \sum_{k \in \Z} \exp(-\frac{(x-x_0+k L_1)^2}{(w L_1)^2}).
\end{eqnarray}
Its height is given in the overall energy unit $\frac{e^2}{\epsilon l_0^2}$ defined by the coulomb interaction and it is centered around $x_0$ in the x-direction in every unit cell. The dimensionless parameters $s$ and $w$ control the height and the width of the potential respectively.
The single-particle matrix elements for the basis functions (\ref{eq:BasisOrt}) in the lowest Landau-level ($n=0$) can be calculated straight forwardly
and are found in appendix \ref{app:MatEls}, Equ. (\ref{eq:MatElsGaussWall}).

These matrix elements are, as already anticipated from the translational invariance in y-direction of $V_{wall}(x)$, diagonal in $k$. Therefore in the many particle system, the potential only causes a mixing of states in subspaces with constant total y-momentum.
Further on, these elements are real, which is an advantage for the numerical diagonalization.

To ``cut'' a notch into this wall, a second potential $V_{notch}$ is superposed. In x-direction it inherits the shape of $V_{wall}$, in y-direction it is similarly Gaussian shaped. It is defined as
\begin{eqnarray}
	\label{eq:Vnotch}
	V_{notch}(x,y) &=& s_{notch} \frac{e^2}{\epsilon l_0^2} \sum_{k \in \Z} \exp(-\frac{(x-x_0+k L_1)^2}{(w L_1)^2}) \times \\ \nonumber
	&\times& \sum_{n \in \Z} \exp(-\frac{(y-y_0+n L_2)^2}{(w_{notch} L_2)^2}). \\ \nonumber
\end{eqnarray}
The matrix elements can again be computed exactly and one ends up with Equ. (\ref{eq:MatElNotch}), appendix \ref{app:MatEls}.

\begin{figure}[htbp]
  \includegraphics[scale=0.4]{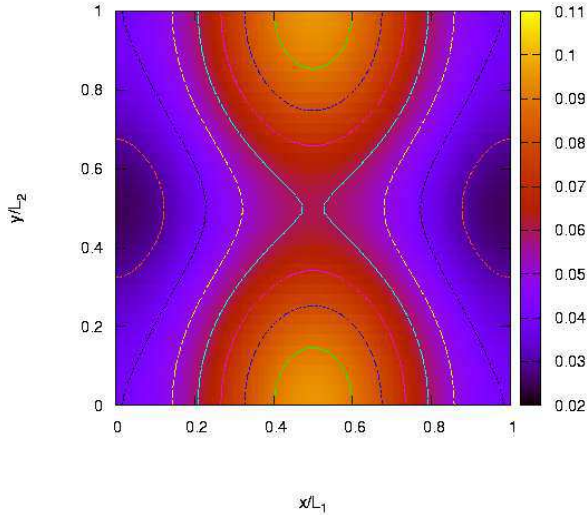}
  \caption{Gaussian shaped potential $V(x,y) = V_{wall}(x) + V_{notch}(x,y)$ of a constriction according to Eq. (\ref{eq:VwallGauss}) and (\ref{eq:Vnotch}). The contour depicts equipotential lines.}
  \label{fig:Potential}
\end{figure}

The shape of a typical potential can be seen in Fig. \ref{fig:Potential}. The colored lines are equipotential lines, the color code represents the height. x- and y-axes are the coordinates in the unit cell.
One technical point is the adjustment of the depth of the notch in the barrier: For a width larger that approximately $0.3 L_2$ the Gaussians of neighboring cells begin to overlap. By this the effective peak height at $y_0$ is higher than defined by $s_{notch}$. This overlap has to be taken into account when setting the parameters (in our program this is done automatically) to keep the potential positive definite.
%

\subsection{Corrections from the next Landau level}
\label{sec:corrections}
As described in section \ref{sec:projection}, the many body problem will be treated in a finite basis containing only states in the lowest Landau level which is equivalent to diagonalizing the projected Hamiltonian $P_{LLL} H P_{LLL}$. While the crucial properties such as the gapped ground state, the incompressibility, the correlations and quasihole excitations were seen to be reproducible in this truncated basis, a problem occurs when evaluating expectation values related to the kinetic momentum operators. This will be described in the following.
\subsubsection{Vanishing kinetic momenta}
\label{sec:PiZero}
In a strong magnetic field an electron is expected to respond to an applied in-plane electric field by a drift perpendicular to $\vec{E}$ and $\vec{B}$.
This E-cross-B-drift velocity $v_{d,y}$ is caused in our case by an external potential $V(x) = \mbox{enu} V_0(x/l_0)$, which is measured in the overall energy unit $\mbox{enu} = \frac{e^2}{\epsilon l_0}$ defined by the Coulomb interaction. So we obtain
\begin{eqnarray}
 	v_{d,y} &=& -\frac{E_x}{B} = -\frac{1}{m \omega_c} \frac{\partial V}{\partial x} \nonumber \\
	        &=& -\frac{e^2}{\epsilon l_0^2 \omega_c m} \frac{\partial V_0}{\partial (x/l_0)} \nonumber \\
		&=& -\frac{e^2}{\hbar} \frac{\partial V_0}{\partial (x/l_0)} \nonumber.
\end{eqnarray}
This drift velocity thus is important if the external potential varies on our energy scale over length scales comparable to the magnetic length.
It will be seen, that for creating effective tunneling barriers, the potential will be of the order of the Coulomb interaction and it varies from zero to its peak value within a few $l_0$. So the drift velocity is not negligible.

On the other hand, calculating the expectation value of the kinetic momenta for a state within the lowest Landau level returns zero. 
To obtain non-vanishing contributions, a correction from higher Landau-levels is necessary, since these operators are identically zero
within the lowest Landau level.

Vanishing kinetic momenta $\Pi_x$ and $\Pi_y$ in the lowest Landau level formally means that $P_{LLL} \Pi_m P_{LLL} = 0$.
Equivalently, this can be expressed as vanishing matrix elements $\bra{0,k} \Pi_i \ket{0,j} = 0$ where $i=x,y$.

Taking the wavefunction from (\ref{eq:PerBasis}) and using that $\Pi_x$ and $\Pi_y$ commute with $t_x(X_p)$ and making use of Equ. (\ref{eq:aOps}) to express the kinetic momenta as $\Pi_x = \sqrt{\frac{m \hbar \omega_c}{2}} i (a^\dagger - a)$ and $\Pi_y = \sqrt{\frac{m \hbar \omega_c}{2}} (a^\dagger + a)$, the statement above is appreciated as follows
\begin{eqnarray}
  \label{eq:PiZero}
  \bra{0,l}_{prb} \Pi_{x/y} \ket{0,j}_{prb} &=& \bra{0,k}_{prb} \Pi_{x/y} \sum_{k \in \Z} \exp(-i k \alpha) \; t_x \big( (2 \pi j + \beta) \frac{l_0^2}{L_2} + k L_1 \big) \; \ket{0,0} \\ \nonumber
  &=& \bra{0,l}_{prb} \sum_{k \in \Z} \exp(-i k \alpha) \; t_x \big( (2 \pi j + \beta) \frac{l_0^2}{L_2} + k L_1 \big) \; \Pi_{x/y} \ket{0,0} \\ \nonumber
  &\propto& \bra{0,l}_{prb} \sum_{k \in \Z} \exp(-i k \alpha) \; t_x \big( (2 \pi j + \beta) \frac{l_0^2}{L_2} + k L_1 \big) \; (a^\dagger \pm a) \ket{0,0} \\ \nonumber
  &=& \bra{0,l}_{prb} \ket{1,j}_{prb} = 0. \\ \nonumber
\end{eqnarray}
This means that the quantum mechanical current operator only yields non-vanishing contributions if the wavefunction has contributions in higher
Landau levels. This will be analyzed more precisely in the following sections.  
\subsubsection{Perturbative contribution of the first Landau level}
\label{sec:Perturb}
To solve the contradiction between the vanishing kinetic momenta and the expected $\vec E \times \vec B$ drift, the amount of admixture
of the second Landau level due to the electric field caused by the potential is calculated.

This is a single particle effect and hence it is sufficient to consider the single particle Hamiltonian which can be separated in the kinetic part like in (\ref{eq:Hamilton}) -- here called $H_0$ -- plus a perturbing external potential $V$, which will be accounted for by perturbation theory. Let $V(x)$ be a potential only depending on the x-coordinate that additionally possesses a periodicity with respect to the unit cell's size.
Thus the wavenumber $k$ remains a good quantum number, meaning that there is no mixing of eigenfunctions with different $k$ caused by $V(x)$.
The eigenfunctions of the unperturbed system $H_0$ are known from (\ref{eq:PerBasis}). The first order correction in perturbation theory now reveals
\begin{eqnarray}
  \label{eq:FirstOrder}
  \ket{k}^1 &=& \sum_{n \neq m} \frac{ \bra{n, k} V(x) \ket{m, k} }{E_n^0 - E_m^0} \; \ket{n, k} \\ \nonumber
  &\simeq& \frac{1}{\hbar \omega_c} \left( -\bra{0,k}V(x)\ket{1,k} \; \ket{0,k} + \bra{1,k}V(x)\ket{0,k} \; \ket{1,k} \right).
\end{eqnarray}
In the second step all higher Landau-levels except for the first two were neglected. The appearing matrix element can be calculated using the basis in direct space from (\ref{eq:BasisOrt}). Due to the diagonality in $k$ all wavefunctions only have strong contributions in the vicinity of $-X_k$. So a Taylor expansion of $V(x)$ around $-X_k$ makes sense. The linear term will cause a transition between neighboring Landau levels. This follows from the orthonormality relation $\int_{-\infty}^{\infty} dx \; \exp(-x^2)H_n(x)H_m(x) = \delta_{n,m} 2^n n! \sqrt{\pi}$ for Hermitian polynomials. Thus, all higher terms of order $n$ produce contributions only between Landau level $m$ and $m \pm n$ and can therefore be skipped without increasing the error of this approximation. The remaining matrix element can be calculated analytically, where the Taylor expansion of $V(x)$ is used.
This leads to the matrix element
\begin{eqnarray}
  \label{eq:PerturbMat}
  \bra{0, k} V(x) \ket{1, k} &=& \frac{l_0}{\sqrt{2}} \left.\frac{\partial V}{\partial x} \right|_{X_k}.
\end{eqnarray}
Using that, the correction of $\ket{0, k}$ becomes 
\begin{eqnarray}
  \label{eq:CorrPsi}
  \ket{k}^1 = \frac{1}{\sqrt{2}} \left. \frac{ \partial V/(\hbar \omega_c) } { \partial x/l_0 } \right|_{X_k} \left( -\ket{0,k} + \ket{1,k} \right).
\end{eqnarray}
The formula (\ref{eq:CorrPsi}) indicates that the correction of the \emph{wavefunction} however is small if $V$ does not vary much in units of $\hbar \omega_c$ over distances of order of the magnetic length $l_0$. To lowest order, the wavefunction remains normalized when adding the correction term. So it is possible to calculate the expectation value of the current in y-direction in this new state as
\begin{eqnarray}
  \label{eq:ExpCurrent}
  \langle \Pi_y \rangle &=& (\bra{0,k} + \bra{k}^1) \Pi_y (\ket{0,k} + \ket{k}^1) \\ \nonumber
  &=& 2 \frac{1}{\sqrt{2}} \left. \frac{ \partial V/(\hbar \omega_c) } { \partial x/l_0 } \right|_{X_k} \bra{0,k} \Pi_y \ket{1,k} \\ \nonumber
  &=& -\frac{1}{\omega_c} \underbrace{\left.\frac{\partial V}{\partial x} \right|_{X_k}}_{=eE} \\ \nonumber
  &=& -\frac{m}{e B} eE_x = m v_{d,y}
\end{eqnarray}
In the first step the vanishing matrix elements of $\Pi_y$ in the same Landau-level was used as well as the hermiticity of this operator. The second step involves the explicit calculation of $\bra{0,k} \Pi_y \ket{1,k} = -\frac{1}{\sqrt{2}}\frac{\hbar}{l_0}$ in direct space. Thus we gain the result for the drift current which was classically anticipated from the $\vec E \times \vec B$ drift. This correction, as stated above, is important if $V$ varies on the energy scale $\mbox{enu}$ over a distance of order $l_0$.

\subsubsection{Hellman-Feynman Theorem}
\label{sec:HellFeyn}
Another way to appreciate the correction from the last section is via the Hellman-Feynman theorem.
This theorem is applicable to parameter dependent eigenvectors of a Hermitian operator depending on the same parameter.
The parameter in question here is the phase factor $\beta$ from the periodic boundary conditions in Equ. (\ref{eq:PRB}), which is
useful when calculating the kinetic momentum.
The parameter appeared originally in the eigenvalue equations for the periodic boundary condition (Equ. \ref{eq:PRB}), but however it is possible
to apply a unitary transformation like in section \ref{sec:AlphaBeta}, Equ. (\ref{eq:UnitBeta}), which causes $\beta$ to reside in the 
Hamiltonian's kinetic y-momentum, as shown in Equ. (\ref{eq:TrxPiy}).
By this transformation, the transformed eigenvectors obtain a fixed phase factor $\beta' = 0$ for the periodic boundary condition in y-direction.
Due to its unitarity, the transformation does not affect any observable, especially all eigenvectors are transformed by the same transformation $U(y)$ to become eigenvectors of the transformed Hamiltonian $H'_\beta = U(y) H U^\dagger(y)$.
After transforming, the Hellman-Feynman theorem (for example from \cite{Lange}) is applicable to this problem.

Now assume we have the same system as in the previous subsection --- namely a single-particle problem with a potential $V(x)$ --- and have calculated the eigenvectors of the Hamiltonian $H'_\beta$ which are denoted by $\ket{k,\beta}$.
Normalized eigenvectors assumed, we have
$\frac{\partial}{\partial \beta} \braket{k,\beta}{l,\beta} = 0$, from which follows $\frac{\partial \bra{k,\beta}}{\partial \beta} \ket{l,\beta} = - \bra{k,\beta} \frac{\partial \ket{l,\beta}}{\partial \beta}$. In our case we yield for $H'_\beta$
\begin{eqnarray}
  \label{eq:FeynmanHellman}
  \bra{k,\beta}H'_\beta\ket{l,\beta} &=& E_{k,\beta} \delta_{k, l} \\ \nonumber
  \frac{\partial E_{k,\beta}}{\partial \beta} \delta_{k, l} &=& \bra{k,\beta}\frac{\partial \ket{l,\beta}}{\partial \beta} (E_{k,\beta} - E_{l,\beta}) + \bra{k,\beta} \frac{\partial H'_\beta}{\partial \beta} \ket{l,\beta}.
\end{eqnarray}
Where $\delta_{k,l}$ denotes the Kronecker symbol. Keeping in mind the $\beta$-dependence of $\Pi_{\beta,y}$, the derivative of $H'_\beta$ can be expressed as $\frac{\partial H_\beta}{\partial \beta} = \frac{\hbar}{m L_2} \Pi_{\beta,y}$. 
Thus for $k=l$ we arrive at a connection between the energy dispersion and the current as
\begin{eqnarray}
  \label{eq:CurrentDisp}
  \frac{\hbar}{m L_2} \langle \Pi_y \rangle_k = \partial_\beta E_{k,\beta}.
\end{eqnarray}
What we actually calculate when restricting the basis to the lowest Landau-level are not the eigenvalues of $H'_\beta$ but those of $P_{LLL}H'_\beta P_{LLL}$ where $P_{LLL}$ is the projection to the lowest Landau-level (see section \ref{sec:Projector}).
Further on, because $V(x)$ only depends on $x$, this potential does not mix any of the eigenstates in the lowest Landau-level.
That's why the basis states are already eigenstates of the system, only their degeneracy is lifted and replaced by the following dispersion
\begin{eqnarray}
  \label{eq:DispersionV}
  E_{k,\beta} &=& \frac{\hbar \omega_c}{2} + \bra{k_\beta} V(x) \ket{k_\beta} \\ \nonumber
  &\simeq& \frac{\hbar \omega_c}{2} + V(-X_{k,\beta}) \\ \nonumber
  \rightarrow \partial_\beta E_{k,\beta} &\simeq& \partial_\beta V(-X_{k,\beta}) = -\frac{L_1}{2 \pi N_s} \left. \frac{\partial V}{\partial x} \right|_{-X_{k,\beta}} \\ \nonumber
&=& -\frac{l_0^2}{L_2} \left. \frac{\partial V}{\partial x} \right|_{-X_{k,\beta}},
\end{eqnarray}
where in the second step the localization of the basis states around $-X_{k,\beta}$ as seen from Equ. (\ref{eq:BasisOrt}) was used together with the assumption of only small variations of the potential $V(x)$ on the length of $l_0$.
In the last line the definition of $X_{k,\beta}$ from the same equation was taken into account.
Merging Equ. (\ref{eq:CurrentDisp}) into (\ref{eq:DispersionV}), we gain the same result as obtained in the previous section by means of perturbation theory. If, on the other hand, this perturbation is not taken into account, $\Pi_y$ will be zero (according to section \ref{sec:PiZero}) which will lead to a contradiction between Equ. \ref{eq:CurrentDisp} and \ref{eq:DispersionV}.
In the case of the current operator, the contributions from the next Landau level are necessary and sufficient to arrive at a consistent picture for the projected system.

\subsubsection{Corrected kinetic momenta}
In the previous sections it was shown that consistency is achieved only if the drift currents due to admixture of the next higher Landau level is
taken into account. The reason for this simply lies in the nature of the perturbation expansion. Restricting the wavefunction to the lowest Landau level means that it is the solution of the perturbation to zeroth order $\ket{0,k}$. On the other hand, the eigenenergies we calculate are eigenvalues of a Hamiltonian containing the external potential $V(x)$ and especially the diagonal terms $\bra{0,k}V\ket{0,k}$, which are already corrections of first order in perturbation theory for the energy. Thus it is not surprising that inconsistency appears if we evaluate an operator ($\Pi_{x,y}$) in a state that was obtained by taking the limit $\hbar \omega_c \gg \mbox{enu}$. The correct procedure is to first take into account mixing of higher Landau-levels by perturbation theory, calculate the expectation value and take the limit at the end. The reason for this is, that the operator itself can have matrix elements whose contribution between adjacent Landau levels diverge as we take the limit.

In the single-particle system it is easy to correct for this lack, as shown in section \ref{sec:Perturb}. Since the external potential is the only
perturbation here, instead of working with the eigenvectors in first order perturbation, we can directly put this correction into the
single-particle operators of which we wish to evaluate the expectation values, namely the current operators.
Reviewing Equ. (\ref{eq:ExpCurrent}), we obtain the same result for the expectation value if we correct the operator $\Pi_y$ with the following additional term 
\begin{equation}
\tilde\Pi_y = \Pi_y - \sum_{k} \frac{1}{\omega_c} \left.\frac{\partial V}{\partial x} \right|_{X_k} \ket{0,k} \bra{0,k}.
\label{eq:CorrectedPiy}
\end{equation}
%
%

\subsubsection{Corrected current density}
The results above were only shown for the y-component of the momentum. But clearly the corrections cannot depend on the gauge, thus they must taken into account also for the x-direction. What we are interested in here is not the expectation value of the momentum but rather the quantum mechanical current density, which is (as a single-particle operator)
\begin{equation}
\label{eq:CurrentOp}
\vec j(\vec r_0) = \frac{1}{2} \left(\vec G \delta(\vec r - \vec r_0) + \delta(\vec r - \vec r_0) \vec G \right),
\end{equation}
with $\vec G$ being the velocity operator. As seen from the sections before due to the projection obviously $\vec G \neq \frac{1}{m} \vec \Pi$ 
as expected from the correspondence principle is not valid.
Also the calculated  correction for $\Pi_y$ in Equ. (\ref{eq:CorrectedPiy}) cannot be used generally, firstly because it depends on the gauge 
and therefore cannot be applied to $\Pi_x$ and secondly it was assumed that the potential only depends on $x$.
A more rigorous approach for deriving the velocity operator just using the features of the projections of $x$ and $y$ onto the lowest Landau level (as used by Shankar in \cite{Shankar}) is found in appendix \ref{app:CorrectedMom}. This approach yields a different kind of current density operator which doesn't have any contributions from the kinetic momenta but is similar to the correction term calculated by perturbation theory
\begin{eqnarray}
\label{eq:CurrentOpCorrected}
\vec j_x(\vec r_0) &=& \delta(\vec r - \vec r_0) \frac{1}{2 m \omega_c} \frac{\partial V(\vec r)}{\partial y} + h.c. \\ \nonumber
\vec j_y(\vec r_0) &=& -\delta(\vec r - \vec r_0) \frac{1}{2 m \omega_c} \frac{\partial V(\vec r)}{\partial x} - h.c. .
\end{eqnarray}
Due to its agreement with the corrections calculated before and its consistency with Ehrenfest's and Hellmann-Feynman's theorem, this current density is believed to be the correct one. Another feature can be directly deduced from its formula (\ref{eq:CurrentOpCorrected}). The current density
is perpendicular to the potential's gradient. By this, the electrons flow along the equipotential lines as expected in high magnetic fields.
This will be seen in Fig. \ref{fig:dnweakbarrier}.

\subsection{Quasihole excitations in inhomogeneous Systems}
\label{sec:inhomogeneous}
After having confirmed the stability of the quasihole states in section \ref{sec:QHcoulomb} for Coulomb interaction
and in section \ref{sec:QuasiHolesHard} for hard core interaction and having seen the additional feature of being a ground state of the
homogeneous system in the latter case, we now want to inject a quasihole into an inhomogeneous system and calculate its time evolution.
First we are considering a weak potential that actually is no real constriction for the electrons but which for sure doesn't destroy the correlated state.
\subsubsection{Weak potentials}
\label{sec:WeakPot}
\begin{center}
\begin{figure}[htbp]          
  \includegraphics[scale=0.4]{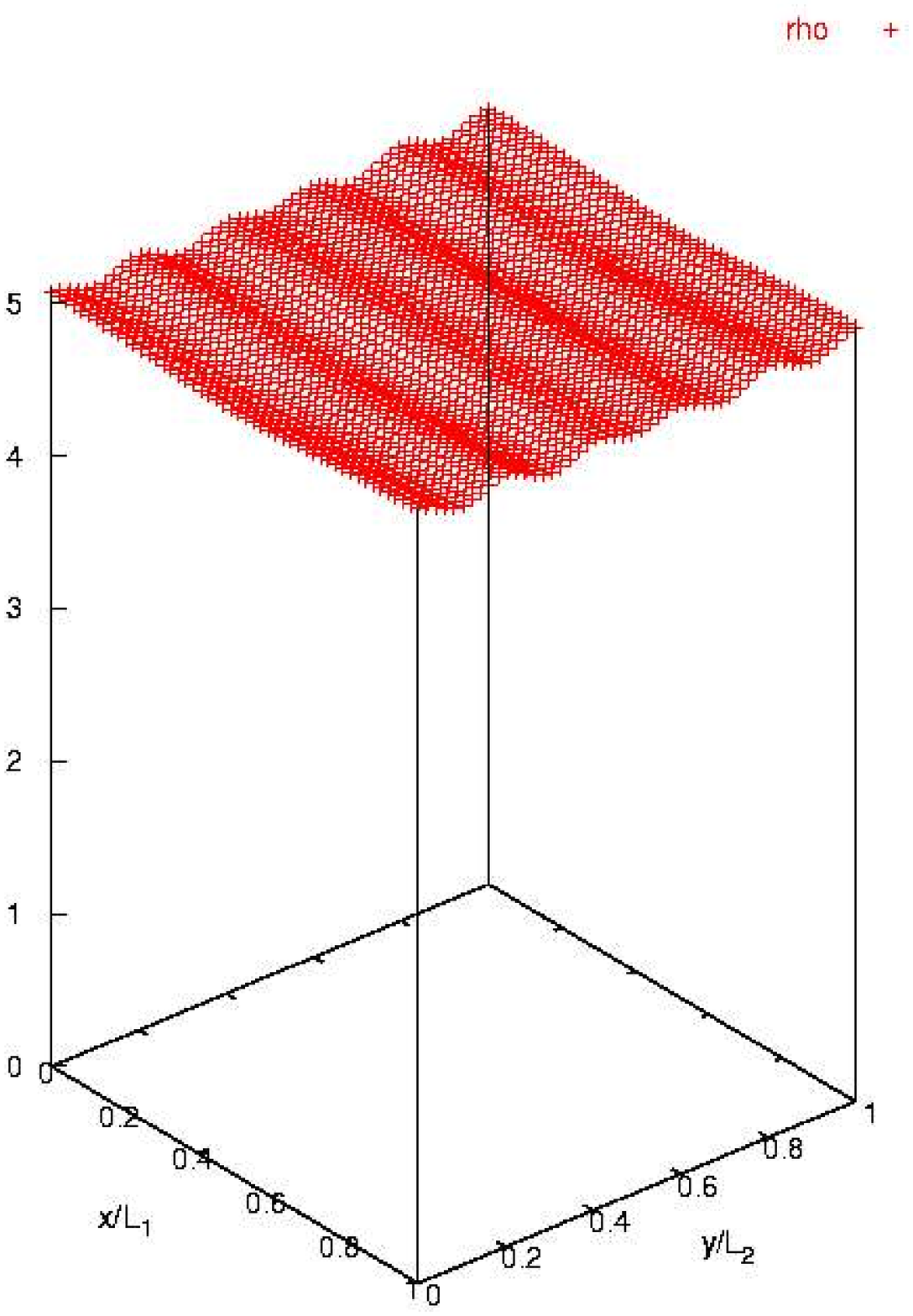}\includegraphics[scale=0.5]{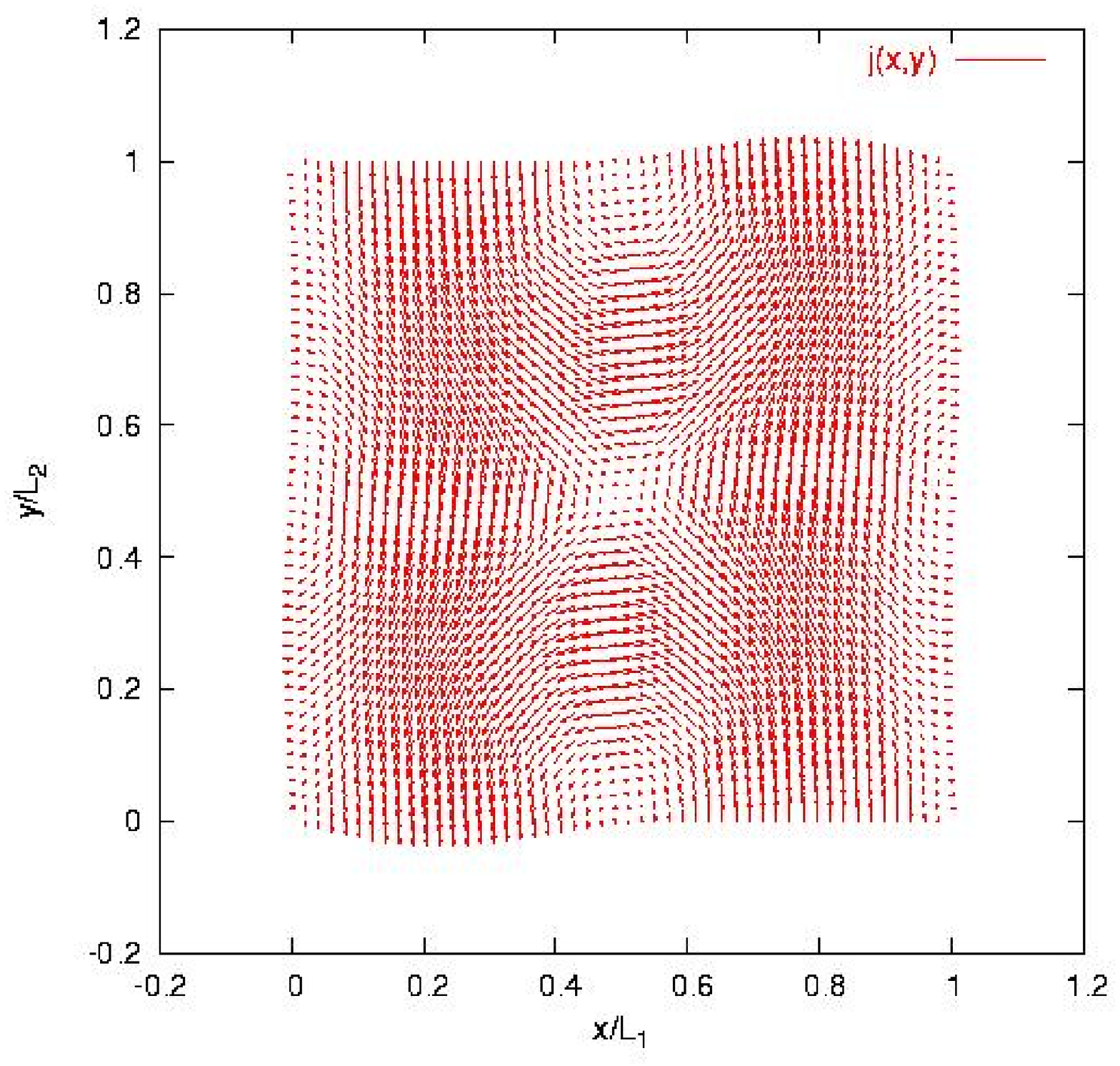}
  \caption{Left: Electronic density of the ground state for the inhomogeneous $5/15$-system. The potential is defined by the values from Tbl.\ref{tab:CommonPars} , the depth of the notch is $0.04 \mbox{enu}$. Due to the weak potential there is hardly a response of the density (incompressibility). Right: Current density for the same state. The current flows along the equipotential lines.}
  \label{fig:dnweakbarrier}
\end{figure}
\end{center}

\begin{center}
\begin{figure}[htbp]          
  \includegraphics[scale=0.4]{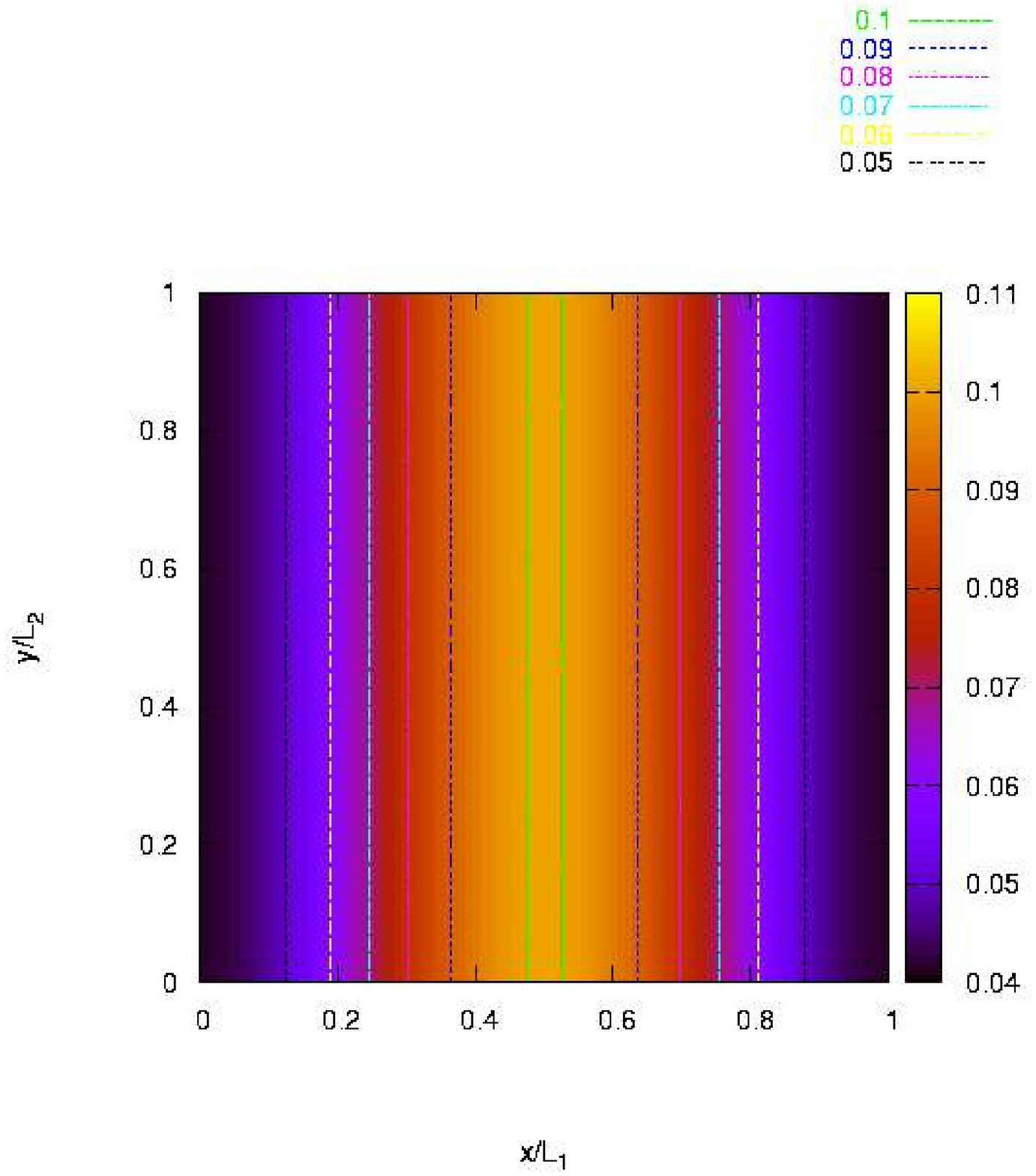}\includegraphics[scale=0.4]{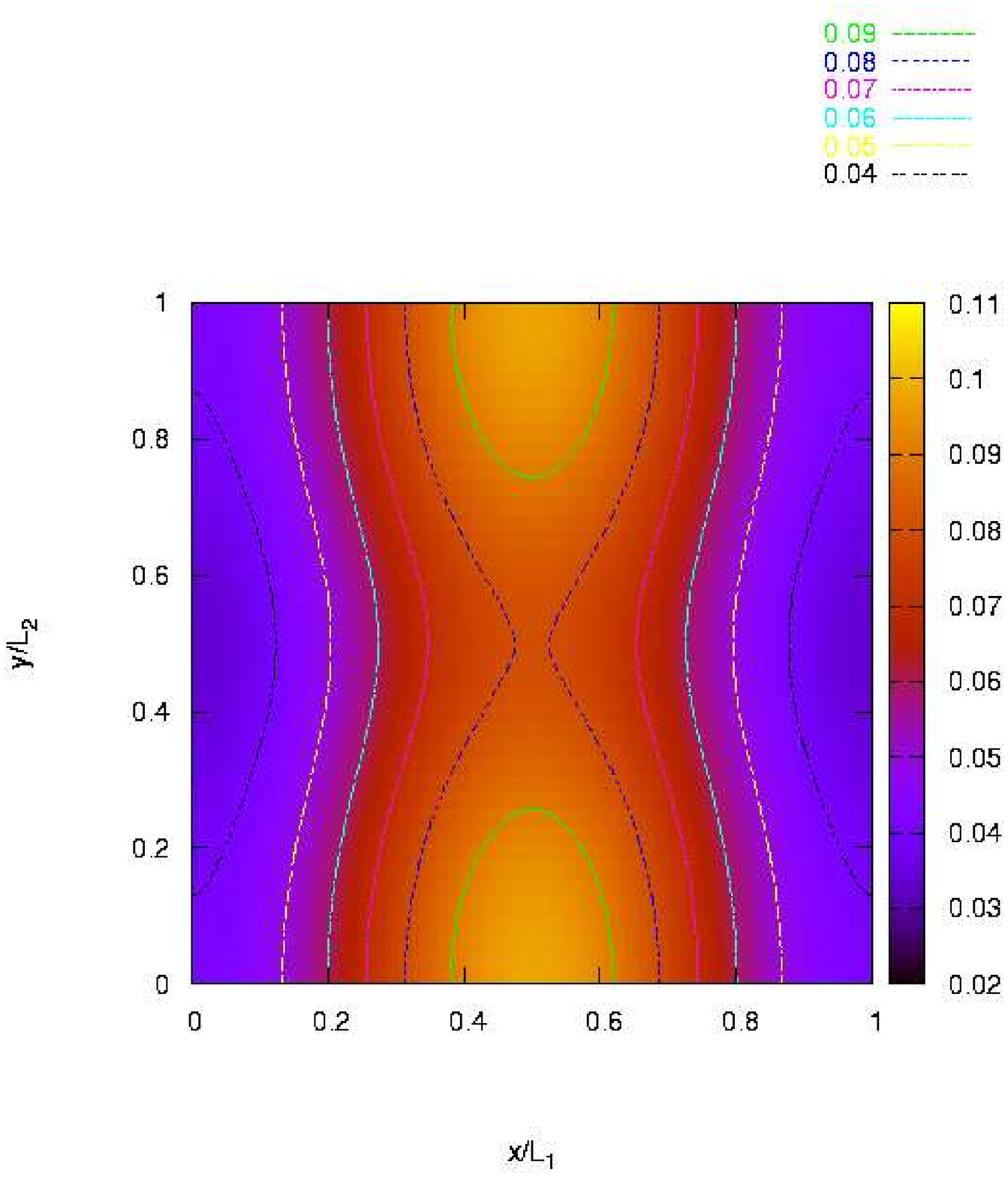}
  \includegraphics[scale=0.4]{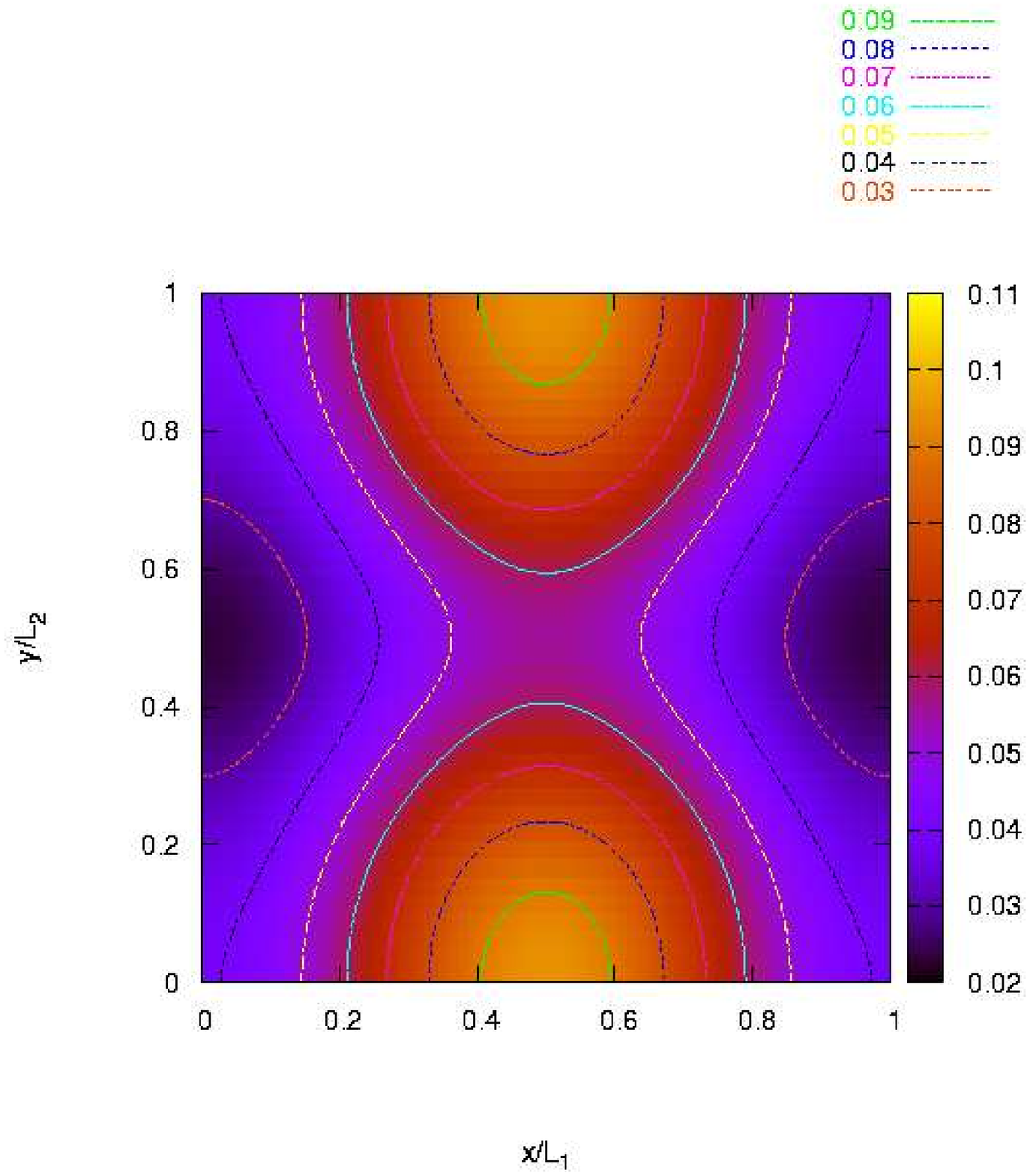}\includegraphics[scale=0.4]{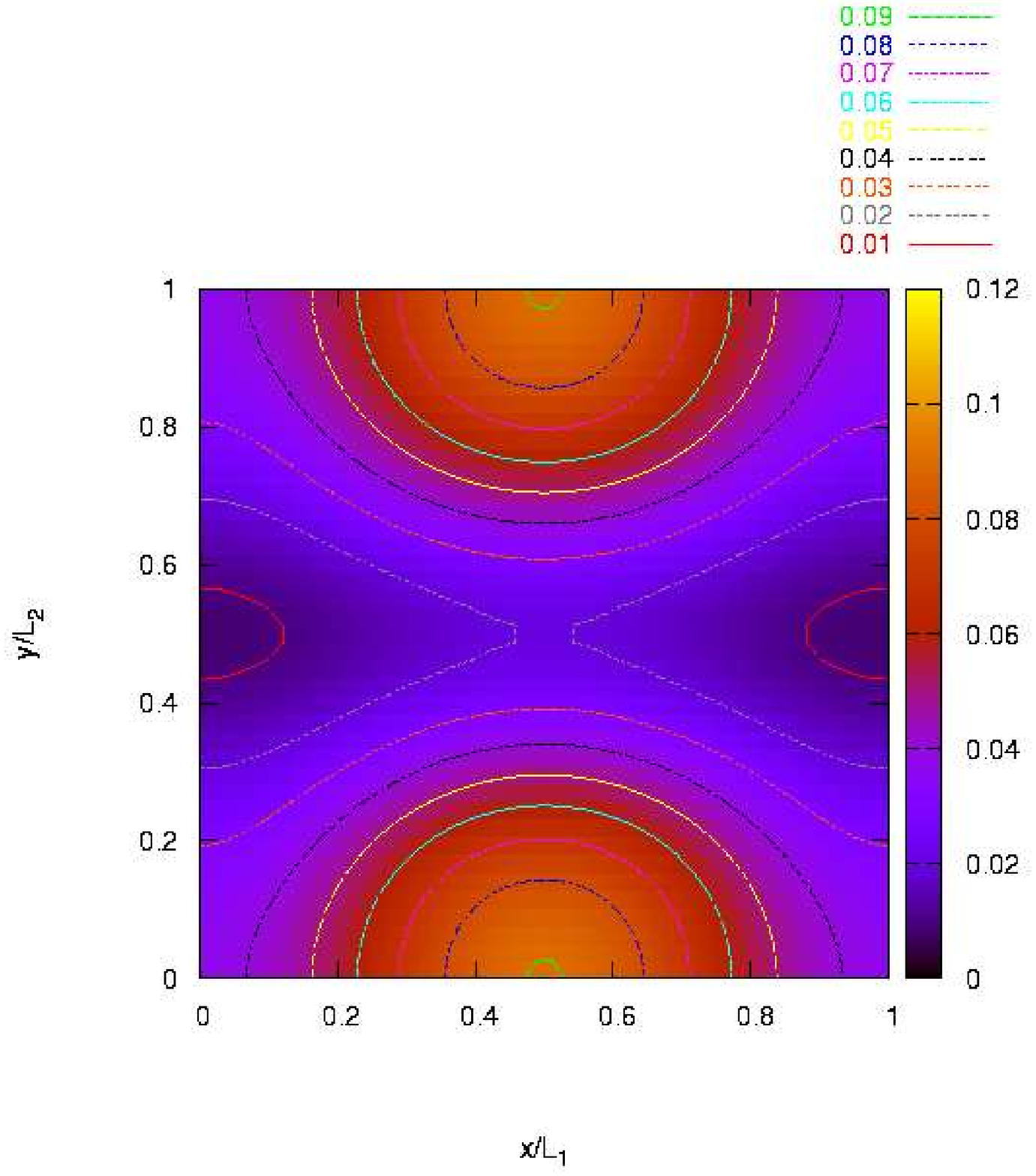}
  \caption{Contour plot of the external potentials. Left up: No notch; right up: Depth of notch $0.02 \mbox{enu}$; left down: Depth of notch $0.04 \mbox{enu}$; right down: Depth of notch $0.08 \mbox enu$.}
  \label{fig:PotentialNoHoleLowHole}
\end{figure}
\end{center}

For a system of 5 electrons and 15 flux quanta with hard-core interaction, different soft-walled potentials are investigated. The shape of the potential is a wall with a notch inside. The depth of this notch is the only parameter which is varied in what follows.
The height of the potential was chosen such that the fractional quantum Hall state of the system is not destroyed. This is confirmed by
comparing the chemical potential for adding an electron (from Tbl. \ref{tab:MuPMHomogeneous}) of $2.91 \frac{e^2}{\epsilon l_0^2}$ to the peak height of the potential (see Tbl. \ref{tab:CommonPars}). The system is in a regime, where electrons can simply overcome this barrier. Therefore,
the density hardly responds to the barrier potential, which can be seen as a sign of incompressibility (compare Fig. \ref{fig:dnweakbarrier}). 
Rather than as a tunneling barrier, the potential should in this case be regarded as a source of a dispersion $E(k)$ for the different single particle states $k$. This in turn causes a dispersion for the many particle states which are serving as a basis. This dispersion should cause a quasihole to move, due to its charge. In Fig. \ref{fig:dnweakbarrier} this motion can also be found in the electronic current density calculated according to Equ. (\ref{eq:CurrentOpCorrected}). The current is due to the electrons drift along the equipotential lines. This can be confirmed by
comparing the current density to Fig. \ref{fig:PotentialNoHoleLowHole}. Although the quasiholes have positive charge a similar motion is to be expected for them since not only the Lorentz force of the $B$-field depends on the charge but also the
force by the constriction potential: For electrons it is repulsive, for quasiholes attractive.
A quasihole is generated at the initial position $(0.2,0.0)$ and its time evolution is calculated. Due to the additional quasihole there
are now 16 flux quanta inside the system.
Four different values for the  depth of the notch are chosen. The potential landscape for those values along with equipotential lines are plotted in Fig. \ref{fig:PotentialNoHoleLowHole}.
\begin{table}[htbp]
\begin{tabular}{|l|l|l|}
\hline
Height of barrier $(\frac{e^2}{\epsilon l_0})$ & Width of barrier $(L_1)$ & Width of hole $(L_2)$\\
\hline	
 0.1 & 0.4 & 0.3\\
\hline
\end{tabular}
\caption{Common parameters used for all barrier potentials. Only the depth of the notch is being varied.}
\label{tab:CommonPars}
\end{table}
Classically we can assume that the quasihole ---treated as a charged particle of charge $q_{qh}=\frac{1}{3}e$--- should follow the equipotential line it was created on. A potential without any notch should lead to a downward drift of the quasihole. Shallow notches in the potential cause a slight deformation of the equipotential lines. As long as the equipotential line the quasihole is starting on does not come close to the saddle point of the potential, no qualitatively different picture from the case without any notch is expected.
In the case of an equipotential line crossing the saddle point of the potential, the quasihole has two possibilities: It can pass through the barrier as well as it can pass by. A yet stronger notch leads to equipotential lines connecting the left and the right part of the system. A quasihole
starting on such a line should therefore pass from the left half to the right half of the system.
The calculation of the density's time evolution for an initial state with a quasihole in the position $x_{hole}=0.2$ and $y_{hole}=0.0$ has been performed to confirm or dismiss these expectations.

\begin{center}
\begin{figure}[htbp]          
  \includegraphics[scale=0.4]{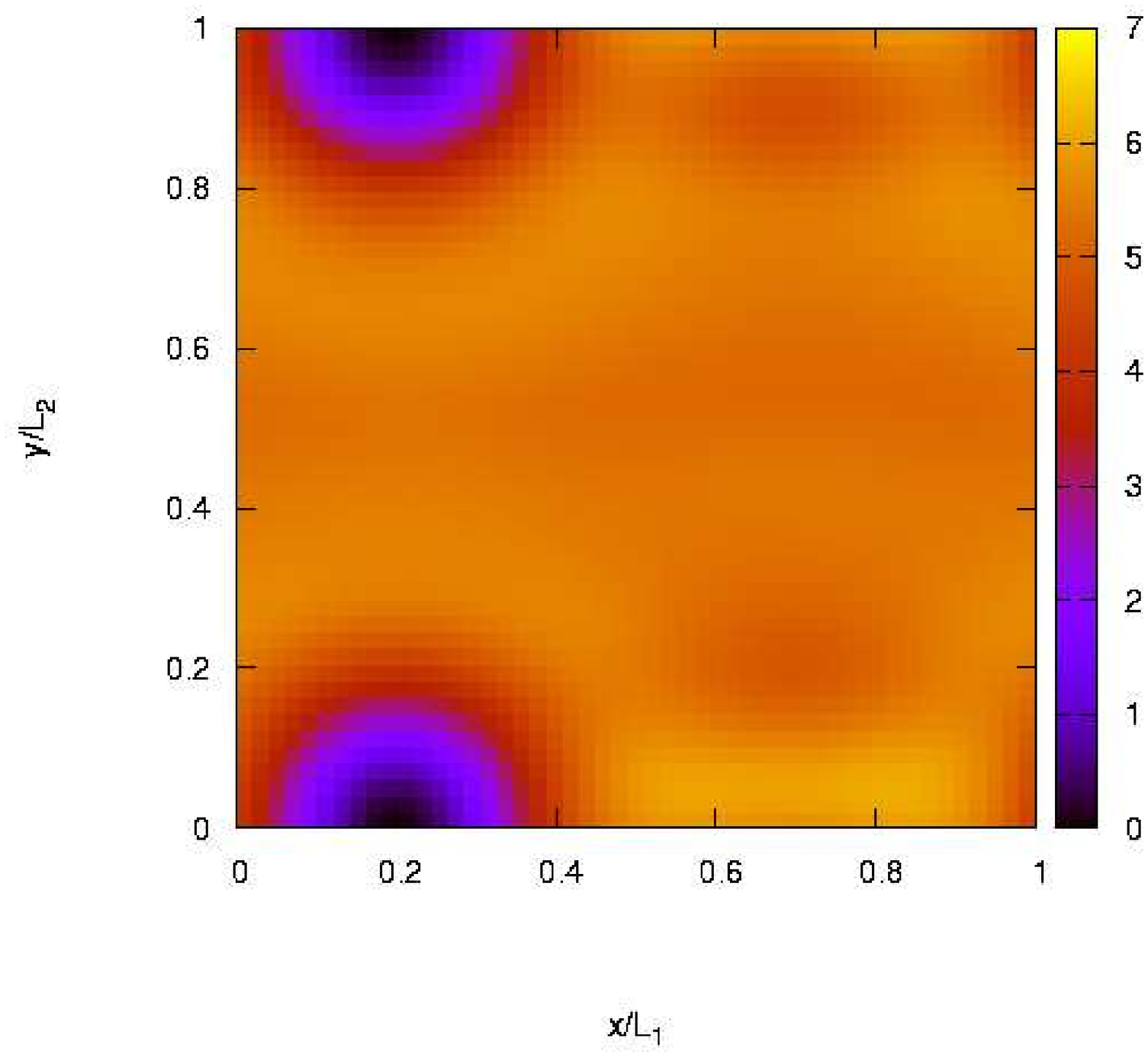}\includegraphics[scale=0.4]{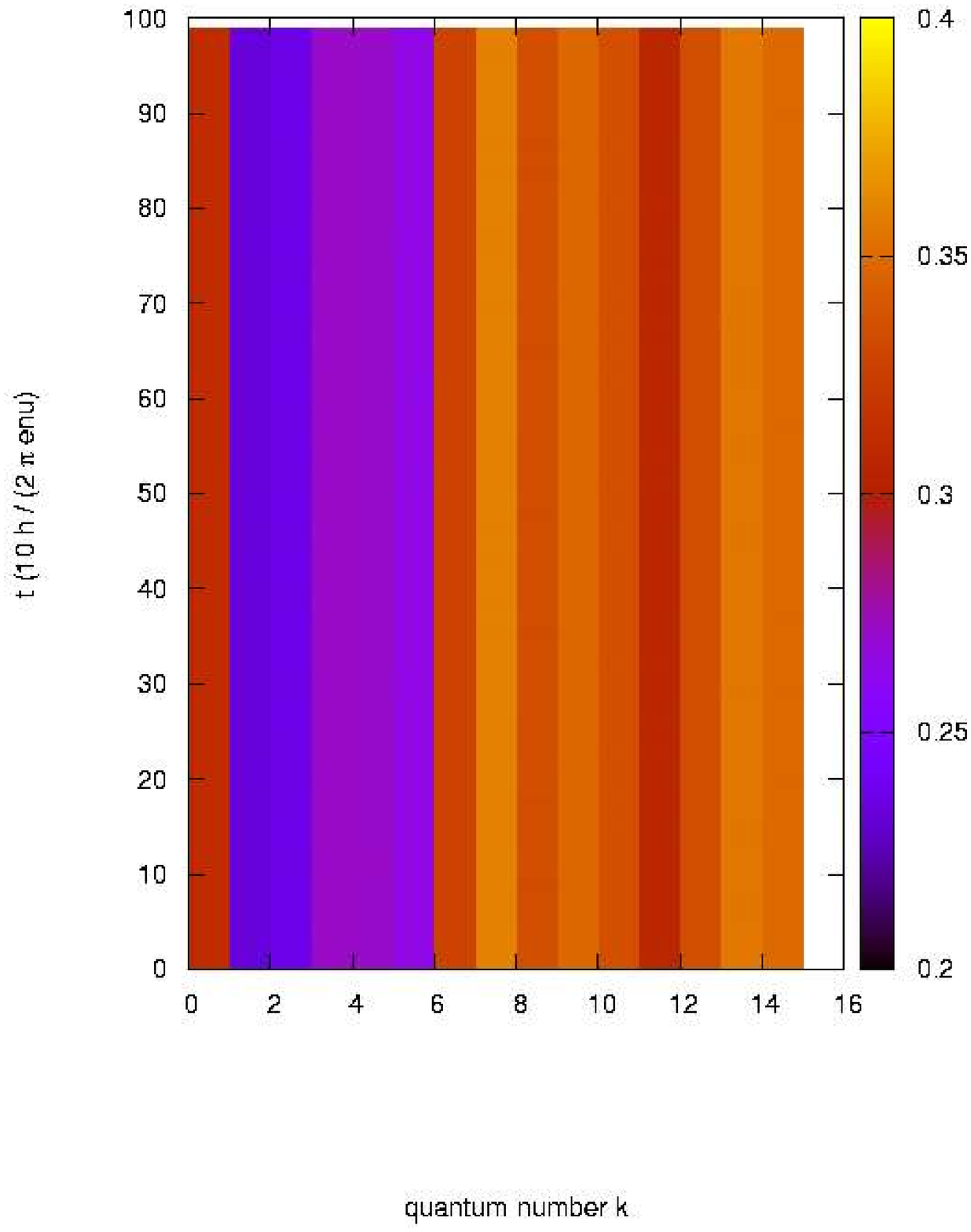}
  \caption{Left: Density of the initial state for $t=0$ with a quasihole at (0.2,0.0) (looking similar for all potentials).
           Right: Single-particle occupation probability $n_k(t)$ depending on time for the system without a notch.}
\label{fig:dnInitial}
\label{fig:occ0.00}
\end{figure}
\end{center}

\begin{center}
\begin{figure}[htbp]
  \includegraphics[scale=0.4]{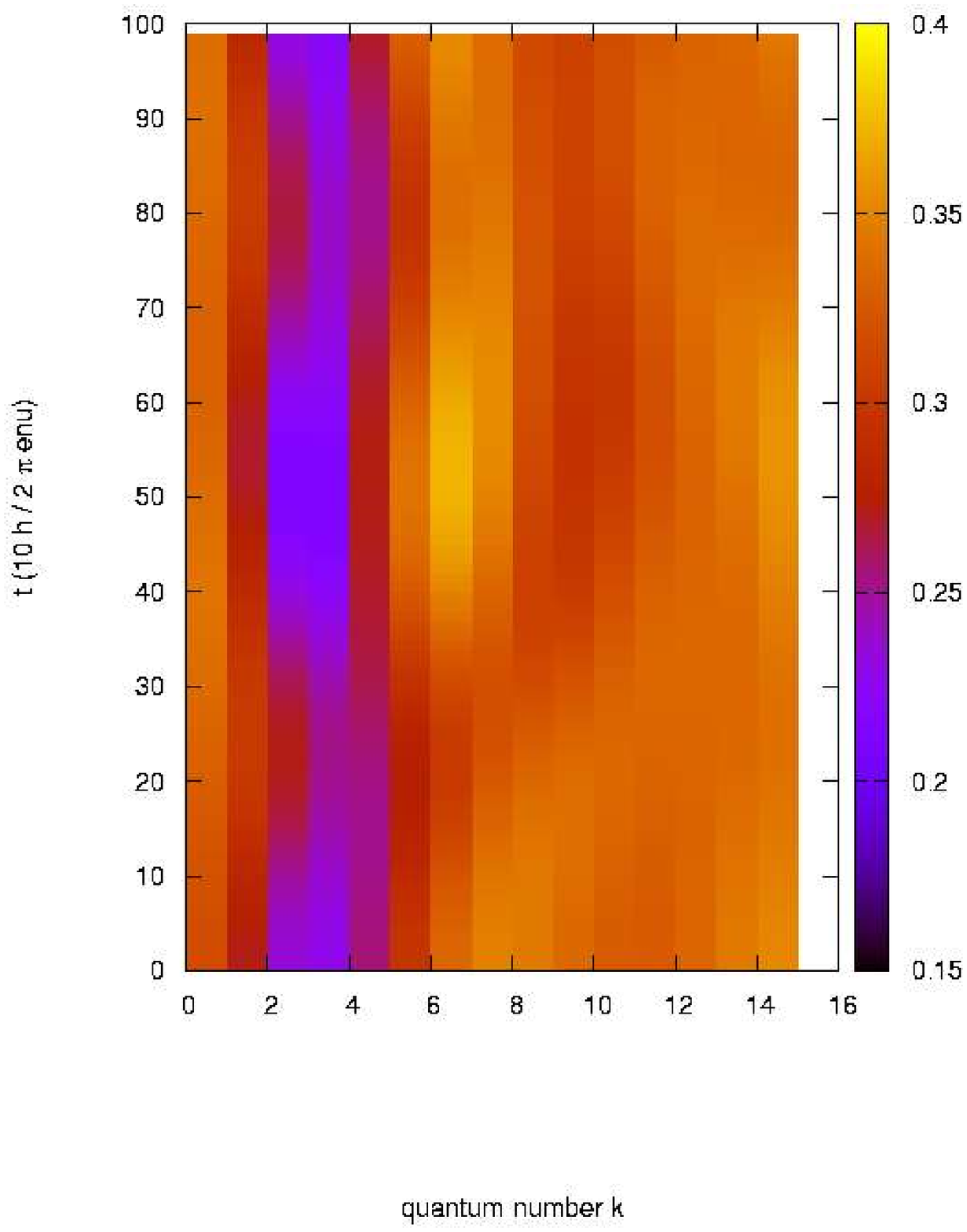}
  \caption{Time evolution of the single particle occupation numbers $n_k(t)$ for a quasi hole initially at $(0.2,0.0)$.
	   Depth of the notch is $0.02 \mbox{enu}$.}
\label{fig:occ0.02}
\end{figure}
\end{center}

\begin{center}
\begin{figure}[htbp]          
  \includegraphics[scale=0.4]{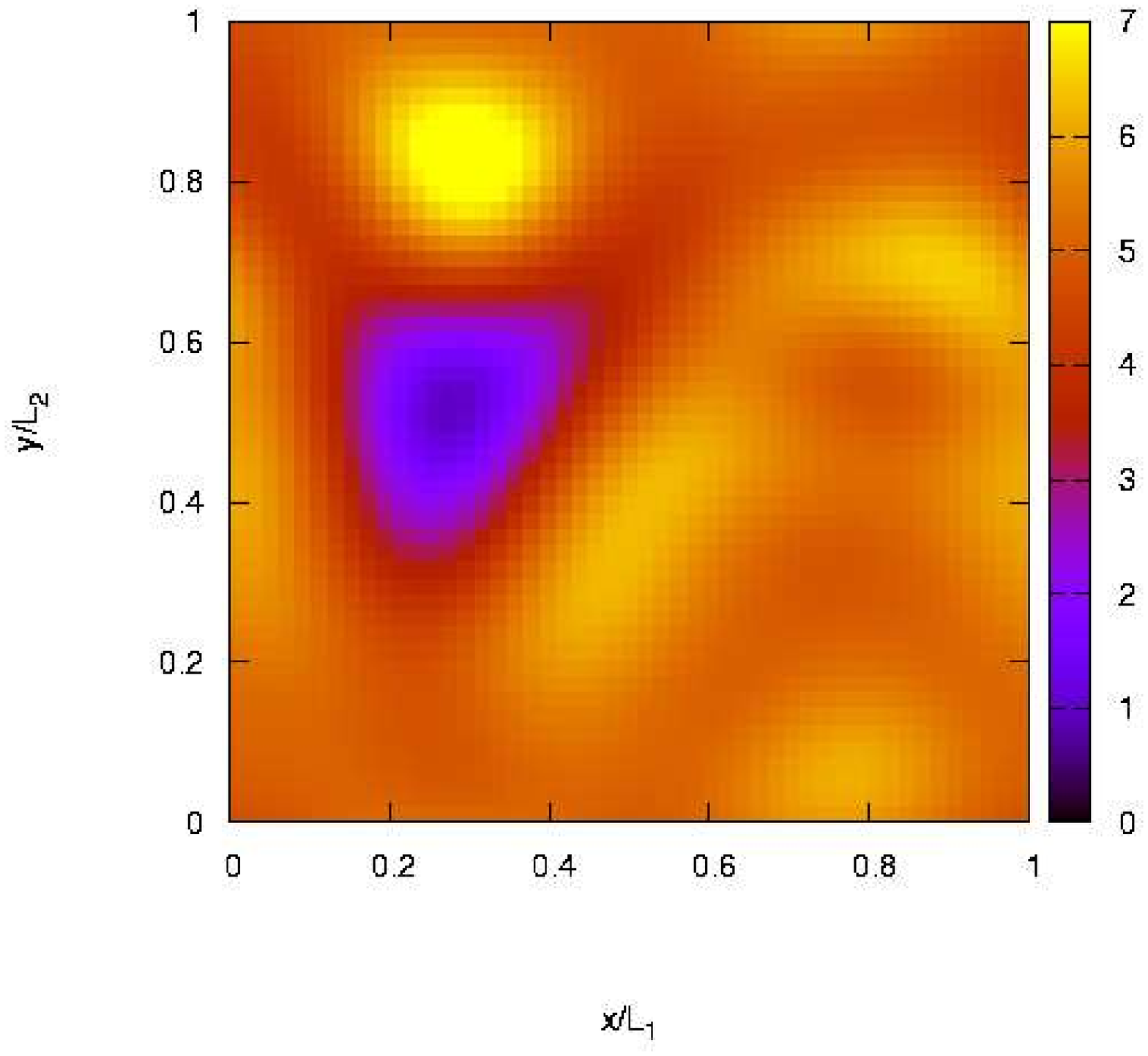}\includegraphics[scale=0.4]{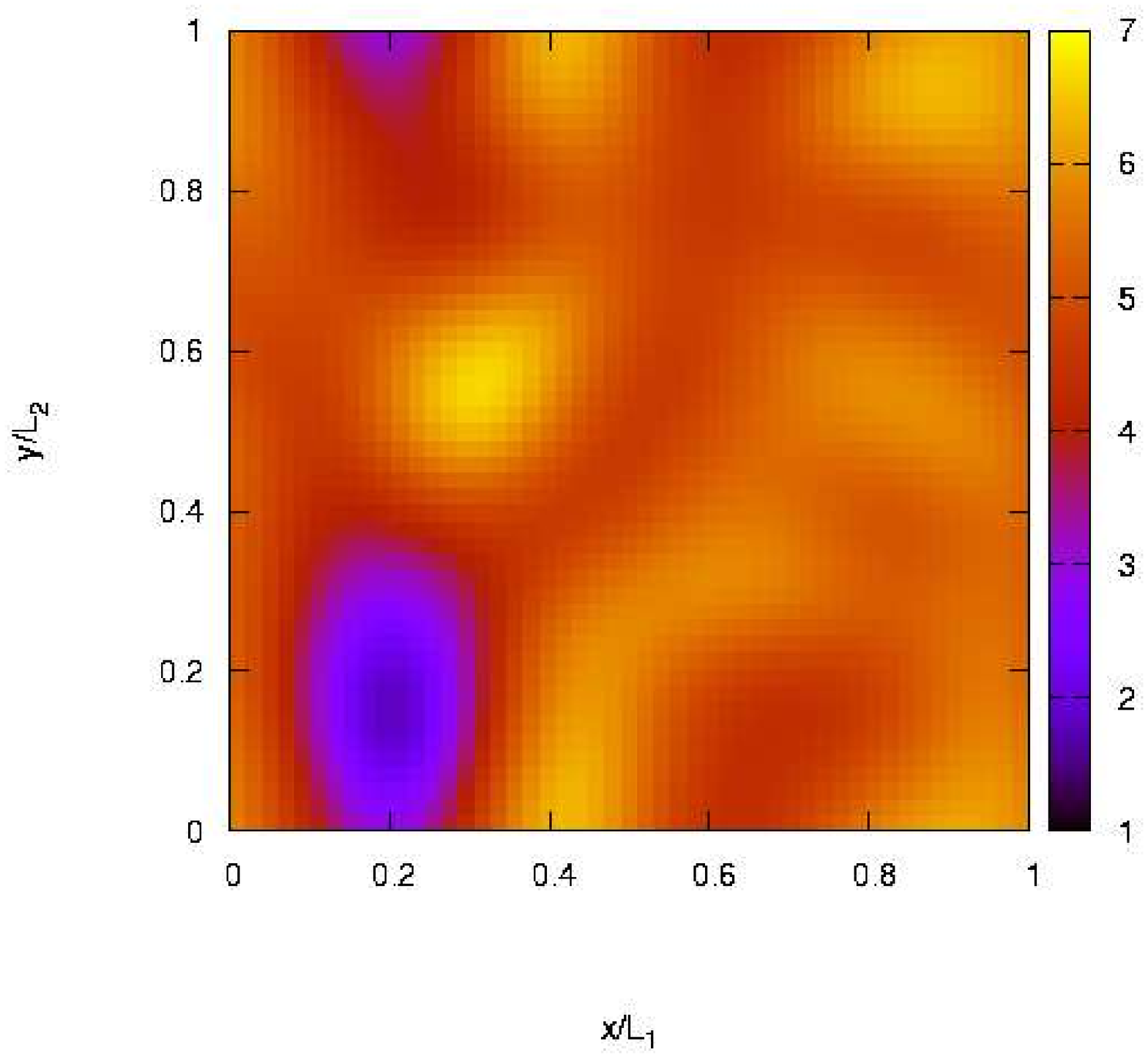}  
  \caption{Left: Density at $t=300 \frac{\hbar}{\mbox{enu}}$ in a system with a shallow notch (depth $0.02 \mbox{enu}$).
	The quasihole is ``chased'' by some charge accumulation centered at $\simeq (0.30,0.83)$
        Right: Density at time $t=500 \frac{\hbar}{\mbox{enu}}$ of the same system. The charge accumulation is decaying again, the quasihole almost rearrived at its starting point.}
\label{fig:dn0.02_t30_t50}
\end{figure}
\end{center}

\begin{center}
\begin{figure}[htbp]          
	\includegraphics[scale=0.4]{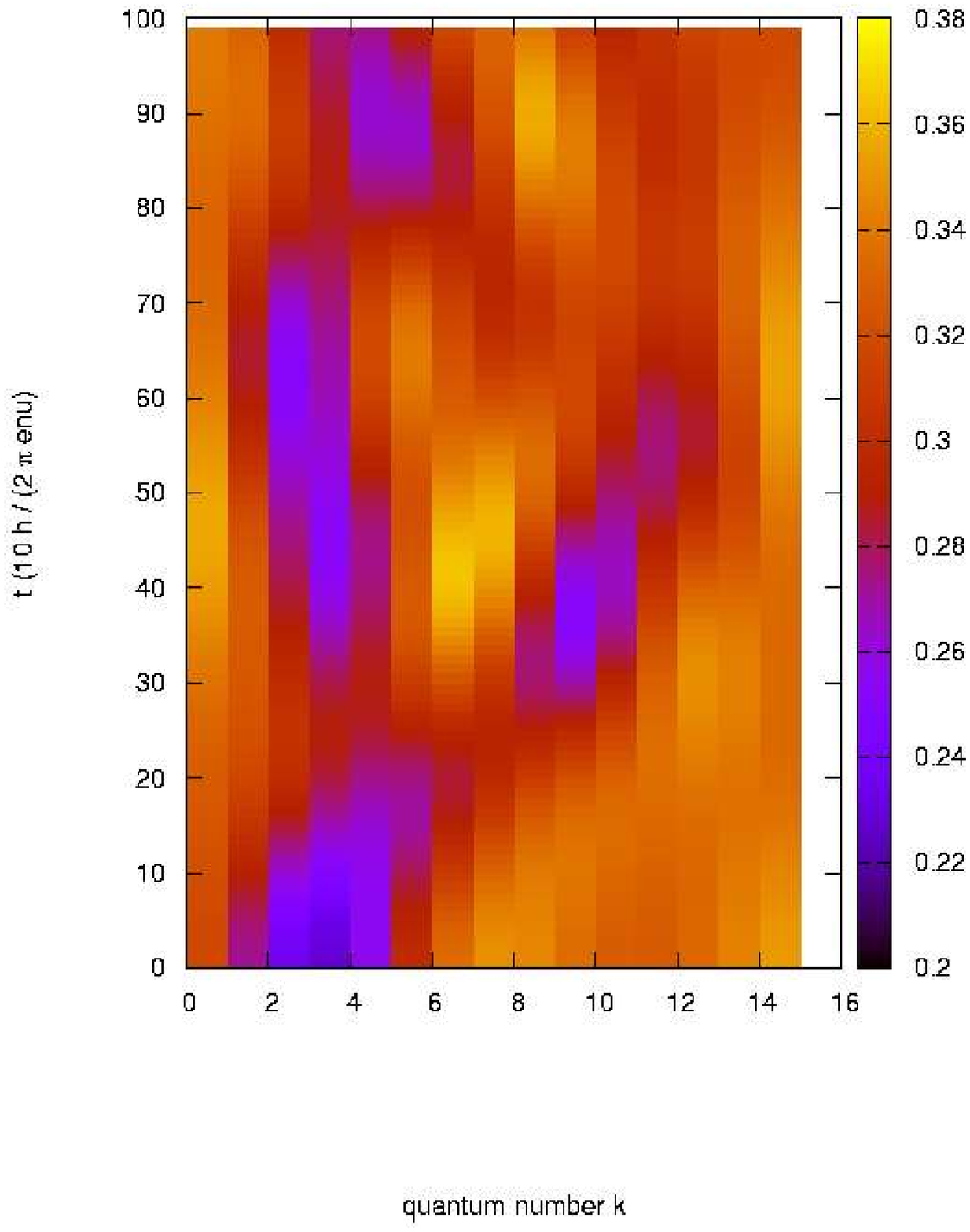}\includegraphics[scale=0.4]{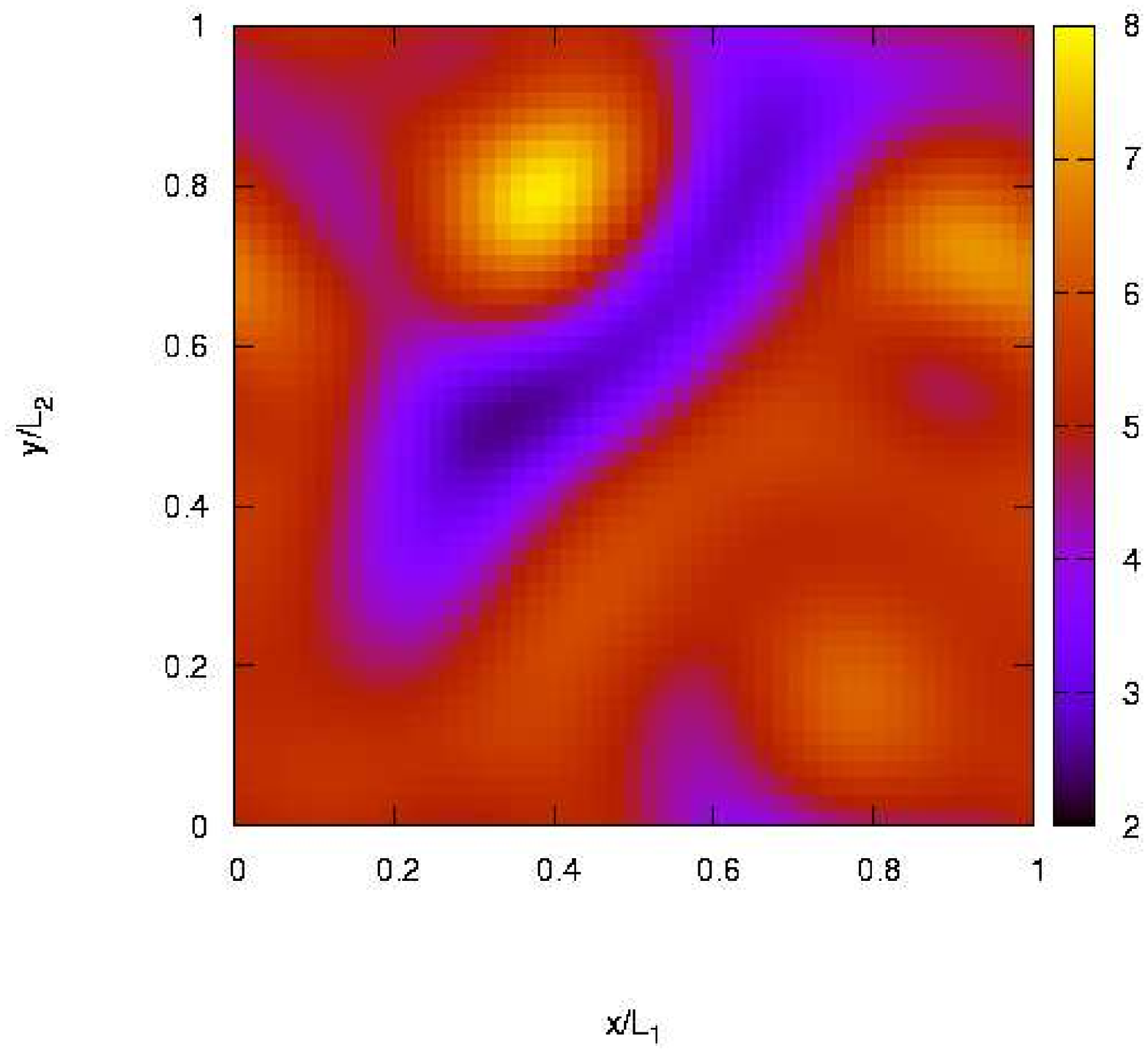}
\caption{Left: Time evolution of the occupation numbers for the quasihole and a depth of notch $0.04 \mbox{enu}$. 
         Right: Density at time $t=380 \frac{\hbar}{\mbox{enu}}$ for the same system: The quasihole is spread out along the equipotential line.}
  \label{fig:dnMidt380}
\label{fig:occ0.04}
\end{figure}
\end{center}

\begin{center}
\begin{figure}[htbp]          
  \includegraphics[scale=0.4]{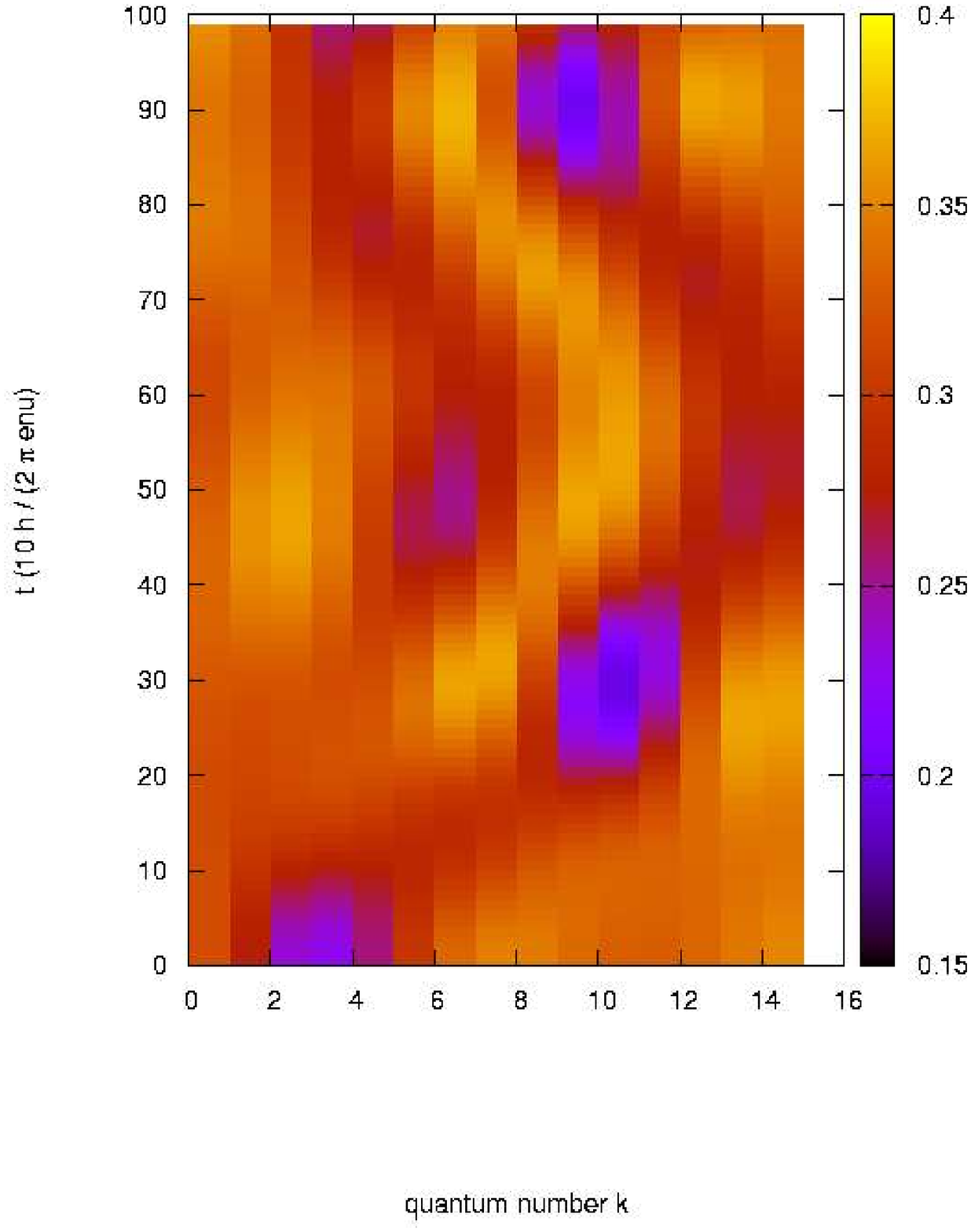}\includegraphics[scale=0.4]{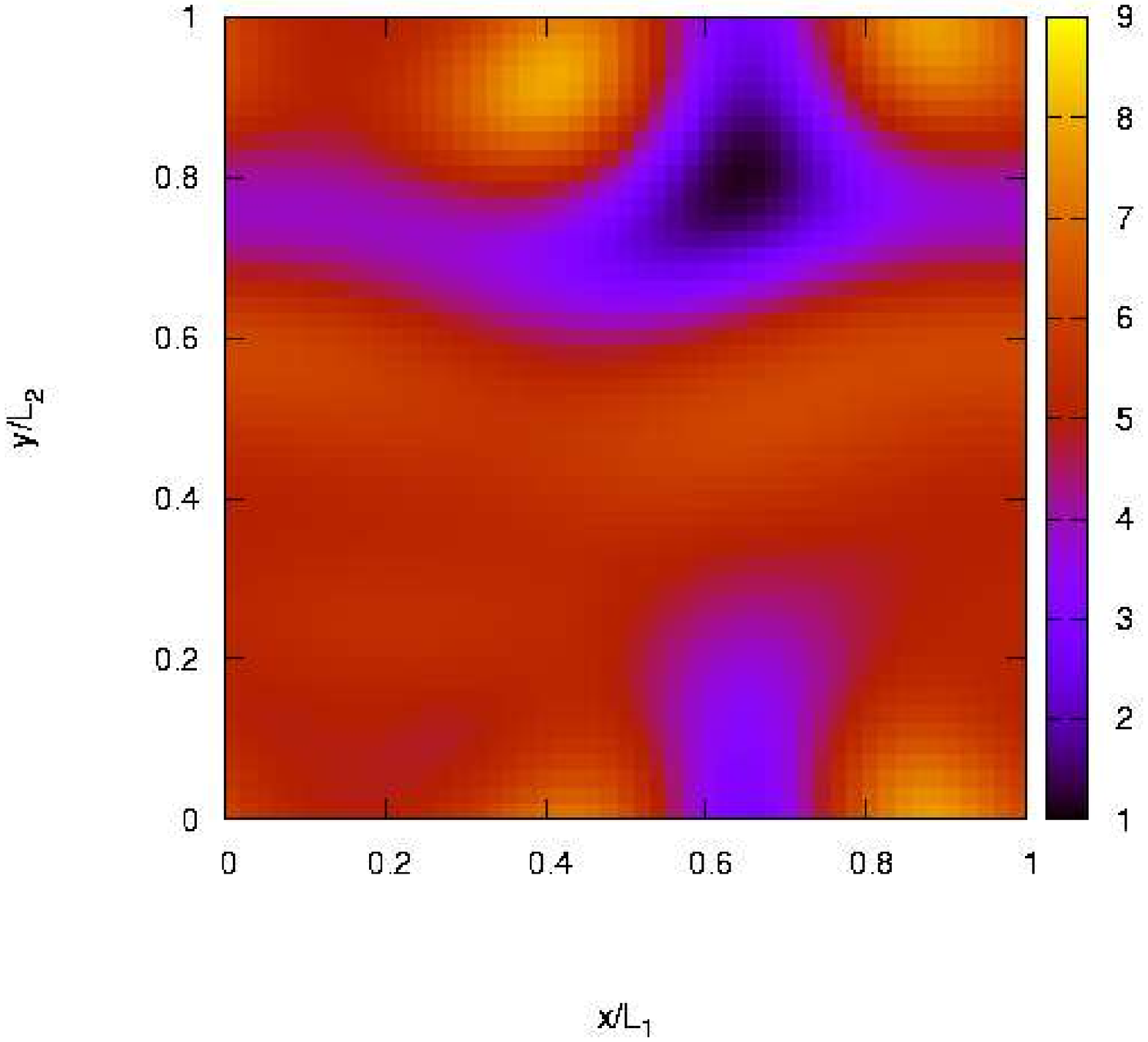}
  \caption{Left: Time evolution of occupation numbers for the quasihole initially at (0.2,0.0) for a notch of depth $0.08 \mbox{enu}$.
	   Right: Density for $t=280 \hbar/\mbox{enu}$ for the same system. The quasihole around $\simeq (0.65,0.80)$ has passed the barrier.}
\label{fig:occ0.08}
\label{fig:dnDeept280}
\end{figure}
\end{center}

The density of an initial state is found in Fig. \ref{fig:dnInitial}. This state is created for each of the potentials separately by application of the quasihole creation operator to the ground state of the $N_e/N_s=5/15$-system with the respective barrier. The densities however all look very similar to Fig. \ref{fig:dnInitial}, which again is due to the small impact of a shallow barrier on the incompressible system.
Like in section \ref{sec:QHcoulomb}, the time evolution of the single particle occupation numbers is computed. As an illustration that goes beyond occupation numbers, for distinct points in time the electronic densities for the systems are calculated. 
At time $t=0$ figures \ref{fig:occ0.00} through \ref{fig:occ0.08} look similar: The quasihole can be identified with the quantum numbers $1\ldots5$ which are less occupied than the average ($\frac{1}{3}$) which results in a density similar to the one in Fig. \ref{fig:dnInitial}.
Fig. \ref{fig:occ0.00} illustrates the occupation numbers in the case of no notch inside the barrier. The barrier thus is translationally invariant along the y-axis. This means that it cannot cause any mixing of states with different $k$, since the crystal momentum in y-direction is conserved. However,
the barrier causes a dispersion in the single particle states $E(k)$ which makes the quasihole drift downward in parallel to the barrier. This drift cannot be infered from Fig. \ref{fig:occ0.00}, but it was confirmed by calculating the density as a function of time. The formal reason for this drift is the phase difference the expansion coefficients acquire due to the dispersion.
If we cut a shallow notch into the barrier (compare Fig. \ref{fig:PotentialNoHoleLowHole}, up right), mixing of single particle states takes place
and causes the occupation numbers to fulfill a more interesting evolution, as found in Fig. \ref{fig:occ0.02}. Starting at the same initial position as before, the quasihole seems to move along a periodic trajectory that has its closest approach to the barrier at approximately $250\ldots300 \hbar / enu$ and reaches its farthest distance at $t=0$ and $t \simeq 550 \hbar / enu$. This is in agreement with the expectation that it should follow the slightly curved, periodic equipotential line it started on. Fig. \ref{fig:dn0.02_t30_t50} (right) is a plot of the density at $t=500 \hbar / enu$, where the quasihole has almost reached its starting point again, due to the periodic boundary conditions. This explains the periodicity in the occupation numbers.

Another interesting finding in Fig. \ref{fig:occ0.02} (left) are the points in time, where the signature of the quasihole becomes clear ($t=0$, $t\simeq550  \hbar / enu$) and those where it appears to be washed out ($t=300 \hbar / enu$). The density for the point $t=300 \hbar/ \mbox{enu}$ where the quasihole is not clearly visible in the occupation numbers reveals Fig. \ref{fig:dn0.02_t30_t50}, left.
The cause for the appearent disappearence is due to an accumulation of charge directly obove the quasihole. This amount of charge obviously compensates locally for the quasihole's charge deficit and therefore makes the dip in the occupation numbers vanish. For later times ($t=500 \hbar/\mbox{enu}$) this accumulation of charge seems to decay again. (see Fig. \ref{fig:dn0.02_t30_t50}, right).
The moving quasihole seems to create some excitations of the system as a side effect. The fact that the charge deficit of the quasihole is almost cancelled out at $t=300 \hbar /\mbox{enu}$ by the charge accumulation renders the speculation temptative to assume the accumulation to be a quasielectron. But apart from this indication there are is no direct verification.
A notch of depth $0.04 \mbox{enu}$ leads to an equipotential line that connects the starting point of the quasihole with the potential's saddle point (see Fig. \ref{fig:PotentialNoHoleLowHole}). The interesting point is the quasihole's behavior when arriving at the saddle point, where it has the possibility to pass by the barrier as well as it can cross it.
In Fig. \ref{fig:occ0.04} this point is reached at about $t=200 \hbar / enu$. Until this time the quasihole clearly moves towards the saddle point. Shortly later the plot suggests that the hole splits or spreads out. At $ t \simeq 380 \hbar / enu$ there are two areas, one to the left the other to the right of the barrier, in which the occupation numbers are lowered. Comparison with the density in Fig. \ref{fig:dnMidt380} confirms the assumption that the formerly spacially localized quasihole is now spread out along the equipotential line. The part left to the barrier is moving downward the right part is moving upward. Due to the periodic boundary conditions it is hard to draw conclusions for later times, since several images of the quasihole, that have passed the borders of the unit cell, are superposed.

Obviously, this kind of potential causes a more complicated behavior. Namely the splitting of the quasihole at the saddle point cannot be understood classically. One might think of it as a superposition of two states, the one of which having a quasihole left to the barrier, the other one one to the right. However, this interpretation was not checked, but could be investigated by projecting the system's wavefunction $\Psi(t)$ to the one or the other final state.
Finally, in the case in which the notch is very deep the behavior resembles qualitatively that of the shallow notch, only rotated by 90 degrees:
The quasihole travels along the equipotential line and arrives on the right side of the (widely opened) barrier. At $t=280 \hbar/enu$ in Fig. \ref{fig:occ0.08} the quasihole can be identified as the dip in the occupation numbers $9\ldots 11$ . Comparison with the density in Fig. \ref{fig:dnDeept280} shows the hole which for the most part is on the right of the barrier. Here again, to the left and to the right of the quasihole the density shows two areas of charge accumulation, similar to case of the shallow notch mentioned earlier.

To end with, the following conclusions can be drawn from this section.
Shallow potentials can be regarded as sources of dispersion, but do not affect the density of the incompressible state to a big extent. This dispersion leads to an $\vec E \times \vec B$-drift of charged quasiholes.
Inserting a quasihole on an equipotential line that doesn't come close to a saddle point of the potential, the motion of the quasihole coincides with the classically expected $\vec E \times \vec B$-drift.

What exactly happens at the crossing point of equipotential lines is unclear, but a superposition of two states both carrying a quasihole on different sides of the barrier seems plausible. As seen from the system of two delta potentials, tunneling of quasiholes is possible. The situation found here is not much different, since classically the quasihole must strictly stay on the equipotential line it started on. If however equipotential lines left and right of the barrier come close to each other near a saddle point of the potential, communication between quasihole states on the left and those on the right takes place and the quasihole can partly pass over to the other side. Following this process, the part of the quasihole resting on the left side travels on in its original direction, while the part on the right moves in the opposite direction along the equipotential line. The quasihole being spread out along the equipotential line (Fig. \ref{fig:dnMidt380}, density) resembles the behavior in the system of two delta potentials, where there was a high occupation probability for the quasihole between the two peaks. Further investigations should focus on the state obtained after the quasihole crossed the saddle point of the potential. This could be done by projecting it to an assumed final state.

In these inhomogeneous systems the quasiholes suffer from dispersion effects due to their spacial extent, which causes a deformation of the quasihole. Nevertheless, in systems where the quasihole does not come close to any saddle point of the potential, it did not decay during the time to cross the unit cell. Larger systems with shallow potentials should lead to less distortion, because the extent of the quasihole will be realtively smaller. A relatively narrower spacial extent also means that the curvature of the dispersion $E(k)$ has a lower effect on the quasihole. This is because the $\vec E \times \vec B$-drift of a single-electron state is proportional to $\frac{\partial E}{\partial k}$ (section \ref{sec:HellFeyn}, Eq. \ref{eq:DispersionV}) and in a bigger system the extent of a quasihole in $k$-space is relatively narrower as well. Thus all single-particle states ``carrying'' the quasihole move with approximately the same drift velocity and hence the deformation of the quasihole is low.

It would also be interesting to have a closer look at the described effect of charge density accumulation along the trace the quasihole moves.
The question here is to check if the maxima in the electronic density can be identified with quasielectrons or if they are charge density waves excited by the charged quasihole moving through the system. They were found to be created spontaneously and also decayed much faster than the quasihole (which is stable, apart from its deformation). A quasielectron therefore seems to be the less plausible explanation, since we would expect its features to be more similar to those of a quasihole.
\subsubsection{Strong potentials}
\label{sec:Strongpot}
The weak potential in the previous section did not force the density of the system to zero inside the barrier. Therefore it is not
an appropriate choice to model a constriction like a point contact. To create an effective barrier for the electrons a stronger potential of the
order of their chemical potential is needed.

\begin{figure}[htbp]          
  \includegraphics[scale=0.4]{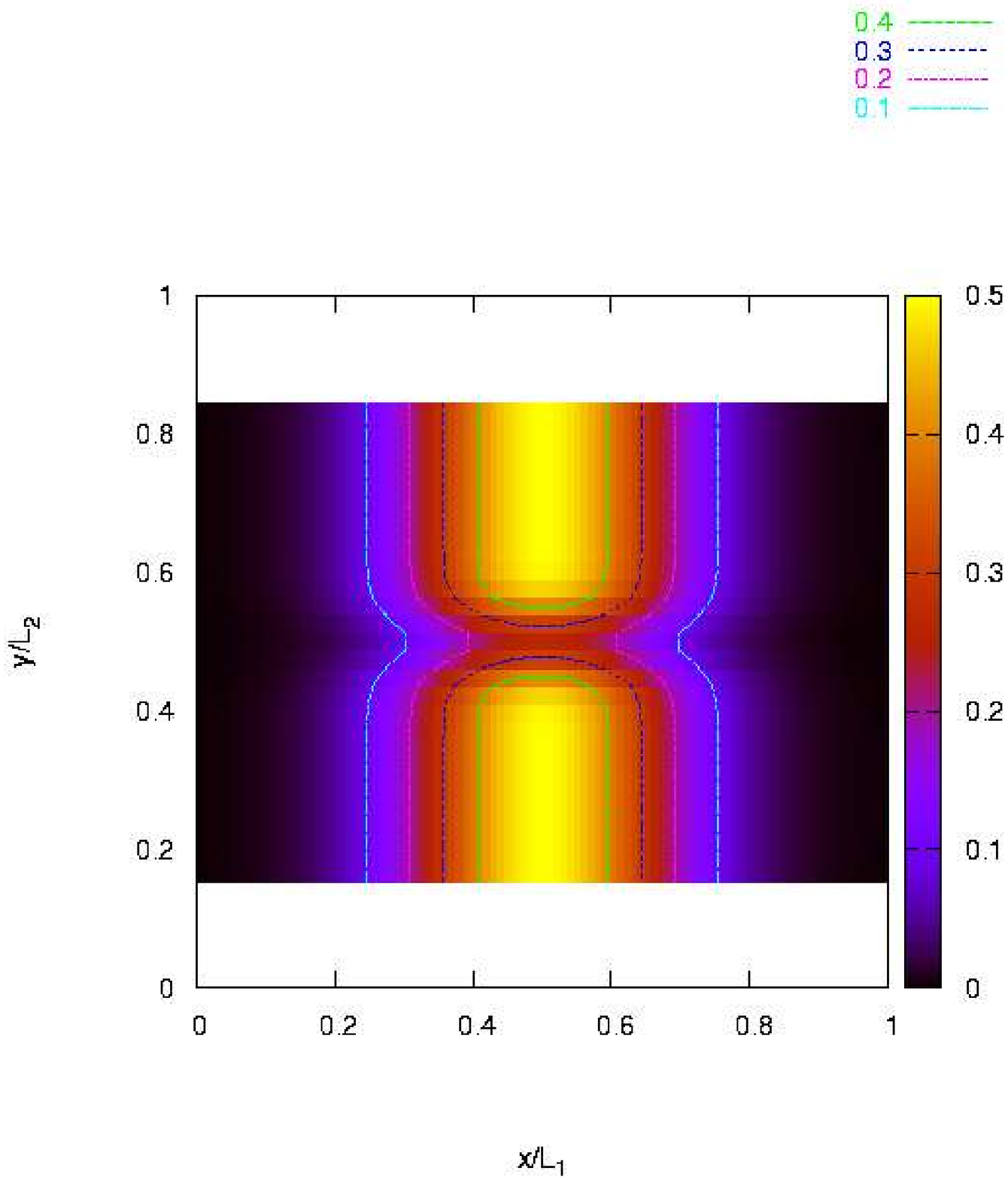}\includegraphics[scale=0.4]{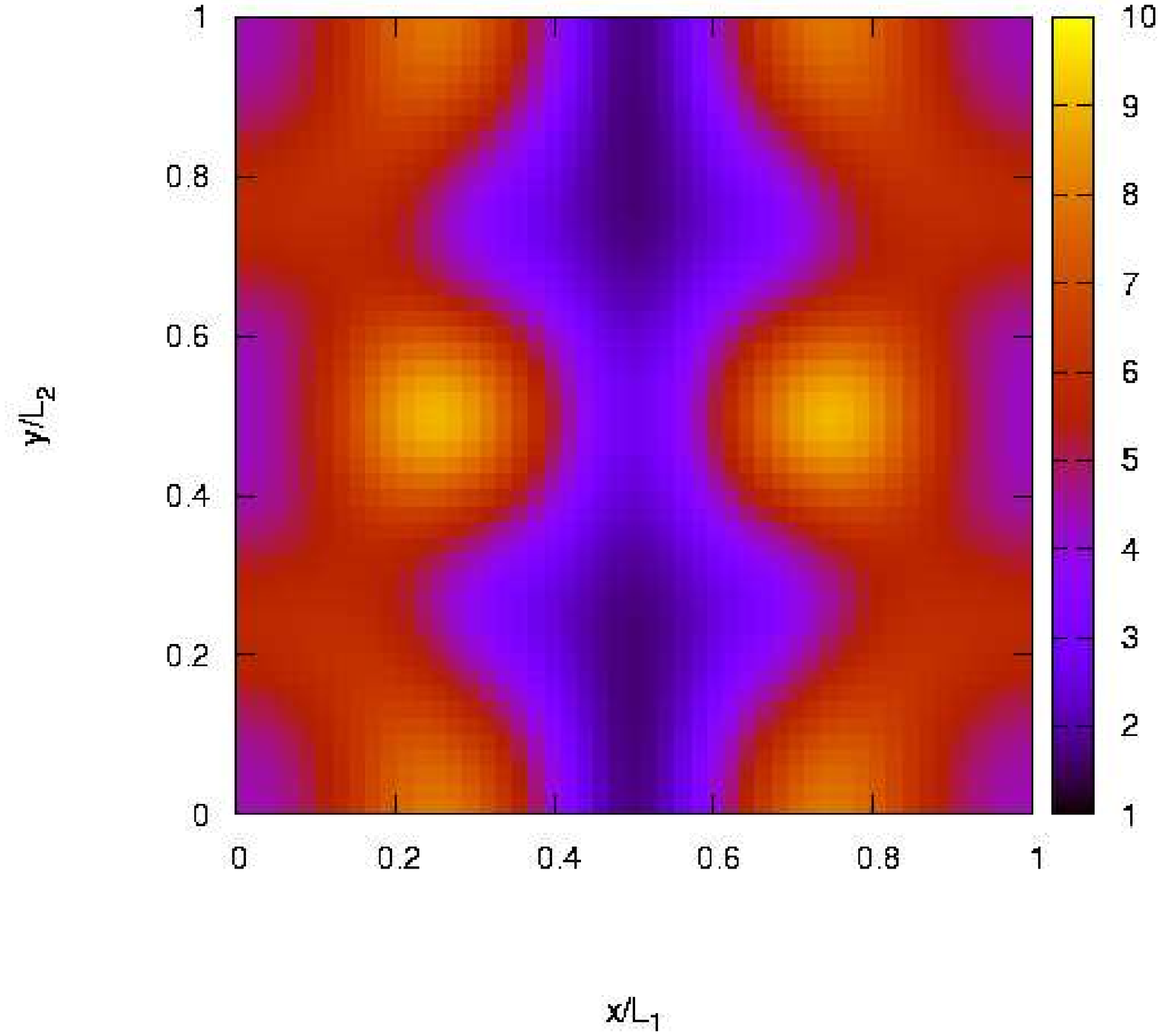}
  \caption{Left: Plot of the potential with parameters given in Tbl. \ref{tab:ParamStrongPot}. Right: Resulting electronic density of the ground state of the 5/15-system with hard-core interaction.}
  \label{fig:PotentialStrong}
  \label{fig:GSbarrier5-15}
\end{figure}

A system with a barrier enforcing a strong dip in the density is investigated here. This barrier can not yet be regarded as a tunneling barrier since the density below it is still too high. Compared to the chemical potential of $2.91 \frac{e^2}{\epsilon l_0}$, its height is roughly $\frac{1}{6}$. The $\nu=\frac{1}{3}$ filled system again has 5 electrons and 15 flux quanta. The electrons interact via hard-core interaction. An impression of the potential can be gained from its contour plot in Fig. \ref{fig:PotentialStrong}.
%
%
\begin{table}[htbp]
\begin{tabular}{|l|l|l|l|}
\hline
Height of barrier $(\frac{e^2}{\epsilon l_0})$ & Width of barrier $(L_1)$ & Depth of hole $(\frac{e^2}{\epsilon l_0})$ & Width of notch $(L_2)$\\
\hline
0.5 & 0.2 & 0.25 & 0.05\\
\hline
\end{tabular}
\caption{Parameters of the potential to create a tunneling barrier. The parameters are defined in Eq. (\ref{eq:VwallGauss}) and (\ref{eq:Vnotch}).}
\label{tab:ParamStrongPot}
\end{table} 
The eigenenergies of the ground states of a system with 15 and 16 flux quanta (see Tbl. \ref{tab:GSenInhomog}), respectively, show that in a system with a barrier it is more favorable to have more flux quanta. This leads to a negative energy per flux quantum (adding a flux quantum lowers the energy).
It is understandable, if one keeps in mind that the incompressible state of a 5/15-system needs all the available zeros of the wavefunction to reduce the hard-core interaction and thus does not have any freedom to react to the barrier. The additional flux quantum in turn allows to fix a zero near the barrier's peak to reduce the potential energy and still has the freedom to reduce the interaction to zero. In the $5/16$-system the expectation value $\langle V_{hard-core} \rangle_{GS}$ of the interaction potential decreases tremendously by a factor of approximately 9 (see Tbl. \ref{tab:GSenInhomog}) compared to the $5/15$-system. So there is a tradeoff between reduction of potential energy and interaction energy.

When looking at the excitation energy of a quasihole, these energies are always given with respect to the ground state energy of the 5/16-system with barrier. Another point to make is that, in contrast to the homogeneous system, the energy of a quasihole state depends on the position where the hole is created. This will be surveyed in the following section.
\begin{table}[htbp]
\begin{tabular}{|l|l|l|}
\hline
System & GS energy $(\frac{e^2}{\epsilon l_0})$ & interaction energy  $\langle V_{\mbox{short-range}} \rangle_{GS} \; (\frac{e^2}{\epsilon l_0})$\\
\hline
5/15 H & 0.803164 & 0.179325\\
5/16 H & 0.676082 & 0.020571\\
\hline
\end{tabular}
\caption{Ground state energies and interaction energy $\langle V_{\mbox{short-range}} \rangle_{GS}$ of systems with 5 electrons and 15 or 16 flux quanta, respectively. The potential is defined by the parameters from Tbl. \ref{tab:ParamStrongPot}. By inserting an additional flux quantum, the interaction energy can be reduced by approximately a factor of 9.}
\label{tab:GSenInhomog}
\end{table} 
\subsubsection{Excitation spectrum in dependence of the quasihole's position}
\label{sec:ExEnDep}
\begin{center}
\begin{figure}[htbp]          
  \includegraphics[scale=0.6]{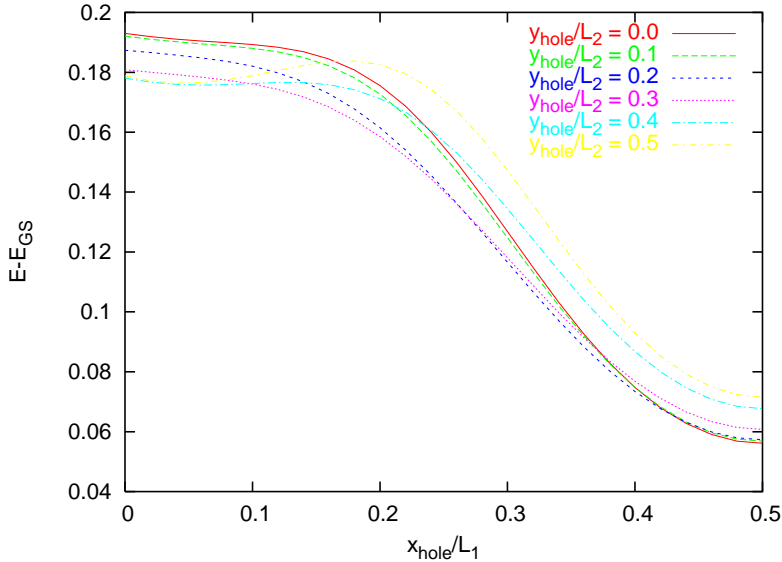}
  \caption{Excitation energy $E$ for quasiholes inserted at different positions $(x_{hole},y_{hole})$ in the unit cell. The quasihole was created by application of the creation operator to the 5/15-system's ground state. The excitation energy is calculated according to equation (\ref{eq:ExcEn}).}
  \label{fig:ExcEnergy}
\end{figure}
\end{center}
Fig. \ref{fig:ExcEnergy} shows the dependence of the excitation energy on the position the quasihole was created. It reveals that
the energy of a quasihole inside the barrier ($x_{hole}=0.5, y_{hole}=0.0$) is close to the groundstate energy of the system.
Holes created further away have higher energies. This implies, that the ground state of the $5/16$-system should be a state, in which the
quasihole is somehow distributed inside the barrier. The variation of the energy with the $y$-coordinate of the quasihole is quite low.
The reason is the relatively narrow notch of $0.05 L_2 \simeq 0.5 l_0$. The quasihole, having a characteristic size of $l_0$, cannot resolve this short ranged potential variation. The structure in the density varying on the scale of $l_0$ in Fig. \ref{fig:GSbarrier5-15} corroborates this.

The density in Fig. \ref{fig:GSbarrier5-15} shows that although the barrier does not yet force the density to zero, it has a strong impact on the areas besides the peak of the potential. Here, the electrons tend to reside in two stripes parallel to the barrier with a highest probability close to the notch. The local filling factor is about 20 to 100 per cent higher than $\frac{1}{3}$. Whether the correlations ---which are crucial for the fractional quantum Hall effect--- still exist in the inhomogeneous system, can be answered by looking at the interaction energy of the hard-core interaction for this state.
According to Tbl. \ref{tab:GSenInhomog}, the interaction energy of $0.179 enu$ takes round about 22 per cent of the ground state energy.
The comparison to the system with 16 flux quanta shows, that here the interaction is only $0.021 enu$, which takes up only 3 per cent of
the total energy. 
This again fortifies the assumption, that the system with one flux quantum in addition has more freedom to react on the barrier than the system at $\nu=\frac{1}{3}$ filling. 
This agrees with the comparison of the densities in Fig. \ref{fig:GS5-16HoleEx} (left) of a system with 16 flux quanta to the one in Fig. \ref{fig:GSbarrier5-15} in the case of 15 flux quanta: The reaction on the barrier potential is much more concentrated near the actual potential in the first case, while in the latter case the potential has a more long-range impact on the density. In addition, the density below the barrier takes lower absolute values in the case of an additional flux quantum. 
In conclusion it can be stated that an additional flux quantum leads to a relaxation of the competition between interaction and potential energy.
\subsubsection{Quasihole excitation by a delta potential}
\begin{figure}[htbp]
\includegraphics[scale=0.4]{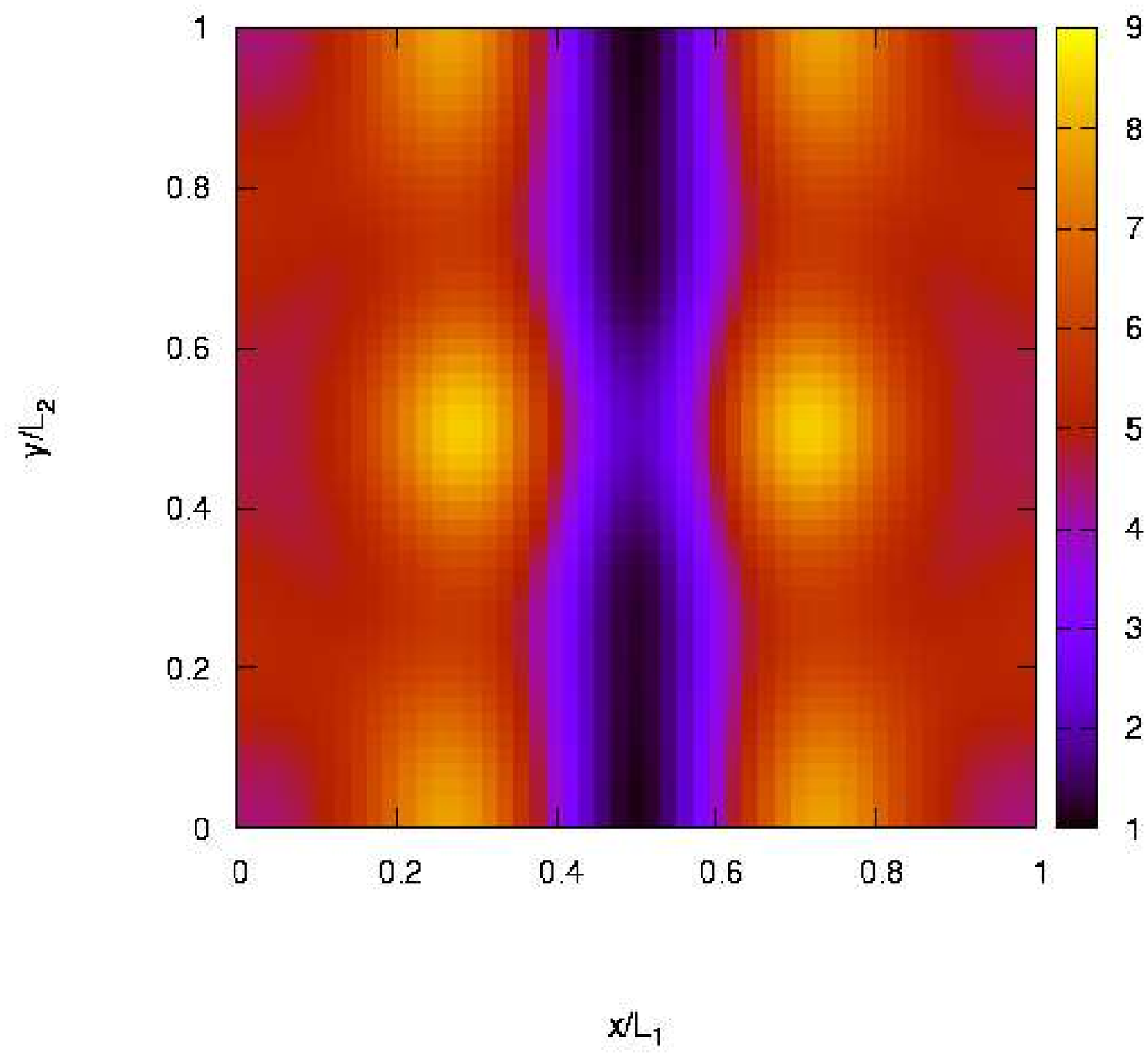}\includegraphics[scale=0.4]{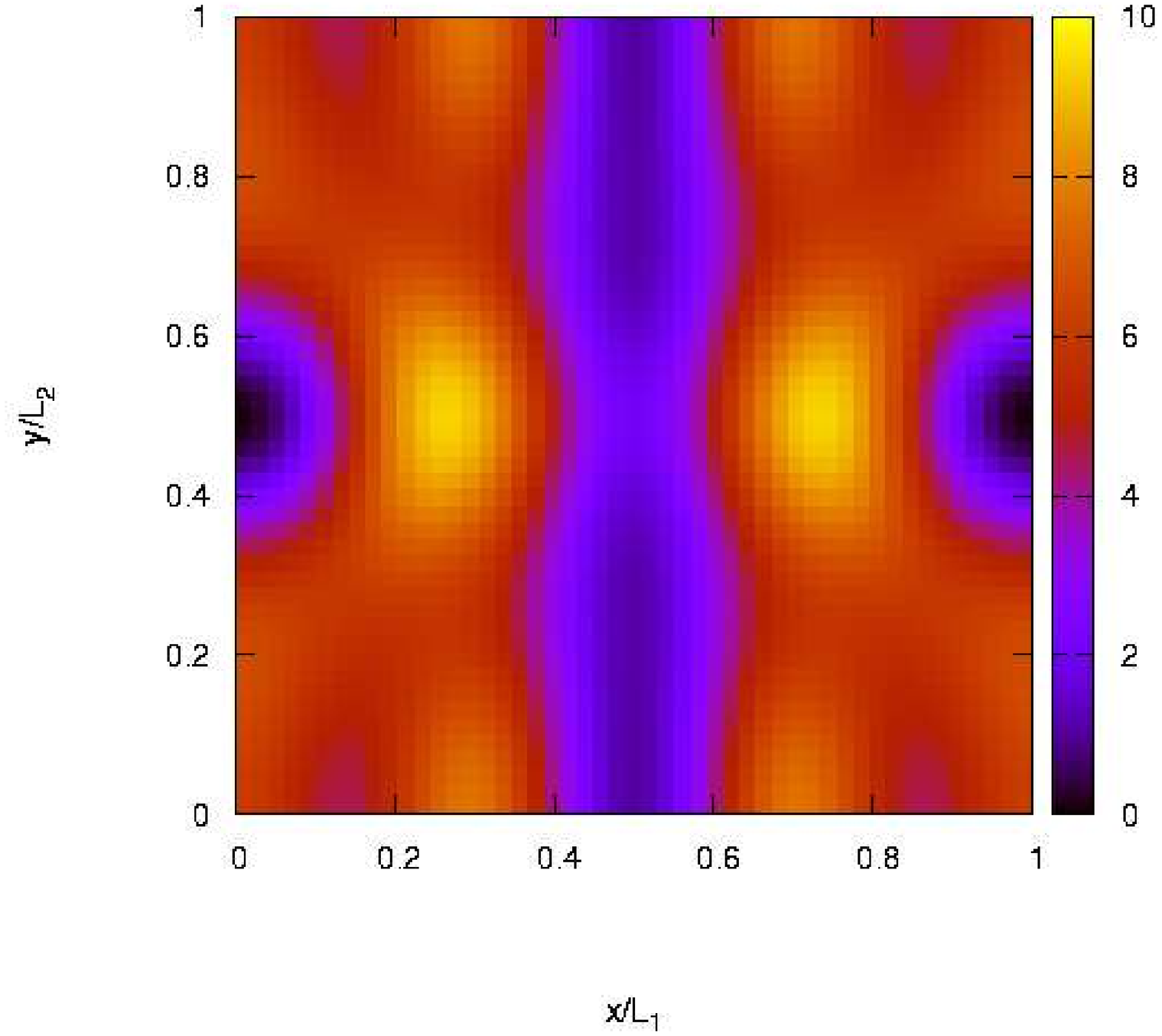}
\caption{Left: Density of the ground state of a $5/16$-system with the potential defined by Tbl. \ref{tab:ParamStrongPot}.
         Right: a quasihole excitation is created by a delta potential at (0.0,0.5) in this system. The Excitation energy is $E-E_{GS}=0.101099 \mbox{enu}$. The profile of the densities near the barrier are the same, only the absolute depth in the right plot is less, because now one quasihole is ``missing''.}
\label{fig:GS5-16HoleEx}
\end{figure}

Applying the method to create a quasihole by means of a delta potential introduced into the inhomogeneous system has the same effect as in the homogeneous case: One zero of the wavefunction is pinned at the delta's position. Similar to the case of homogeneous systems with Coulomb interaction, here the quasihole state created by this method results in a lower energy. As an example, a hole was created at ($x_{hole} = 0.0, y_{hole} = 0.5$). Its excitation energy above the ground state is $0.101099 enu$. In contrast the energy obtained for the operator-generated hole as shown in Fig. \ref{fig:ExcEnergy}, yellow plot at $x=0$, is about $0.18 enu$ above ground state.
Again we find the delta potential method resulting in a lower excitation energy compared to the operator generated state.
The reason for this is the quasihole creation operator that was deduced from the trial-wavefunctions for the homogeneous system. 
Applying it to the ground state of an inhomogeneous system results in a more or less good approximation for the quasihole state.
In contrast, diagonalizing with an additional flux quantum and the constraint of a fixed zero in the wavefunction allows the system to benefit
from its degees of freedom optimally.
Fig. \ref{fig:GS5-16HoleEx}, (right) shows the density for the state obtained in this way. 
In the vicinity of the barrier it almost looks identical to the state without a quasihole (left). One difference is the absolute value of the density below the barrier: The density rose in the case of the injected quasihole. This is in agreement with the explanation, that to minimize the potential energy, quasiholes are pinned below the barrier. Fixing one away from the potential must increase the density below the barrier.

\subsection{Overcoming the incompressibility at the constriction}
The incompressible nature of the ground state in the fractional quantum Hall regime has an impact on the system's reaction to an introduced 
barrier potential.
By comparing the density profile of ground states of systems with a barrier, one of which
having a filling factor of $\nu=\frac{1}{3}$ the other owning an additional flux quantum, we can see a different behavior.
It can be traced back to the incompressibility at $\frac{1}{3}$ filling where the chemical potential of a homogeneous system has a step. It is the reason for a finite amount of energy needed to create a pair of quasiparticles (quasihole and -electron) (see section \ref{sec:gsmu}). Introducing an external potential into this system, these particles are created and allow the system to adapt (if the potential energy is higher than the creation energy of a quasiparticle pair). But, due to the finite energy to pay, the system is rigid.
Moving off the fraction of $\frac{1}{3}$ by increasing the number of flux quanta, the excessive quasiholes are available in the system ``for free''. They make the system become more adaptive to the external potential. The densities for such two systems were already investigated in section \ref{sec:inhomogeneous}. Figure \ref{fig:GSbarrier5-15} depicts the case where the filling is $\frac{1}{3}$, whereas the left plot in figure \ref{fig:GS5-16HoleEx} shows the same system with an additional flux quantum. In the latter system, the area in which the density is lowered by the barrier is much more localized around the barrier potential. The system at $\frac{1}{3}$ in contrast shows a more long range impact of the barrier.

Another way to explain these different reactions of the systems can be attributed to the concurrence between the interaction energy and potential energy of the electrons. Thinking in terms of the wavefunction, we can quantify the interaction in case of a hard-core pseudopotential by the positions of the wavefunction's zeros in the relative electron coordinates. We have a vanishing hard-core interaction energy whenever it is possible to have a threefold zero in the wavefunction upon approach of two electrons. Clearly, an additional potential tends to attract zeros of the wavefunction in order to minimize the potential energy. In the case of a delta-peak potential it was shown that a single zero is pinned at the position of the delta peak. In turn, there is one zero missing to diminish the interaction to zero.
But, on the other hand, if there is one flux quantum above $\frac{1}{3}$ inside the system, this will be pinned by the delta inhomogeneity, leaving $3$ zeros per relative coordinate to minimize the interaction to zero.

In the following three subsections different approaches are tested for their applicability to create an effective constriction while maintaining
the correlations of the incompressible fractional Quantum Hall state.
\subsubsection{Pinning flux quanta by delta potentials}
\label{sec:deltapeakbarrier}
The idea arising from the previous observations is to build up a constriction out of several Dirac delta functions, each of which pinning one additional flux quantum. The density inevitably goes to zero at the positions of the delta potentials. By placing the delta potential adequately, it is possible to construct a wall-shaped barrier with a notch in it. Since there are 3 unbound flux quanta per electron remaining, the system can still exhibit Laughlin-like correlations with a vanishing interaction.

This was done for a system of 5 electrons interacting via the hard-core pseudopotential.
Three delta potentials were located at $(0.5,0.0)$, $(0.5,0.25)$ and $(0.5,0.75)$ (in units of cell size). The number of flux quanta is varied from 15 corresponding to a filling of $\nu = \frac{1}{3}$ in case of the homogeneous system till 18. If the idea is correct, in the case of 18 flux quanta the system will show properties of a fractional $\frac{1}{3}$-state.
\begin{figure}
\includegraphics[scale=0.5]{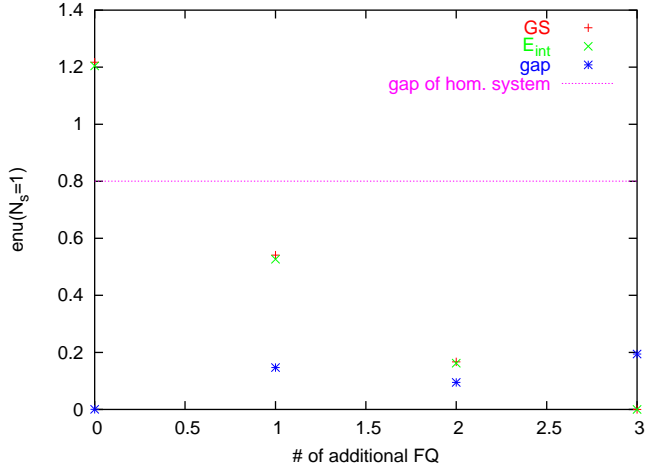}
\caption{Dependence of ground state energy, interaction energy $\langle V_{hard-core} \rangle_{GS}$ and gap 
on the number of additional flux quanta ($N_s-15$) for a system of 5 electrons interacting via hard-core interaction.}
\label{fig:GS3delta}
\end{figure}
To verify this, the eigenenergy of the groundstate, the interaction energy $E_{int} = \langle V_{hard-core}\rangle_{GS}$ and the gap is plotted against the number of additional flux quanta in Fig. \ref{fig:GS3delta}. As expected, the ground state energy is reduced with increasing number of flux quanta. The change in energy is mostly carried by the interaction part. So the delta potentials enforce zeros in the wavefunction at their respective position on costs of the interaction. In the case of 18 flux quanta the eigenenergy vanishes again, indicating Laughlin-like correlations that minimize the interaction. Also the degeneracy of the ground state becomes threefold, as in the case of a filling of $\frac{1}{3}$. This was already expected from section \ref{sec:QuasiHolesDelta} for the case of one delta potential and one additional flux quantum and also holds in this case. The gap between the ground state and the first excited state reaches its maximum in the case of 18 flux quanta, although this increase is not monotonically. This may be due to an odd/even symmetry in the number of flux quanta. The absolute value of the gap is still only approximately 25 per cent of that in a homogeneous system, which is $0.800236 \mbox{enu}(N_s=1)$ (see Tbl. \ref{tab:GShomogeneous}). ($\mbox{enu}(N_s=1)$ denoting the energy unit in the case of 1 flux quantum).
\begin{figure}
\includegraphics[scale=0.4]{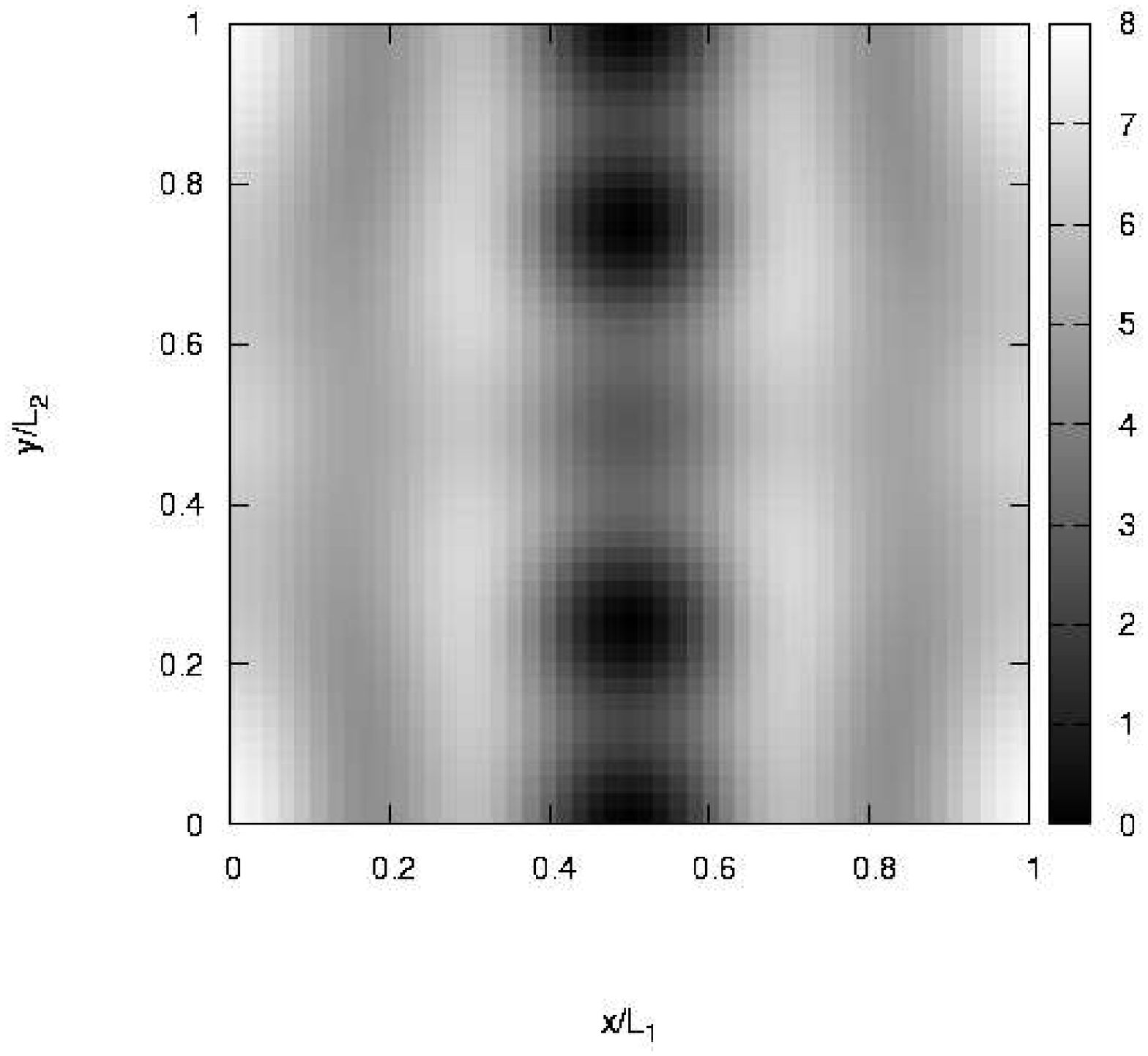}\includegraphics[scale=0.4]{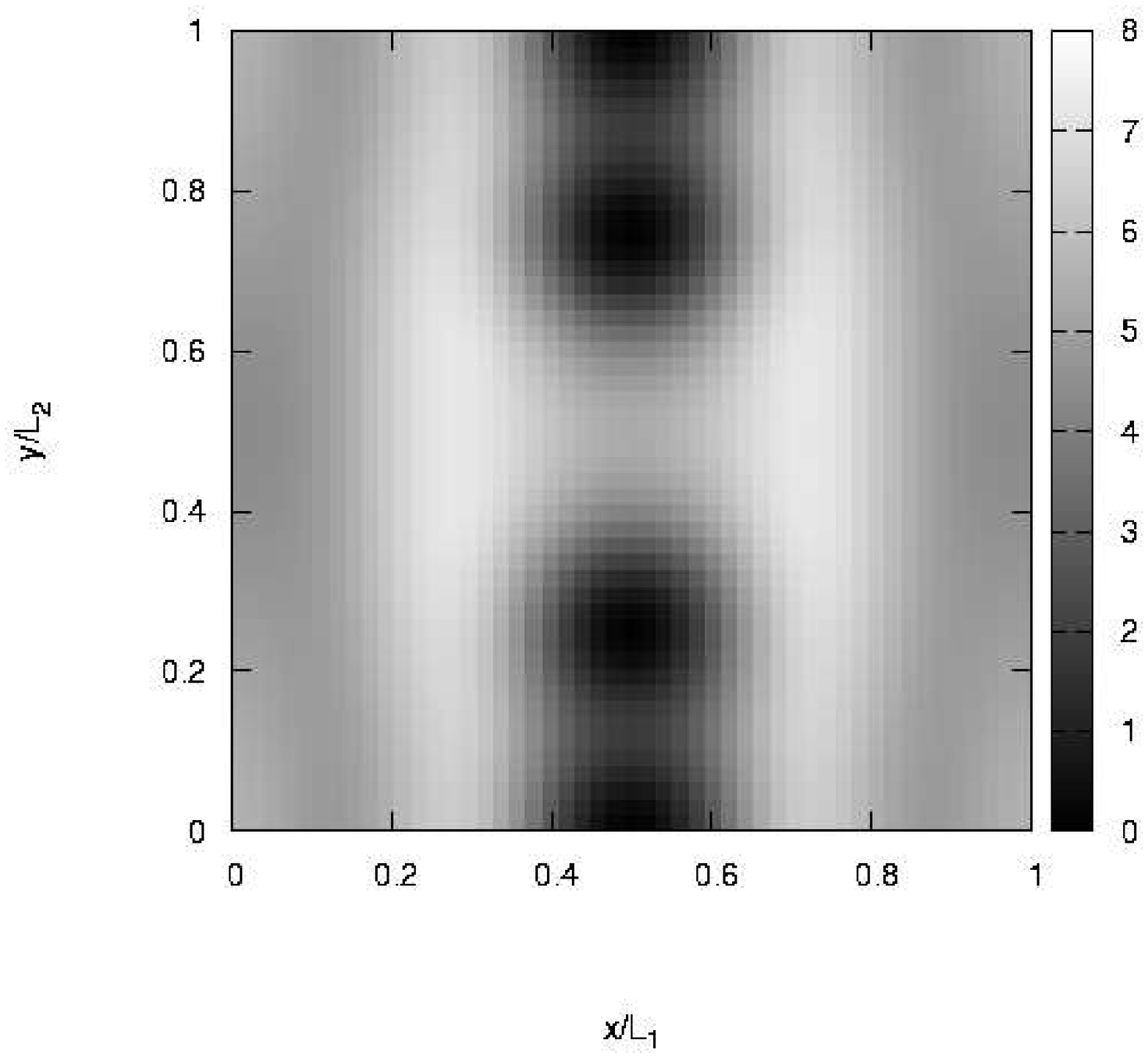}
\includegraphics[scale=0.4]{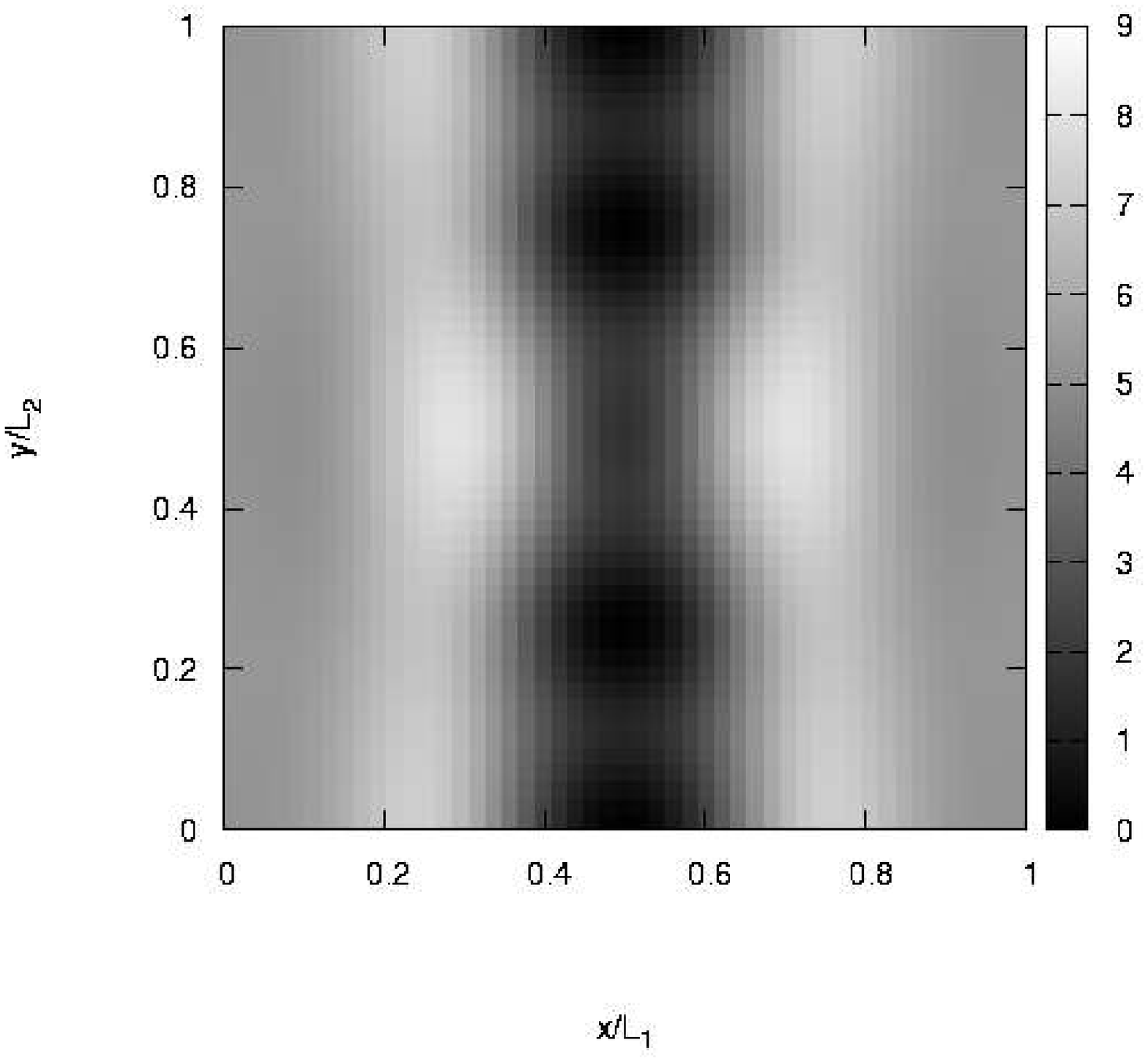}\includegraphics[scale=0.4]{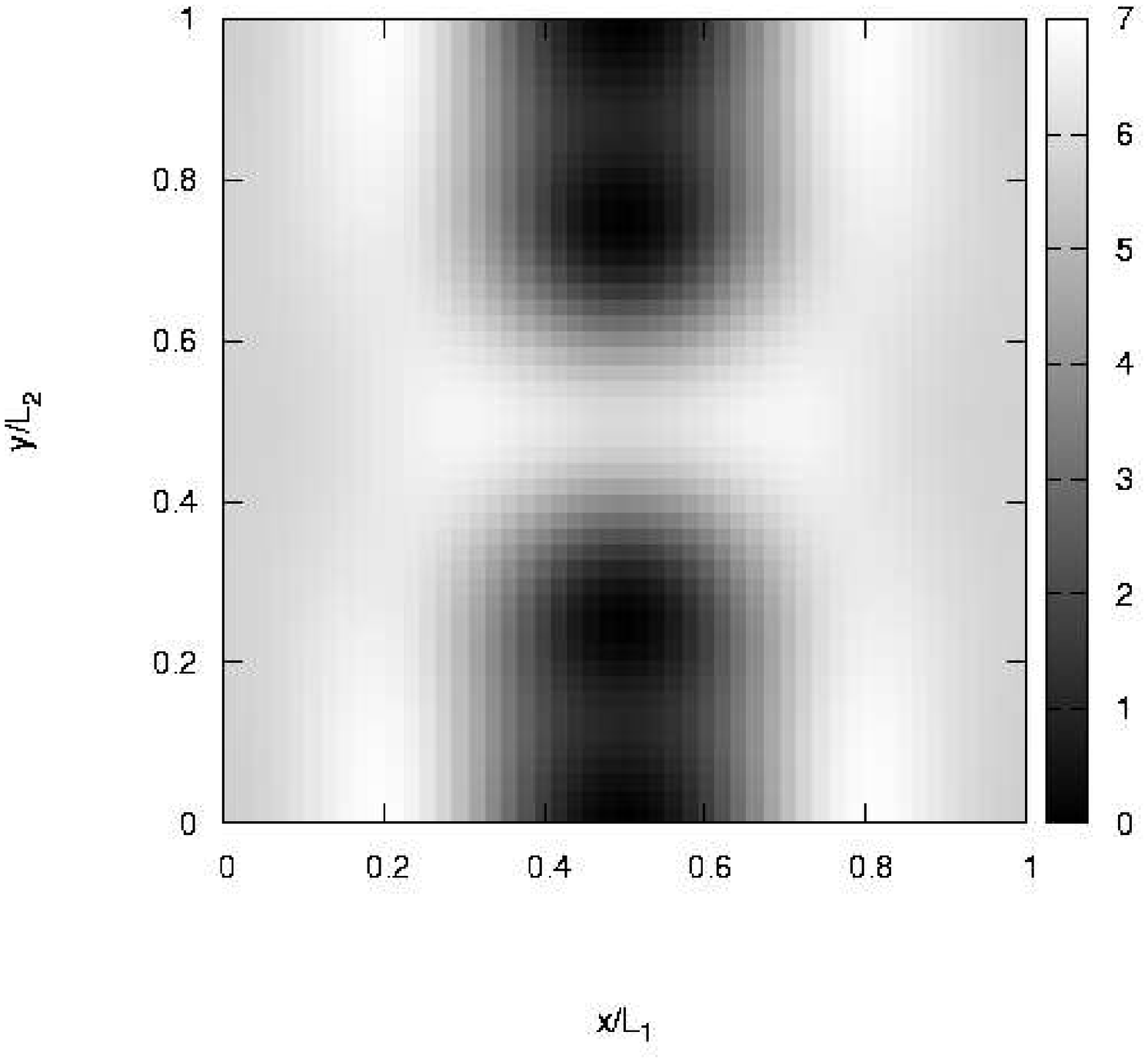}
\caption{Density for 5 electrons interacting via hard-core interaction in a system with three delta potentials at $(0.5,0.0)$, $(0.5,0.25)$ and $(0.5,0.75)$. The number of flux quanta is increased: Top left: $N_s=15$; top right: $N_s=16$; bottom left: $N_s=17$; bottom right: $N_s=18$.}
\label{fig:DensDeltaBarr}
\end{figure}
In figure \ref{fig:DensDeltaBarr} the densities of the ground states for the four different systems are shown.
The top left plot depicts the case of 15 flux quanta, top right contains 16, bottom left 17 and finally bottom right has 18 flux quanta. 
A qualitative difference seems to exist between the systems with an even number of flux quanta (right) and an odd number (left). In case of odd numbers the density inside the notch is a bit lower. The density of the system with three additional flux quanta (bottom right) shows the most homogeneous distribution in the region away from the delta peaks, while the cases with a lower amount of flux quanta show some stripes parallel to the ``barrier''. Also, in the case of three additional flux quanta, the density reaches the value of approximately 6 in the area far away from the potentials, which equals a local filling factor of $\frac{1}{3}$ again.
\begin{figure}
\includegraphics[scale=0.4]{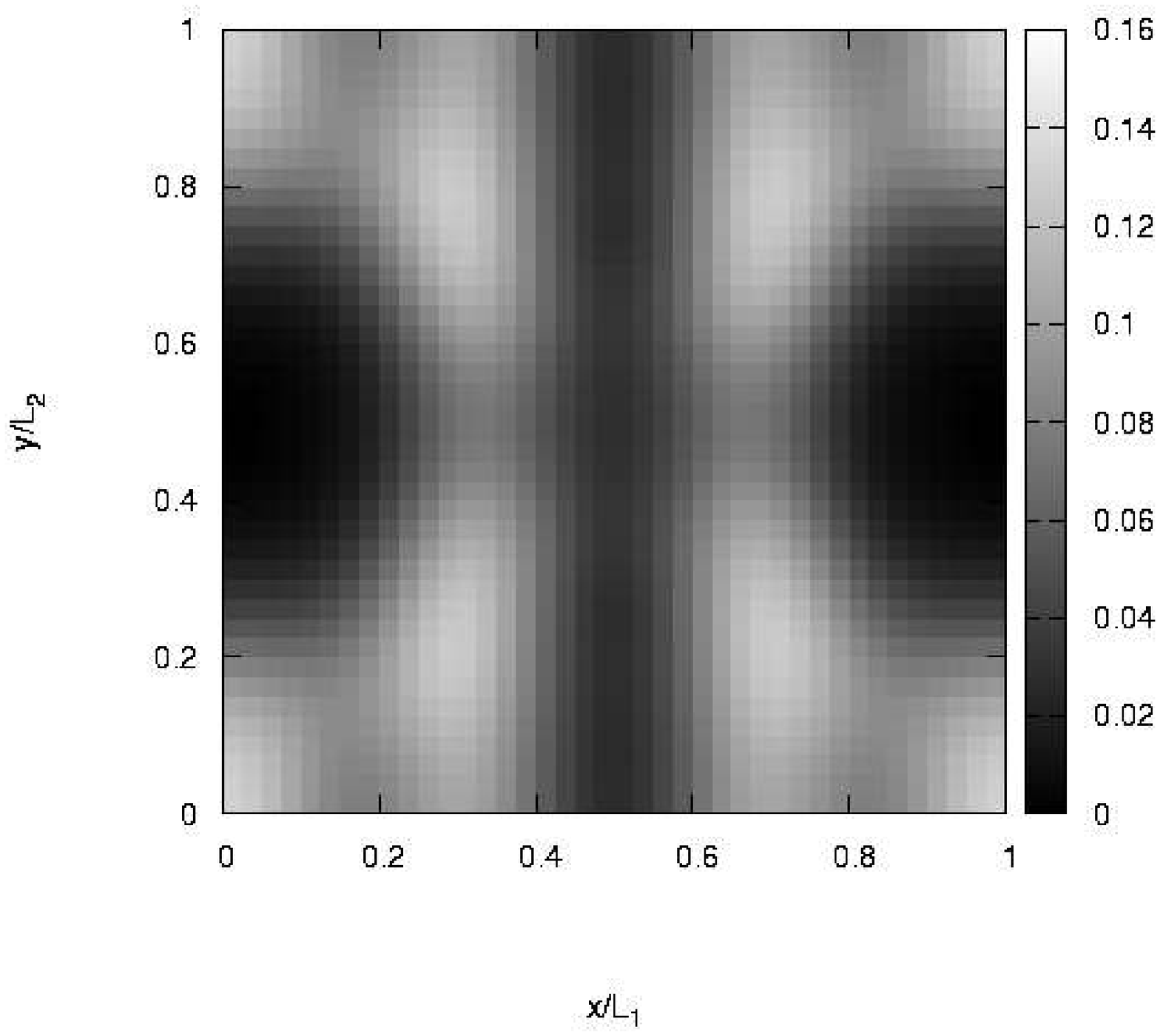}\includegraphics[scale=0.4]{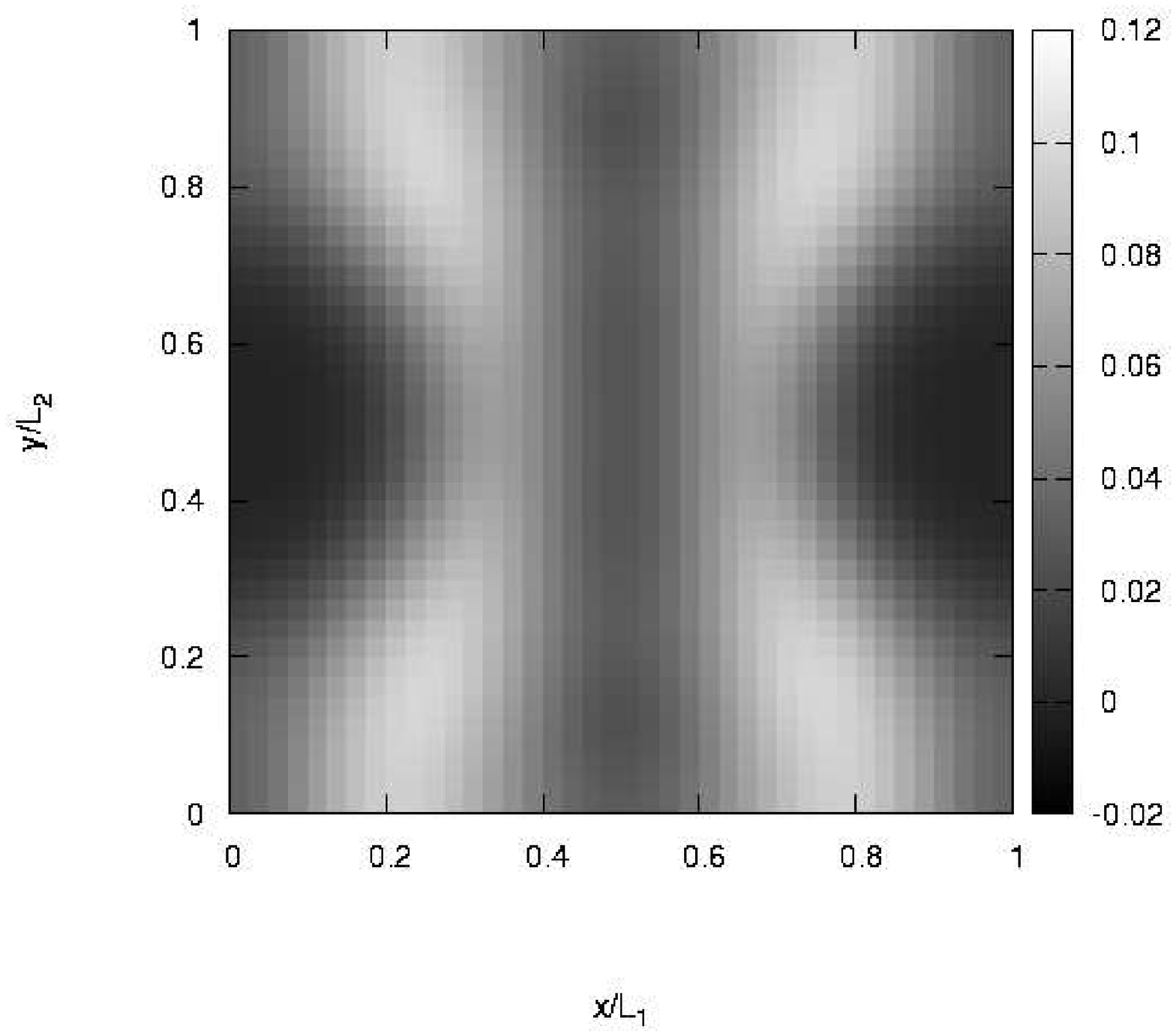}
\includegraphics[scale=0.4]{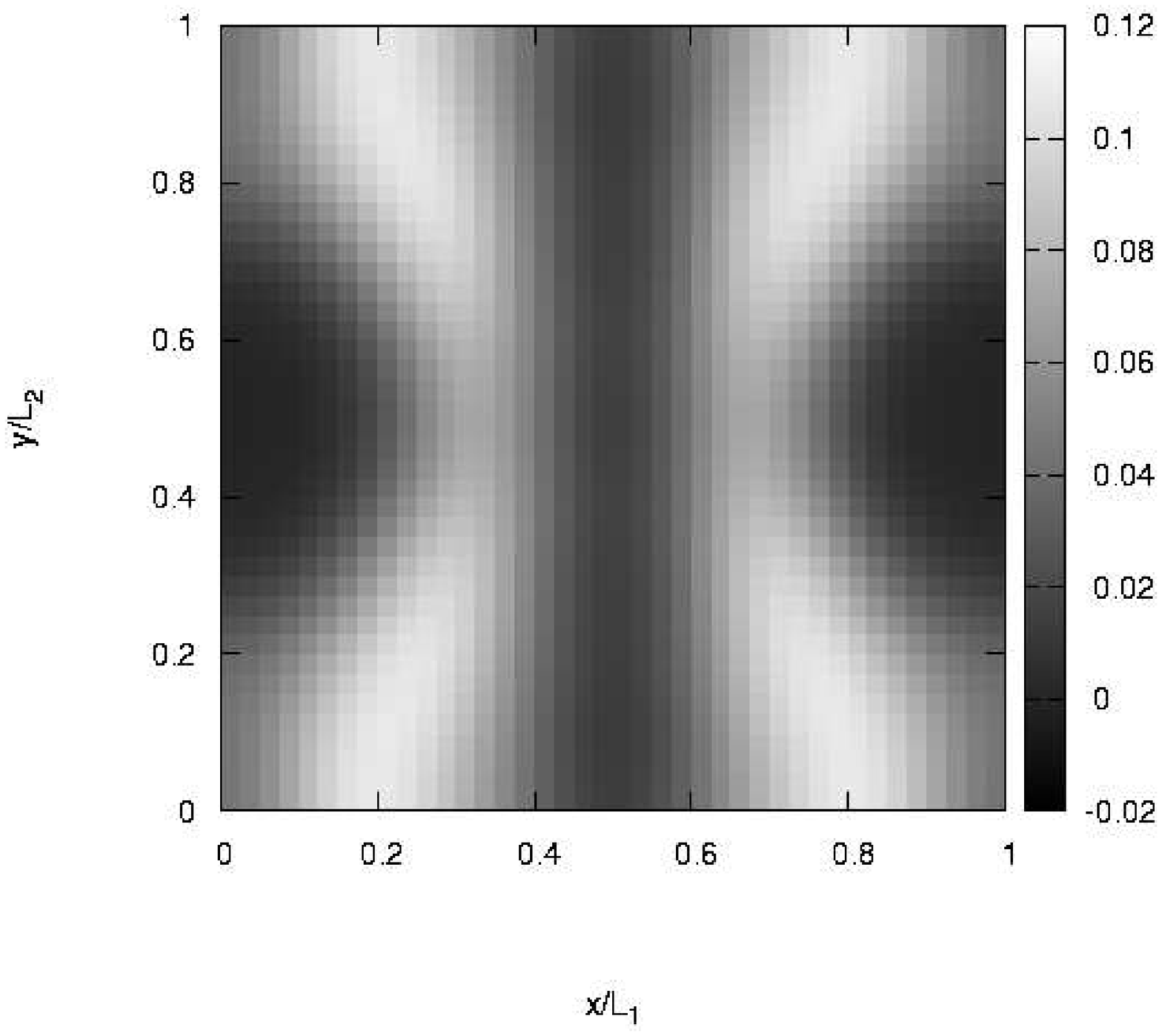}\includegraphics[scale=0.4]{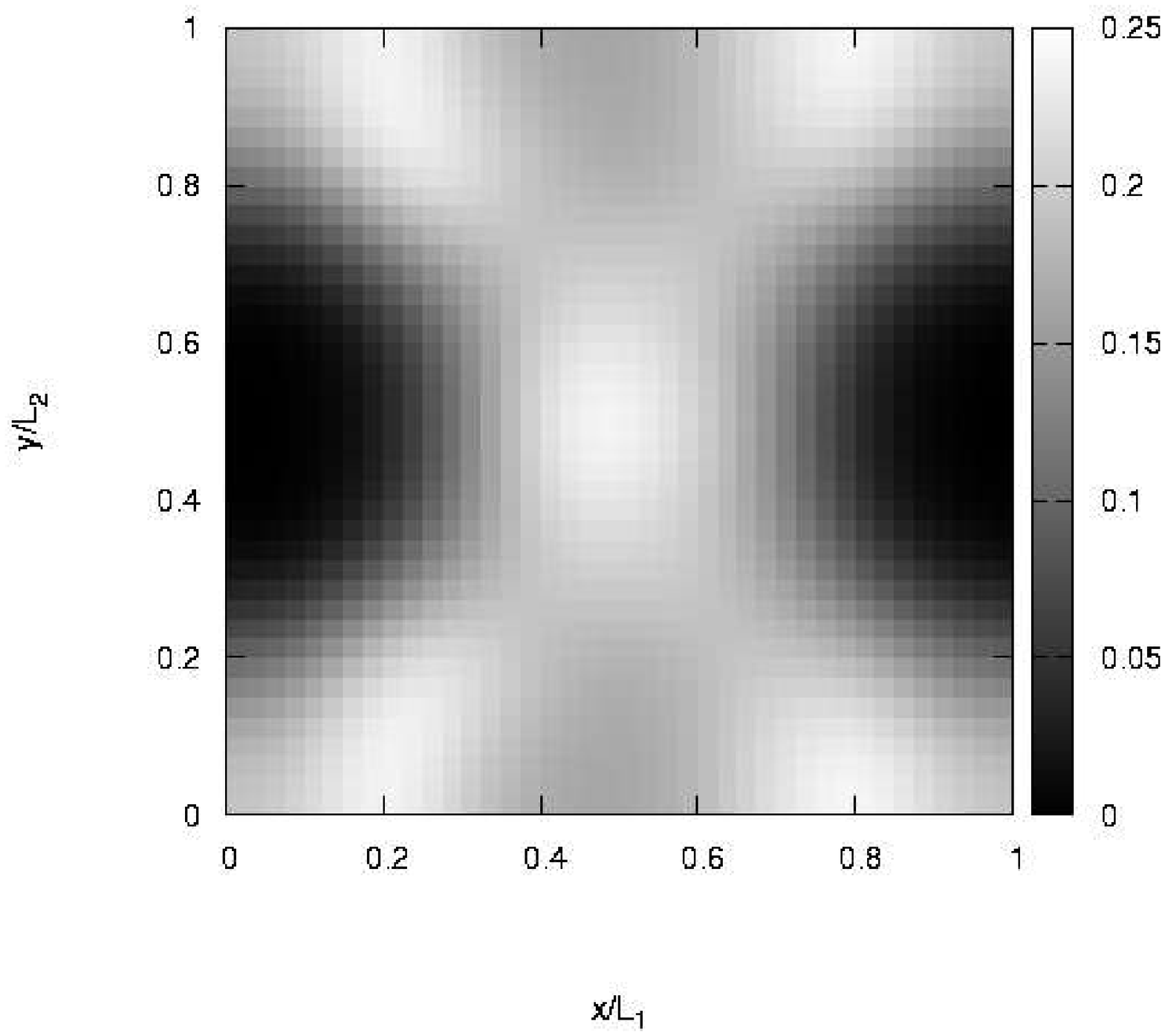}
\caption{Two-point correlation for an electron at $R_1 = (0.0,0.5)$ in a system of 5 electrons interacting via hard-core interaction and three delta potentials at $(0.5,0.0)$, $(0.5,0.25)$ and $(0.5,0.75)$. The number of flux quanta is increased: Top left: $N_s=15$; top right: $N_s=16$; bottom left: $N_s=17$; bottom right: $N_s=18$.}
\label{fig:corrDeltaBarr}
\end{figure}
Finally the two-electron correlation functions defined by equation (\ref{eq:2PointCorr}) are calculated for one electron being positiond sitting at$R_1 = (0.0, 0.5)$,
\begin{equation}
\label{eq:2PointCorr}
g(\vec R_1, \vec R_2) = \frac{(L_1 L_2)^2}{N_e(N_e-1)}\sum_{i \neq j} \bra{\Psi} \delta(r_i - R_1)\delta(r_j - R_2)\ket{\Psi}.
\end{equation}
 Figure \ref{fig:corrDeltaBarr} shows the results. As expected from the densities, again the system with three additional flux quanta shows the most liquid-like correlations. This correlation function is essentially structureless in the areas away from the barrier, whereas the systems with a fewer amount of flux quanta still show some stripe-like structures well off the barrier.

\subsubsection{Gaussian constrictions and additional flux quanta}
\label{sec:barraddflux}
The approach in the last section showed, that a number of additional flux quanta enables the system to react to the potential and to
keep the Laughlin-like correlations. A drawback of this method was that the density does not vanish completely below the barrier, since the
zeros are fixed at certain positions. Another point is, that an additional quasihole in such a system will again lie completely inside
the ground state sector and thus does not show any non-trivial time evolution. The delta potentials that create the barrier obviously
don't produce any dispersion.

Here we will combine some additional number of flux quanta with a potential to create the barrier. As seen from the previous section,
three additional flux quanta seem to be enough to enable the system to exhibit a density profile that reflects the barriers potential.
Therefore we will stick here to fillings of 5 electrons in 18 flux quanta.

\begin{figure}
\includegraphics[scale=0.45]{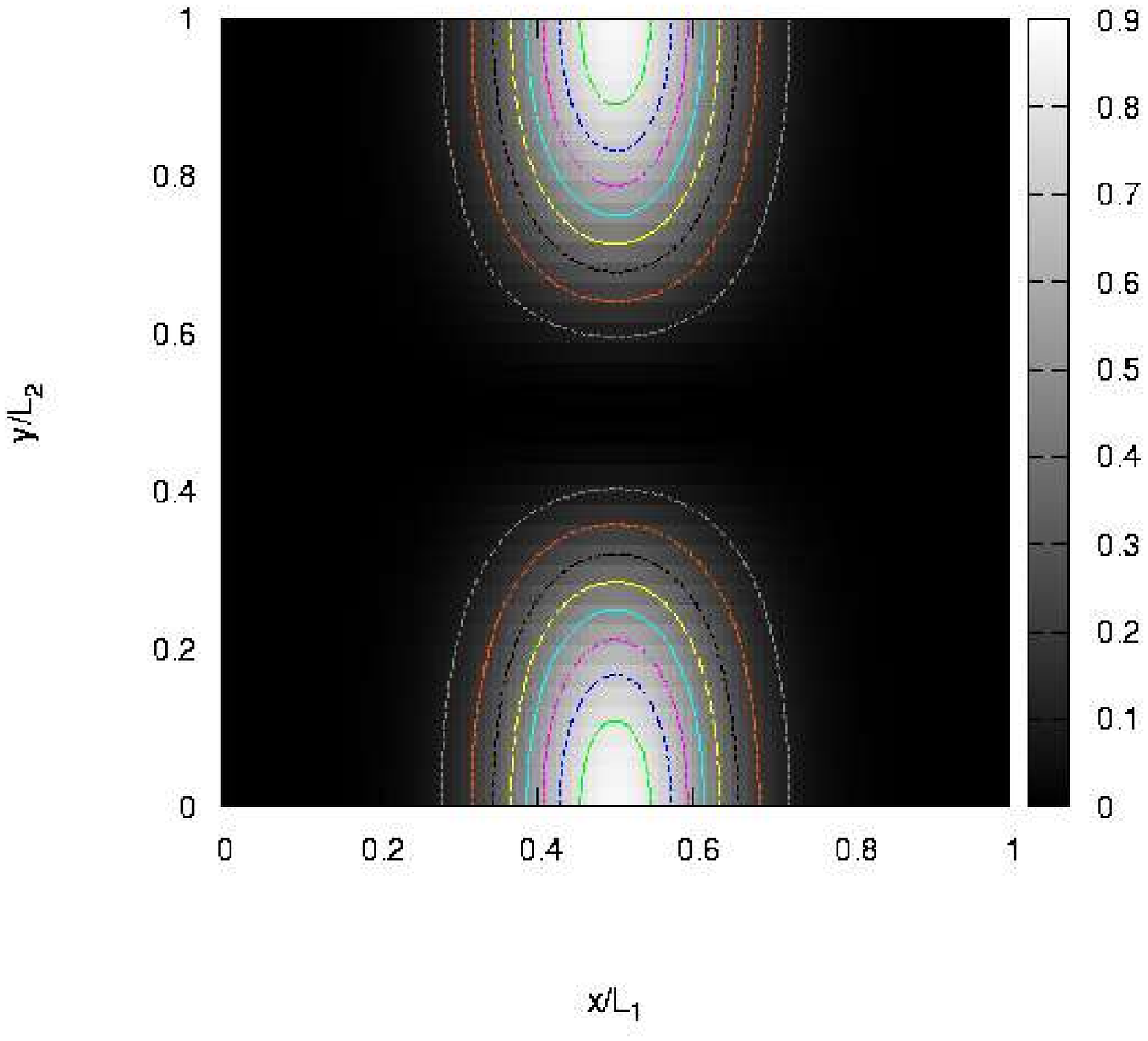}\includegraphics[scale=0.45]{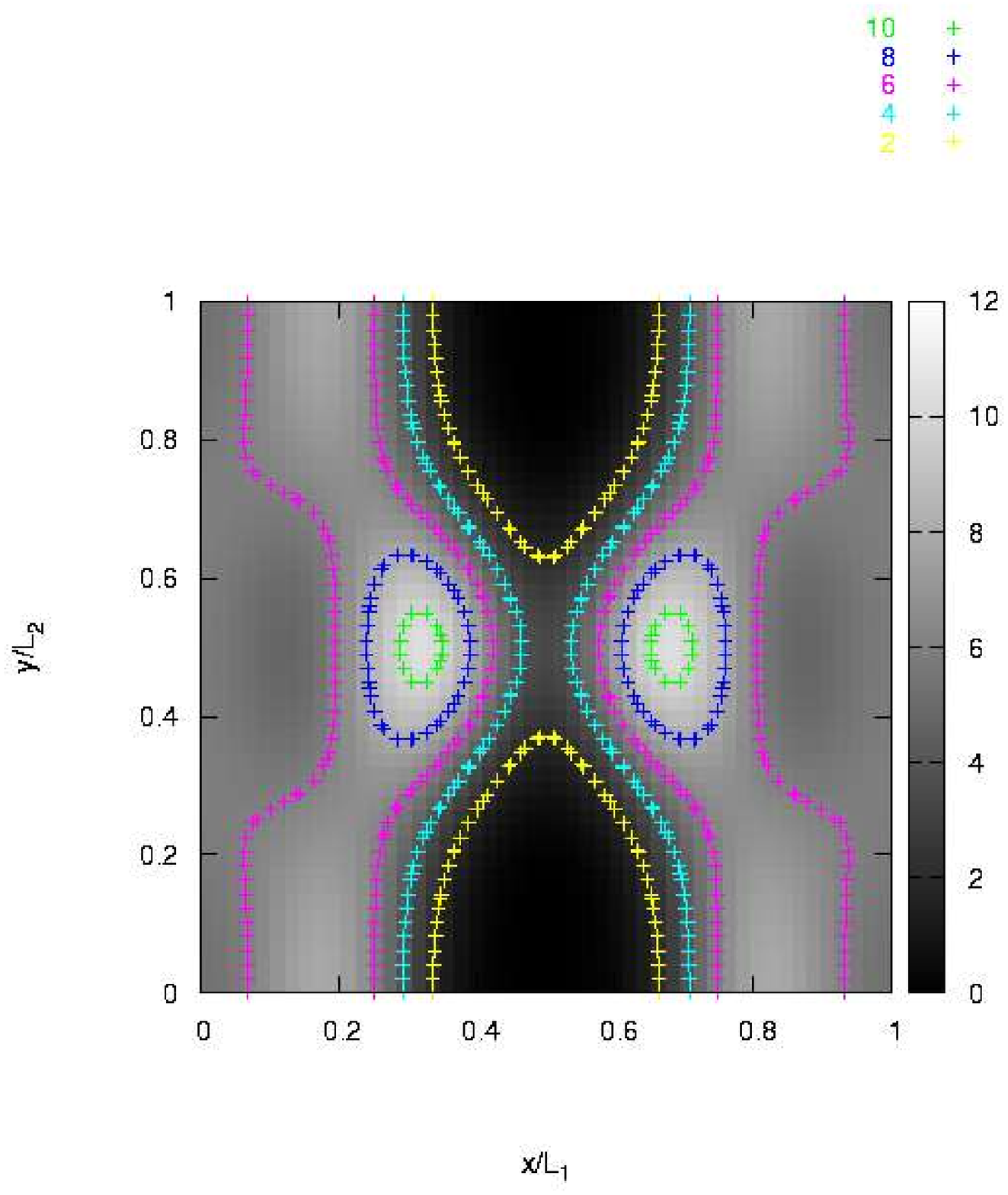}
\caption{Left: Potential used in the 5/18-system with the parameters from Tbl. \ref{tab:ParamAddBarr}.
         Right: Density of the 5/18-system with this barrier and hard-core interaction. A local filling of $\frac{1}{3}$ amounts to a density of 6 in the units used here. The parts of the system away from the barrier exhibit this filling (see equidensity line 6).}
\label{fig:PotDenBarrAdd}
\end{figure}

The barrier used here has the parameters from Tbl. \ref{tab:ParamAddBarr}. The equipotential lines of this potential can be found
in Fig. \ref{fig:PotDenBarrAdd} (left plot) along with the electronic density of the system (right plot). The density is sufficiently low below the barrier and also quite homogeneous away from the barrier. The purple colored isodensity lines ($\rho=6$) indicate a local filling of $\frac{1}{3}$. Although the notch inside the barrier is quite large, the electrons hardly populate this area, instead they tend to localize more strongly left and right of the notch. Similar effects were found in the work of Krause-Kyora \cite{Krause-Kyora} for large notches.
\begin{table} 
\begin{tabular}{|l|l|l|l|} 
\hline
height of wall (enu) & width of wall ($L_1$) & depth of notch (enu) & width of notch ($L_2$)\\
\hline
1.0 & 0.15 & -1.0 & 0.3\\
\hline
\end{tabular}
\caption{Parameters of the Gaussian barrier used in the 5/18-system.}
\label{tab:ParamAddBarr}
\end{table}
\begin{figure}
\includegraphics[scale=0.6]{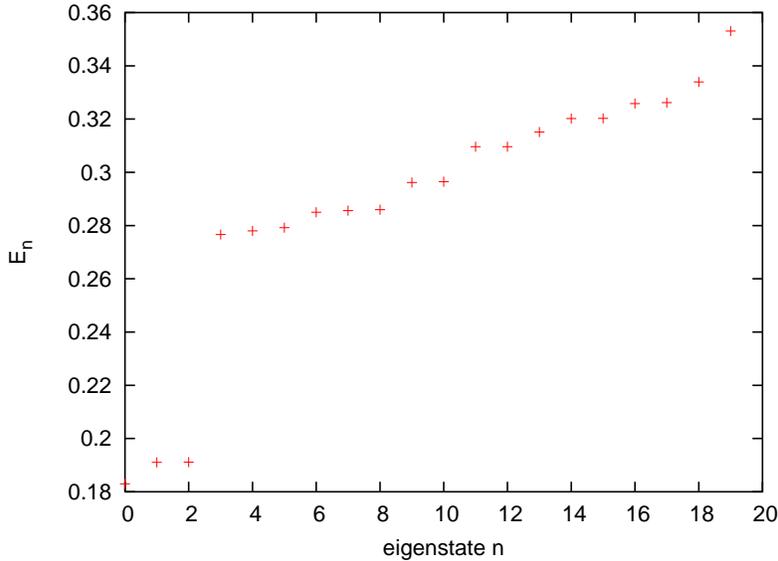}
\caption{Spectrum of the 5/18-system with a Gaussian barrier (parameters from Tbl. \ref{tab:ParamAddBarr}) and hard-core interaction. The ground state is threefold quasi-degenerate and separated by a gap from the excitations.}
\label{fig:spBarrAddFlux}
\end{figure}
The ground state of this system is non-degenerate, but looking at the spectrum in Fig. \ref{fig:spBarrAddFlux} shows three states that are
close in energy and which are separated by a gap from the excitations. It resembles very much the spectrum of a homogeneous $\nu = \frac{1}{3}$ system. The interaction energy from Tbl. \ref{tab:GSIntBarrAdd} confirms that the short-range interaction is effectively minimized such that it only
takes up $6.4$ per cent of the ground state energy.
\begin{table}
\begin{tabular}{|l|l|l|} 
\hline
ground state (enu) & gap (enu) & interaction energy $\langle V_{hard-core}\rangle_{GS}$ (enu)\\
\hline
0.18297 & 0.08552 & 0.01177 (6.4 per cent of GS)\\
\hline
\end{tabular}
\caption{Ground state energy, gap (between the highest of the three quasi-deg. GS and the first excited state) and interaction energy for the 5/18-system with hard-core interaction and a potential defined by the parameters from Tbl. \ref{tab:ParamAddBarr}.}
\label{tab:GSIntBarrAdd}
\end{table}

These observations corroborate that the system is in a fractional quantum Hall state. Now, a quasihole can be inserted. As a starting position we chose $(0.32,0.0)$ which lies near the edge of local filling of $\frac{1}{3}$. A starting point of $(0.2,0.0)$ resulted in similar behavior as described below.
If we assumed that the quasihole should just follow the equipotential lines, we would expect it to appear on the other side of the barrier after some time. It was found to be true for shallow potentials in section \ref{sec:WeakPot}.
After creating the initial state by diagonalizing the $5/19$-system with a delta potential at $(0.32,0.0)$,
the calculation of the time evolution for this initial state is performed and the density of the system is evaluated for every time step.
In figure \ref{fig:dnt0_t15} through \ref{fig:dnt90} the difference of the density of the system containing a quasihole and the density of the
inhomogeneous 5-18-system's ground state is shown for ascending time steps.
The first of those plots shows the quasihole at its initial position $(0.32,0.0)$. Along with a dip in the density at this position, there is
a lesser occupied region right of the notch. Left to it there is a small accumulation of electrons. For $t=15 \hbar/\textsf{enu}$, the quasihole moves downward and, as also seen from prior calculations, is smeared out a bit. The darker region on the right side remains stationary. Until this time both sides of the system still appear to be decoupled. At $t=30 \hbar / \textsf{enu}$ and $t=45 \hbar / \textsf{enu}$ the dip in the density right of the barrier starts moving up. This can be caused by a part of the quasihole following the equipotential line to the right half of the system (compare the potential in Fig. \ref{fig:PotDenBarrAdd}). But at the same time the quasihole's larger part stays on the left side and continues traveling downward. Looking at the density's values at this time shows that the dips on both sides became shallower. This is what we would expect, if there was tunneling between the two edges left and right of the barrier which would cause the quasihole state to become a linear combination of two quasihole states each of which located on one edge. This is in agreement with the results from section \ref{sec:tunnel}. Beginning at $t=45 \hbar / \textsf{enu}$ and being more pronounced at $t = 60 \hbar/\textsf{enu}$ the density starts becoming more inhomogeneous with several hills and valleys. This effect was already seen in section \ref{sec:WeakPot} for weak potentials.
At this time, in the lower left corner a dip appears which looks again like a quasihole and is more clearly visible for later times.
For $t = 75 \hbar /\textsf{enu}$ the quasihole on the left side reaches its initial position again. This is the time, where periodic boundary conditions will definitely have a big impact on the results. Simultaneously, the stripe of lower density which was found for earlier times right
to the barrier concentrates in one point near the notch and reaches its maximum depth around $t = 90 \hbar / \textsf{enu}$.
One very striking discovery was made by comparison of the density for $t= 60 \hbar/\textsf{enu}$ with that of $t = 90 \hbar / \textsf{enu}$.
These plots are approximately the inverse of each other. As said before this may be another effect caused by the periodic boundary conditions.
\begin{center}
\begin{figure} 
\includegraphics[scale=0.4]{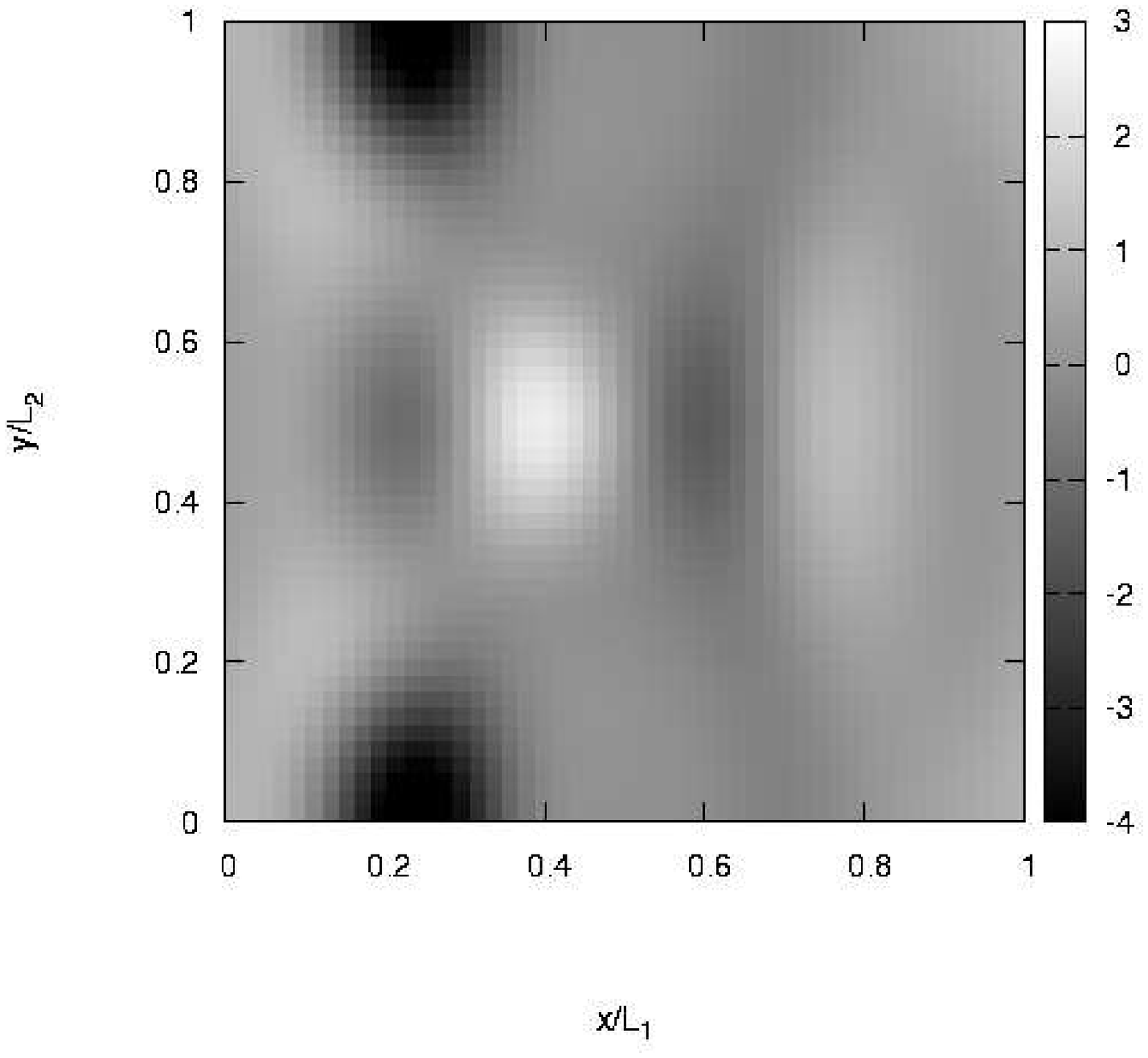}\includegraphics[scale=0.4]{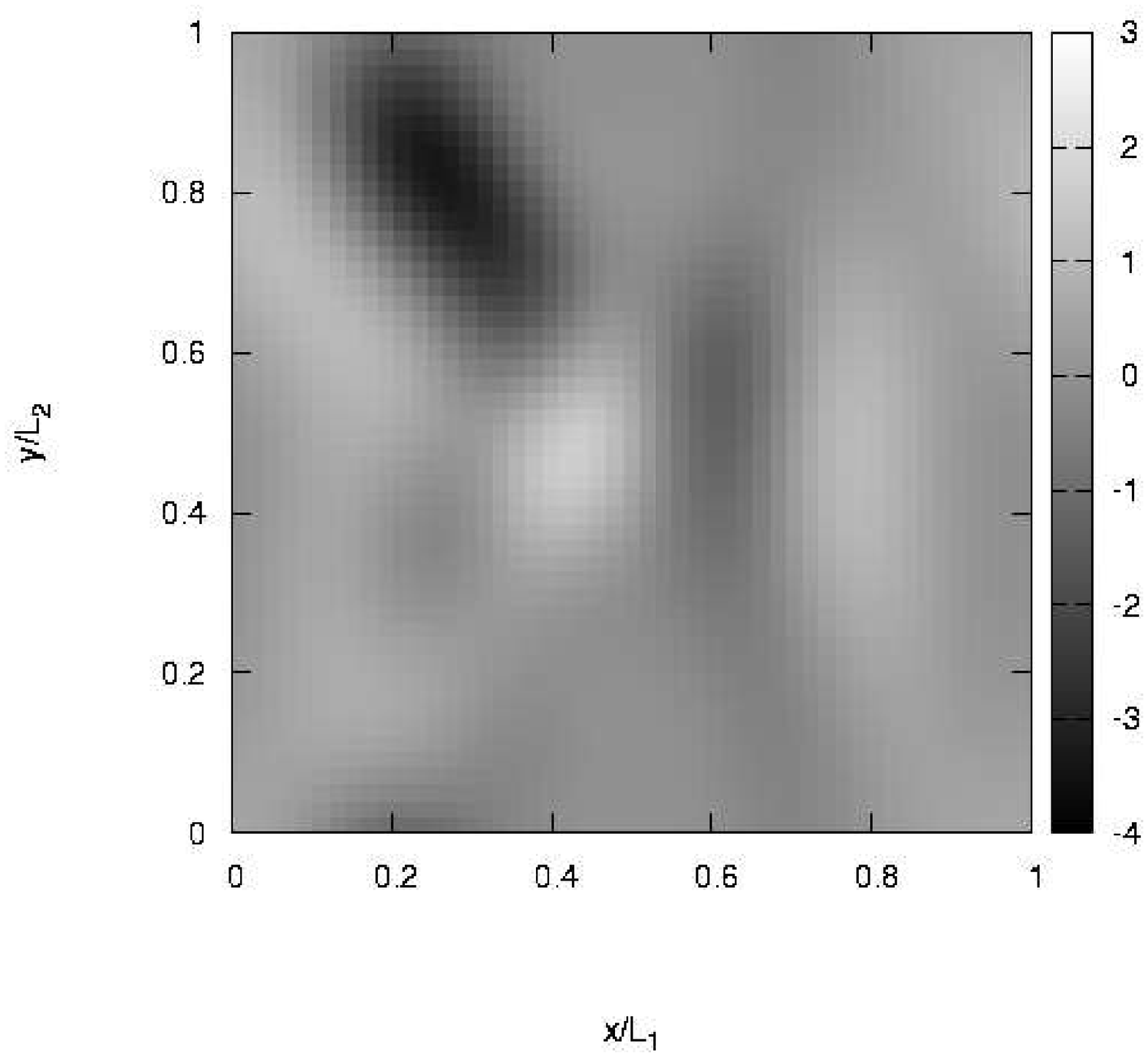}
\includegraphics[scale=0.4]{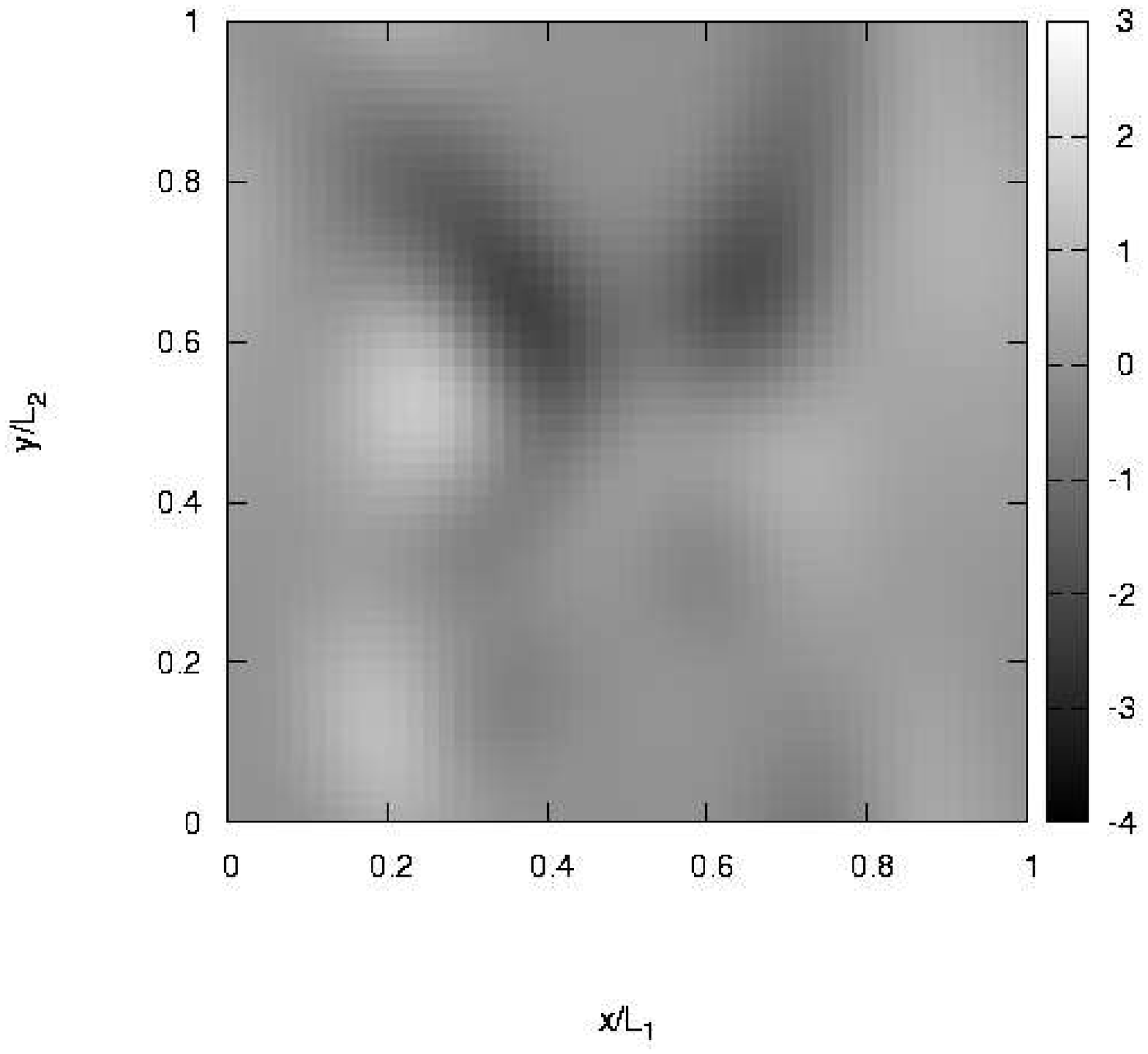}\includegraphics[scale=0.4]{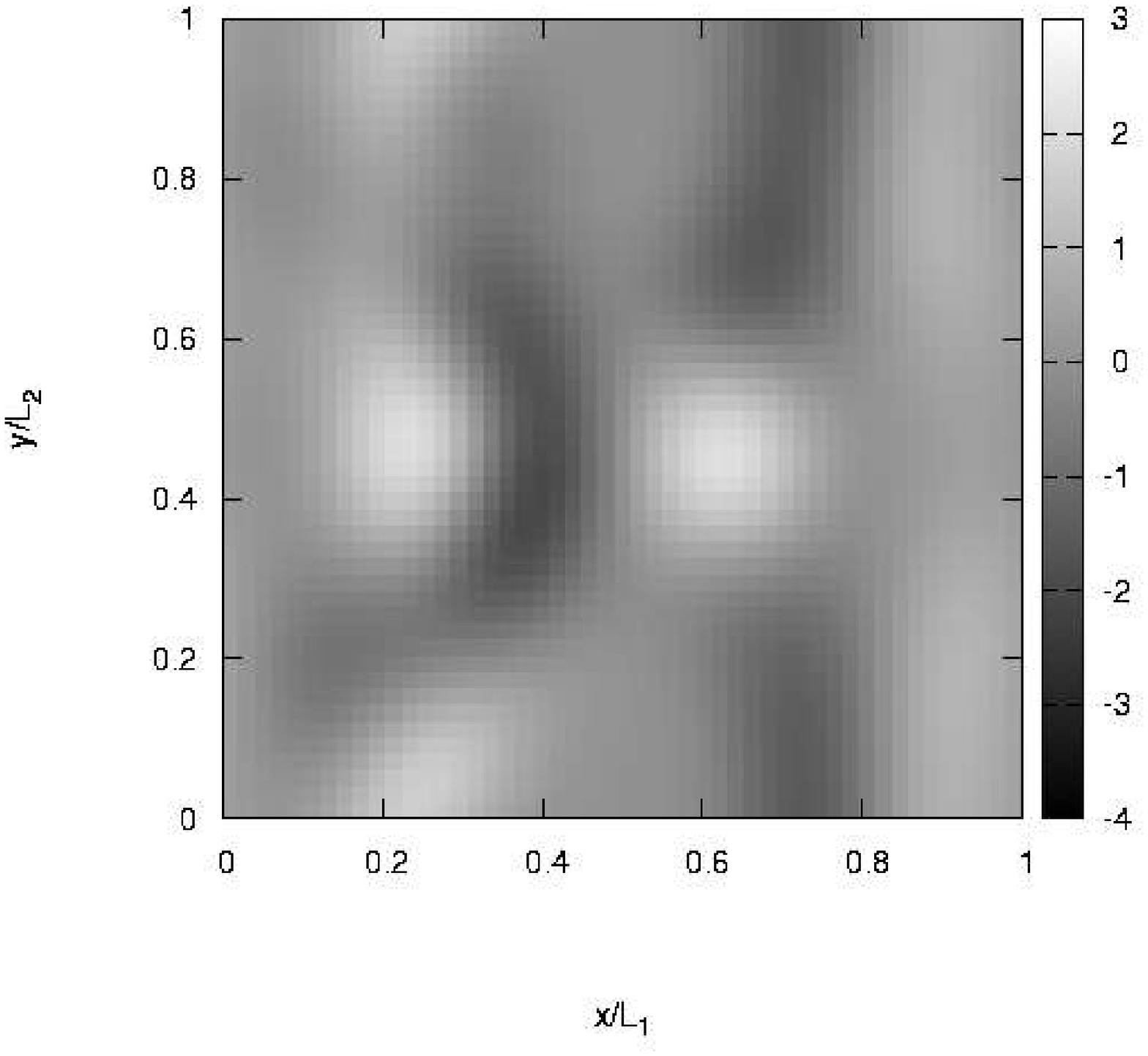}
\includegraphics[scale=0.4]{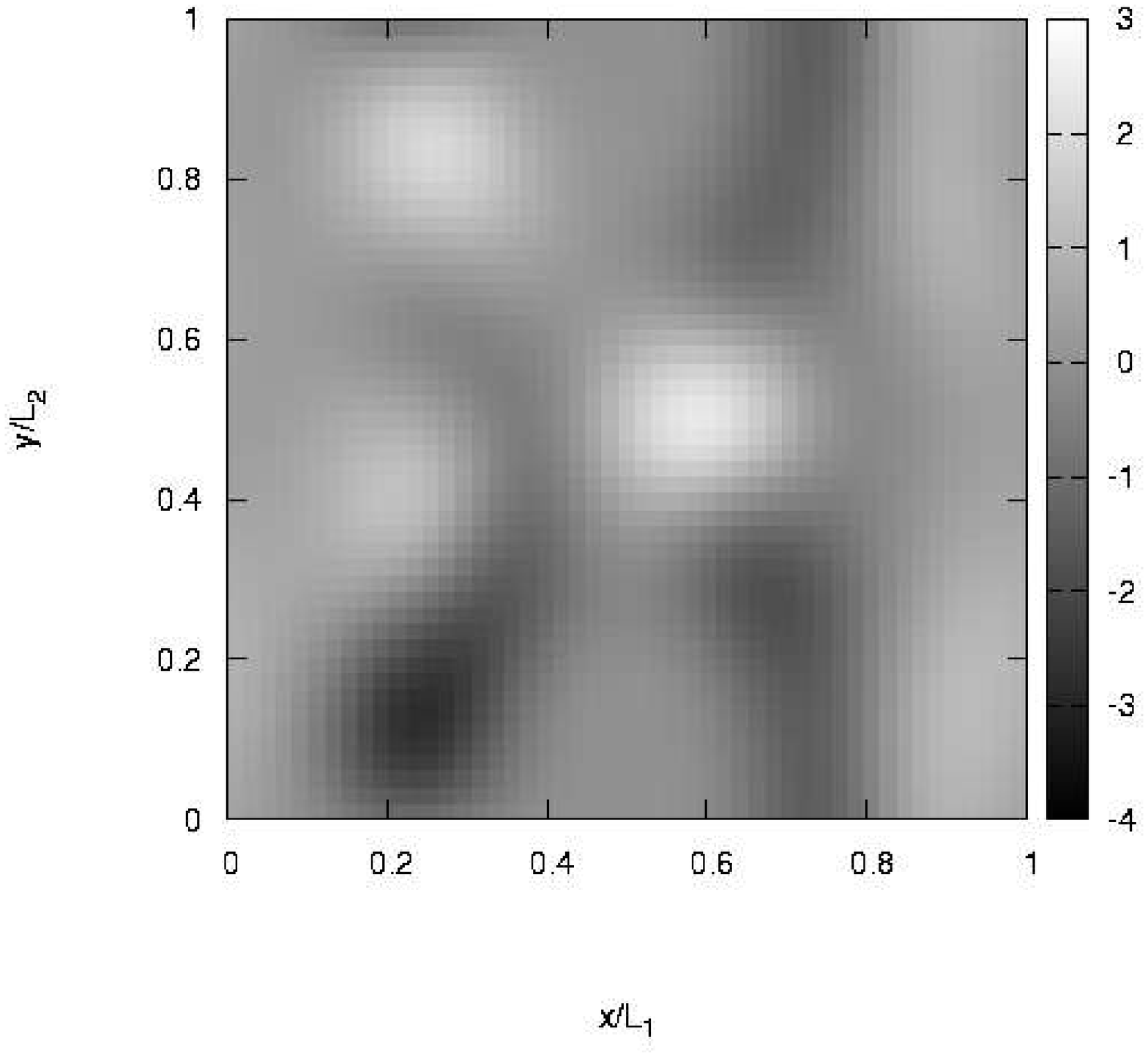}\includegraphics[scale=0.4]{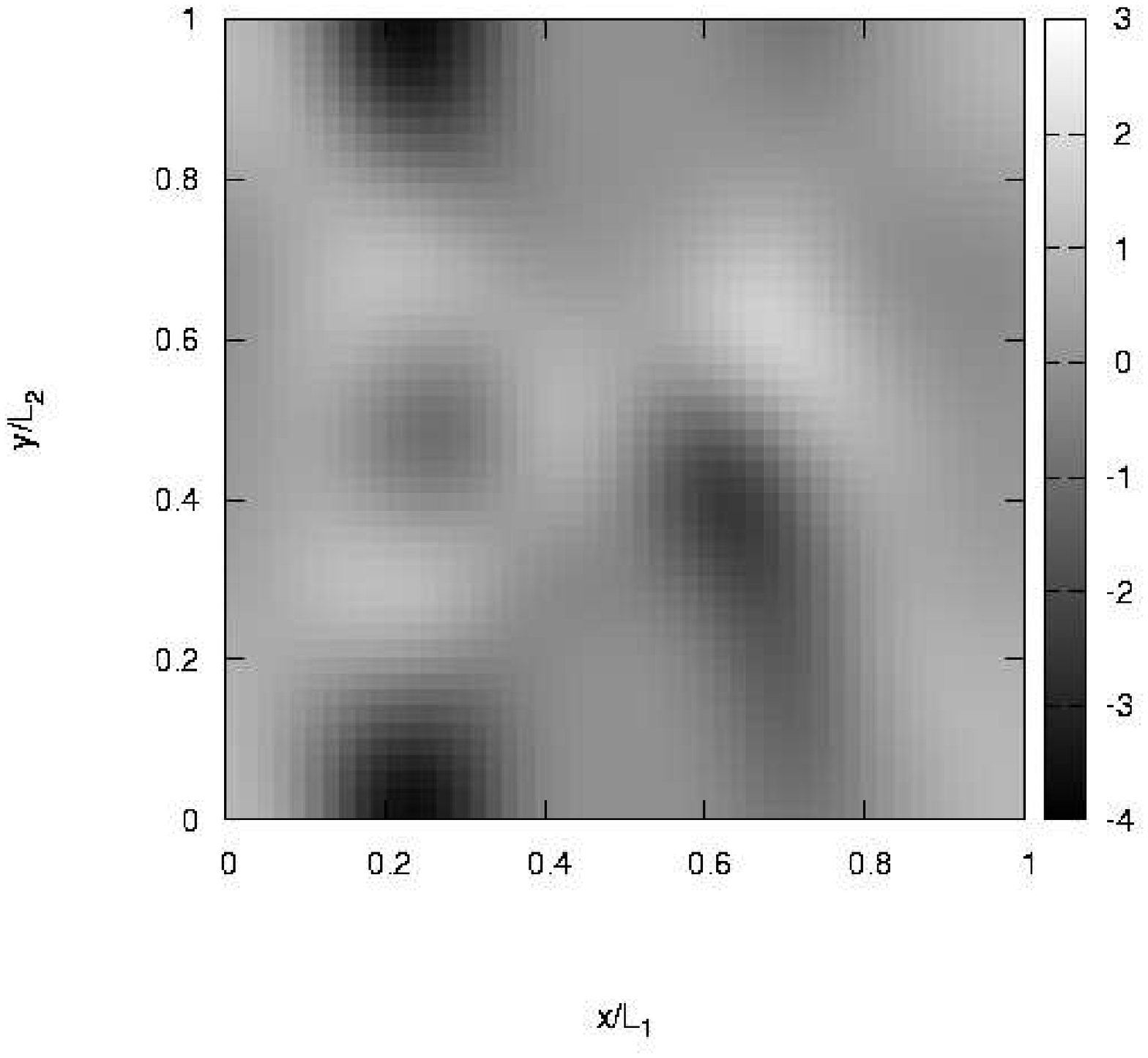}
\caption{Time evolution of the density of a quasihole injected at $t=0$ into the inhomogeneous 5/18-system. The parameters of the potential are given by Tbl. \ref{tab:ParamAddBarr}. From top left to bottom right: $t = 0, 15, 30, 45, 60, 75 \hbar / \textsf{enu}$. The density of the inhomogeneous $5/18$-system (Fig. \ref{fig:PotDenBarrAdd}) was subtracted. Initially the quasihole is located at $0.32,0.0)$.}
\label{fig:dnt0_t15}
\label{fig:dnt30_t45}
\label{fig:dnt60_t75}
\end{figure}
\begin{figure}
\includegraphics[scale=0.4]{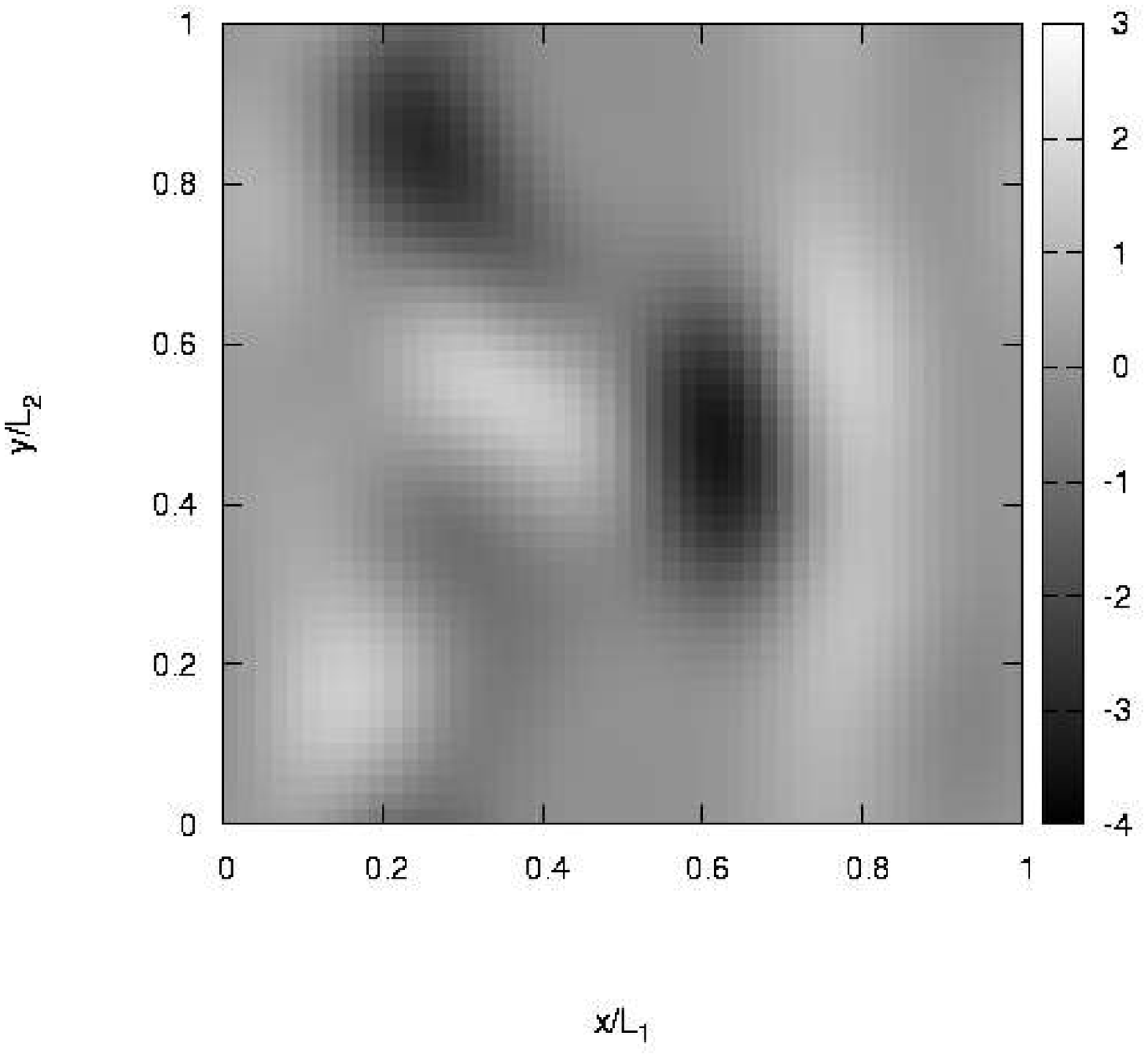}
\caption{Density of the quasihole state for $t=90 \hbar / \textsf{enu}$.}
\label{fig:dnt90}
\end{figure}
\end{center}
To end this section with, some concluding remarks on the observed effects will be given. 
Developed from the previous sections, the approach to overcome the incompressibility by inserting additional flux quanta into an inhomogeneous system appears to be a suitable way to create an effective tunneling barrier in the fractional quantum Hall regime. The ground state of this system shows features (gap, threefold degeneracy, low interaction energy) of the homogeneous $\nu = \frac{1}{3}$ system and its density exhibits a local filling near $\frac{1}{3}$ in about 40 per cent of the system's area well off the barrier. 
So the insertion of a quasihole makes sense since the system is in a fractional quantum Hall state.

Although the density in Fig. \ref{fig:PotDenBarrAdd} pretends an effective separation of both parts of the system,
the time evolution shows a strong coupling of the left and the right part of the system.
The periodic boundary condition in x-direction of course provides a connection between left and right but reviewing the densities in the time evolution they do not show any dynamics around $x = 0$. This is because the quasihole is attracted by the barrier and the important effects take place in the vicinity of the notch.
For sure at $t = 30 \hbar / \textsf{enu}$ there is a communication between the left and the right edge causing oscillations on the right side for later times. The density distribution for this time looks quite symmetric and resembles a superposition of two quasihole states, each having the hole on one side of the barrier. As found in the tunneling setup with two delta potentials, tunneling of quasiholes is possible. The symmetric state that forms at $t = 30 \hbar / \textsf{enu}$ could be due to tunneling of the quasihole between the two edges. 
Also the  structures appearing on the right half of the system at $t = 90 \hbar / \textsf{enu}$ look quasihole-like and make quasihole tunneling plausible but have to be treated with some care, since at that time the periodic boundary conditions might already have a big impact.

An interesting feature of the quasihole is that it first smears out during its time evolution, but for later times it tends to reshape.
Since this smearing-out is most strongly pronounced at the time the quasihole is near the saddle point of the potential, it could be understood by thinking of the quasihole to distribute between the two edges due to tunneling. This is also supported by
comparing the densities for $t=0$ and $t=75 \hbar/\textsf{enu}$, the quasihole on the left side is located at its starting position $(0.32,0.0)$ (or at least near to it). For $t=75 \hbar / \textsf{enu}$ the right side of the system exhibits a dip in the density near the notch that was absent for $t=0$. This is another indication that at $t=75 \hbar / \textsf{enu}$ the system is in a superposition of states in one of which the quasihole is right of the barrier. This it what would be expected for tunneling.
  
Not only the quasihole itself undergoes a process of decay and reformation but it is accompanied by the appearance of charge accumulations which were already found in the system with a weak potential (section \ref{sec:WeakPot}). The startling resemblance of the density for $t=60 \hbar/ \textsf{enu}$ with the inverted density for $t=90 \hbar / \textsf{enu}$ suggests that the quasiholes and these charge accumulations should be treated equivalently.
To check if these accumulations are quasielectrons of the system they could be compared against the trial wavefunction given by  Laughlin \cite{Laughlin}. Since these excitations appeared in systems of moving quasiholes only, the simplest case of a quasihole driven by a homogeneous electric field (which can be realized according to section \ref{sec:efield}) in a homogeneous system should be investigated.

\subsubsection{Delta constriction in k-space}
\label{sec:deltapot}
Already in the work of Krause-Kyora \cite{Krause-Kyora} it was seen that Gaussian shaped barriers tend to cause vast areas of depletion in the vicinity of the constriction and only small areas of the system remain homogeneous. This observation led to the request of finding an effective constriction that forces the electronic density to zero while having only small impact on the ``bulk'' area of the system far away from it.
In the case of Gaussian barriers we tried to compensate for this by additional flux quanta which are trapped by the potential and were shown to help maintaining the correlations. Increasing the number of flux quanta also increases the basis' size and thus puts a stronger limit on the number of electrons we can treat. So it would be desirable to have a method which is less ``flux-consumptive''.

The ansatz used here is defining the constriction directly by its matrix elements. Due to our gauge it is easily possible to create a localized
wall by this. The matrix elements of this wall are such that they lift the energy of one selected single-particle state $\ket{0, j_0}$ in the lowest Landau level, whereas the energies of the others remain the same. This ``potential'' is actually the projector onto this one lifted state, $V_{\delta} = s_{\delta} \frac{e^2}{\epsilon l_0^2} \ket{0,j_0}\bra{0,j_0}$. Hence it is rather a pseudopotential as there is no local representation in x-space since the projector is non-local: There are two coordinates to integrate over to calculate its matrix elements.
The diagonal terms of this ``nonlocal potential'' are given by a Gaussian shaped potential centered around $x \simeq -X_{j_0}$ with $X_{j_0} = j_0 \frac{L_1}{N_s} + \beta \frac{L_1}{2 \pi N_s}$. The intention of this procedure is to create the narrowest possible barrier. Due to the localization of the single particle states
on a range of the cyclotron radius $l_0$, this gives a lower limit for structures' sizes the system can resolve. Thus even narrower barriers than this ``delta''-barrier cannot be constructed. Its matrix elements are obviously
\begin{eqnarray}
  \label{eq:MatElDeltaWall}
  \bra{0,j} V_{\delta} \ket{0,k} &=& s_{\delta} \frac{e^2}{\epsilon l_0^2} \delta_{j,j_0} \delta_{k,j_0}.
\end{eqnarray}
To create a notch inside this barrier, we take the matrix elements of a potential wall parallel to the $y$-direction as in equation (\ref{eq:WallY}) and multiply them symmetrically with the matrix elements from equation (\ref{eq:MatElDeltaWall})
\begin{equation}
  \label{eq:WallY}
  V_{Wallx} = s_{Wallx} \frac{e^2}{\epsilon l_0^2} \sum_{n \in \Z} \exp \left( -\frac{(y-y_0+n L_2)^2}{(w_{Wallx} L_2)^2} \right).
\end{equation}
Its width and depth is defined by the width and (negative) strength of this wall. The resulting matrix elements are found in appendix \ref{app:MatEls}, equation (\ref{eq:MatElDeltaNotch}).

\begin{figure}
\includegraphics[scale=0.4]{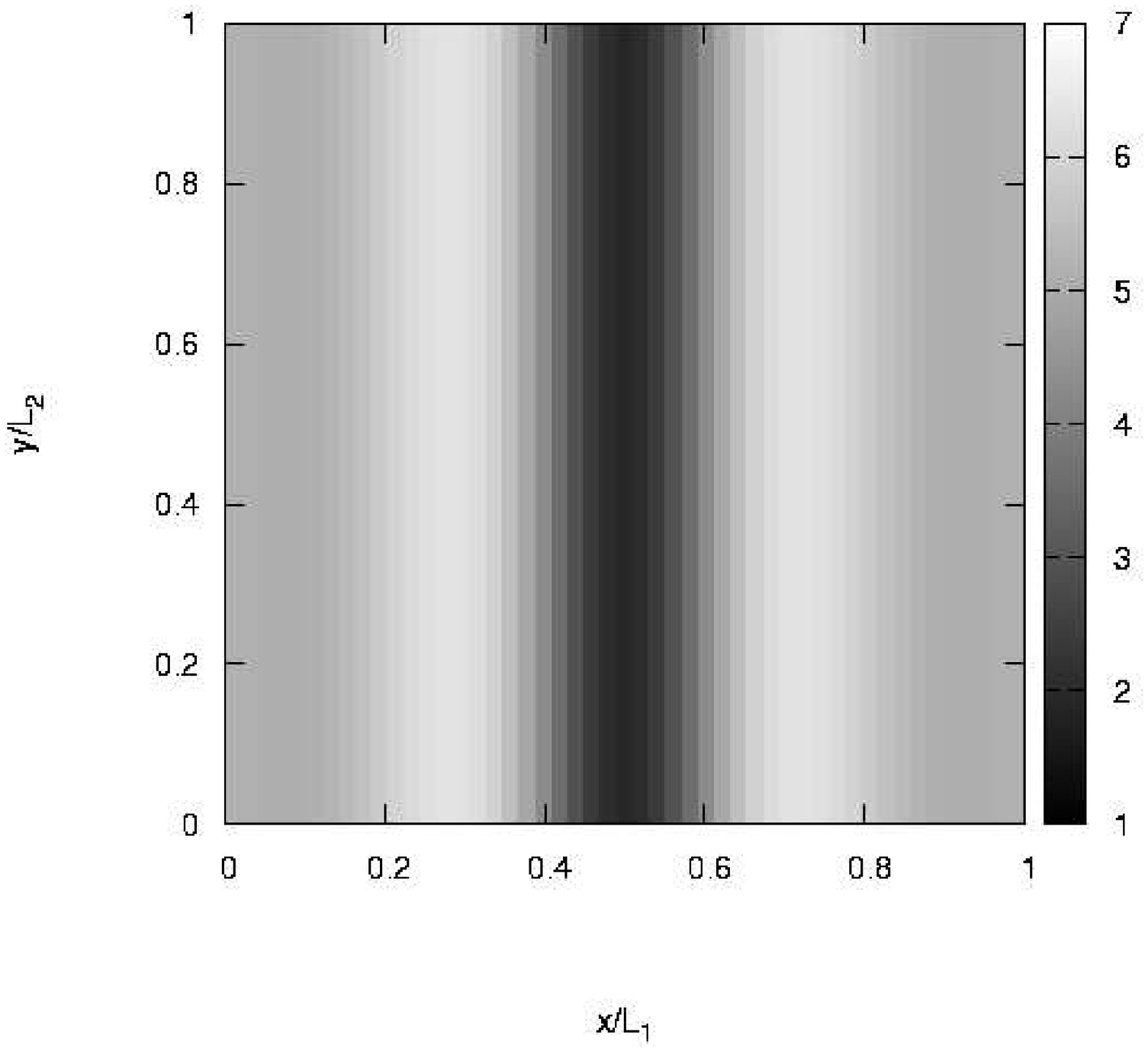} \includegraphics[scale=0.4]{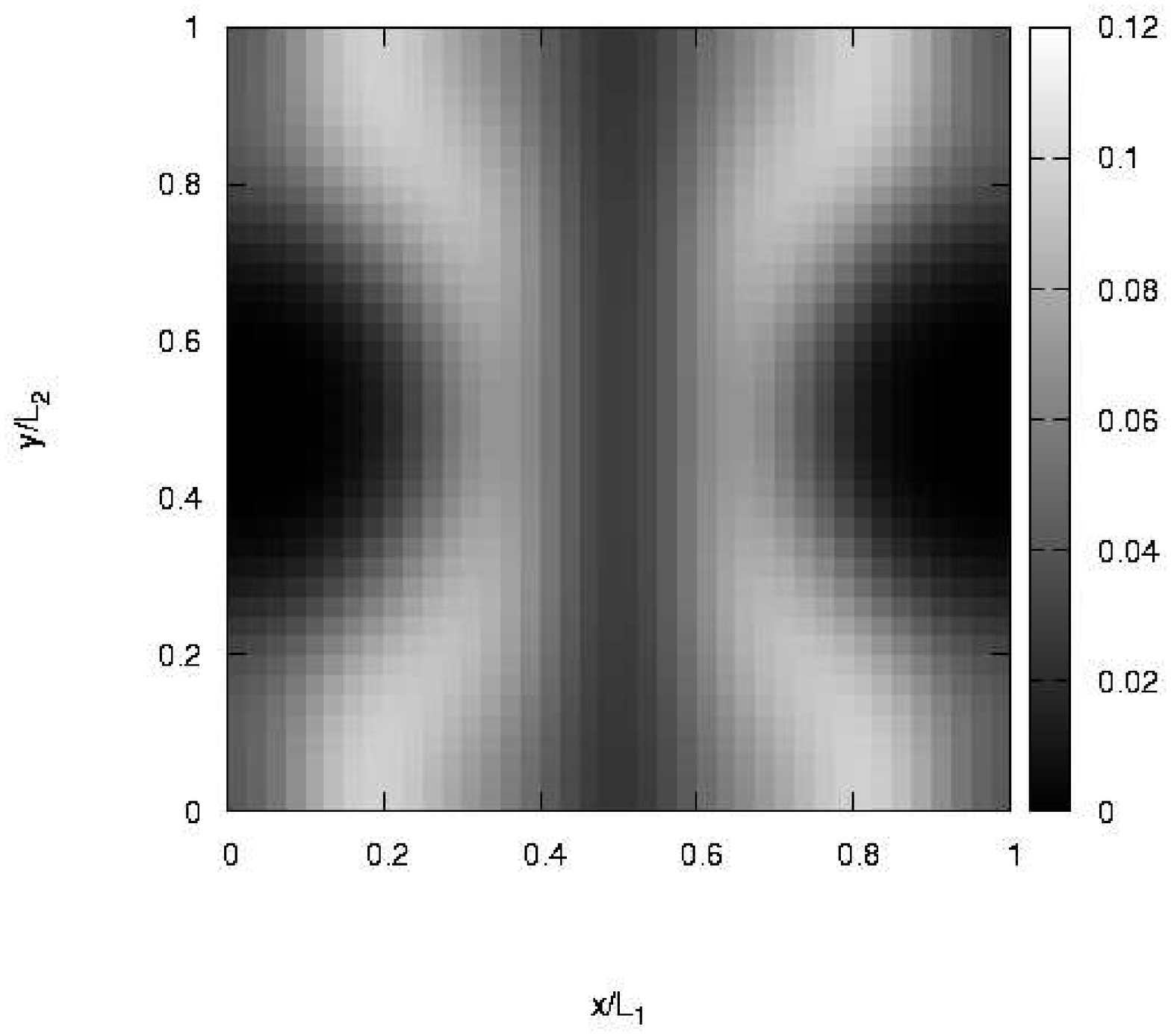} 
\caption{Left: Electronic density, right: 2 point correlation function for a 5/16 system with a delta barrier}
\label{fig:DeltaBarrDNCorr}
\end{figure}

%
%
A system with 5 electrons and 16 flux quanta was diagonalized using a delta barrier as a constriction.
Its height was chosen to be $1 \mbox{enu}$.
In figure \ref{fig:DeltaBarrDNCorr} the density of the system is depicted. The area between $x=0.4 L_1$ and $x=0.6 L_1$ is depleted which is a width of $2 l_0$. In the area $x > 0.8 L_1$ and $x < 0.6 L_1$ which makes up 40 per cent of the whole system the density is rather homogeneous and at a local filling factor of $\frac{1}{3}$. The two point correlation function for one electron fixed at the point of largest distance from the barrier is shown in the right plot of this figure, too. The correlation hole around the electron is well established and the other electrons are most likely to be found in the region approximately $2 l_0$ and further away from the first one. Apart from the dip at the barriers position this correlation function is
structureless --- as it has to be for a liquid-like state. The ground state energy is $0.00311 \mbox{enu}$ of which are $0.00204 \mbox{enu}$ interaction energy which is quite low.
%
%
These findings make this type of barrier appear to be another adequate system for inserting quasiholes.
From a point of view concerning the numerics, compared to the Gaussian barrier in the previous section where three additional flux quanta were needed, one additional flux quantum ($\equiv$ one additional single-electron state) is sufficient here to obtain similar features, which of course reduces the basis' size and makes larger systems possible to be calculated.
\subsection{Conclusions on the inhomogeneous system}
The inhomogeneity induced by an external potential demands a careful treatment in taking the limit of high magnetic field.
Specifically, corrections to the current operators are crucial to obtain a consistent treatment of the system within the lowest Landau level (section \ref{sec:corrections}).

The first part of section \ref{sec:inhomogeneous} dealt with weak inhomogeneities, where the incompressibility of a $\nu=\frac{1}{3}$ system counteracts the response of the system to the imposed potential.
Such a potential can be regarded as a source of dispersion causing the $\vec E \times \vec B$ drift of quasiholes injected into the system.
The time evolution of a quasihole state allows for the following conclusions:
\begin{itemize}
\item Far away from the saddle point of the potential the quasihole behaves like a classical particle in a strong $B$-field. In the electric field
generated by the potential it shows the $\vec E \times \vec B$ drift. Although it spreads out a bit while it moves along an equipotential line it remains identifiable as a dip in the electronic density.
\item Equipotential lines with opposite direction of propagation meet at the saddle point of the potential. At this point the quasihole is smeared out strongly. It can be interpreted as a superposition of a quasihole moving along the original equipotential line and one having passed to the other side of the potential's peak moving in opposite direction.
\item While the quasihole moves through the system it induces excitations that become apparent as accumulations of electrons. The amount of charge accumulated just compensates the charge of a quasihole. This suggests the question whether these excitations could be quasielectrons; however their short lifetime (compared to the quasihole) rather contradicts to this interpretation.
\end{itemize}

On the other hand strong potentials, that affected the density of the system and created an effective barrier for electrons, destroyed the correlations of the homogeneous $\nu = \frac{1}{3}$ state that are essential for quasiholes. Much of the subsequent work was dedicated to finding a proper means of introducing a barrier into the system without destroying the correlations. Doing so, some results of Krause-Kyora \cite{Krause-Kyora} were helpful.
The position-dependent excitation energy of quasiholes confirmed these particles to be attracted by the barrier. Additional flux quanta were found to be bound inside the barrier. A comparison of properties of an inhomogeneous system at the filling factor $\nu =\frac{1}{3}$ to the case of having an amount of excess flux quanta, revealed the latter system to be more desirable: Its ground state shows more signs of a fractional quantum Hall ground state than the one of the system at fractional filling. In conclusion one can state that the excess flux quanta relax the competition between correlations (interaction energy) and potential energy.

In section \ref{sec:deltapeakbarrier} the previous observations were developed further by constructing a barrier through pinning of excess flux quanta at several delta potentials. Although the properties (gap, threefold degeneracy, liquid-like correlation, $E_{GS}$ = 0) of this system's ground state reproduced the features of the homogeneous system's $\nu=\frac{1}{3}$ state, it is not believed to be suitable for investigating quasihole tunneling because of the ``pathologic'' barrier: It does not cause any drift for the charged quasiholes.

Section \ref{sec:barraddflux} uses the best realization of a tunneling barrier found so far: A combination of a Gaussian barrier in a system with a relaxed incompressibility due to additional flux quanta leads to a ground state of the system that resembles very much a fractional quantum Hall state.
The time evolution of a quasihole injected into this system could be studied.
Although the quasihole's deformation was found to be more severe than for shallow potentials, it was still possible to identify it as a dip in the electronic density most of the time.
Near the meeting point of the edges the quasihole --- initially traveling downwards in the left half --- favored a symmetric state between the left and the right edge. For later times the impact on the right half of the system evolved into a quasihole-like dip in the density.
A tunneling process between the edges seems to be a plausible explanation for this behavior.
To corroborate this, it would be possible to project the state of the time evolution onto a final state with a quasihole on the right side.

Excitation of charge accumulations moving through the system, as observed earlier in the case of shallow potentials, were also found here. A striking symmetry between the quasiholes and these ``hills'' in the density was observed but cannot be understood so far.
To check whether these excitations are quasielectrons would be possible by comparing them to the quasielectron wavefunction of Laughlin \cite{Laughlin}.
Charge density waves excited by the moving charged quasihole may be an alternative explanation.
To clarify this, a homogeneous system with a quasihole driven by an electric in-plane field (realizable according to section \ref{sec:efield}) would be a system to focus on.
 
Finally in section \ref{sec:deltapot} an alternative approach to realize an effective tunneling barrier in the fractional quantum Hall regime by a delta-barrier in $k$-space is presented. It has promising features and tunneling of quasiparticles could be investigated in this system in a future work.

\newpage
\section{Conclusions and perspectives}
The motivation and the aim of this work was to consider single-quasiparticle tunneling at a quantum point contact (QPC) in the fractional quantum Hall regime. Focussing on a finite electronic system the behavior of inserted quasiholes near this constriction should be studied by treating the many-particle Hamiltonian by exact numerical diagonalization.

Initially preparatory work was necessary to create a suitable system in which quasihole tunneling would be possible to observe. The preparations targeted in two directions: Firstly, a proper means of describing an injected quasihole in our system had to be found and its validity checked. Secondly, a system with a quantum point contact had to be realized. For the second problem it was possible to resort to results of a preceding work by Krause-Kyora \cite{Krause-Kyora}.

Finally the combination of these efforts allowed the main question of quasihole tunneling through a QPC to be addressed and results corroborating the tunneling process were found.
\medskip

In order to relate and to compare the numerical approach to known results for homogeneous systems,
in section \ref{sec:hardcore} Laughlin's trial wavefunctions were introduced and their uniqueness in a system with a short-ranged interaction was pointed out. Handling a homogeneous system numerically in section \ref{sec:gsmu} revealed the expected gapped ground state and its incompressibility for both kinds of interactions -- Coulomb and short-ranged.
Confirming Laughlin's correlations in section \ref{sec:correlations} for our system with short-range interaction showed the vanishing interaction energy $\langle V_{\mbox{short-range}} \rangle_{GS}$ to be an indicator for these correlations.

Addressing the injection of a quasihole in our framework, (section \ref{sec:QuasiHoles}) a quasihole creation operator was derived based on Laughlin's trial wavefunctions. Excitations it produces in homogeneous systems were verified to have the desired properties such as fractional charge, low excitation energy and stability for the short-ranged and for Coulomb interaction (section \ref{sec:QuasiHolesHard} and \ref{sec:QHcoulomb}).
An alternative approach to create quasiholes proved to be advantageous for Coulomb interaction, where it resulted in a lower excitation energy and improved stability. A side effect of this method was the finding of a localized ground state at a delta potential in a system with one excess flux quantum, i.e. $N_s = 3 N_e + 1$.

Treating such a system with one excess flux quantum and two delta potentials in section \ref{sec:tunnel} in an effective single-quasiparticle picture was qualitatively in agreement with the numerically obtained results. Section \ref{sec:twodeltas} analyzed the ground state of this system. It is predominantly a symmetric superposition of the quasihole's ground state at the one delta or the other, respectively, and complies with the simple picture of a single quasihole tunneling between bound states.
Efforts were made in section \ref{sec:DistanceDepTunnel} and \ref{sec:AsymmDelta} to extract the tunneling strength $t$ from the distance dependence of the tunnel-splitting. However, only the increase of $t$ with decreasing separation between the potentials could be understood qualitatively. Quantitative statements can possibly be drawn from a similar analysis with different parameters (see conclusion on section \ref{sec:tunnel}).
As an alternative treatment of tunneling it would be interesting to prepare this system in an initial state with the quasihole at one delta and perform its time evolution. Localization of the quasihole at either potential alternating in time can be expected.

Turning to inhomogeneous systems, the projection to the lowest Landau level, considered as taking the limit $\hbar \omega_c \gg \frac{e^2}{\epsilon l_0}$, demands a careful treatment of evaluating the expectation value of the kinetic momentum operators.
A consistent treatment is achieved in section \ref{sec:corrections} making a redefinition of these operators necessary which is equivalent to taking into account mixing with the next Landau level.

Inhomogeneous systems were treated in two different limits: Weak potentials that leave the incompressible system's density nearly homogeneous and strong potentials (comparable with $\mu_{\mbox{electron}}$) that create abrupt edges in the system and are more problematic \cite{Krause-Kyora}.
Section \ref{sec:WeakPot} deals with quasiholes in systems with shallow potentials which are easy to handle. Being the source of a dispersion for the single particle states, they cause charged particles to undergo an $\vec E \times \vec B$ drift.
Due to their charge, quasiholes follow the classical motion along equipotential lines as long as they are far away from a saddle point of the potential. They remain relatively stable during time spans of interest.
At the saddle point equipotential lines with different direction of propagation meet.
The quasiholes' quantum mechanical behavior becomes apparent here and it is interpretable as a superposition of two states:
One, carrying the quasihole along its original equipotential line, and a second where the quasihole moves on the corresponding equipotential line on the other side of the saddle point.

In the limit of strong potentials, the competition between correlations and potential energy prohibits creating an effective tunneling barrier while simultaneously maintaining the correlations of a homogeneous system. This can be understood by thinking of the constriction to bind vortices which are in turn ``missing'' to establish Laughlin's correlations. Section \ref{sec:Strongpot} corroborated this intuitive picture. Having more flux quanta than needed for realizing a filling factor of $\frac{1}{3}$, these ``free'' vortices are bound by the potential.
In section \ref{sec:deltapeakbarrier} the necessary amount of excess flux quanta to create a QPC was estimated.

Applying the idea of excess flux quanta to a system with a Gaussian barrier in section \ref{sec:barraddflux} resulted in a usable setup to obtain abrupt edges while conserving the correlations: A threefold quasidegenerate gapped ground state was obtained similar to the homogeneous $\nu=\frac{1}{3}$-system.
Inserting a quasihole into the left edge of this system and performing the time evolution allows the following conclusions to be drawn:
The quasihole suffers more severely from distortion compared to the case of a weak potential.
Similar to the shallow potential, the motion follows the equipotential lines. When the quasihole reaches the point where both edges are close to each other a symmetric state between the edges is adopted. It is interpretable as a superposition of two states, each having a quasihole on one edge.
For later times there is evidence for having a quasihole in both halves of the system. 
In order to support this suggestion, in a future work the final state obtained by performing the time evolution should be analyzed by projecting it to the supposed final state.
These observations corroborate tunneling of the quasihole through the constriction and are thus in contradiction with recent theoretical results \cite{KaneFisher2} that found the tunneling process in the limit $T \rightarrow 0$ to be forbidden.

As a perspective for further work, an approach of creating the narrowest barrier possible in our system revealed promising results in section \ref{sec:deltapot}. This system would be suitable for quasihole injection as well.

\newpage
\section{Deutsche Zusammenfassung}
Die vorliegende Arbeit beschäftigt sich mit dem Tunneln von Quasilöchern im Fraktionalen-Quanten-Hall-Regime.
Die Fragestellung erwuchs aus den theoretischen Beschreibungen des fraktionalen Quanten Hall Effekts und führte kürzlich experimentell auf überraschende Ergebnisse \cite{Heiblum2}.
Kurz nach der Entdeckung des fraktionalen Quanten Hall Effekts (FQHE) identifizierte Laughlin \cite{Laughlin} die elementaren niedrigenergetischen
Anregungen des Systems als fraktional geladene Quasiteilchen und Quasilöcher und konnte damit die Inkompressibilität des Zustands erklären.
Auch in darauf aufbauenden Entwicklungen \cite{Haldane} spielten diese Quasiteilchen eine zentrale Rolle und ihre Eigenschaften wurden in
Experimenten \cite{Simmons,Saminadayer,Heiblum1,Heiblum2} und theoretischen Arbeiten \cite{HaldaneRezayi,Pokrovsky,KaneFisher1,KaneFisher2}
untersucht. Einige dieser Arbeiten basieren auf der Beschreibung des FQHE mit Hilfe von Randzuständen \cite{MacDonald,Wen} die sich auf deren
Erfolg in der Erklärung des ganzzahligen Quanten Hall Effekts \cite{Laughlin1, Halperin} stützt. 
In diesem Zusammenhang werden Tunnelexperimente von Quasiteilchen (und -löchern) in Quanten Punkt Kontakten als probates Mittel zur Verifikation des Randzustands-Bildes angesehen.
\par
Ziel dieser Diplomarbeit ist es, das Tunneln von Quasilöchern von einer anderen Seite her zu beleuchten. Dazu werden Quasilöcher in einem endlichen System von Elektronen im fraktionalen Quanten Hall Regime beschrieben wobei die Frage des Quasiloch-Tunnelns insbesondere in der zeitlichen Entwicklung inhomogener Systeme untersucht wird. Hierfür wird die Methode der exakten numerischen Diagonalisierung verwendet.
Im \ref{sec:intro}. Kapitel wird eine Einführung in das Thema gegeben und die Fragestellung in den Kontext bestehender und aktueller
experimenteller und unabhängiger theoretischer Ergebnisse gestellt.
Die für die Beschreibung des Systems nötige Basis zusammen mit den gewählten Randbedingungen wird im \ref{sec:one}. Kapitel abgeleitet.
Anhand homogener Systeme werden im \ref{sec:two}. Kapitel die Rechnungen anhand bekannter Ergebnisse (Inkompressibilität) verifiziert
und eine kurzreichweitige Elektron-Elektron Wechselwirkung eingeführt. Zwei Möglichkeiten zur Erzeugung von Quasiloch-Anregungen, basierend auf 
bekannten Versuchswellenfunktionen \cite{Laughlin}, werden abgeleitet. Die Stabilität der Anregungen wird sowohl für Coulomb- als auch für die kurzreichweitige Wechselwirkung untersucht.
Ein System mit einem lokalisierten Zustand eines Quasilochs bietet eine erste Möglichkeit zur Untersuchung von Quasiloch-Tunneln.
Dazu befasst sich das \ref{sec:three}. Kapitel mit der in diesem Rahmen einfachsten Realisierung eines Systems, das Quasiloch-Tunneln zwischen zwei
lokalisierten Zuständen zeigt. Die Abhängigkeit des Tunnelns von dem Abstand der bindenden Systeme ist qualitativ nachvollziehbar.
Zur Untersuchung eines Quanten Punkt Kontaktes (QPC) werden im \ref{sec:InhomSystems}. Kapitel inhomogene Systeme betrachtet.
Beiträge des nächsten Landauniveaus werden hier wichtig um ein kosistentes Bild zu erhalten; sie führen zu Korrekturen der Stromoperatoren.
Für schwache Potentiale (verglichen mit der Anregungslücke) zeigt die Zeitentwicklung quasiklassisch erwartete $\vec E \times \vec B$-Drift des Quasilochs entlang der Äquipotentiallinien, doch auch eine teilweise Transmission des Quasilochs zwischen Äquipotentiallinien kann beobachtet werden. Die Inkompressibilität des Systems steht der Realisierung eines QPCs durch ein stärkeres Potential im Weg. 
Eine Möglichkeit, dennoch ein System mit einer Tunnelbarriere in einem inkompressiblen Zustand zu erzeugen wird gefunden und die Zeitentwicklung eines Quasiloch-Zustands in diesem System berechnet. Es findet eine Kommunikation zwischen den beiden Rändern über die Barriere hinweg statt, die sich durch Tunneln des Quasilochs erklären lässt.
Das letzte Kapitel enthält eine zuammenfassende Betrachtung und einen Ausblick auf Fragen, die in dieser Arbeit auftraten, aber nicht erschöpfend geklärt werden konnten.

\newpage
\begin{appendix}
\section{Derivation of a quasihole creation operator}
\label{app:DerivQHOp}
In section \ref{sec:QHOpIdea} the trial wavefunction (Equ. (\ref{eq:HoleEx})) of a quasihole excitation in a fractional quantum Hall state at a filling factor $\frac{N_e}{N_s} = \frac{1}{m}$ was given along with the trial wavefunction of the ground state, both for homogeneous systems.
Basing on these wavefunctions, an operator shall be derived that, applied to the ground state wavefunction, results in the quasihole excited state with a quasihole inserted at a position $z_0$ inside the unit cell. This operator will be used as a generalization to create trial wavefunctions for arbitrary systems (i.e. with inhomogeneities).
The quasihole is manifest in the wavefunction as a single zero with respect to the electrons' coordinates $z_j$. Since adding a zero to the wavefunction amounts to inserting an additional flux quantum, the final state belongs to a system with $N_s+1$ flux quanta.
This operator will then be applied to the many-particle basis states of the system with $N_s$ flux quanta. The result can be expressed in the many-particle basis with $N_s+1$ flux quanta. Once knowing this basis transformation, the operator can act on arbitrary initial states to obtain trial wavefunctions for quasihole excitations.

As seen in section \ref{sec:QHOpIdea} Equ. (\ref{eq:HoleEx}), the relative zero between $z_j$ and $z_0$ is caused by multiplication with a factor $\vartheta_1\Big( \pi \frac{-i (z_j^*-z_0^*)}{L_2} | i\frac{L_1}{L_2} \Big)$
for every electron $j$. The additional flux quantum needs to be accounted for by the center of mass wavefunction $F_{N_s}$, too. It must be replaced by a solution $F_{N_s+1}$ of the single-particle Schrödinger equation given by Equ. (\ref{eq:Fcm}) for $N_s+1$ flux quanta.
This wavefunction can be obtained from the center of mass part of the ground state by applying transformations that only act on the center of mass
coordinate.
More concrete, given a center of mass wavefunction for $N_s$ flux-quanta it is seen from (\ref{eq:Fcm}) that we can construct a valid one 
for $N_s+1$ flux quanta simply by a spatial shift of $Z\ \rightarrow Z + \Delta Z$ and a momentum shift $\exp(i \Delta K Z)$.
From the two conditions in (\ref{eq:Fcm}) connecting the wavenumber $K$ and the zeros $Z_\nu$ of the center of mass function follows, that this can be done by $Z_\nu \rightarrow Z_\nu + \frac{L_1}{2 m}n + \frac{L_2}{2 m}i k$ and $K \rightarrow K + \frac{\pi}{2 L_2} n$ for odd integers $n, k$. This transformation can be expressed as a shift of the center of mass $Z$ and an accompanying shift of the center of mass momentum. Additionally, $N_s$ in the Gaussian must be replaced by $N_s+1$.
Applying this transformation on the whole wavefunction will not affect its relative part.
Thus the ansatz for the operator belonging to the insertion of a quasihole at position $z_0$ looks like
\begin{eqnarray}
  \label{eq:OperHole}
  O_{hole}(z_0) &=& \Pi_{j=1}^{Ne} \vartheta_1 \Big( \pi \frac{-i (z_j^*-z_0^*)}{L_2} | i\frac{L_1}{L_2} \Big) \\ \nonumber
  && \underbrace{ \exp( -\frac{\pi (N_s+1) \sum_{j=1}^{Ne} x_j^2}{L_1 L_2} ) T^{c.m.}(\frac{z_0}{m}) \exp(i \Delta K Z) T^{c.m.}(\Delta Z) \exp( \frac{\pi N_s \sum_{j=1}^{Ne} x_j^2}{L_1 L_2} )}_{ \rightarrow f_{rel}({z_i}) F_{N_s+1}(Z + \frac{z_0}{m}) \exp( -\frac{\pi (N_s+1) \sum_{j=1}^{Ne} x_j^2}{L_1 L_2})}, 
\end{eqnarray}
where $\Delta Z = -\frac{L_1}{2 m}n - \frac{L_2}{2 m}i k$ and $\Delta K = \frac{\pi}{2 L_2} n$, $n,k$ odd integers.
Here we choose $n = 1, k = 1$ because we want to construct a state that does not differ much from the state we started with. Therefore, the smallest possible center of mass and momentum translations are chosen.
As indicated by the curly bracket, the latter part of the operator is intended to fix the boundary conditions and yield the part of the trial
wavefunction (\ref{eq:HoleEx}) that is written below this bracket.

It is obvious that this wavefunction has got the desired zeros at $z_0$ produced by the first factor in \ref{eq:OperHole}.
In the following calculation we will apply this operator to the many-particle basis states of the system with $N_s$ flux quanta.
It can be calculated separately, but it is also evident from this calculation, that the obtained wavefunction with a quasihole will
obey the correct boundary conditions, since they will be shown to be expandable in the many-body basis of the $N_e/N_s+1$-system.
\subsection{Application of $O_{hole}$ to the basis states}
The many-body wavefunctions in the finite system with $N_s$ flux quanta are represented as linear combinations of Slater determinants of 
one-particle wavefunctions from Equ. (\ref{eq:BasisOrt}). Therefore, if one wants to insert a quasihole into such a state, given its
decomposition into these basis states, the operator $O_{hole}$ can just be used to act on these basis states and the result again has to 
be expressed in the many particle basis states of the system with $N_s+1$ flux quanta. If once this basis transformation is calculated, also
the transformation of arbitrary states expanded in this basis is known. In the following this basis transformation is calculated explicitly.

Since each product in the slater determinant contains $N_e$ single particle wavefunctions as factors,
a Gaussian factor can be split off the many-particle basis wavefunction, like
\begin{eqnarray}
  \label{eq:DiagGS}
  \Psi(z_1,...,z_{Ne}) = f(z_1,...,z_{Ne}) \exp( -\frac{\pi N_s \sum_{j=1}^{Ne} x_j^2}{L_1 L_2} ).
\end{eqnarray}
This is true for arbitrary wavefunctions expressed in this basis.
Applying the operator $O_{hole}$ to this state yields
\begin{eqnarray}
  \label{eq:HoleDiag}
    \Psi_{Hole}(z_1,...,z_{N_e}; z_0) &=& O_{hole}(z_0) \Psi(z_1,...,z_{Ne}) \\ \nonumber
    &\propto& \Pi_{j=1}^{Ne} \vartheta_1( \pi \frac{-i (z_j^*-z_0^*)}{L_2} | i\frac{L_1}{L_2})\\ \nonumber
    && \exp(K Z^*) f(\{ z_i +\frac{z_0}{N_s} + \frac{\Delta Z}{N_e} \} ) \exp(-\frac{\pi (N_s+1) \sum_{j=1}^{Ne} x_j^2}{L_1 L_2})
\end{eqnarray}
It was used, that the center of mass translation, due to being sandwiched between the two Gaussian factors, only affects the coordinates of $f$ and that a translation
of the center of mass by $\frac{z_0}{m}$ is equivalent to a translation of each electrons' coordinate by $\frac{z_0}{N_s N_e}$. A factor $\exp(i K \frac{z_0}{m})$ was omitted because it is just a c-number which will be absorbed by the normalization.

Every element of the many particle basis is a sum of products of $N_e$ single-particle wavefunctions (\ref{eq:BasisOrt}), each of which depends on one of the coordinates $\{z_1,...,z_{Ne}\}$.
Expanding the Gaussian factor's exponent allows to separate a Gaussian factor proportional to $\exp(-x^2)$ and leads to the following form
for $\phi_{0,j}$

\begin{eqnarray}
  \label{eq:Separ1T}
  \phi_{0,j}(x,y) &=& \frac{1}{\sqrt{2 \pi} l_0} \sum_k \exp \Big( \frac{X_j + k L_1}{l_0^2}(-x+iy) - i k \alpha \Big) \; \exp \Big( - \frac{(X_j+k L_1)^2}{2 l_0^2} \Big) \\ \nonumber
  && \exp \Big( -\frac{x^2}{2 l_0^2} \Big) \\ \nonumber
  &=& f_{j}(z) \; \exp \Big( -\frac{x^2}{2 l_0^2} \Big).
\end{eqnarray}
An arbitrary wavefunction expanded in the many-body-basis can be written explicitly as 
\begin{eqnarray}
  \label{eq:MPBasis}
  \Psi(z_1,...,z_{Ne}) &=& \bra{z_1,...,z_{Ne}} \sum_{j_1<...<j_{Ne}} C_{j_1,...,j_{Ne}} \ket{j_1,...,j_{Ne}} \\ \nonumber
  \mbox{where} \quad \braket{z_1,...,z_{Ne}}{j_1,...,j_{Ne}} &=& det( \{f_{j_k}(z_l)\}_{k,l=1..N_e} ) \exp(-\frac{\pi N_s \sum_{j=1}^{Ne} x_j^2}{L_1 L_2})
\end{eqnarray}
and $C_{j_1,...,j_{Ne}}$ are the coefficients in this basis. Now it is possible to apply the operator (\ref{eq:OperHole}) to the state in (\ref{eq:MPBasis}).
This done, one arrives at
\begin{eqnarray}
  \label{eq:PsiHole}
  \Psi_{hole} &=& O_{hole}(z_0) \sum_{j_1<...<j_{Ne}} C_{j_1,...,j_{Ne}} \ket{j_1,...,j_{Ne}} \\ \nonumber
  &=& \sum_{j_1<...<j_{Ne}} C_{j_1,...,j_{Ne}} det( \{ \vartheta_1( \pi \frac{-i (z_{j_k}^*-z_0^*)}{L_2} | i\frac{L_1}{L_2})  exp(K z_l^*) f_{j_k}(z_l+\frac{z_0}{N_s}+\frac{\Delta Z}{N_e}) \}_{k,l=1..N_e} ) \\ \nonumber
 && \exp(-\frac{\pi (N_s+1) \sum_{j=1}^{Ne} x_j^2}{L_1 L_2}).
\end{eqnarray}
This expression has to be written as a linear combination in the many-particle basis of a system with $N_s+1$ flux quanta. This is possible, since the wavefunction (\ref{eq:PsiHole}) satisfies
the periodic boundary conditions. Using the periodicity of the $\theta$-function (for example from \cite{Gradstein}) by
\begin{eqnarray}
  \label{eq:ThetaPBC}
  \vartheta_1(u+\pi|\tau) &=& -\vartheta_1(u|\tau) \\ \nonumber
  \vartheta_1(u+\pi \tau) &=& - \exp(-i \pi \tau) \exp(-2iu) \vartheta_1(u|\tau) \\ \nonumber
  \mbox{where} \quad \tau &=& i\frac{L_1}{L_2},
\end{eqnarray}
it is possible to show, that the wavefunction $\Psi_{hole}$ satisfies modified periodic boundary conditions like
\begin{eqnarray}
  \label{eq:PBCPsihole}
  \Psi_{hole}(z_1,...,z_i+L_1,z_{i+1},...,z_{N_e}) &=& \exp(i\alpha) \Psi_{hole}(z_1,...,z_{Ne}) \exp(-\frac{2 \pi}{L_2}i (N_s+1) y_i) \\ \nonumber
  \Psi_{hole}(z_1,...,z_i+iL_2,z_{i+1},...,z_{N_e}) &=& \exp(i\beta) \Psi_{hole}(z_1,...,z_{Ne}). \\ \nonumber
\end{eqnarray}
Stated differently, equations (\ref{eq:PBCPsihole}) show, that the wavefunction $\Psi_{hole}$ is an eigenfunction of the magnetic translation operators for a system with $N_s+1$ flux quanta, since the appearing exponential in the first line is compensated by the one appearing in the magnetic translation from Equ. (\ref{eq:MagTrans}).

According to \cite{Gradstein} the odd elliptic theta function of first order is defined as
\begin{eqnarray}
  \label{eq:ThetaFunc}
  \vartheta_1(u|\tau) &=& \frac{1}{i} \sum_{n=-\infty}^\infty (-1)^n \exp(i \pi (n+\frac{1}{2})^2 \tau + i (2n+1) u) .
\end{eqnarray}
Using that, one can express the factor $\vartheta_1( \pi \frac{-i (z^*-z_0^*)}{L_2} | i\frac{L_1}{L_2}) \; f_{j}(z)$ appearing in Equ. (\ref{eq:PsiHole}) as a linear combination of basis functions, namely
\begin{eqnarray}
  \label{eq:OneFactor}
  \vartheta_1( \pi \frac{-i (z^*-z_0^*)}{L_2} | i\frac{L_1}{L_2}) \; f_{j}(z) &=&
  -i \sum_{d \in \Z} (-1)^d \exp \left( - z_0^* \frac{2 \pi}{N_s L_2}(j+ \frac{\beta}{2 \pi} + (d+\frac{1}{2})N_s) \right) \times \\ \nonumber
  && \times \exp \left( -\pi \frac{L_1}{L_2 (N_s+1)N_s} ( (j+\frac{\beta}{2 \pi} + N_s d + \frac{N_s+1}{2})^2 - \frac{N_s+1}{4} ) \right) \times \\ \nonumber
  && \times \exp (- \frac{i \pi}{N_s}(j + \frac{\beta}{2 \pi}) ) \ \phi^{N_s+1}_{j-d}(z) \\ \nonumber
  &=:& \sum_{d \in \Z} G(d,j,z_0) \ \phi^{N_s+1}_{j-d}(z) \\ \nonumber
  &=& \sum_{h=0}^{N_s} \underbrace{\left( \sum_{n \in (N_s+1) \Z} G(j-h-n, j, z_0) \exp(-i \frac{n}{N_s+1} \alpha) \right)}_{=: K(j,h,z_0)} \ \phi^{N_s+1}_{h}(z)
\end{eqnarray}
where $\phi^{N_s+1}_{h}(z)$ names the h-th single-particle basisvector of the system with $N_s+1$ flux quanta.

Now the expression (\ref{eq:PsiHole}) can be rearranged such that $\Psi_{hole}$ is expressed as a linear combination of many particle basis functions.
To accomplish that, one element of the matrix under the determinant can be expressed as a product of two matrices by inserting (\ref{eq:OneFactor}) into (\ref{eq:PsiHole}) and applying the identity $det (A\cdot B) = det(A) det(B)$ on the determinant,
\begin{eqnarray}
  \label{eq:matmult}
  \Psi_{hole} &=& \sum_{h_1<\ldots<h_{N_e}} \sum_{j_1<\ldots<j_{Ne}} C_{j_1,\ldots,j_{Ne}} \det \left( \{ \sum_{m=1}^{N_e} K(j_k,h_m,z_0) \phi^{N_s+1}_{h_m}(z_l) \}_{k,l=1,\ldots,N_e} \right) \\ \nonumber
  &=& \sum_{h_1<\ldots<h_{N_e}} \underbrace{\sum_{j_1<\ldots<j_{Ne}} C_{j_1,\ldots,j_{Ne}} \det \left( \{ K(j_k,h_m,z_0) \}_{m,k=1\ldots N_e} \right)}_{C^{hole}_{h_1,\ldots,h_{N_e}}} \times \\ \nonumber
  && \times \det \left( \phi^{N_s+1}_{h_m}(z_l) \}_{m,l=1,\ldots,N_e} \right).
\end{eqnarray}
Finally, the new coefficients $C^{hole}_{h_1,\ldots,h_{N_e}}$ in the last equation can be calculated explicitly as a matrix multiplication. $G(d,j,z_0)$ is used from (\ref{eq:OneFactor}). This results in
\begin{eqnarray}
  \label{eq:RealFinalPsiHole}
  \ket{\Psi_{hole}} &=& C^{hole}_{h_1,\ldots,h_{N_e}} \; \ket{h_1,\ldots,h_{N_e}} \\ \nonumber
  \mbox{with} \quad C^{hole}_{h_1,\ldots,h_{N_e}} &=& \sum_{j_1<\ldots<j_{N_e}} K^{j_1,\ldots,j_{N_e}}_{h_1,\ldots,h_{N_e}} C_{j_1,\ldots,j_{N_e}} \\ \nonumber
  \mbox{and} \quad K^{j_1,\ldots,j_{N_e}}_{h_1,\ldots,h_{N_e}} &=& det \Big( \{ K(j_p,h_m,z_0) \}_{p,m=1\ldots N_e} \Big) \\ \nonumber
  K(j,h,z_0) &=& \sum_{n \in (N_s+1) \Z} G(j-h-n, j, z_0) \exp(-i \frac{n}{N_s+1} \alpha) \\ \nonumber
             &=& -i Q(x_0) \sum_{n \in \Z} (-1)^{j-h} \exp(-i n \alpha) \exp(-i \frac{\pi}{N_s}(j+\frac{\beta}{2 \pi}+n N_s(N_s+1))) \times \\ \nonumber
             && \times \exp(2 \pi i \frac{y_0}{L_2} \frac{N_s+1}{N_s}( j-N_s n + (\frac{1}{2}-h)\frac{N_s}{N_s+1} + \frac{\beta}{2 \pi (N_s+1)} )) \times \\ \nonumber
             && \times \exp ( -\pi \frac{L_1}{L_2} \frac{N_s+1}{N_s} ( j-N_s n + (\frac{1}{2}-h)\frac{N_s}{N_s+1} + \frac{\beta}{2 \pi (N_s+1)} + \frac{1}{2 (N_s+1)} + \frac{x_0}{L_1})^2 ).
\end{eqnarray}
Here $Q(x_0)$ is a real function of $x_0$. It doesn't have to be computed, since the resulting vector will be normalized anyway.
\section{Single particle matrix elements}
\label{app:MatEls}
\subsection{The constriction}
All single-particle operators are evaluated by aid of second quantization, thus it is sufficient to calculate the single particle matrix elements
to construct the many particle operator $\hat O = \sum_{i,j} O_{i,j} a^\dagger_i a_j$ from it.

A straight forward calculation yields the single-particle matrix elements of the Gaussian constriction potential given in Equ. (\ref{eq:VwallGauss}) for the basis functions from Equ. (\ref{eq:BasisOrt}) in the first Landau level.
As seen from the potential only depending on $x$ the matrix elements conserve the momentum in y-direction.
\begin{eqnarray}
	\label{eq:MatElsGaussWall}
	 \bra{0,j} V_{Wall} \ket{0,k} &=& s \frac{e^2}{\epsilon l_0^2} \frac{w}{\sigma} \sum_{m \in \Z} \exp \left(-\frac{1}{\sigma^2} \Big( \frac{j+\beta/2\pi}{N_s}+m-x_0/L_1 \Big)^2 \right) \; \delta_{j,k}, \\ \nonumber
	\mbox{where} \; \sigma &=& \sqrt{w^2+\frac{1}{2 \pi N_s L_1/L_2}}.
\end{eqnarray}
The calculation of the matrix elements for the ``notch'' inside the constriction, the potential of which is given in (\ref{eq:Vnotch}), can be carried out analytically as well. A rather lengthy calculation yields
\begin{eqnarray}
	\label{eq:MatElNotch}	
	\bra{0,j} V_{notch} \ket{0,k} &=& s_{notch} \frac{e^2}{\epsilon l_0^2} \; w \;w_{hole} \; \frac{\sqrt{\pi}}{\sigma} \sum_{l \in 2 \Z, m \in 2 \Z} \left( A(l) B(m) + A(m+1) B(l+1) \right) \\ \nonumber
	\mbox{where} \; A(l) &=& \exp \left( -\frac{1}{\sigma^2} \Big( l/2 - x_0/L_1 - \frac{1}{2}(\frac{j+k}{N_s} + \frac{\beta}{\pi N_s}) \Big)^2 \right) \\ \nonumber
	\mbox{and} \; B(m) &=& \exp \left( 2 \pi i (k-j + N_s l) \frac{y_0}{L_2} - i \alpha l - (\frac{\pi L_1}{2 N_s L_2}+ \pi^2 w_{Hole}^2)(k-j + N_s l)^2 \right).
\end{eqnarray}
As a remark, in the case that $y_0 = 0.5 L_2$, which is a notch in the center of the unit cell (with respect to the y-direction),
and $\alpha=0$ or $\alpha=\pi$, the matrix elements for the notch become real. This is of course a benefit for numerical calculations since
only a real symmetric instead of a complex Hermitian matrix has to be diagonalized.

The matrix elements for the notch inside the delta-barrier are constructed by symmetric multiplication of those of the delta wall and those of
a Gaussian wall parallel to the x-direction defined by Equ. (\ref{eq:WallY}). Calculating them yields
\begin{eqnarray}
\bra{0,j} V_{\mbox{wall}_x} \ket{0,k} = \frac{e^2}{\epsilon l_0^2} s_{\mbox{wall}_x} w_{\mbox{wall}_x} \sqrt{\pi}
 \sum_{l \in \Z} \exp \left(-2 \pi i (j-k + N_s l) \frac{y_0}{L_2} + i \alpha l \right)\\ \nonumber
 \times \exp \left( -(\pi^2 w_{\mbox{wall}_x}^2 + \frac{\pi L_1}{2 N_s L_2})(j-k + N_s l)^2 \right).
\end{eqnarray}
Therefore the matrix elements of the notch turn out to be
\begin{eqnarray}
  \label{eq:MatElDeltaNotch}
  \bra{0,j} V_{\mbox{deltanotch}} \ket{0,k} &=& \frac{1}{2} \Big( \bra{0,j} V_{\mbox{wall}_x} \ket{0,j_0} \delta_{k,j_0} + \bra{0,j_0} V_{\mbox{wall}_x} \ket{0,k} \delta_{j,j_0} \Big),
\end{eqnarray}
where the depth of the notch is given by the negative strength of $V_{\mbox{wall}_x}$ while $w_{\mbox{wall}_x}$ width defines the width of the notch.
\subsection{Kinetic momenta in the lowest Landau level}
\label{app:CorrectedMom}
To obtain a gauge invariant expression for the kinetic momentum operator we state that Ehrenfest's theorem  must be fulfilled
\begin{equation}
\frac{1}{m} \langle \tilde \Pi_y \rangle = \langle \dot y \rangle = \frac{i}{\hbar} \langle [H,y] \rangle.
\end{equation}
We use this as a definition for the operator $\Pi_y$ or respectively $v_y$
\begin{equation}
\label{eq:DefVel}
\tilde v_y \equiv \dot y = \frac{i}{\hbar} [H,y].
\end{equation}
It is sufficient to restrict ourselves to the single particle problem given by the Hamiltonian
\begin{eqnarray}
H &=& H_{kin} + V(x,y) \\ \nonumber
  &=& \frac{1}{2 m} \left( \Pi_x^2 + \Pi_y^2 \right) + V(x,y),
\end{eqnarray}
where $H_{kin}$ is defined like in Equ. (\ref{eq:Hamilton}). The problem shall be solved after projecting it to the lowest Landau level.
This projection will also be done for the $x$ and $y$ operator following \cite{Shankar}. The ``projected versions'' of these operators will appear
to obey a conical commutation relation.

Defining the operators $k_x$ and $p_y$ by
\begin{eqnarray}
k_x &=& -i\hbar \partial_x + e B y \\ \nonumber
p_y &=& -i\hbar \partial_y
\end{eqnarray}
and comparing them with Equ. (\ref{eq:MagTrans}) respectively (\ref{eq:MagTransY}) from section \ref{sec:magtrans} identifies them
as the generators of magnetic translations. $x$ and $y$ can now be expressed in these new operators and the kinetic momenta (found in Equ. (\ref{eq:Hamilton})). The projection to the lowest Landau level is performed by $P_{LLL}$ defined by Equ. (\ref{eq:PLLL}) in section \ref{sec:Projector} and results in
\begin{eqnarray}
\label{eq:ProjXY}
x &=& \frac{1}{e B} \left( \Pi_y - p_y \right) \\ \nonumber
y &=& \frac{1}{e B} \left( k_x - \Pi_x \right) \\ \nonumber
x_{LLL} &\equiv& P_{LLL} x P_{LLL} = -\frac{1}{e B} p_y \\ \nonumber
y_{LLL} &\equiv& P_{LLL} y P_{LLL} = \frac{1}{e B} k_x. \\ \nonumber
\end{eqnarray}
As shown in section \ref{sec:PiZero} the projection of $\Pi$ onto the lowest Landau level is zero.
Since $k_x$ and $p_y$ commute with the Hamiltonian $H_{kin}$, $k_x$ and $p_y$ don't cause mixing of different Landau levels and we can omit the surrounding projectors in Equ. (\ref{eq:ProjXY}).

Now, $x_{LLL}$ and $y_{LLL}$ inherit the canonical commutation relation of $k_x$ and $p_y$
\begin{equation}
[x_{LLL},y_{LLL}] = i l_0^2.
\end{equation}
The velocity operator can be calculated according to our definition from Equ. (\ref{eq:DefVel}). This yields
\begin{eqnarray}
\tilde v_y &=& \frac{i}{\hbar} \left( [H_{kin},y_{LLL}] + [V(x_{LLL},y_{LLL}), y_{LLL}]\right)\\ \nonumber
           &=& \frac{i}{\hbar}[V(x_{LLL},y_{LLL}), y_{LLL}].\\ \nonumber
\end{eqnarray}
The kinetic part commutes with $y_{LLL}$, the potential yields the expected correction term. It can be calculated using 
$[x_{LLL}^k, y_{LLL}] = i l_0^2 \; k \; x_{LLL}^{k-1}$ and a Taylor expansion of $V(x,y)$ in $x$ from which follows
\begin{eqnarray}
  [V(x,y), y] &=& \sum_{k=0}^\infty \left. \frac{1}{k!} \frac{\partial^k V(x,y)}{\partial x^k}\right|_{x=0} [x^k_{LLL},y_{LLL}]\\ \nonumber
	      &=& \sum_{k=0}^\infty \left. \frac{1}{k!} \frac{\partial^k V(x,y)}{\partial x^k}\right|_{x=0} i l_0^2 \; k \; x_{LLL}^{k-1} \\ \nonumber
	      &=& i l_0^2 \frac{\partial V(x,y)}{\partial x}.
\end{eqnarray}
This results in the corrected velocity operator
\begin{equation}
  \tilde v_y = -\frac{1}{m \omega_c} \frac{\partial V(x,y)}{\partial x}.
\end{equation}
An analogous treatment of the velocity in $x$-direction therefore results in
\begin{equation}
  \tilde v_x = \frac{1}{m \omega_c} \frac{\partial V(x,y)}{\partial y}.
\end{equation}
This definition leads to a different velocity operator than the naive approach of taking the kinetic momenta. It not just gives us a correction
but states that the kinetic momentum $\vec \Pi$ doesn't cause any current in the lowest Landau level at all.
\section{Two-particle matrix elements}
\subsection{General two-particle operator depending on $\vec r_i - \vec r_j$ only}
Most of the two particle operators in this work only depend on the difference coordinate $\vec r_1-\vec r_2$ between the two particles.
For those cases it is sufficient to calculate the two-particle matrix element of the operator $\exp(i \vec q (\vec r_1 - \vec r_2))$.
The matrix element of the Coulomb interaction, the short range interaction and the correlation functions can be obtained simply from their Fourier transforms.
Since we use generalized boundary conditions we cannot just reuse the results of Yoshioka from \cite{Yoshioka} which are valid only for boundary conditions of Equ. (\ref{eq:PRB}) with $\alpha=\beta=0$.
The wavevector $\vec q$ is restricted to values lying on a lattice $\vec q = \left(\frac{2 \pi}{L_1}s,\frac{2 \pi}{L_2}t \right)$ where $s,t \in \Z$ since all the operators have to be periodic in the unit cell.
Calculating this operator's matrix elements in the lowest Landau level spanned by our basis given by Equ. (\ref{eq:BasisOrt}) yields
\begin{eqnarray}
\label{eq:MatElExp}
\bra{j_1,j_2} \exp \left(i \vec q (\vec r_1 - \vec r_2)\right) \ket{j3,j4} &=& \sum_{l \in Z} \exp \left(-\frac{1}{2}l_0^2 q^2 + i q_x \frac{L_1}{N_s}(j_2-j_4) + i \alpha l \right) \\ \nonumber
&\times& \delta_{j_1-j_4, j_3-j_2+N_m l} \; \delta^\prime_{j_1-j_4,\frac{q_y L_2}{2 \pi}}, \\ \nonumber
\mbox{where} \; \vec q &\in& 2 \pi \frac{\Z}{L_1} \times \frac{\Z}{L_2}
\end{eqnarray}
and $\delta^\prime$ is the Kronecker delta modulo $N_s$.
To evaluate a general two particle operator $\hat O$ with the Fourier transform $V(\vec q)$, it can be written in second quantization which then results in
\begin{equation}
\label{eq:TwoPartGeneral}
\hat O = \frac{1}{2} \sum_{j_1,j_2,j_3,j_4=1}^{N_s} \sum_{q_x \in \frac{2 \pi}{L_1} \Z,\; q_y \in \frac{2 \pi}{L_2}\Z} V(\vec q) \; \bra{j_1,j_2} \exp \left(i \vec q (\vec r_1 - \vec r_2)\right) \ket{j3,j4} \; a^\dagger_{j_1} a^\dagger_{j_2}  a_{j_4}  a_{j_3}.
\end{equation}
\subsection{Coulomb interaction operator}
The Coulomb matrix elements follow from Equ. (\ref{eq:TwoPartGeneral}) by using the Fourier transform in two dimensions for periodic functions in the unit cell. The calculation of its Fourier transform is carried out here because it was not found explicitly in the cited papers.
By taking into account the interaction between all the images of one electron in neighboring cells, the Coulomb interaction is periodic in the unit cell. Alternatively we can think of the $N_e$ electrons residing on a topological, flat torus and interacting with each other via all possible ways,
\begin{equation}
V^{per}_{Coul} = \frac{e^2}{\epsilon} \sum_{k,l \in \Z} \frac{1}{\vec r - k L_1 \vec e_x - l L_2 \vec e_y}.
\end{equation}
Its Fourier transform clearly only has components for $\vec q$-vectors satisfying $\vec q \in (\frac{2 \pi}{L_1}\Z, \frac{2 \pi}{L_2} \Z)$.
It can be calculated as follows
\begin{eqnarray}
V^{per}_{Coul}(\vec q) &=& \frac{1}{L_1 L_2} \frac{e^2}{\epsilon} \int_0^{L_1} \; dx \; \int_0^{L_2} \; dy \; \sum_{k,l \in \Z} \frac{1}{|\vec r - k L_1 \vec e_x - l L_2 \vec e_y|} \exp(-i \vec q \vec r)\\ \nonumber
 &=& \frac{1}{L_1 L_2} \frac{e^2}{\epsilon} \underbrace{\int_{-\infty}^{\infty} \; dx \; \int_{-\infty}^{\infty} \; dy \; \frac{\exp(-i \vec q \vec r)}{|\vec r|}}_{\frac{2 \pi}{|\vec q|}}.
\end{eqnarray}
Writing the periodic Coulomb interaction in its Fourier representation then leads to the expression found in \cite{Yoshioka}
\begin{eqnarray}
\label{eq:VCoulPer}
V_{Coul}(\vec r) &=& \frac{1}{L_1 L_2} \frac{2 \pi e^2}{\epsilon} \sum_{\vec q \in (\frac{2 \pi}{L_1}\Z,\frac{2 \pi}{L_2}\Z)} \frac{1}{|\vec q|} \exp(i \vec q \vec r) \\ \nonumber
	&=& \frac{e^2}{\epsilon l_0} \frac{1}{N_s} \sum_{\vec q \in (\frac{2 \pi}{L_1}\Z,\frac{2 \pi}{L_2}\Z)} \frac{1}{|\vec q| l_0} \exp(i \vec q \vec r).
\end{eqnarray}
\subsection{Short range interaction operator}
In the case of short range interaction the Fourier transform is according to Equ. (\ref{eq:FourierShort}) proportional to $-q^2$ and we just replace $\frac{1}{l_0 q}$ in Equ. (\ref{eq:VCoulPer}) by $l_0^2 q^2$, thus
\begin{equation}
V_{short}(\vec r) = \frac{-e^2}{\epsilon l_0} \frac{1}{N_s} \sum_{\vec q} (|\vec q|l_0)^2 \exp(i \vec q \vec r).
\end{equation}
\subsection{Pair distribution function}
To obtain the pair distribution function used in section \ref{sec:correlations}, Equ. (\ref{eq:CorrAvgd}), we need the two particle matrix element of $\delta^{per}(\vec r + \vec r_1 - \vec r_2)$ where the argument of the delta is taken modulo cell size, i.e. periodic as $\delta^{per}(\vec r) = \sum_{k,l \in \Z} \delta(\vec r + k L_1 \vec e_x + l L_2 \vec e_y)$.
It can again be calculated by aid of Equ. (\ref{eq:MatElExp}). This is done by writing this delta function in Fourier space as 
\begin{equation}
\delta^{per}(\vec r_0 + \vec r) = \frac{1}{L_1 L_2} \sum_{\vec q \in (\frac{2 \pi}{L_1}\Z,\frac{2 \pi}{L_2} \Z)} \exp(i \vec q \vec r_0) \exp(i \vec q \vec r).
\end{equation}
\subsection{Two particle correlation function}
For inhomogeneous systems the two-particle correlation function in Equ. (\ref{eq:2PointCorr}) is used. Its two-particle matrix elements are clearly simply a product of wavefunctions from Equ. (\ref{eq:BasisOrt}), thus
\begin{equation}
\bra{j_1,j_2} \delta(\vec r_1 - \vec R_1) \delta(\vec r_2 - \vec R_2) \ket{j_3,j_4} =  \phi^*_{0,j_1}(\vec R_1)\phi^*_{0,j_2}(\vec R_2) \; \phi_{0,j_3}(\vec R_1)\phi_{0,j_4}(\vec R_2).
\end{equation}
Its representation in second quantization is obvious then.
\end{appendix}
\newpage
%
%
%

%
%
\newpage
\thispagestyle{empty}
\flushleft {\bf \huge Danksagung}

\vspace{1cm}

Zuerst möchte ich meinen Eltern danken, die mir ein sorgenfreies Studium ermöglichten und mich in jeder Weise unterstützten.

\vspace{1cm}

Der weitere Dank geht an die ganze Gruppe ``\verb+th1_km+'' für ihr angenehmes und freundschaftliches (Arbeits-)Klima.

\vspace{1cm}

{\bf \large Danke !}
\newpage
\thispagestyle{empty}
\flushleft Hiermit versichere ich, die vorliegende Diplomarbeit selbstständig verfasst und nur die angegebenen Hilfsmittel verwendet zu haben.
\vspace{3cm}
\begin{tabbing}
Hamburg, 28.11.2003 \= \\
 \\
 \\
\>............................................\\
\>\phantom{............}Moritz Helias\\
\end{tabbing}
\newpage

\end{document}